\documentclass{aa}
\usepackage{graphicx}
\usepackage{mwe}
\usepackage{longtable}
\usepackage{siunitx}
\usepackage[table]{xcolor}
\usepackage{chemist}
\usepackage{multirow}
\usepackage{caption}
\usepackage[flushleft]{threeparttable}
\usepackage{subcaption}
\usepackage{float}
\usepackage{txfonts}
\titlerunning{}

\usepackage{natbib,twoopt}
\usepackage{comment}
\usepackage[colorlinks=true, pdfstartview=FitV, linkcolor=blue, citecolor=blue, urlcolor=blue, breaklinks=true]{hyperref} %% to avoid \citeads line fills
\bibpunct{(}{)}{;}{a}{}{,}             %% natbib format for A&A and ApJ
\makeatletter
  \newcommandtwoopt{\citeads}[3][][]{\href{http://adsabs.harvard.edu/abs/#3}%
    {\def\hyper@linkstart##1##2{}%
     \let\hyper@linkend\@empty\citealp[#1][#2]{#3}}}
  \newcommandtwoopt{\citepads}[3][][]{\href{http://adsabs.harvard.edu/abs/#3}%
    {\def\hyper@linkstart##1##2{}%
     \let\hyper@linkend\@empty\citep[#1][#2]{#3}}}
  \newcommandtwoopt{\citetads}[3][][]{\href{http://adsabs.harvard.edu/abs/#3}%
    {\def\hyper@linkstart##1##2{}%
     \let\hyper@linkend\@empty\citet[#1][#2]{#3}}}
  \newcommandtwoopt{\citeyearads}[3][][]%
    {\href{http://adsabs.harvard.edu/abs/#3}
    {\def\hyper@linkstart##1##2{}%
     \let\hyper@linkend\@empty\citeyear[#1][#2]{#3}}}
\makeatother

\usepackage{color}
\newcommand{\Rsolar}{\mbox{$R_{\odot}\,$}}

\usepackage{color}

\usepackage[capitalise]{cleveref}

\usepackage{booktabs}% http://ctan.org/pkg/booktabs

\begin{document}
\title{Pulsating chromosphere of classical Cepheids}
\subtitle{Calcium infrared triplet and H$\alpha$ profile variations\thanks{Based on observations made with ESO telescopes at Paranal observatory  under  program  IDs: 098.D-0379(A), 0100.D-0397(A) and 0101.D-0551(A)}.}

\titlerunning{Calcium infrared triplet and H$\alpha$ profile variations}
\authorrunning{Hocd\'e et al. }

\author{V. Hocd\'e \inst{1} 
\and N. Nardetto \inst{1}
\and S. Borgniet\inst{2}
\and E. Lagadec \inst{1}
\and P. Kervella \inst{2}
\and A. M\'erand \inst{3}
\and N. Evans \inst{4}
\and D. Gillet\inst{5}
\and Ph. Mathias\inst{6}
\and A.~Chiavassa\inst{1}
\and A.~Gallenne\inst{1}
\and L.~Breuval\inst{2}
\and B.~Javanmardi\inst{2}
}

\institute{Universit\'e Côte d'Azur, Observatoire de la C\^ote d'Azur, CNRS, Laboratoire Lagrange, France,\\
email : \texttt{vincent.hocde@oca.eu}
\and LESIA (UMR 8109), Observatoire de Paris, PSL, CNRS, UPMC, Univ. Paris-Diderot, 5 place Jules Janssen, 92195 Meudon, France
\and European Southern Observatory, Karl-Schwarzschild-Str. 2, 85748 Garching, Germany
\and  Smithsonian Astrophysical Observatory, MS 4, 60 Garden St. Cambridge MA 02138
\and Observatoire de Haute-Provence – CNRS/PYTHEAS/Université d’Aix-Marseille, 04870 Saint-Michel l’Observatoire, France
\and IRAP, Université de Toulouse, CNRS, CNES, UPS. 14 Av. E. Belin, 31400 Toulouse, France
%\and European Southern Observatory, Alonso de C\'ordova 3107, Casilla 19001, Santiago 19, Chile   
}

\date{Received ... ; accepted ...}

\abstract{It has been shown recently that the infrared emission of Cepheids, which is constant over the pulsation cycle, might be due to a pulsating shell of ionized gas of about 15\% of the stellar radius, which could be attributed to the chromospheric activity of Cepheids.} {The aim of this paper is to investigate the dynamical structure of the chromosphere of Cepheids along the pulsation cycle and quantify its size. %and also evaluate what is the impact of  chromospheric activity on the radial velocity measurements from the Radial Velocity Spectrometer (RVS) on board \textit{Gaia}.
}  {We present H$\alpha$ and Calcium Near InfraRed triplet (Ca IR) profile variations using high-resolution spectroscopy with the UVES\thanks{Ultra Violet and Echelle Spectrograph project developed by the European Southern Observatory.} spectrograph of a sample of 24 Cepheids with a good period coverage from $\approx$ 3 to 60 days. After a qualitative analysis of the spectral lines profiles, we quantify the Van Hoof effect (velocity gradient between the H$\alpha$ and Ca IR) as a function of the period of the Cepheids. Then, we use the Schwarzschild mechanism (a line doubling due to a shock wave) to quantify the size of the chromosphere.}{We find a significant Van Hoof effect for Cepheids with period larger than $P=10$ days, in particular H$\alpha$ lines are delayed with a velocity gradient up to $\Delta v \approx$30 km/s compared to Ca IR. By studying the shocks, we find that the size of the chromosphere of long-period Cepheids is of at least $\approx$ 50\% of the stellar radius, which is consistent at first order with the size of the shell made of ionized gas previously found from the analysis of infrared excess. Last, for most of the long-period Cepheids in the sample, we report a motionless absorption feature in the H$\alpha$ line that we attribute to a circumstellar envelope that surrounds the chromosphere.}{Analyzing the Ca~IR lines of Cepheids is of importance to potentially unbias the period-luminosity relation from their infrared excess, particularly in the context of forthcoming observations of radial velocity measurements from the Radial Velocity Spectrometer (RVS) on board \textit{Gaia}, that could be sensitive to their chromosphere.}

\maketitle

%synchronization

\section{Introduction}\label{s_Introduction}
%Atmosphere of Cepheids are
Cepheids are milestones of the extragalactic distance scale since their period and luminosity are correlated, which is known as the Leavitt law \citep{leavitt08} also called the period-luminosity relation (hereafter PL relation). These variable stars have provided among the most essential advances in the history of astronomy from the discovery of galaxies to the expansion of the Universe \citep{HUBBLE26,hubble29}. Still today, the discovery of the accelerated expansion of the Universe \citep{riess98} has demonstrated the central importance of Cepheids in modern astronomy. However, uncertainties on both zero point and slope of the PL relation are today one of the largest contributors to the error on the extragalactic distance ladder and therefore on the determination of H0, the Hubble-Lemaître constant \citep{riess2019}. 

A plausible source of uncertainty could be due to InfraRed (IR) excess emitted by CircumStellar Envelope (CSE) such as the ones discovered using long-baseline interferometry \citep{kervella06a,merand06} in the K-band. However the origin and the nature of these CSEs are still debated. In particular, while CSE emission is explained by dust emission in some cases \citep{Gallenne2012,gallenne13b,Gro2020}, it fails to reproduce the IR excess in other studies \citep{Schmidt2015}.

The extended and dynamical atmosphere of Cepheids could be at the origin for the observed IR excess. Recently, \cite{hocde2020} analytically modeled free-free and bound-free emission from a thin shell of ionized gas to explain the near and mid-IR excess of Cepheids. This shell is modeled with a size of about 15\% of the star radius, whatever the pulsation phase, which corresponds to the region of the lower chromosphere. 
%\textbf{This shell is modeled \N{with} a thickness of about 15\% of the star radius, whatever the pulsation phase,} which corresponds to the region of the lower chromosphere. 
In this model the ionized material could be provided by periodic shocks occuring in the atmosphere, which heat up and ionize the gas.

Shocks in radially pulsating atmosphere of Cepheids have been largely studied in the H$\alpha$ Balmer line, providing valuable insight to the atmosphere dynamics  \citep{breitfellner93a,breitfellner93c,breitfellner93b,nardetto08b,gillet14}. These studies have also been supported by radiative hydrodynamic models \citep{fokin91,fokin96,Fadeyev2004} which have demonstrated how the  shock waves are generated then propagated through the atmosphere during a pulsation cycle. These studies emphasized the fundamental differences on the dynamics depending on short, medium or long pulsation period of Cepheids. %Also, from H$\alpha$ observations of T~Mon (27.02d) \cite{wallerstein1972} suggests H$\alpha$ emission could originates from the chromosphere. 
In addition, the H$\alpha$ emissions reported in the latter studies, in particular in upper atmosphere of long-period Cepheids, could indicate a chromospheric emission of pulsational origin.

Chromospheric activity of Cepheids has been first probed using optical Ca II K line \citep{wilson1957,kraft57} which showed transitory emissions after minimum light, with an increasing duration and strength with pulsation period. Later, upper chromospheric emission was detected on $\beta$~Dor using ultraviolet Mg II \textit{h} and \textit{k} \citep{schmidt1979} followed by \cite{SP82I,SP84II} who found heterogeneous emissions in the chromosphere with both rising and falling materials traveling through tenth of stellar radii. 

Then, in the most outer part of the chromosphere, \cite{sasselov94b} observed a steady outflow of infrared HeI $\lambda$10830 absorption line, while time-dependent and non-LTE hydrodynamic modeling constrained by previously published observational data \citep{sasselov94a,sasselov94c} show that the upper chromosphere should be permanently heated by an acoustic or magneto-hydrodynamic energy provided by mean of convection. Hence, a combination of both low-frequence excitation provided by pulsational shocks and persistent high-frequency acoustic heating from turbulent convection cells seem to be responsible for the chromospheric dynamics. Others mechanisms leading to high temperature plasmas ($10^6$K) are not excluded.
Indeed \cite{bohm1983} possibly detected X-ray emission in the spectra of $\zeta$~Gem, probably due to a chromospheric activity, which has been confirmed later, around $\phi=0.5$ in the case of $\delta$~Cep and $\beta$~Dor \citep{engle2017}. Although several heating mechanisms to produce such amount of energy are considered, for example fast-moving shocks or magnetic reconnections, a coherent physical explanation is still to be provided.

Alternative lines for probing upper atmosphere dynamic
and chromospheric activity are provided by the calcium infrared triplet ($\lambda$8498.018, 8542.089 and 8662.140\AA). Indeed, \cite{Linsky1970} have shown first that Ca IR lines from the Sun are formed in the lower chromosphere. These lines are also sensitive to the temperature since they are collisionally controlled like Ca H and K lines, between which an empirical correlation has been determined \citep{Martin2017}. These properties make  Ca IR a suitable indicator of chromospheric activity \citep{linsky1979a,foing89,Chmielewski2000,busa07}. While the emission in the core is generally weaker than other chromospheric indicators these lines have the advantage not to be blended by circumstellar or interstellar absorption contrary to Ca H and K and Mg II \textit{h} and \textit{k}. A series of paper on the solar chromosphere has also highlighted the importance of studying Ca IR complementary to H$\alpha$ for probing the chromospheric activity \citep{Cauzzi2008I,vecchio09II,reardon09III,cauzzi09IV}. Indeed, as shown by \cite{Vernazza1973}, the formation region of H$\alpha$ and Ca IR covers most of the low chromosphere in the Sun. Therefore, it is interesting to use modern high-resolution instrument capabilities to observe Ca IR complementary to H$\alpha$ line for studying the pulsating chromosphere of classical Cepheids.
 
In addition, Ca IR lines were recently used to measure radial velocities (RVs) in Cepheids' atmospheres \citep{wallerstein2015,wallerstein2019}. Indeed, as stated by \cite{wallerstein2019}, since the Radial Velocity Spectrometer (RVS) of ESA \textit{Gaia} survey observes in the same wavelength range \citep{Munari1999,Sartoretti2018,Katz2019}, it is of prime importance to understand the dynamic of these lines for Cepheid RV measurement accuracy, in particular in the context of the use of the Baade-Wesselink method for extragalactic distance scale measurement. 
The paper is structured as follows. 
%This paper aims at better understanding the dynamics and the structure of the chromosphere of Cepheids by analysing simultaneously H$\alpha$ and Ca IR profile variations. 
We first present the UVES high-resolution profiles of H$\alpha$ and Ca IR  for the 24 Cepheids in our sample in Sect.~\ref{sect2:UVES}. We analyze H$\alpha$ and  Ca IR profiles variations in Sect.~\ref{sect3:Ha} and Sect.~\ref{sect4:ca_ir}, respectively. In Sect.~\ref{sect6:vanhoof}, we study the dynamics of the chromosphere from the Van Hoof effect. We finally estimate the size of the chromosphere from the Schwarzschild mechanism observed in Ca IR in Sect.~\ref{sect5:schwarz}. We then discuss our results in Sect.~\ref{sect:discuss} and conclude in \ref{sect:conclusion}.

\section{UVES observations and Data reduction \label{sect2:UVES}}
\subsection{Observations and data reduction}
\begin{figure*}[h]
     \centering
         \includegraphics[width=\textwidth]{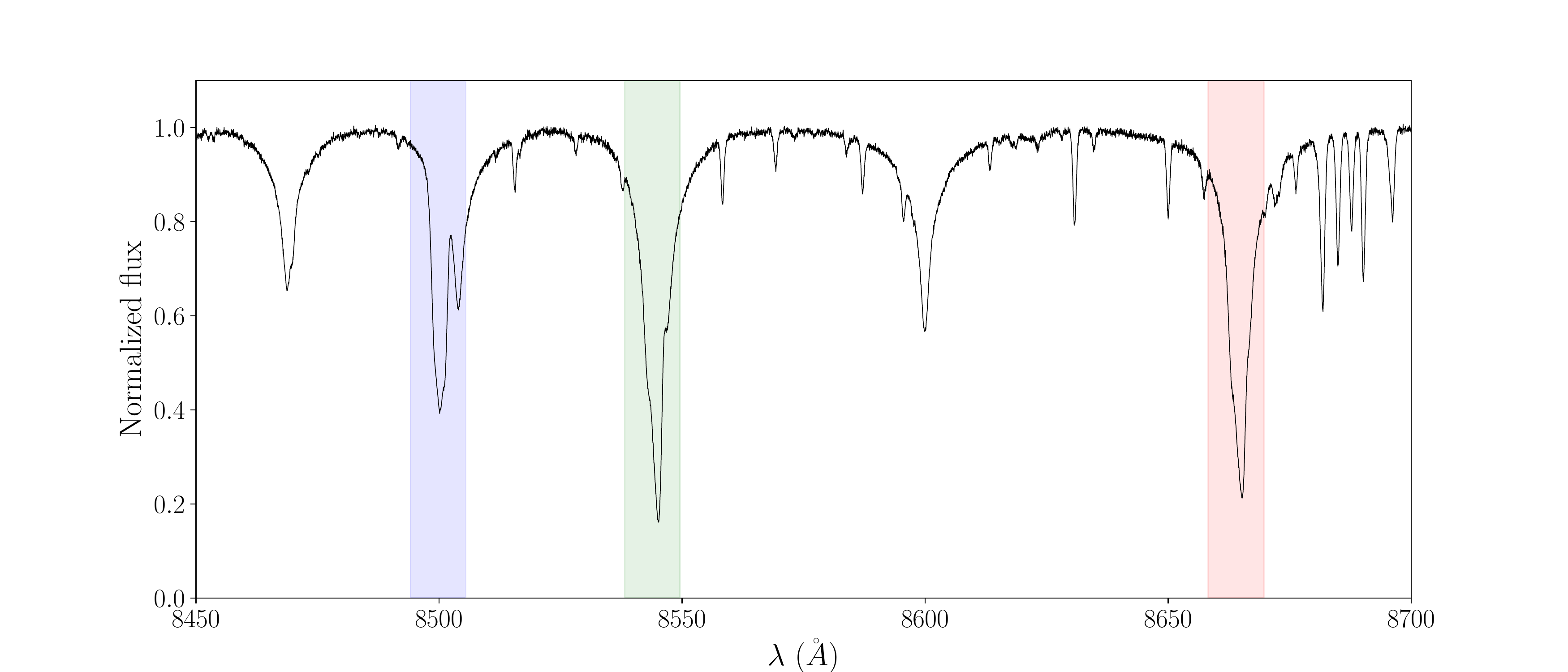}
     \caption{\small Typical normalized UVES spectrum plotted over the RVS spectral window. Calcium infrared triplet at 8498, 8552 and 8662 angstroms for VZ Pup at $\phi=0.91$ is presented. In the following the spectral line profiles are plotted over the [\text{-}200,\text{+}200] km/s velocity range, represented by vertical strips around each line.}\label{fig:calcium_exemple}
\end{figure*}

\begin{figure*}
     \centering
         \begin{subfigure}[b]{0.24\textwidth}
         \centering
         \includegraphics[width=\textwidth]{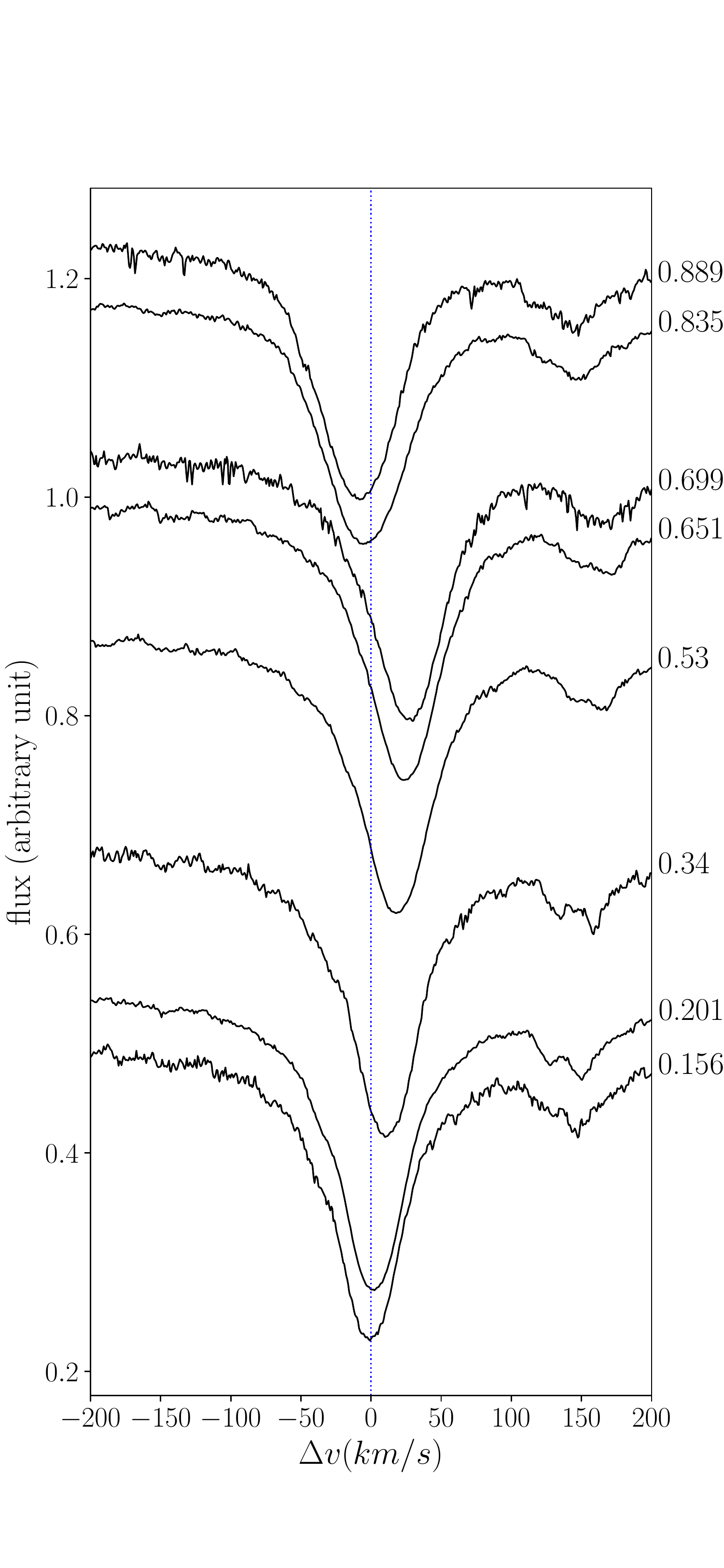}
         \caption{$\lambda$8498}
     \end{subfigure}
     \hfill
     \begin{subfigure}[b]{0.24\textwidth}
         \centering
         \includegraphics[width=\textwidth]{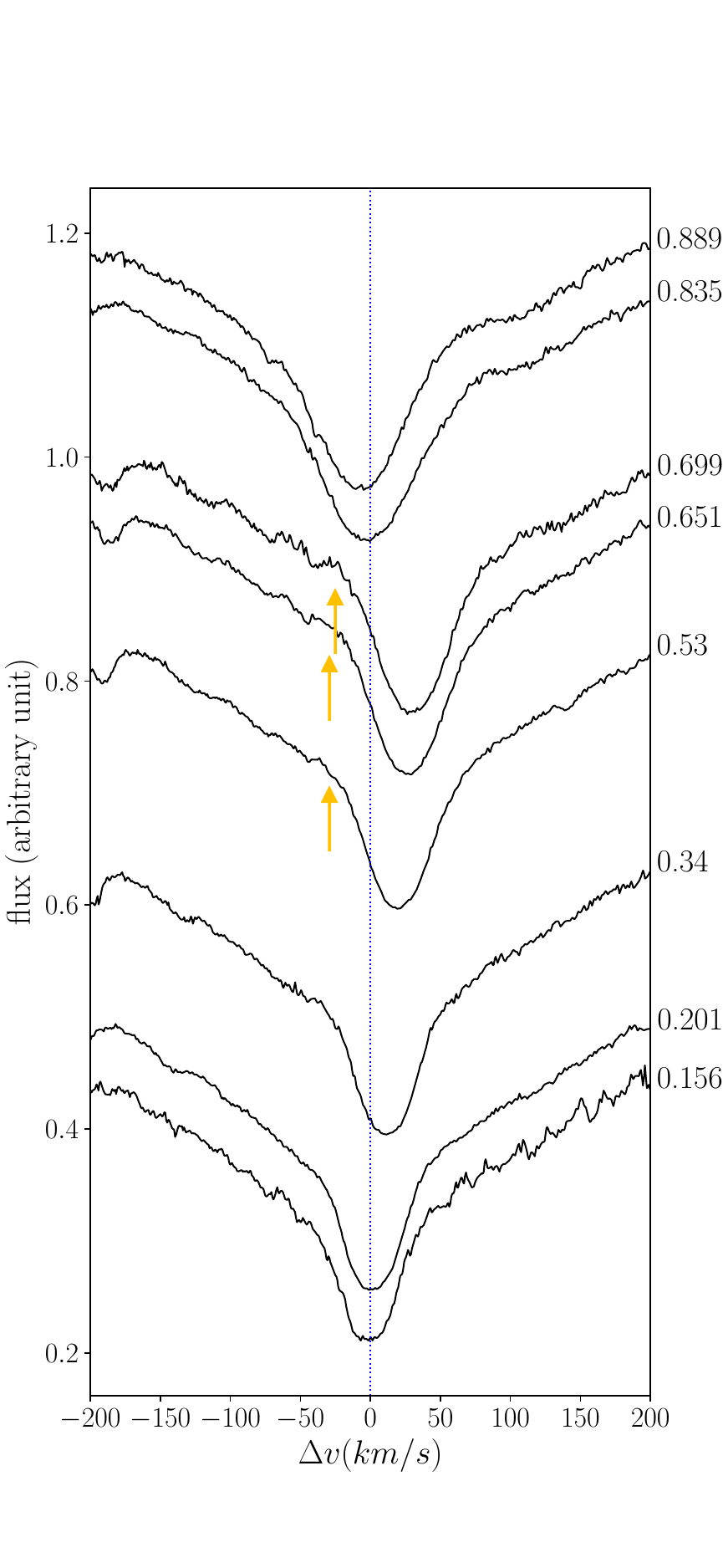}
         \caption{$\lambda$8542}
     \end{subfigure}
     \hfill
     \begin{subfigure}[b]{0.24\textwidth}
         \centering
         \includegraphics[width=\textwidth]{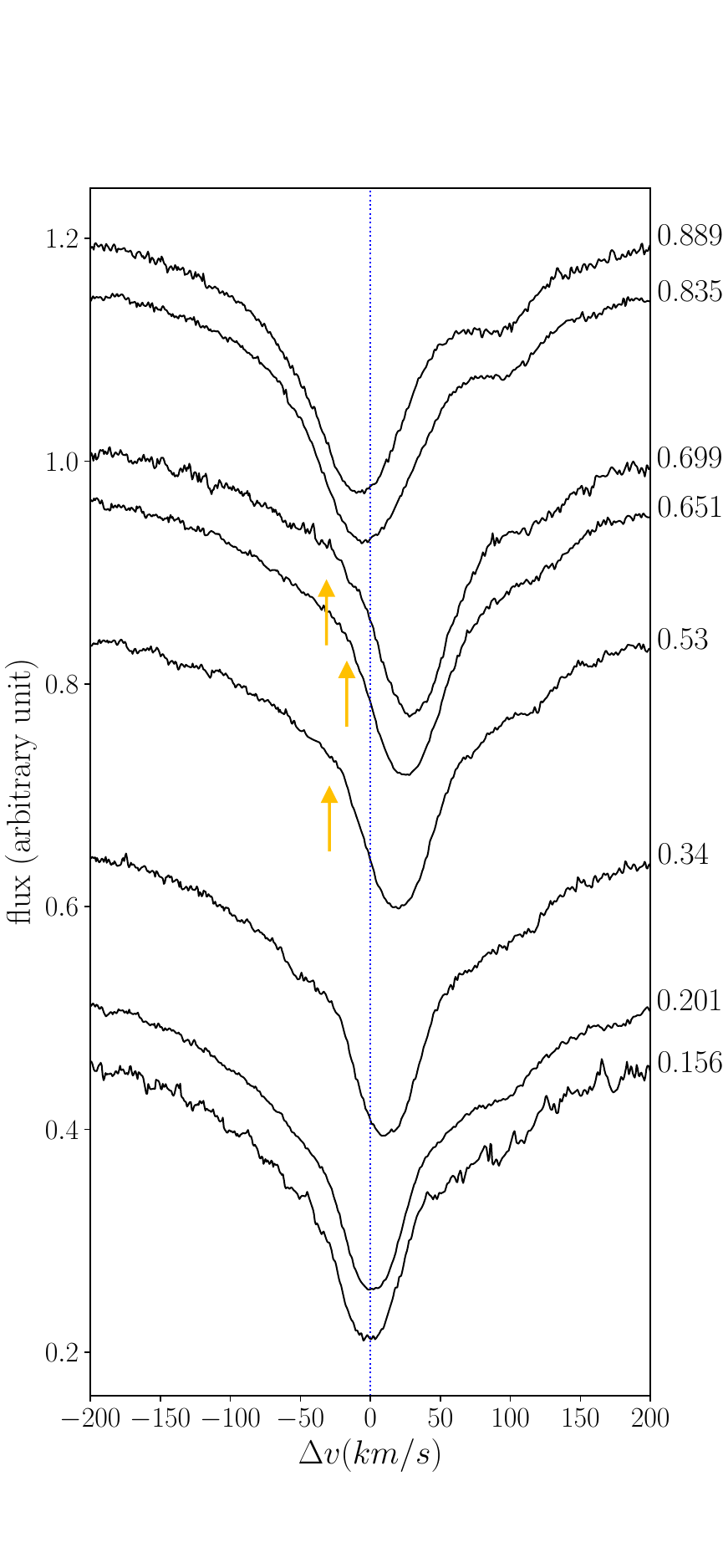}
         \caption{$\lambda$8662}
     \end{subfigure}
     \hfill
     \begin{subfigure}[b]{0.24\textwidth}
         \centering
         \includegraphics[width=\textwidth]{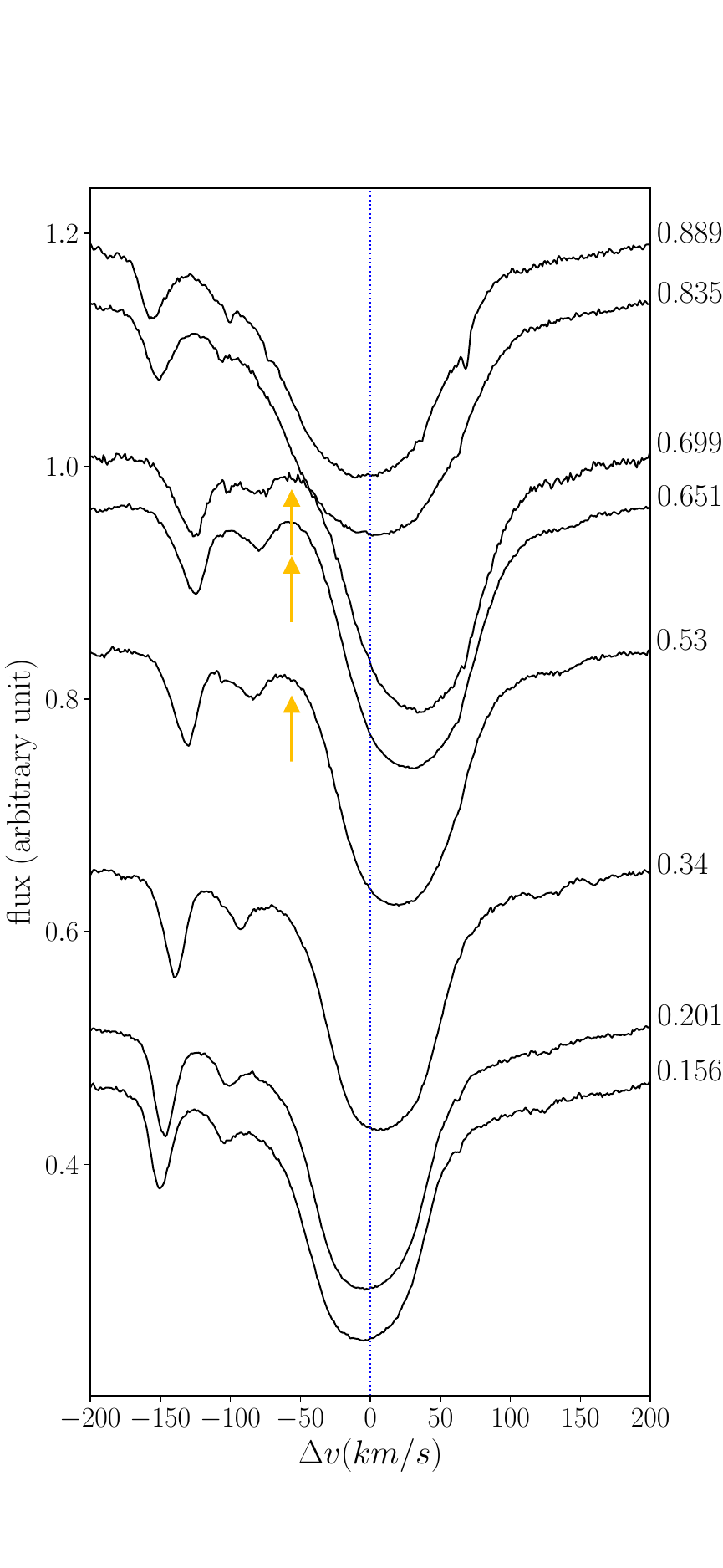}
         \caption{H$\alpha$}\label{fig:ax_cir_Ha}
     \end{subfigure}
        \caption{AX Cir, 5.27d. Colored arrows refer to remarkable features in these profiles and are also associated to the color from Fig.~\ref{fig:std_model}. Orange arrow is the blue-shifted emission due to a blended inverse P Cygni profile. \label{fig:ax_cir}}
\end{figure*}

\begin{figure*}
         \begin{subfigure}[b]{0.24\textwidth}
         \centering
         \includegraphics[width=\textwidth]{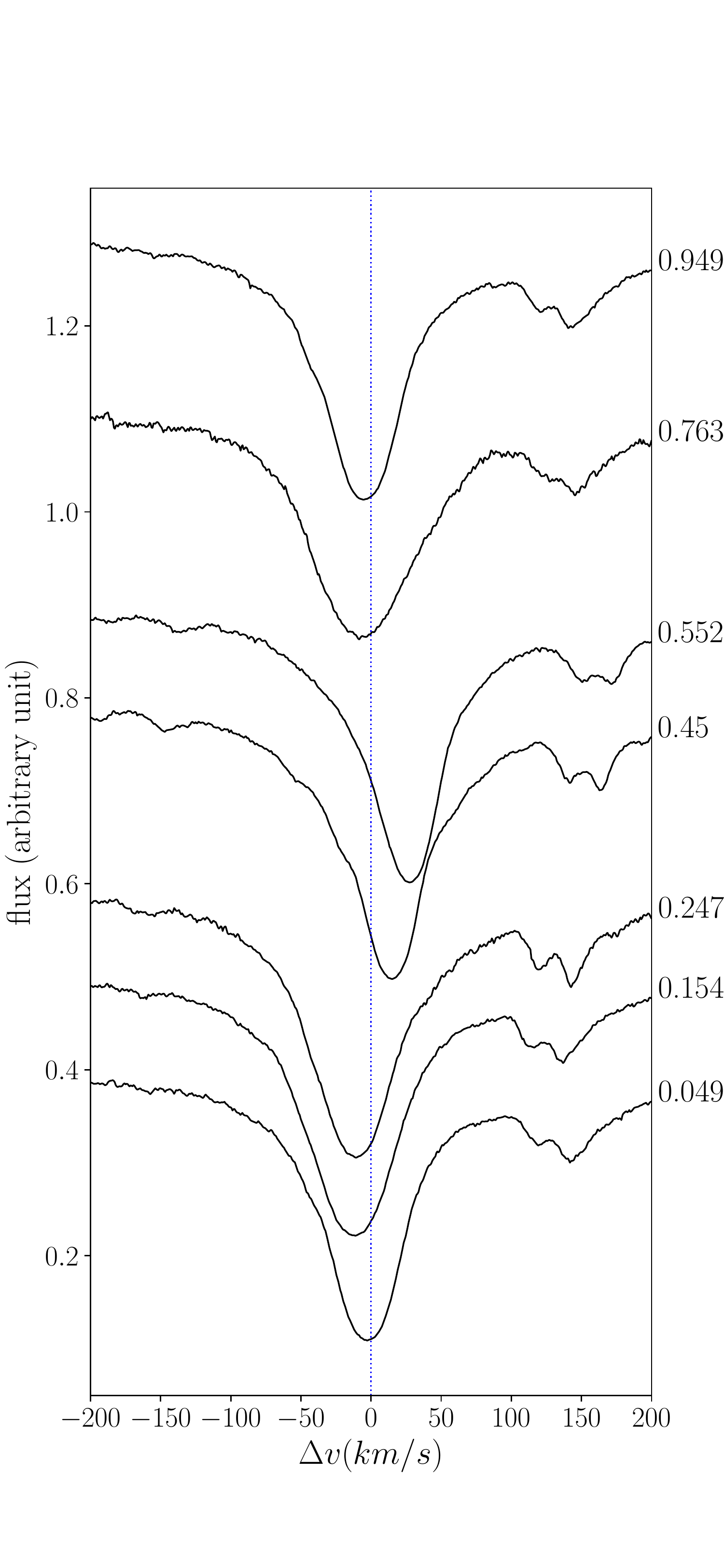}
        \caption{$\lambda$8498}
     \end{subfigure}
     \hfill
     \begin{subfigure}[b]{0.24\textwidth}
         \centering
         \includegraphics[width=\textwidth]{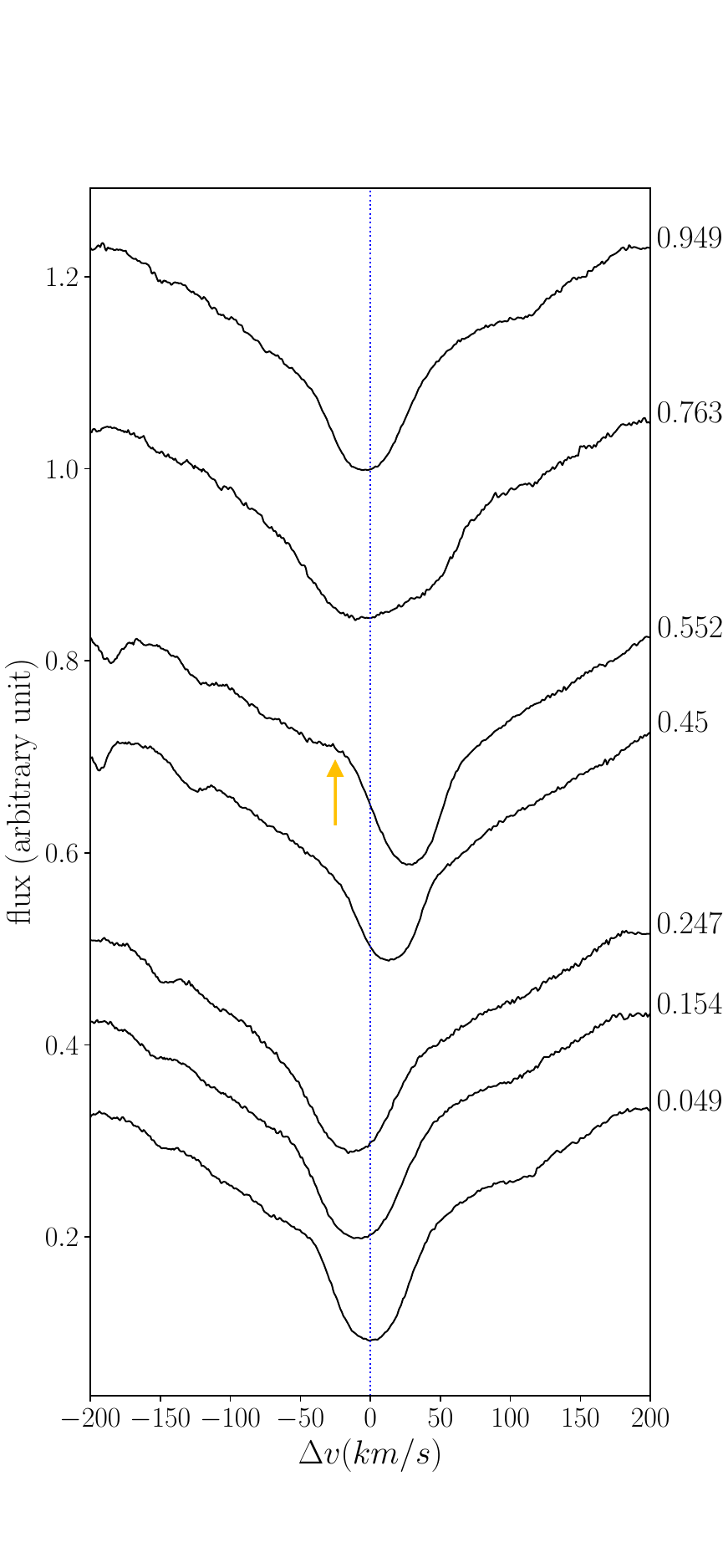}
         \caption{$\lambda$8542  \label{fig:s_nor_8542}}

     \end{subfigure}
     \hfill
     \begin{subfigure}[b]{0.24\textwidth}
         \centering
         \includegraphics[width=\textwidth]{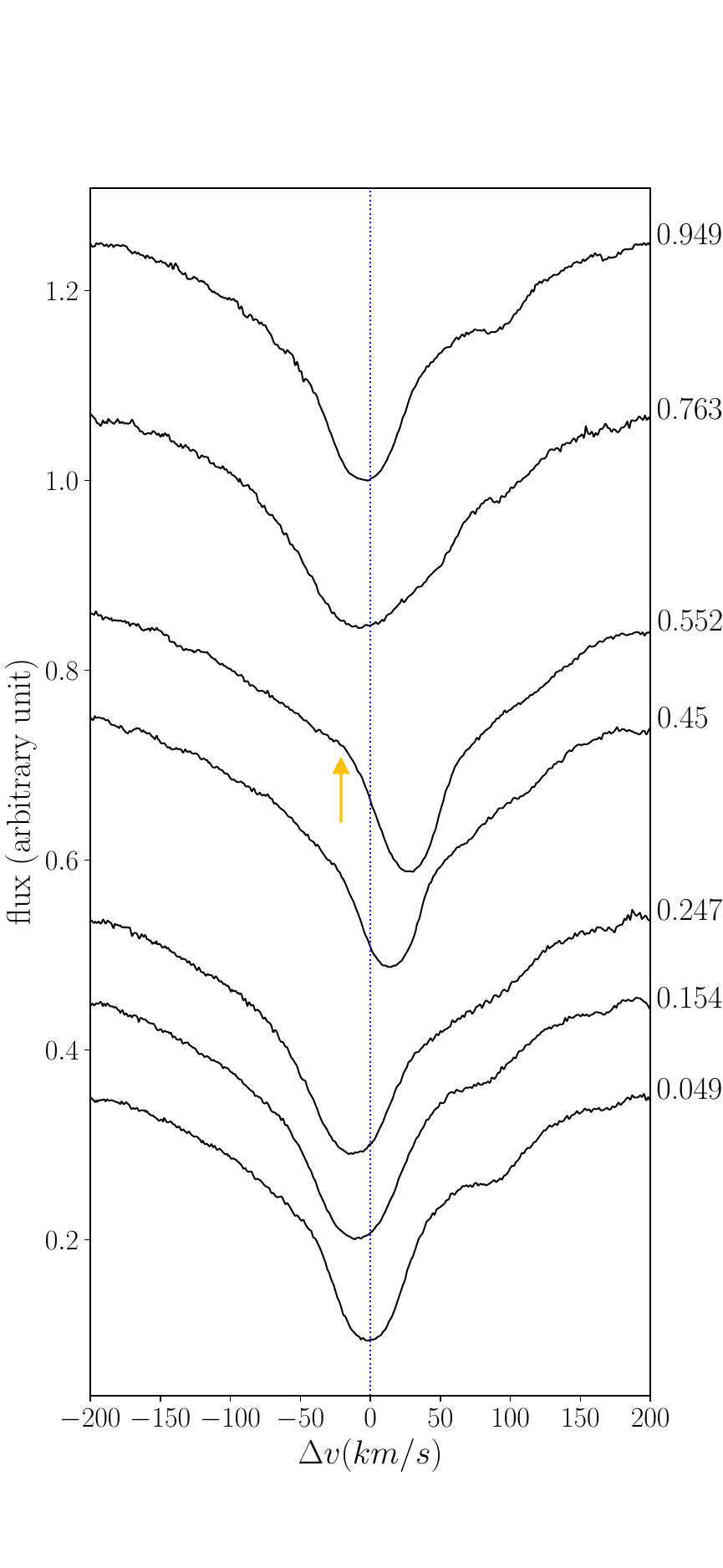}
         \caption{$\lambda$8662  \label{fig:s_nor_8662}}
     \end{subfigure}
     \hfill
          \begin{subfigure}[b]{0.24\textwidth}
         \centering
         \includegraphics[width=\textwidth]{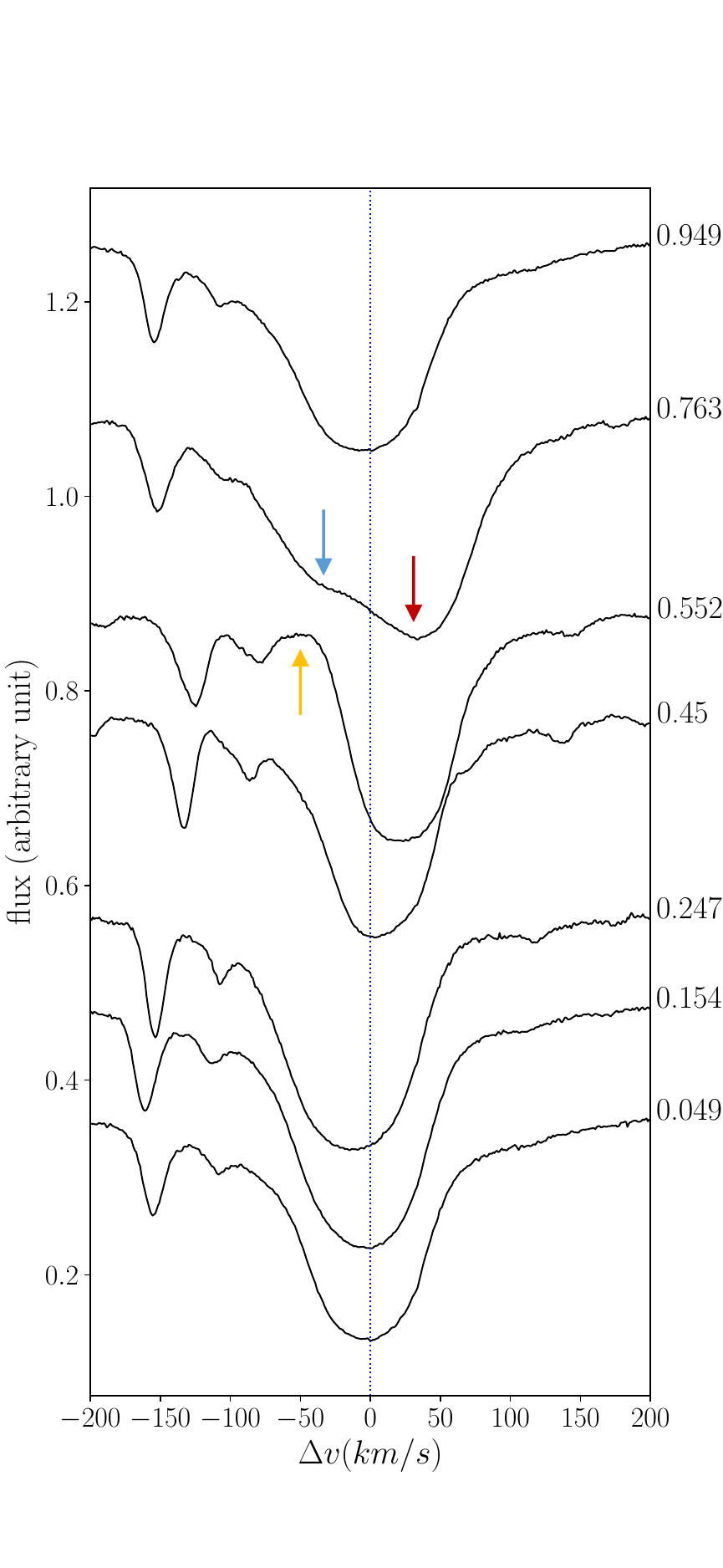}
         \caption{H$\alpha$}\label{fig:Ha_s_nor}
     \end{subfigure}
        \caption{S Nor, 9.75d. Blue arrow is the blue-shifted absorption due to the main shock. Red arrow is the red-shifted absorption of the in-falling atmosphere. Orange arrows are blue-shifted emission due to a blended inverse P Cygni profile. \label{fig:s_nor}}
\end{figure*}

\begin{figure*}
     \centering
          \begin{subfigure}[b]{0.24\textwidth}
         \centering
         \includegraphics[width=\textwidth]{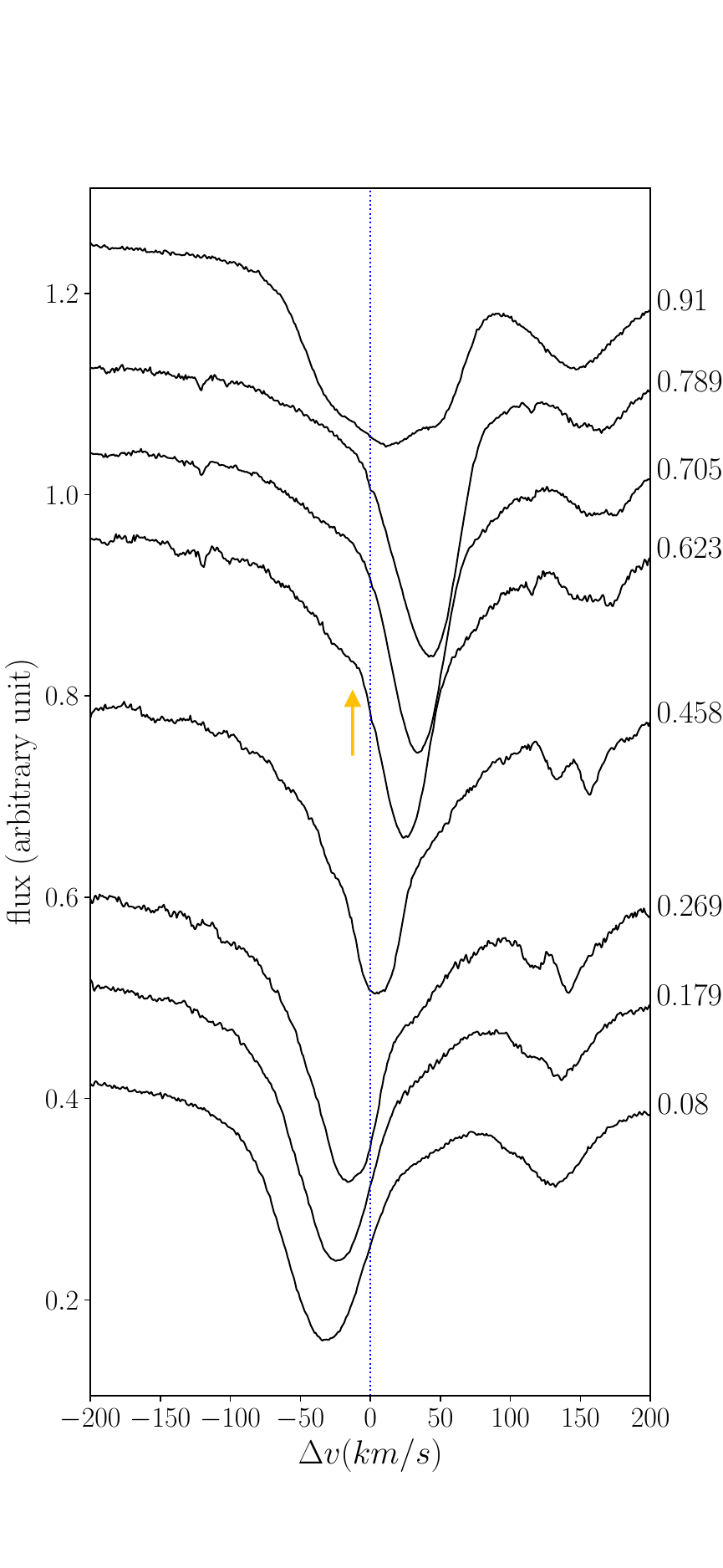}
         \caption{$\lambda$8498}\label{fig:vz_pup_all_1}
     \end{subfigure}
\hfill
     \begin{subfigure}[b]{0.24\textwidth}
         \centering
         \includegraphics[width=\textwidth]{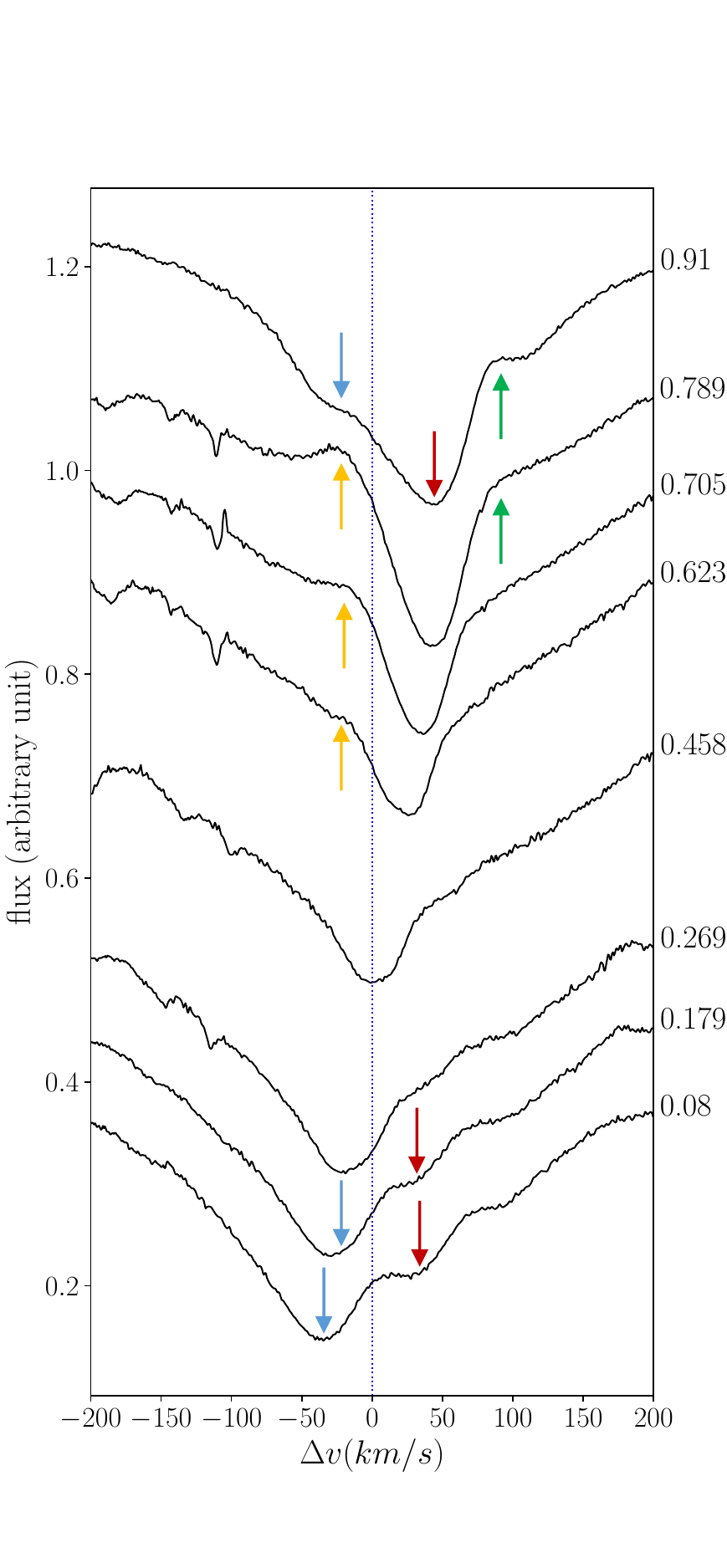}
        \caption{$\lambda$8542 }\label{fig:vz_pup_all_2}

     \end{subfigure}
\hfill
     \begin{subfigure}[b]{0.24\textwidth}
         \centering
         \includegraphics[width=\textwidth]{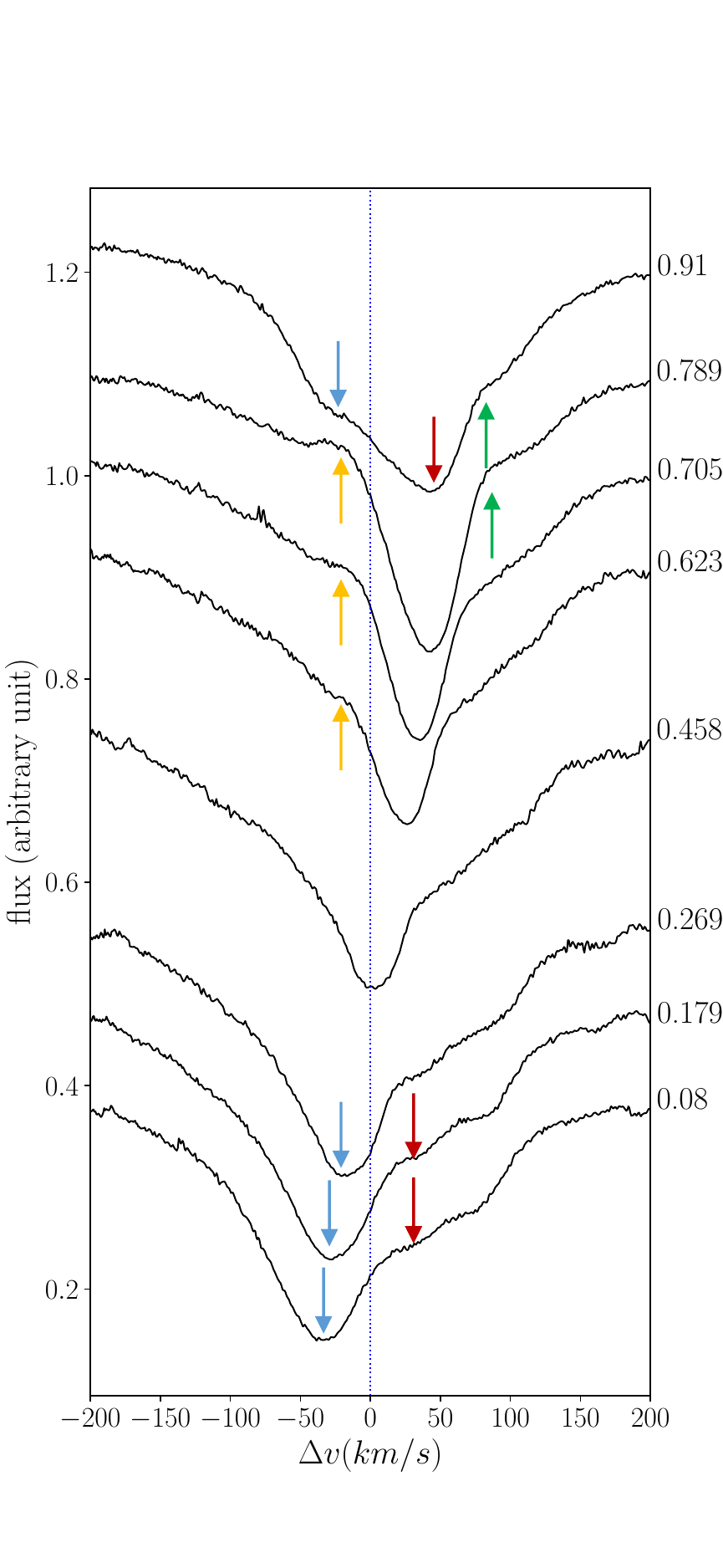}
         \caption{$\lambda$8662}\label{fig:vz_pup_all_3}

     \end{subfigure}
     \hfill
          \begin{subfigure}[b]{0.24\textwidth}
         \centering
         \includegraphics[width=\textwidth]{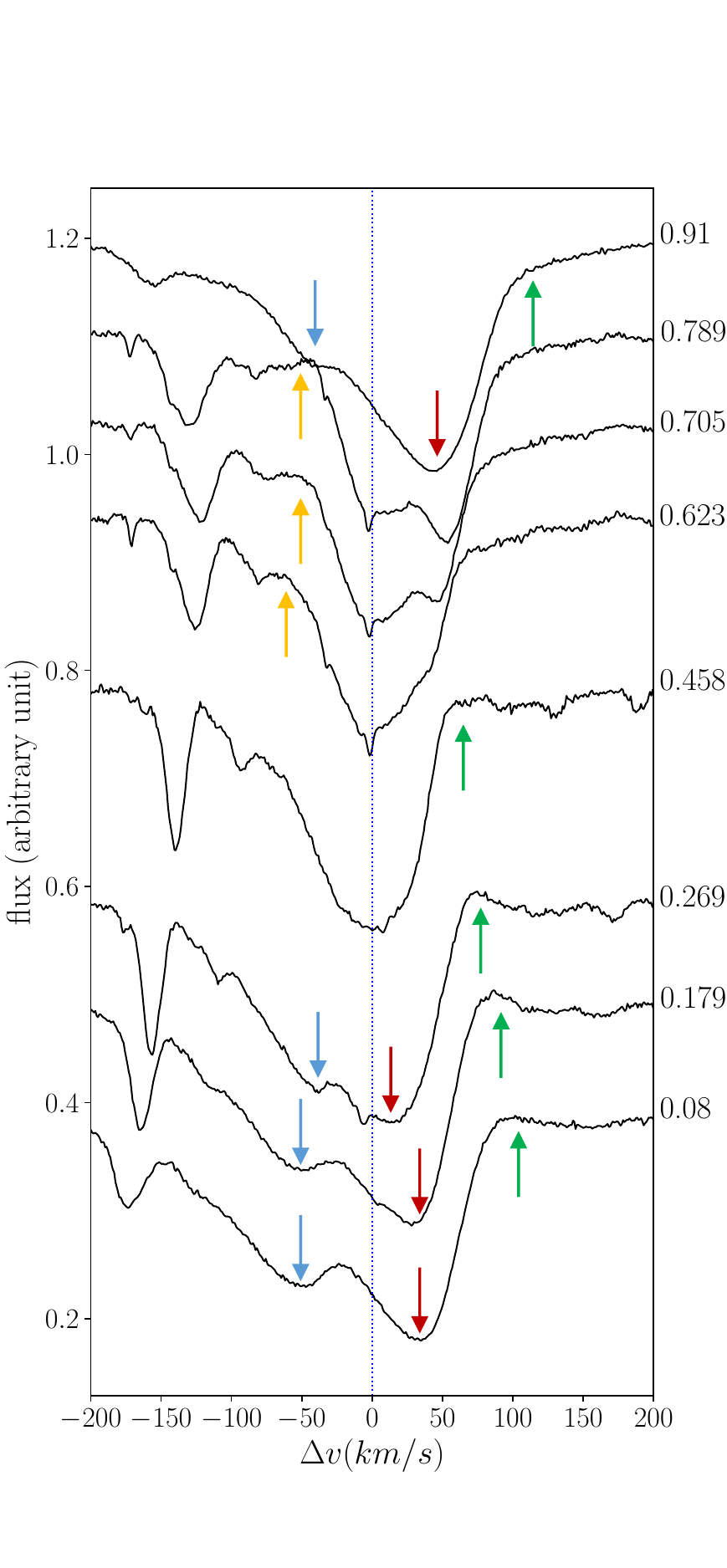}
         \caption{H$\alpha$}\label{fig:vz_pup_Ha}
     \end{subfigure}
        \caption{VZ Pup, 23.17d. Blue arrows are blue-shifted absorption due to the main shock. Red arrows are red-shifted absorption of the in-falling atmosphere. Green arrows are red-shifted emission due to a P Cygni profile. Orange arrows are blue-shifted emission due to an inverse P Cygni profile.} \label{fig:vz_pup}
\end{figure*}

We gathered 1350 high-resolution ($R \sim 75000$) spectra from 24 Cepheids acquired with the red arm ($\sim$570 to $\sim$940 nm wavelength range) of the UVES spectrograph \citep{dekker2000} mounted on the UT2 telescope at the Very Large Telescope (VLT). The processing and normalization of these spectra are detailed in \cite{Borgniet2019}. We acquired several consecutive spectra (up to a dozen) at each observing epoch. 
%\N{For each Cepheid, we summed the UVES spectra observed consecutively (Sect.~\ref{sect2:UVES}) to merge them into a single average spectrum per observing epoch. This allows us to increase the SNR on the Ca IR profiles while keeping the same precision in terms of shape and Doppler shift. There are indeed no significant changes to the line profiles due to the short UVES exposure times on such bright targets.}
The total number of distinct observing epochs is 193 over the whole data sample, with an average number of 8 epochs per target. The total number of epoch per target is indicated in Table~\ref{Tab.vrad}. The average SNR per single spectrum is $\approx$100. In order to discuss qualitatively the spectral line profiles, at a given epoch, we consider only the first snapshot of a series, without averaging the consecutive spectra.
We consider the different observing epochs to be part of a unique pulsation cycle for each target, despite possible cycle-to-cycle variations \citep{anderson2016}.
%The average SNR \SIMON{per single} spectrum is $\approx$100, thus, for each consecutive spectra, we retrieved only the first snapshot so that we can use the specific observation date without averaging the consecutive data. We have also removed spectra for which pulsation phases are very close to each other. 
 %We consider the different observing epochs to be part of a unique pulsation cycle for each target, for convenience. Thus, the spectral line profile variability from one epoch to another has to be interpreted with caution regarding possible cycle-to-cycle variations \citep{anderson2016}.
 To study the line profile variations, we focus on the core of each line, hence we consider the same RV window ranging from -200 to 200 km/s and centered on the line rest wavelength corrected by the star center-of-mass velocity (also called $\gamma$-velocity, $V_{\gamma}$). The $V_{\gamma}$ values we consider in the following of the analysis are listed in Table~\ref{Tab.vrad}. We present an UVES spectrum centered on the Ca~IR triplet in Fig.~\ref{fig:calcium_exemple}. AX~Cir (5.27d), S~Nor (9.75d) and VZ~Pup (23.17d) are considered as prototypes of small-, mid-, and long-periods respectively in the following of the paper, because of their good phase coverage (see Figs.~\ref{fig:ax_cir}, \ref{fig:s_nor} and \ref{fig:vz_pup}) and recent ephemeris (see Table \ref{Tab.vrad}) presented in next section.

\begin{table*}
\caption{\label{Tab.vrad} \small The $\gamma$-velocities of the stars in the sample are indicated together with the corresponding references. The ephemeris (T$_0$, P) are retrieved either from \cite{gaia2018} (G18), \texttt{SPIPS} software or the GCVS catalogue. Otherwise, the ephemeris of EW Sct and V1496 Aql are retrieved from AAVSO and \cite{berdnikov2004}, respectively. The number of UVES epoch of observations is indicated in the last column. The Cepheids that we consider as prototypes in the following of the paper are indicated in bold. }
\begin{center}
\begin{tabular}{ccccccc}
\hline
\hline
Star	&	$\gamma$ (km/s)	&	Ref. & P (days) &T$_0$ (MJD)& Ephemeris & UVES epoch	\\
\hline					
AV Cir	&	8.7$\pm$1.6&	(1)	&  3.065255  &   56890.9571   &  G18& 10\\
				
BG Cru	&	-19.3$\pm2.2$	&	(2)	&      3.342477  &   56965.8567 & G18   &6\\

RT Aur	&	20.3$\pm$0.3	&	(2)	&      3.728313  &   47956.8877  & SPIPS   &7\\

AH Vel	&	26.0$\pm$2.9	&	(2)	&    4.226461  &   56915.1433 &G18  &7\\
					
\textbf{AX Cir}	&	-20.9$\pm4.6$	&	(2)	&   5.275967  &   56887.7452   & G18  &8\\
					
MY Pup	&	11.0$\pm$2.9	&	(2)	&      5.692962 &   56948.3850 &  G18  &7\\
				
EW Sct	&	-18.6$\pm$0.3	&	(2)	&       5.82363 &    49705.23    & AAVSO  &9\\
% AAVSO catalogue of variable stars: http://vizier.u-strasbg.fr/viz-bin/VizieR?-source=B/vsx
				
U Sgr	&	2.8$\pm$0.3	&	(2)	& 6.745332   &   48336.5001 & SPIPS&10\\%(!) \SIMON{12?}\\

V636 Sco	&	9.1$\pm$0.17	&	(3)	&      6.79671  &40364.392& GCVS    &7\\

R Mus	&	3.8$\pm$2.9	&	(2)	&  7.510276   &   56912.6852   & G18&6\\
				
%(!) \SIMON{10?}\\
				
S Mus	&	-1.9$\pm$0.4	&	(3)	& 9.658900    & 56867.6654 & G18&7\\%(!) \SIMON{8?}\\
				
$\beta$ Dor	&	7.2$\pm$0.7	&	(2)	&    9.842661    & 50274.9261    & SPIPS& 5\\
				
{\bf S Nor}	&	5.6$\pm$0.05	&	(4)	&      9.753759   & 56874.8021   & SPIPS   &7\\
				
$\zeta$ Gem	&	2.8$\pm$0.2	&	(2)	&      10.149857 &   48708.0588& SPIPS   &6\\
				
TT Aql	&	3.0$\pm$0.3	&	(2)	&  13.754750   & 48308.5570  & SPIPS  & 10\\
				
RU Sct	&	-4.8$\pm$0.3	&	(2)	&   19.70445    &    48335.5908   & SPIPS  &7 \\%\SIMON{8?}\\

RZ Vel	&	24.1$\pm$2.4	&	(2)	&      20.497635   & 56875.8093&  G18 &5\\% \SIMON{6?}\\
				
WZ Car	&	-14.7$\pm$2.7	&	(2)	&     23.01759  &    53418.78& GCVS   &10\\
				
{\bf VZ Pup}	&	63.3$\pm$2.7	&	(2)	&     23.172844   &   56904.7167&  G18   &8\\%(!) \SIMON{11?}\\23.172844 2456905.2167
				
T Mon	&	30.4$\pm$0.2	&	(2)	&      27.029570    &    43783.7905   & SPIPS   &6\\%(!) \SIMON{7?}\\
				
$\ell$ Car	&	3.3$\pm$0.7	&	(2)	&    35.55783 &   50583.7427& SPIPS   &6\\
				
U Car	&	1.7$\pm$2.3	&	(2)	&      38.717914 &    56849.1763   & G18  &8\\
					
RS Pup	&	24.6$\pm$0.5	&	(2)	&      41.464114   &   56872.3158 & G18   &  5\\%(!) \SIMON{6?}\\
				
V1496 Aql	&	71.8$\pm$5.8	&	(1)	&      65.3679  & 51736.0660&   (5) &8 \\%\SIMON{10?}\\
\hline																					
\end{tabular}
\end{center}
\begin{tablenotes}
\small
\item (1): \cite{gaia2018}\label{gaia2018},
\item (2):	\cite{gontcharov2006},
\item (3):    \cite{pourbaix2004},
\item (4):    \cite{mermilliod2008},
\item (5):    \cite{berdnikov2004}
\end{tablenotes}
\normalsize

\end{table*}
%Since We limited our study to a qualitative approach except for

%We present EW~Sct (5.82d), S~Nor (9.75d) and VZ~Pup (23.17d) in the corpus of this paper that we have chosen to be the prototypes of small-, mid-, and long-periods respectively, because of their better phase coverage (see Figs.~\ref{fig:ew_sct}, \ref{fig:s_nor} and \ref{fig:vz_pup}).

\subsection{Ephemeris}
% J'ai remis cette phrase elle est importante dans le raisonnement (et appréciée par le referee voir premier rapport) de plus SPIPS permet justement d'atteindre cette justesse ainsi que les récentes observations de gaia
It is essential to use accurate ephemeris to analyze the profile variations of Cepheids along their pulsation cycle with precision. 
Ephemeris of Cepheids are characterized by their pulsation period $P$ and their referenced epoch $T_0$ given at maximum light. However, since Cepheids undergo period change with time depending on evolutionnary state \citep{Turner2006}, it is necessary to use ephemeris with reference epoch which is close to UVES date of observation. Modified Julian Date (MJD) of the UVES observations is approximately 58000, which renders  suitable ephemeris obtained from \cite{gaia2018}. 
When these epheremis are unavailable we use
SpectroPhoto-Interferometric modeling of Pulsating Stars (\texttt{SPIPS}) which is a model-based parallax-of-pulsation code. \texttt{SPIPS} gathers multiple different data sets available in the literature from photometric, interferometric, effective temperature and radial velocity measurements in a robust model fit \citep{merand15}. \texttt{SPIPS} was already extensively introduced and used in several studies \citep{merand15,breitfelder16,kervella17,Gallenne2017,hocde2020}. 
When stars are not found in the precedent methods, which is the case for four stars, we retrieve data from the General Catalogue of Variable Stars \footnote{\url{http://www.sai.msu.su/gcvs/gcvs/}} \citep[GCVS,][]{GCVS2017} or the American Association of Variable Star Observers  \footnote{\url{http://vizier.u-strasbg.fr/viz-bin/VizieR?-source=B/vsx}} \citep[AAVSO,][]{aavso2006}. We note that older ephemeris could introduce a phase shift in the radial velocity curves, in particular in the case of the long period V1496 Aql which is known to present a significant stochastic period change \citep{berdnikov2004}. 
The ephemeris are presented in Table~\ref{Tab.vrad} while profile variations are shown in Appendix~\ref{Fig.profiles}.

%We did not corrected ephemeris from period changes since it does not shift the pulsation phase significantly.

%The phase of each spectrum using the ephemeris obtained with the SPIPS algorithm are thought to be more reliable than the ephemeris from General Catalogue of Variable Stars (GCVS) \citep{GCVS2017}. \V{Indeed, we have checked showing a perfect consistency between the phases of the different spectra obtained with SPIPS whereas those derived with GCVS presented more scattered results with less consistency between the spectra.}
 
\paragraph{}
In the next sections, we proceed in two steps to analyze the data: we first describe the H$\alpha$ profiles as an observational basis, then, we analyze Ca IR profiles.

\section{Observational basis: H$\alpha$ profile variations\label{sect3:Ha}}
In this section, we mostly confirm the H$\alpha$ line behaviour which has been found in previous studies, but for a larger sample of 24 Cepheids with a good period coverage from P$\approx$3 to 60 days. 

%We classify these stars into three groups depending on their pulsation period. We present the star sample from short (10 stars), medium (4 stars), to long (10 stars) period Cepheids. These results are used as a solid observational basis for comparing with Ca IR profiles in the next section.

\subsection{Short-period cepheids P<10d \label{sect3:Ha_small}}
\begin{figure}
          \begin{subfigure}[b]{0.5\textwidth}
         \centering
         \includegraphics[width=\textwidth]{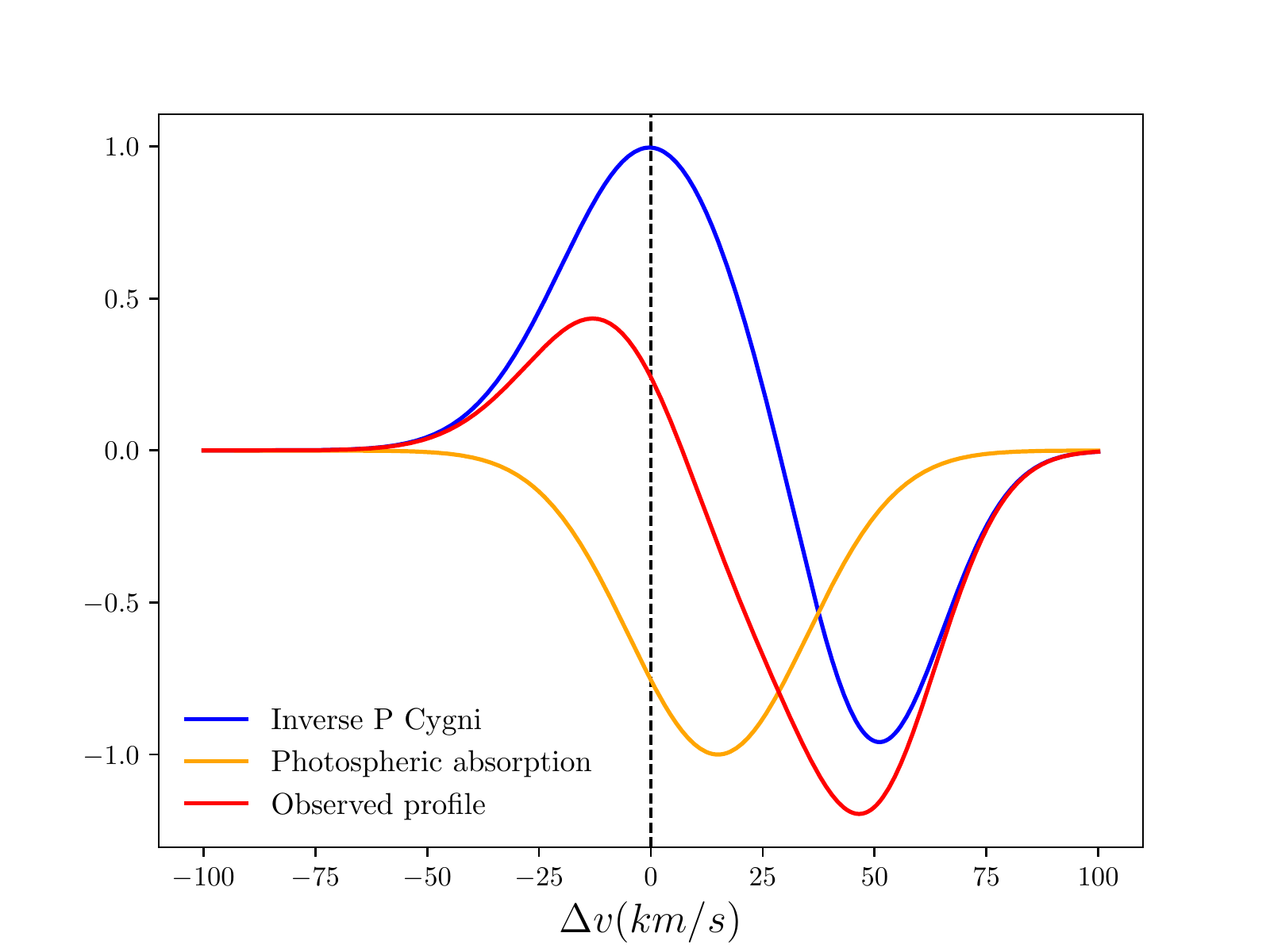}
         \caption{Inverse P Cygni blended with photosphere}\label{fig:inverse_p_cyg}
     \end{subfigure}
%\vskip
     \begin{subfigure}[b]{0.5\textwidth}
         \centering
         \includegraphics[width=\textwidth]{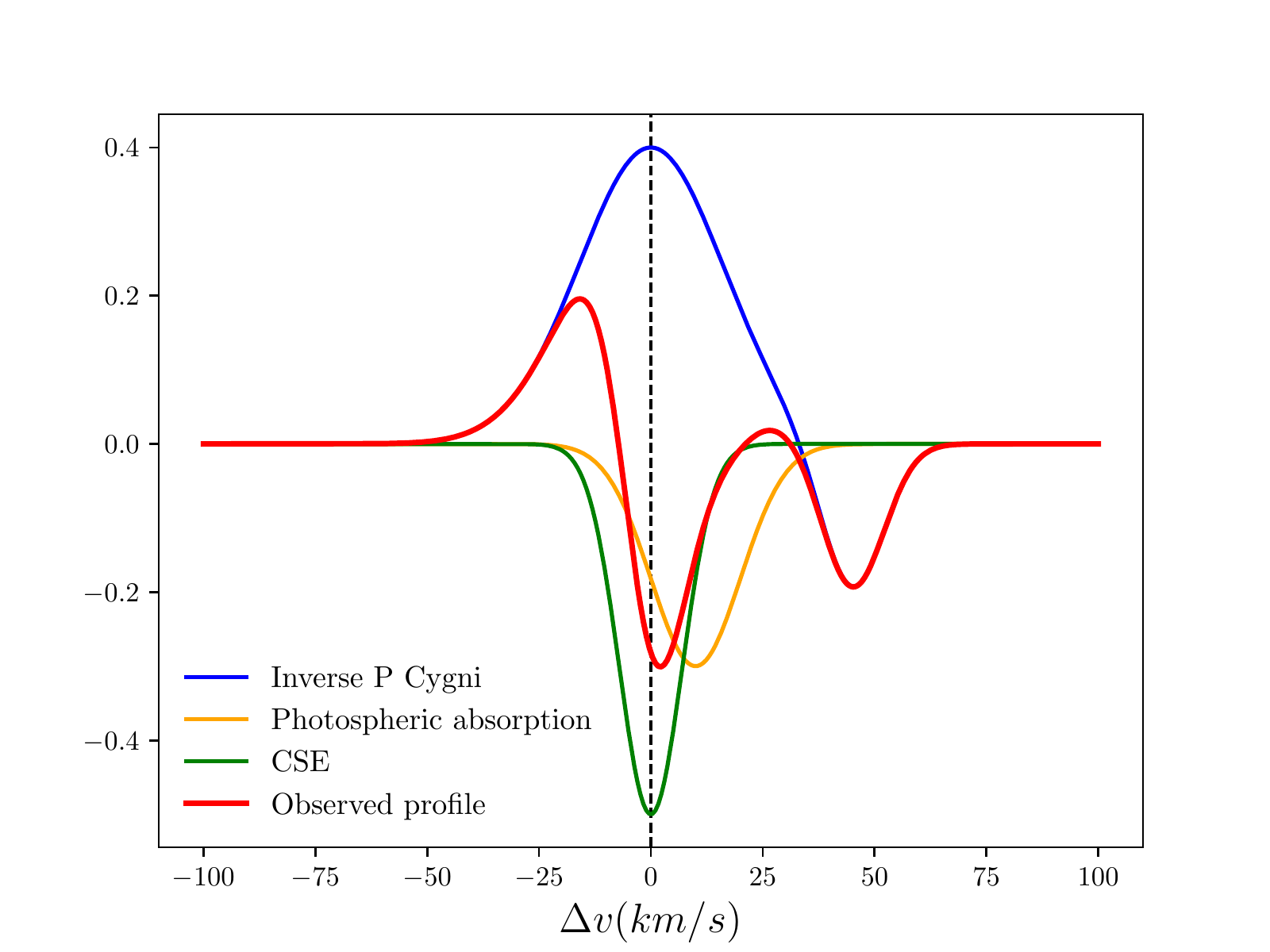}
        \caption{Inverse P Cygni blended with CSE + photosphere}\label{fig:inverse_p_cyg_ha}
     \end{subfigure}
        \caption{\small Schematic profile formation of a blended inverse P Cygni profile. (a) The supersonic ($\approx$ 50 km/s) descending atmospheric layers produce a radiative emission centered on 0 km/s (blue curve). This profile is blended with photospheric red-shifted ($\approx$15 km/s) absorption (orange curve) resulting in a an apparent weak blue-shifted emission. (b) In long-period Cepheids a component centered on 0km/s is attributed to a CSE (green) and causes the appearance of a double absorption profile.}\label{fig:inverse_p_cygni}
\end{figure}

Several studies have reported the quiescent behaviour of H$\alpha$ profiles of short-period cepheids which indeed present similar behaviour than other metallic absorption lines \citep{schmidt1970,nardetto08b}, while some authors have found differences at some phases \citep{Jacobsen1981}. The UVES observations of the 10 short-period Cepheids in the sample confirm this general trend, with quiescent profiles during most of the pulsation cycle. At some phases, half of the short-period Cepheids present more disturbed features, especially a weak blue-shifted emission is observed between phase $\phi$=0.6-0.9 (see orange arrow in Fig.~ \ref{fig:ax_cir_Ha}).  Because of the gravitational acceleration, upper atmosphere descending layers can reach supersonic velocities, producing an inward radiative shock. An inverse P Cygni profile could appear when the column density is important enough to contribute to the emission. However, the classical 0 km/s emission in P Cygni profile is blended with deep photospheric absorption, resulting in an apparent blue-shifted emission (see Fig. \ref{fig:inverse_p_cyg}).  
%The same phenomenon, while stronger, is observed in mid- and long-period cepheids of the star sample. 
This emission in short-periods is followed by an important enlargement of the line profile, which is usually attributed to an increase of the turbulence during atmosphere compression \citep{breitfellner93a,fokin96}.

\subsection{Medium-period cepheids P$\approx$10d \label{sect3:Ha_medium}}
Medium-period Cepheids differ from the short- and long-period Cepheids by the shape of their radial velocity curve due to the Hertzsprung progression \citep{hertzsprung1926}. These Cepheids are indeed close to the $\omega_2$/$\omega_0$=0.5 resonance, where $\omega_0$ and $\omega_2$ are the periods corresponding to the fundamental and the second overtone modes respectively \citep{kovacs90}.
The UVES sample of medium-period Cepheids contains four stars, namely S~Nor (our prototype), S~Mus, $\zeta$~Gem and $\beta$~Dor, with a rather poor pulsation coverage. These stars have been studied by several authors \citep{Bell1967a,schmidt1970,Jacobsen1982} who have found substantial disturbed profiles and large displacement from metallic lines.

In the star sample, blue-shifted H$\alpha$ emissions are observed in S~Nor, S~Mus and $\beta$~Dor around $\phi=0.5-0.6$ (see orange arrow in Fig.~\ref{fig:Ha_s_nor}). Since the atmosphere is in-falling at these phases for resonance stars, an inverse P~Cygni profile could appear if the gas is high enough and reaches supersonic velocities as it is the case for short- and long-period cepheids. 

 In addition, S Nor and $\beta$ Dor profiles are importantly wider around $\phi=0.75-0.80$ (see Fig.~\ref{fig:Ha_s_nor}  and Fig.~\ref{fig:beta_dor} respectively) with the appearance of a double absorption profile. These features are not observed on S~Mus at phase $\phi$=0.9 and there is no phase coverage beyond $\phi=0.5$ for $\zeta$~Gem. The red-shifted absorption in the profile (see red arrow in Fig.~\ref{fig:Ha_s_nor}) is caused by the infalling motion of the atmosphere while the blue-shifted absorption is likely due to the main shock produced by the $\kappa$ mechanism in the star interior, which is later propagating outward in the star atmosphere. Then, we expect the collision between these in-falling layers and the main shock, which could result in a line enlargement feature around $\phi$=0.8.
 % doubling 

\subsection{Long-period cepheids P>10d \label{sect3:Ha_long}}
For long-period Cepheids ($P > 10$d) we confirm the global atmosphere dynamics from H$\alpha$ profiles presented by \cite{gillet14} in the case of X~Cygni (16.3d). This author described three different H$\alpha$ stages per cycle which are in agreement with the UVES observations of the 10 long-period Cepheids presented in this paper:

\begin{enumerate}
    \item From $\phi$ = 0.9 to 0.3 when the photosphere is expanding, we observe the presence of a blue-shifted H$\alpha$ absorption (see blue arrows in Fig.~\ref{fig:vz_pup_Ha}) which is identified as a gas flow at the rear of the main shock due to $\kappa$-mechanism when emerging from the photosphere at $\phi \approx $0.85 (see blue zone in Fig.~\ref{fig:model_0_03}). This feature is accompanied by a red-shifted absorption due to the infalling atmosphere layers.

    \item Between phase 0.9 and 0.5, a P~Cygni profile is observed (see green arrows in Fig.~\ref{fig:vz_pup_Ha}). This P~Cygni profile is interpreted as H$\alpha$ emission at the rear of the main shock front propagating outward. It appears as a P Cygni profile when it is sufficiently detached from the photosphere (see green zone in Fig.~\ref{fig:model_0_03}). This feature progressively disappears between $\phi$ = 0.3 to 0.6 and is no longer visible around $\phi \approx$ 0.6 (see \cref{fig:vz_pup_Ha,fig:model_03_05}).
    
   \item From $\phi$ = 0.7 to 0.9 a double absorption profile appears with a blue-shifted emission. This double absorption is composed of an absorption centered on the stellar rest frame and a red-shifted absorption at $\approx$50km/s. The motionless absorption feature have been attributed to a CSE by several authors, not only in the case of $\ell$ Car \citep{Rodgers1968,Baldry1997,nardetto08b}, but also in the case of the Type II long-period Cepheid W Virginis \citep{Kovtyukh2011}. On the other hand, the red-shifted absorption is due to hydrogen in-falling layers during the ballistic motion. As a result, the blue-shifted emission (orange arrows in Fig.~\ref{fig:vz_pup_Ha}) is a blend between an inverse P Cygni profile caused by the supersonic infalling layers, and both a photospheric and a CSE absorption (see Figure \ref{fig:inverse_p_cyg_ha}).
   
     \item From $\phi$ = 0.9 to 1.0 the inverse P Cygni profile progressively disappears. Two effects could contribute together to explain this transition. First, since the layers are infalling, at a certain point they are not high enough in the atmosphere to produce a P Cygni profile. Secondly, the main shock emerges from the photosphere at phase $\phi \approx$0.85, then a blue-shifted absorption could progressively blends the P Cygni emission.
\end{enumerate}

\section{Ca IR profile variations \label{sect4:ca_ir}}
\begin{figure*}[h!]
     \centering
     \begin{subfigure}[b]{0.33\textwidth}
         \centering
         \includegraphics[width=\textwidth]{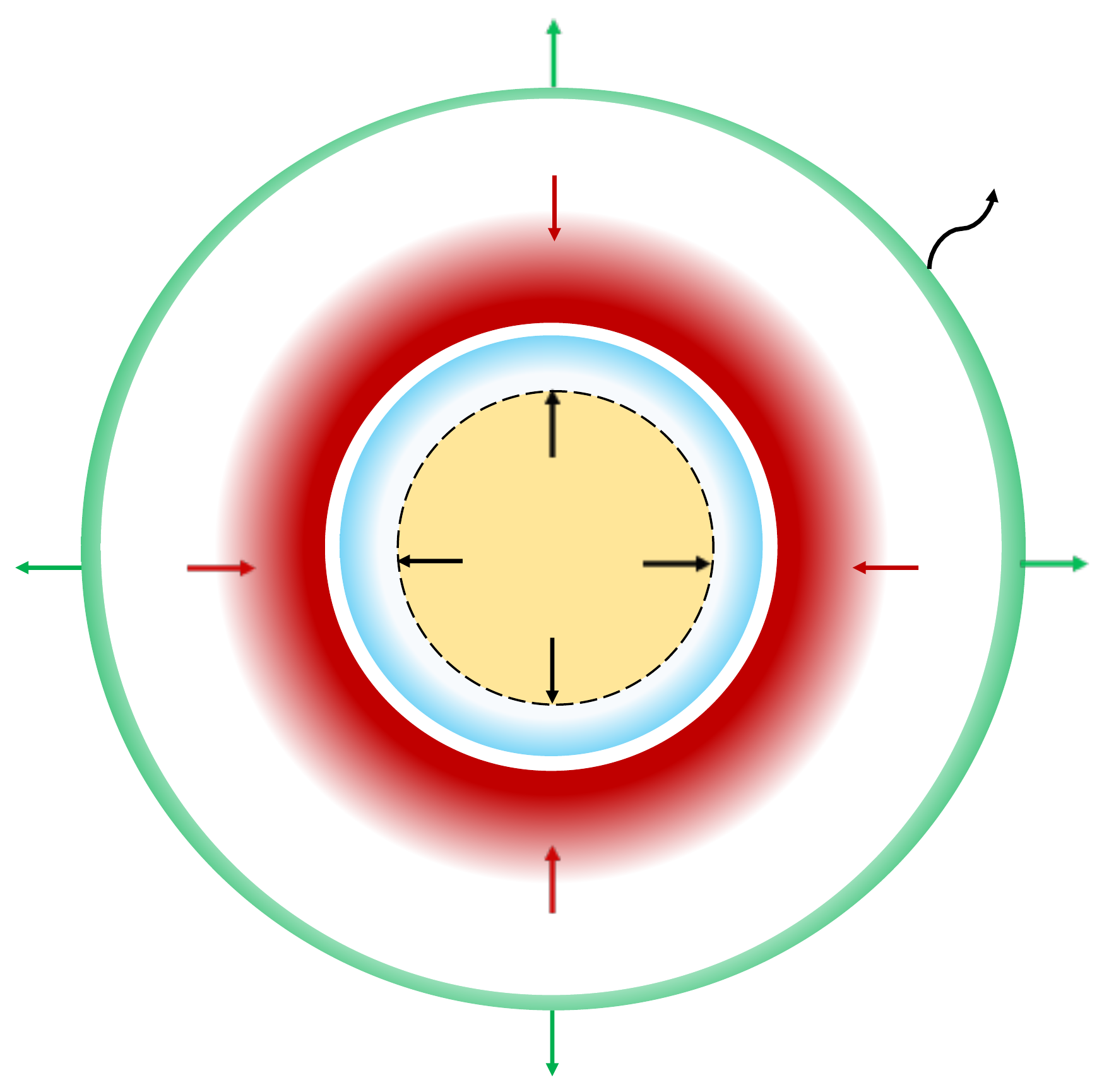}
         \caption{$\phi$=0.9-0.3 : Schwarzschild scenario}\label{fig:model_0_03}
     \end{subfigure}
\hfill
     \begin{subfigure}[b]{0.33\textwidth}
         \centering
         \includegraphics[width=\textwidth]{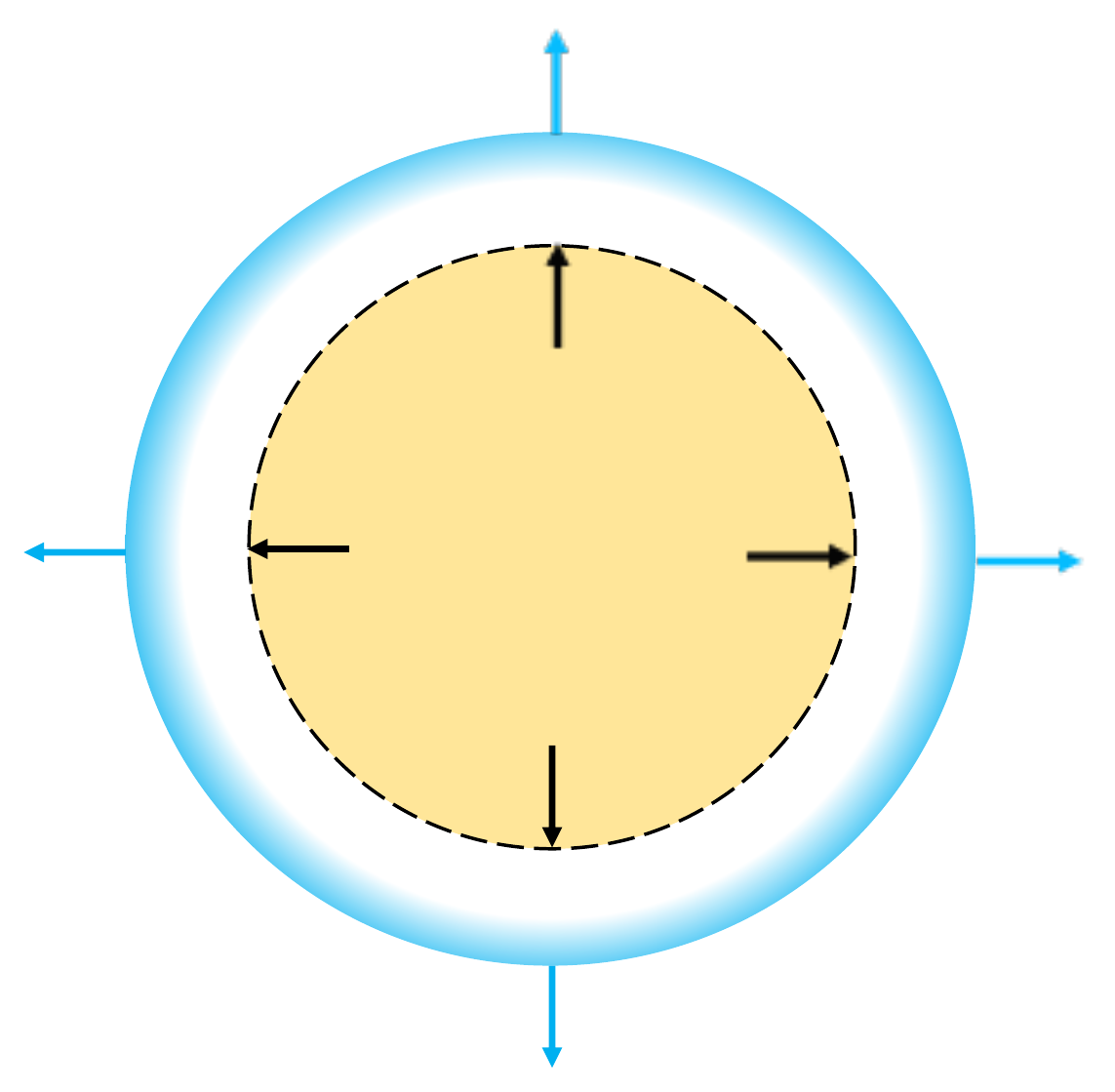}
         \caption{$\phi$=0.3-0.6 : Quiescent profiles}\label{fig:model_03_05}
     \end{subfigure}
\hfill
          \begin{subfigure}[b]{0.30\textwidth}
         \centering
         \includegraphics[width=\textwidth]{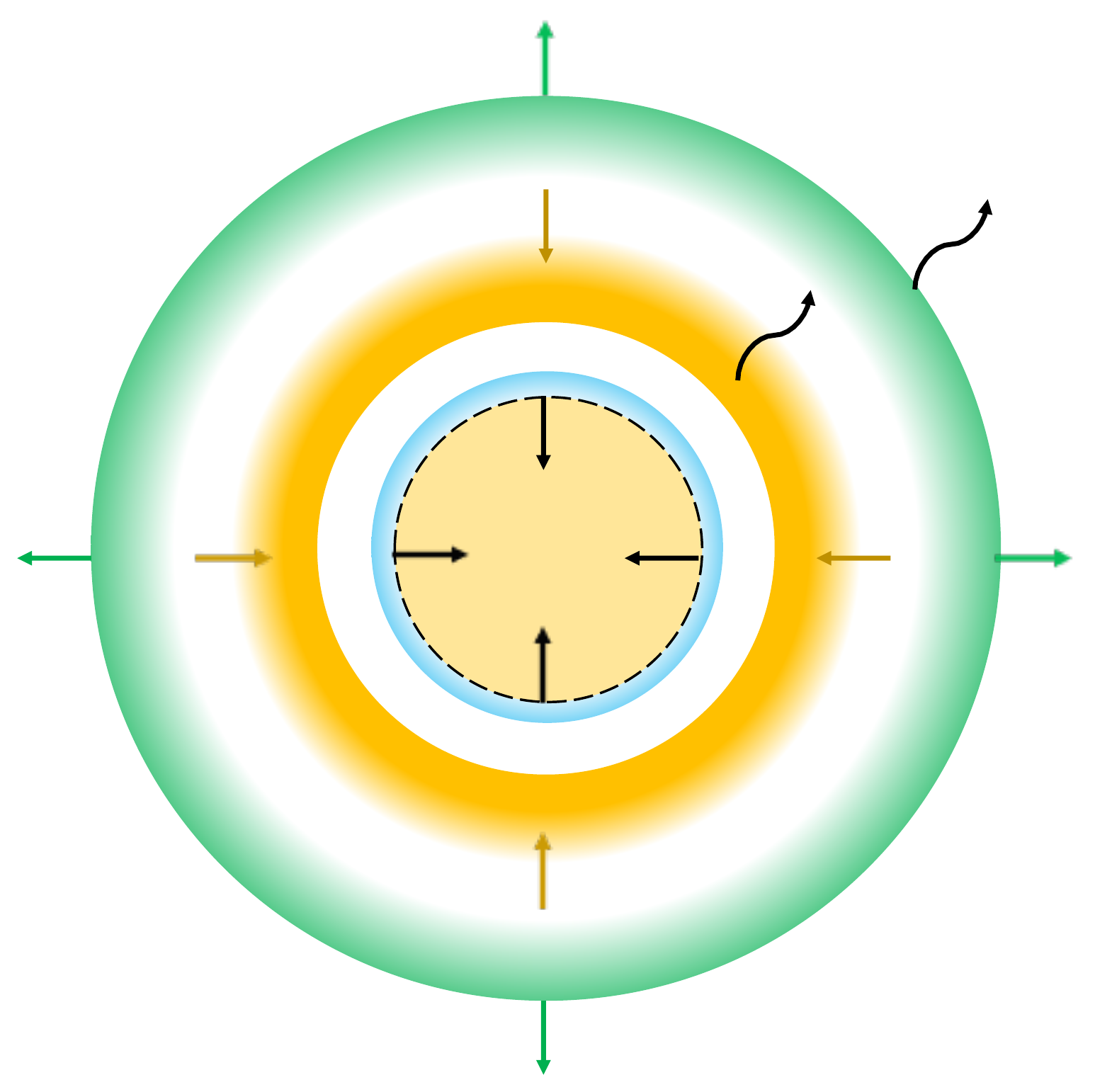}
         \caption{$\phi$=0.6-0.9 : Inverse P Cygni}\label{fig:model_06_09}
     \end{subfigure}
     \caption{General scheme of the pulsating chromosphere dynamics from Ca IR variations. \textbf{(a)}~As the main shock continue to propagate outward causing a radiative wake (black arrow emitted by the green zone), H$\alpha$ exhibits a blended P Cygni profile. At the same time a blue-shifted absorption appears from the emerging main shock from the $\kappa$ mechanism (blue zone). This shock collides the Ca~IR in-falling layers (red zone) which produces a Schwarzschild profile (Sect. \ref{sect5:schwarz}).  \textbf{(b)} Quiescent phase: The main shock has reversed the movement of the atmosphere which is now entirely propagating outward. H$\alpha$ P Cygni profile progressively disappears around $\phi$=0.5. \textbf{(c)} Following a ballistic motion, layers of the atmosphere fall on the star (orange area). Supersonic velocities are reached if the acceleration is strong enough and thus an inverse P Cygni profile appears for H$\alpha$, Ca $\lambda$8542 and $\lambda$8662. The main shock continues to propagate outward and a Ca IR P Cygni profile could appear due to a radiative wake emission from chromosphere (black arrow emitted by the green zone)}\label{fig:std_model}
\end{figure*}

In this section, we qualitatively describe and physically interpret Ca IR profile variations. We further show that Ca IR profile features are similar to H$\alpha$ profile variations since they are induced by the same phenomena. We synthetized our findings in Fig.~\ref{fig:std_model} and Table \ref{tab:summary}.

\begin{table*}[]
\caption{Observational features of calcium infrared triplet in the star sample. The Table is divided into pulsation periods (columns) and phase interval (rows). In Fig.~\ref{fig:std_model} we represent the mean features observed in each phase interval.}\label{tab:summary}
\begin{center}
\begin{tabular}{c|p{5cm}|p{5cm}|p{5cm}}
\hline
\hline
$\phi$ interval	& Small (P<10d)&	Medium (P$\approx$10d)	&	Long (P>10d) \\
\hline	
\hline							
0.9-0.3 	& \textbf{- Quiescent profiles:} \par The atmospheric layers are moving outward. & \textbf{- Quiescent profiles} 	&\textbf{- Schwarzschild mechanism:}\par In-falling chromosphere layers are \par collapsing onto the emerging shock \par causing a double absorption profile \par \textbf{- P Cygni profile:} \par The main shock front is still propagating and progressively leaves Ca IR layers, entering in outermost H$\alpha$ layers.\\
\hline
0.3-0.6	&\textbf{- Quiescent profiles}	&	\textbf{- Inverse P Cygni profile:}\par Emission from supersonic infalling layers.		&\textbf{- Quiescent profiles}\\
\hline
0.6-1.0	&\textbf{- Inverse P Cygni profile:}\par A weak emission is observable in several short-periods	 \par \textbf{- Profile enlargement:}\par The profiles are importantly wider due to turbulences during \par atmosphere collision.	& \textbf{- Line doubling:} \par Enlargement of the line profile \par  which is possibly a double profile due to a Schwarzschild scenario. 	& \textbf{- P Cygni profile:} \par It appears when the main shock \par from the previous cycle is high \par enough in the chromosphere. \par \textbf{- Inverse P Cygni profile:}\par Emission from supersonic infalling layers. \par \textbf{- Transition phase:} \par From 0.9 to 1.0 the latter emission progressively disappears. \par Schwarzschild mechanism initiates.   	\\
\hline							
\hline
\end{tabular}

\end{center}
\end{table*}

\subsection{Short-period cepheids P < 10d \label{sect4:short}}
 Similarly to H$\alpha$ lines presented in Sect~ \ref{sect3:Ha_small}, Ca IR profiles are almost quiescent during the pulsation cycle. However, a finer inspection reveals a weak blue-shifted emission between $\phi \approx$0.6 - 0.9 which is always synchronized with H$\alpha$ blue-shifted emission (see orange arrows in Fig.~\ref{fig:ax_cir}). Hence, we can infer that Ca IR blue-shifted emission has the same origin same origin as a H$\alpha$, the in-falling upper atmosphere layers are causing a radiative shock, resulting in an inverse P~Cygni profile. The line $\lambda$8498 is less disturbed than the two others lines, a result that we find also for medium and long-period Cepheids. According to \cite{Linsky1970} this line is formed lower in the chromosphere than the two others, as a result, the acceleration is less important during ballistic movement of the atmosphere for line $\lambda$8498 and could not be high enough in the atmosphere to produce a P Cygni profile.

\subsection{Medium-period cepheids P$\approx$ 10d \label{sect4:mid}}
In the case of medium-period Cepheids we also find the two profile features identified in the case of H$\alpha$ profiles, that is a blue-shifted emission followed by an enlargement of the lines (see \cref{fig:s_nor_8542,fig:s_nor_8662}). However in this case it is less clear if we observe a simple enlargement due to turbulence or a double absorption profile which would be not resolved. Indeed, medium-period Cepheids are an intermediate case between short-period Cepheids which are dominated by the turbulence during the in-falling motion, and a double absorption profile as it is the case in long-period Cepheids.

\subsection{Long-period cepheids P > 10d \label{sect4:long}}
 For all the profile variations presented in this paper we report the following Ca~IR profile sequence over the pulsation period:
\begin{enumerate}
    \item From $\phi$ = 0.9 to 0.3 there are two important spectral features. First, we observe a double absorption profile centered on 0~km/s velocity. The same features were recently observed by \cite{wallerstein2019} who published Ca~IR profiles of X~Cygni with a good phase coverage. We interpret this feature as a Schwarzschild mechanism \citep{schwarz1952}. Briefly, this phenomenon can be produced when two layers of an atmosphere with an opposite velocity field are colliding on each other. In the present case it is caused by the collision between the main shock expanding outward and the lower chromosphere region descending toward the photosphere (see Fig.~\ref{fig:model_0_03}). Thus, both a blue- and a red-shifted absorption component symmetrically centered on the stellar restframe are observed (see blue and red arrows respectively in \cref{fig:vz_pup_all_2,fig:vz_pup_all_3}). This phenomenon is weaker or absent for the line $\lambda$8498. We study in detail this behaviour in Sect.~\ref{sect5:schwarz}. Secondly, we also observe a red-shifted emission (see green arrows in \cref{fig:vz_pup_all_2,fig:vz_pup_all_3}). We attributed this feature to a P~Cygni profile due to the radiative wake at the rear of the main shock (initiated at the end of the previous cycle) which is still propagating outward.
    
   \item  From $\phi$ = 0.3 to 0.6 the main shock has reversed the atmosphere movement which is now expanding. The main shock continues to propagate, as shown in Fig.~\ref{fig:model_03_05} and all Ca IR profiles are passing through a quiescent phase. No particular features are observed. 
    
    \item Between $\phi$ = 0.6 and 0.9 a blue-shifted emission is observed (see orange arrows in \cref{fig:vz_pup_all_1,fig:vz_pup_all_2,fig:vz_pup_all_3}). In the case of VZ~Pup this emission is seen from $\phi$=0.623 to 0.789 with an increasing intensity, which is almost 20\% of the pulsation cycle. Similarly to H$\alpha$, we attribute this feature to an inverse P~Cygni profile due to the supersonic in-falling movement of the atmosphere (see Fig.~\ref{fig:model_06_09}). However, contrary to H$\alpha$ it is not a double absorption profile. Indeed since the abundance of calcium is lower than hydrogen, we expect a weaker CSE absorption from Ca IR in long-period Cepheids. We suggest it could contribute enough in some cases to blend a part of the Ca IR emission as it is the case in the sample for U Car and $\ell$ Car in \cref{fig:u_car,fig:l_car}.%\cite{wallerstein2019} found a blue-shifted emission from $\phi$=0.63 to 0.71 in the case of X Cyg. %This emission correlates perfectly with H$\alpha$ double absorption profiles. \cite{gillet14} has interpreted H$\alpha$ double absorption profiles as an inverse P~Cygni profile blended with a wide photospheric absorption feature. 

    %Indeed, during the in-falling motion, the atmosphere reaches supersonic velocity close to 40 km/s, producing a radiative wake H$\alpha$ line. Thus an inverse profile is likely to appear, in this case an inverse P~Cygni profile should be observable, with an emission centered on 0 km/s velocity. However, a photospheric absorption component centered on the stellar restframe blends this emission. We suggest the same mechanism for producing Ca IR blue-shifted emission. 

%Hydrogen H$\alpha$ line can be used as a chromosphere diagnostic, but since it can be excited by different mechanisms, this indicator remains uncertain \citep{Linsky2017}. However, Ca IR lines are collisionnally controled, and are therefore an unambigous indicator of chromospheric activity. 
\end{enumerate} 
%As a conclusion of the Ca IR profile variations of the star sample, we find similar Ca IR profiles than H$\alpha$ whatever the pulsation period of the Cepheids. 
As a conclusion from the analysis of the Ca IR profiles in the entire star sample, the presence of Ca IR emissions indicate a chromospheric activity in the Cepheid upper atmosphere \citep{busa07}. Hence, the chromosphere is activated either by shocks during the infalling movement of the atmosphere or by the main shock propagating outward. In case of long-period Cepheids this emission is also observed for a significant part of the pulsation cycle ($\Delta \phi \approx0.2$). 

%In the next section, we analyze the Schwarzschild feature in Ca IR in order to estimate the size of the lower chromosphere. Then, we study the dynamics of the chromosphere by studying the velocity lag of Ca IR respected to H$\alpha$. In the last section, we degrade the UVES spectra down to the RVS resolution to compare the chromospheric motion derived from the simulated RVS observations (based on Ca IR lines) with the atmospheric radial velocity determined from the cross-correlated metallic line profiles.

\section{Chromosphere kinematics using Van Hoof effect \label{sect6:vanhoof}}
\begin{figure}[h!]
     \centering
         \includegraphics[width=0.45\textwidth]{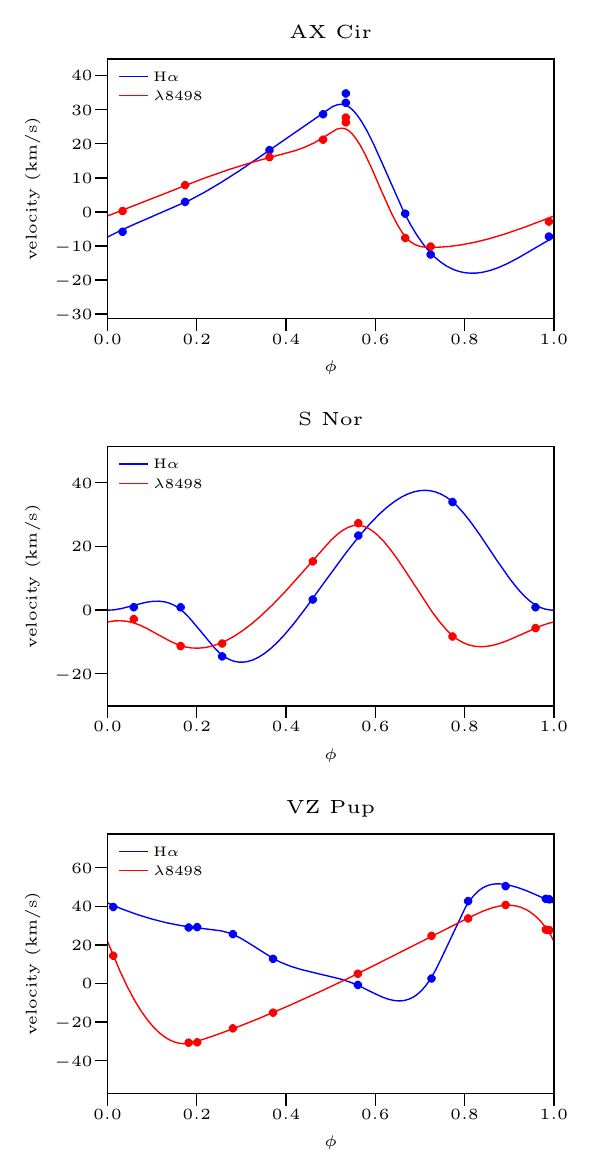}
     \caption{\small \label{fig:vanhoof} Radial velocity of line Ca $\lambda$8498 and H$\alpha$ line along the pulsation cycle $\phi$ for prototype stars. The solid curves are built from a spline fit to the RV data to guide the reader’s eye.}
\end{figure}

By studying solar chromosphere, \cite{Vernazza1973} derived a H$\alpha$ core formation height higher than Ca IR. In addition, according to \cite{Linsky1970}, the core of Ca~$\lambda$8498 is formed lower than the other calcium lines in the chromosphere. Hence, by comparing Ca~$\lambda$8498 and H$\alpha$ core formation layers from Cepheids, we expect to directly obtain information on the velocity gradient between the bottom and the top of the chromosphere. In order to study the motion of the chromospheric layers we derive heliocentric RVs curves of Ca~$\lambda$8498 and H$\alpha$ by measuring the minimum of the line. Therefore, we probe the RV associated to the core-forming region. When a double profile absorption is observed, as it is the case in VZ~Pup, we consider only the red-shifted absorption since it is produced by the atmospheric layers. We present the results in Fig.~\ref{fig:vanhoof}.

From these figures, H$\alpha$ line and Ca~$\lambda$8498 are well sychronized in the case of the short-period AX Cir, whereas the variation of H$\alpha$ is late respected to Ca~$\lambda$8498 in S~Nor and VZ~Pup. This de-synchronization is similar to the Van Hoof effect \citep{VanHoof1953} in the case of hydrogen and metallic lines. This phenomenon is interpreted as an outward propagating wave all along the pulsation cycle \citep{mathias1993}. We demonstrate the global de-synchronization of the chromospheric layers due to the propagation of waves for mid- and long-period Cepheids. 
%We also conclude that the core of Ca IR is formed at lower altitude than the core of H$\alpha$ in the chromosphere.
 Acceleration of Ca IR and H$\alpha$ are almost identical in the case of AX~Cir, which is likely a sign of a compact chromosphere due to the higher gravity field in short-period Cepheids. For both S Nor and VZ Pup we observe an in-falling movement which accelerates up to about 30km/s. Then, a sudden reverse from in-falling (> 0km/s) to outward motion (< 0km/s) is observed around 0.7 in case of S~Nor while it appears later for VZ~Pup at $\phi \approx$0.9. For long-period Cepheids we can hypothesize that reversal motion of the atmosphere occurs later due to larger ballistic motion in Cepheids of lower gravity field. 
Then, Ca IR lines reverse first their movement followed by H$\alpha$ lines (see red and blue curves in Fig.~\ref{fig:vanhoof}). The larger de-synchronization for longer period Cepheids is likely responsible of the Schwarzschild mecanism in Ca IR that we study in next Sect.~\ref{sect5:schwarz}.

\begin{figure}[h!]
     \centering
         \includegraphics[width=0.45\textwidth]{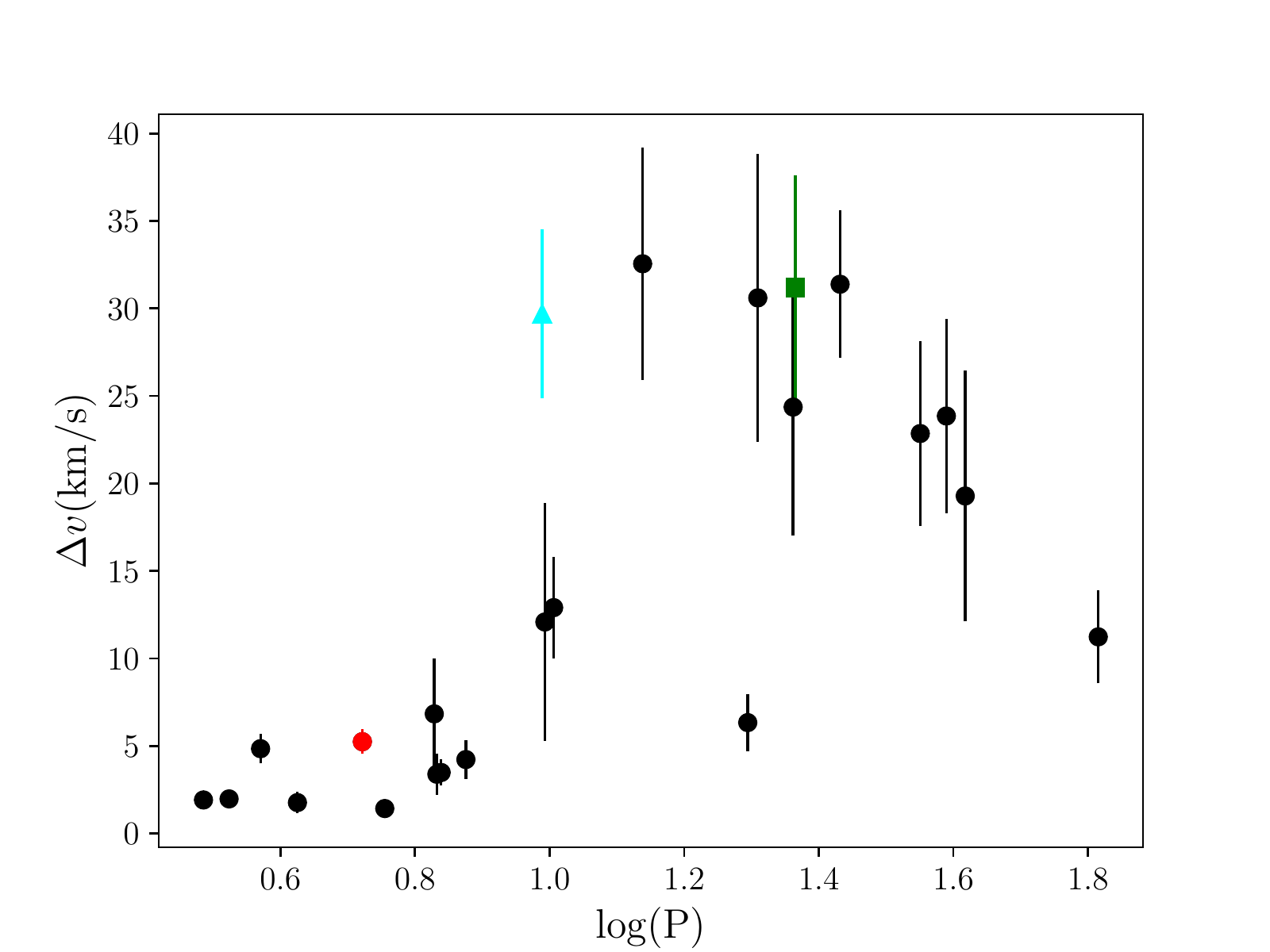}
     \caption{\small Norm of the velocity gradient averaged over the pulsation cycle estimated using Eq~(\ref{eq:p_vs_amp})  versus logarithm of the pulsation period. Red, cyan and green marks are AX Cir, S Nor and VZ Pup respectively.} %S Nor is also comparable to long-periods and RU Sct (19.70d) is comparable to short-periods.         
     \label{fig:p_vs_amp}
\end{figure}

In order to explore the link between period of Cepheids and upper atmosphere de-synchronization in the entire star sample, we derive the norm of the velocity gradient between Ca~IR and H$\alpha$ spectral lines, averaged over the pulsation cycle:

 \begin{equation} \label{eq:p_vs_amp}
    \Delta v=\frac{1}{N} \sum_{i=1}^{N}|V_{\lambda8498}(\phi_i) - V_{\mathrm{H}\alpha}(\phi_i)|
\end{equation}
where $V_{\lambda8498}$ and $V_{\mathrm{H}\alpha}$ are radial velocities of Ca IR and H$\alpha$ measured using the minimum of the line profile, and $N$ is the total number of pulsation phase. We also adopted the standard error of the mean $s$ for $N$ epoch in a pulsation cycle:

 \begin{equation}
    s=\frac{\sigma(\Delta v)}{\sqrt{N}}
\end{equation}
where $\sigma(\Delta v)$ is the standard deviation of the mean.
The result is presented in Fig.~\ref{fig:p_vs_amp}. 

Since the pulsation cycle coverage is relatively poor in the star sample, the $\Delta v$ quantity (Eq. 1) do not cover the full pulsation cycle and depends on the specific phases of observation. However, from Fig.~\ref{fig:p_vs_amp} we can clearly observe a tendency with two distinct regimes:

\begin{enumerate}
    \item For Cepheids with period $P < 10$d, the layers associated to the top of line-forming regions of Ca IR and H$\alpha$, respectively, are  synchronized (with $\Delta v$ < 5 km/s.)
    \item A sudden increase appears for P$\approx$10d with a velocity gradient of the H$\alpha$ layer compared to Ca IR rising up to 30 km/s. %This velocity seems to be maintained for periods between 10 and 30d.
    %\item \V{For Cepheids with long-period $P > 30$d, we observe a decrease of the de-synchronization down to about 10 km/s for V1496~Aql (65.36d).}
\end{enumerate}

\section{The size of the chromosphere estimated from the Schwarzschild mechanism of long period Cepheids\label{sect5:schwarz}}
We observe the Schwarzschild phenomenon for 8 out of 10 long-period Cepheids \citep{schwarz1952} between $\phi$=0.9 and 0.3. Indeed, the observed profiles follow the classical picture of the Schwarzschild scenario in which the center-of-mass velocity falls in between the blue and red peaks. Moreover, since \cite{Linsky1970} found that Ca $\lambda$8498 line is formed lower in the chromosphere than Ca $\lambda$8542 and $\lambda$8662 spectral lines, we can infer that Ca $\lambda$8498 line is colliding first onto the main shock, which is confirmed by the fact that Ca $\lambda$8498 line is always in a more advanced stage in the Schwarzschild scenario than Ca $\lambda$8542 and $\lambda$8662 lines, respectively (see for example Fig.~\ref{fig:vz_pup}). Although we cannot firmly conclude because of possible cycle-to-cycle variation effects \citep{anderson2016}, the Schwarzschild mechanism could be a common feature in long-period Cepheids.
 
\begin{figure}[h!]
     \centering
         \includegraphics[width=0.45\textwidth]{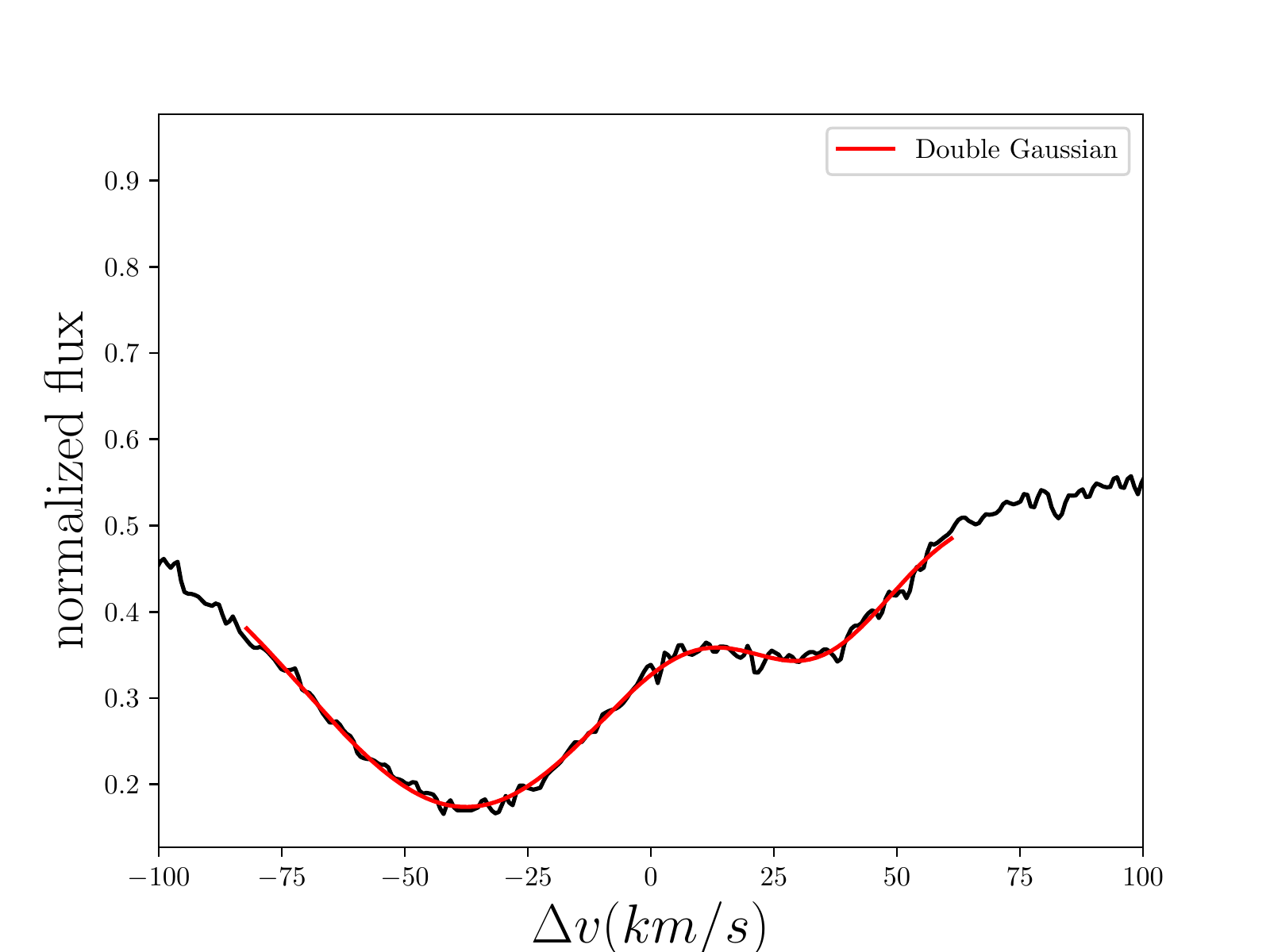}
     \caption{\small Example of bi-Gaussian fit in the case of VZ Pup for line $\lambda$8542 at the pulsation phase $\phi$=0.08. The parameters associated to the bi-Gaussian fits are listed for all stars in Table~\ref{Tab.shock}, together with the derived shock amplitude.}\label{fig:double_gauss}
\end{figure}

The Schwarzschild mechanism itself occurs during the expansion of the Cepheids photosphere around $\phi$=0.9. It is explained by a collision between the in-falling layers of the lower chromosphere on denser region of the ascending photosphere. For each long-period Cepheids presenting double line features we derive RVs from both blue-shifted $V_\mathrm{blue}$ and red-shifted $V_\mathrm{red}$ components using a bi-Gaussian fitting. We provide an example of this fitting in Fig.~\ref{fig:double_gauss}. 
 %We also derive the Gaussian widths $\sigma_\mathrm{blue}$ and $\sigma_\mathrm{red}$ corrected from instrument resolution. 
 From this fitting we obtain both mean and standard deviation values of blue and red absorption components. We also estimate the shock amplitude, by simply assuming $A_\mathrm{shock}=V_\mathrm{red}-V_\mathrm{blue}$ and constant velocities during the in-falling motion. Results are presented in Table~\ref{Tab.shock}.

%\V{Should I let standard deviations ? because I do not discuss these data and I have no interest in it, but may be its interesting for dicussing turbulences ?}} 
% SIMON: Turbulence may be beyond the scope of this paper ? Anyway, this article on turbulence might interest you:  http://arxiv.org/abs/1707.00738

\begin{table}[]
\caption{\label{Tab.rad_chromo} Chromospheric radius $R_\mathrm{chromo}$ estimated from shock amplitude listed in Table~\ref{Tab.shock} in the case of the calcium line $\lambda8542$. $R_\mathrm{chromo}$ is given relatively to the mean stellar radius provided by PR relation from
\cite{Gallenne2017}$^1$. Calculation assumes $\Delta \phi =0.250$ for Schwarzschild mechanism duration.}
\begin{center}
\begin{tabular}{c|c|c|c}
\hline
\hline
Star	&	$A_\mathrm{shock}$(km/s) &$R$(\Rsolar) 		&	$R_\mathrm{chromo}$(\%$R$) \\[0.15cm]
\hline	
\hline							
TT Aql	&	82.9$\pm1.1$    &   82.0$\pm1.0$			&	43.2$\pm1.1$	\\
		
RZ Vel	&	80.1$\pm1.3$     &   107.4$\pm1.6$			&	47.2$\pm1.5$	\\
				
WZ Car	    &   83.2$\pm0.4$ &	116.7$\pm1.8$		& 50.9$\pm1.0$		\\
					
VZ Pup	&   69.3$\pm0.6$     &	117.2$\pm1.8$			&	42.5$\pm1.0$	\\
	
T Mon	&	82.4$\pm0.8$     &   130.2$\pm2.2$		&		53.1$\pm1.4$	\\
				
U Car	&   59.1$\pm0.4$    &	167.0$\pm3.2$			&	42.7$\pm1.1$	\\
					
RS Pup	&   81.3$\pm1.7$    &	174.5$\pm3.5$			&	60.0$\pm2.5$	\\
						
V1496 Aql	&	56.0$\pm1.3$ &238.3$\pm5.5$		& 47.7$\pm2.2$	\\
\hline							
\hline
\end{tabular}
\begin{tablenotes}
(1): Mean radius from PR relation:\par $\mathrm{log}\, R=(0.684\pm0.007)(\,\mathrm{log}\,P-0.517)+(1.489\pm0.002)$
\end{tablenotes}
\normalsize
\end{center}
\end{table}
 We obtain shock amplitudes between 50 and 80 km/s. Although these profiles belong to different pulsation phases we observe no trend between the shock amplitude and the period of the Cepheids.
 %from Table~\ref{Tab.shock}. 
 In the case of VZ~Pup the Schwarzschild mechanism is initiated at the end of the pulsation cycle ($\phi=0.91$) for Ca $\lambda$8542 and $\lambda$8662. The double line feature disappears after $\phi$=0.179. In that case, it took at least a quarter cycle ($\Delta \phi \approx$ 0.250) for Ca $\lambda$8542 and $\lambda$8662 layers  to collapse entirely on the photospheric denser region, resulting in a single blue-shifted component. If we assume a constant shock amplitude during the collapsing of the chromosphere, we can approximate the maximum height of Ca $\lambda$8542 and $\lambda$8662 core formation regions. Indeed this is simply the atmosphere distance covered in a quarter period by the relative velocity of the collapsing layers $R_\mathrm{chromo}=P\Delta \phi A_\mathrm{shock}$. We thus estimate this distance assuming $\Delta \phi=0.250$ and using the shock amplitude from Ca $\lambda$8542 derived in Table~\ref{Tab.shock}. In order to derive the chromospheric radius expressed by the mean stellar radius, we used the Period-Radius (PR) relation from \cite{Gallenne2017}.
Applying the same methodology for the long-period Cepheids in the sample, we find an extension for the 
%lower 
chromosphere of Cepheids from about 40 to 50\% of the mean stellar radius.  The results are presented in Table \ref{Tab.rad_chromo}. This order of magnitude is in agreement with \citep{wallerstein1972} who estimated a Ca II fallen distance of 50\% the radius of the star in the case of T~Mon.

\section{Discussion \label{sect:discuss}}

%A plausible source of uncertainty could be due to InfraRed (IR) excess emitted by CircumStellar Envelope (CSE) such the ones discovered using long-baseline interferometry \citep{kervella06a,merand06} in the K-band. However the origin and the nature of these CSEs is still debated. In particular, while the CSE emission is explained by dust emission in some cases (\cite{Gallenne2012},\cite{gallenne13b}), it fails to reproduce the IR excess in other studies (\cite{Schmidt2015}, Hocdé et al. 2019). 

%\subsection{Multiple origin of chromospheric activation}

 %and analysis of Ca IR below TiO absorption band
 %Ca IR is collisionnally excited \cite{vernazza76} demonstrated that the IRT is formed in the lower chromosphere

 %On the other hand, we demonstrated in Sect.~\ref{sect6:blue_emission} that Ca IR blue-shifted emission originates from an inverse P Cygni profile blended with photospheric absorption. This is effect is caused by the ballistic movement of the atmosphere which is accelerated during the in-falling motion, just before maximum light.
 
 %Also the strong mass transfer in long-period Cepheids provided by mean of Schwarzschild mechanism suggests a massive ionization of the atmosphere during a third of the period.

 \subsection{Pulsating chromosphere as the origin of an ionized shell}
 
A central question raised in the introduction is whether or not the pulsating chromosphere has the potential to ionize the gas in the environment to produce IR excess such as the one that would be caused by a shell of ionized gas, which represents  15\% of the photosphere radius \citep{hocde2020}. In this paper, we studied the lower chromosphere of Cepheids for which we estimated a size to be at least 50\% of the star radius.
%, thus the ionized gas shell is constrained to the low chromosphere region. 

Several processes are able to provide ionized material in the lower chromosphere. We have analyzed three episodes which have the potential to heat up and massively ionize the upper atmosphere: (1) the main shock traveling through the upper atmosphere whose front amplitude can reach 100km/s (2) a shock caused by supersonic descending layers of the atmosphere revealed under the form of an inverse P~Cygni profile, and (3) a Schwarzschild mechanism caused by the main shock which is reversing the movement of the descending layers of the atmosphere. 

 However, according to shock theory \citep{fokin2000,fokin2004}, shocks alone seem unlikely to provide a permanent free-free continuum which could explain a constant IR excess. These authors have shown that cooling by FeI and FeII lines in the post-shock region is more effective than cooling by free-free and bound-free continuum. Moreover these authors have also estimated a size of the post-shock region of about 100 km, and, given that shock velocities are 10 to 100 km/s, we deduce a cooling time of the order of tens of seconds which is short compared to a pulsation period. 
 %Nevertheless, the complexity of all the mechanisms to take into account, and the continuous presence of a shocked atmosphere during a pulsation cycle, lead us to think that there is no clear consensus about the origin of a free-free continuum from an ionized shell gas.
 
 %On the other hand, 
 
 Besides, the main shock traversing the photosphere rules a lot of processes in the behaviour of Cepheids. Indeed, it is already hypersonic when it emerges from the surface of the star, with a shock velocity of about 100 km/s or more, that is a Mach number between 10 and 20. In this case the convection zone could be strongly disrupted or inhibited at least for a fraction of the pulsation cycle.
%a while. 
We propose, therefore, that this shock could be strong enough to bring hot and dense material in the upper atmosphere. Interestingly, \cite{kraft57} also suggested that the transitory development from Ca II K emission in classical Cepheids is associated with the presence of hot material low in the atmosphere. He further discusses that this material could have been convected upward from hydrogen convection zone below the surface.

%\textbf{As an alternative to an IR excess from chromospheric origin, we have found}
 
\subsection{A plausible source of X-ray emission in the upper chromosphere}
Another interesting question is to understand the origin of periodic X-ray emissions around phase $\phi$=0.45 at maximum radius that have been observed by \cite{engle2017}. Such X-ray emissions could stem from a high temperature plasma in the higher chromosphere. Indeed at this phase, the shock is well detached from the photosphere. When the P-Cygni appears in H$\alpha$ line at phase 0.0, the shock has already travelled $\approx$15\% of the photospheric radius with a speed of 100 km/s or more. Then, the shock decelerates below 100 km/s around $\phi$=0.5, thus the shock is no longer able to produce hydrogen emission.  At this time, the shock traveled a distance close to the photospheric radius and is now propagating in the upper atmosphere where the density is 3 or 4 orders of magnitude weaker ($10 ^{- 11}$ - $10^{- 12}$ g / cm$^3$) than that in the photosphere ($10^{- 8} $g / cm$^3$). In this case, according to \cite{Fadeyev2004}, with such pre-shock density in the upper atmosphere, the thickness of the radiative wake increases by almost 3 orders of magnitude while the temperature just after the shock front changes relatively little. Part of the thickness of the radiative shell will be a few or at least thousands of kilometers. The plasma behind the shock is also strongly compressed, proportional to the square of the Mach number (given by Hugoniot-Rankine equation in isothermal condition), and so does the amplification rate of the magnetic field, preexisting or generated by the plasma of the radiative wake, which may easily reach a factor of 100. Therefore, in the upper chromosphere, part of the energy dissipation of the shock may be in non-collisional form. As a conclusion, we suggest that the energy producing X-ray flux between $\phi \approx $ 0.4-0.5 could be deposited by the main shock in the outermost layers of the chromosphere.

\subsection{Is the Gaia RVS sensitive to the chromosphere dynamics ?}
The {\it Gaia}~RVS has a narrow wavelength range centered on the Ca IR triplet \citep[from $\sim$845 to $\sim$872~nm, see][]{Sartoretti2018}. Thus, we can expect the RV time series produced by the RVS to be dependent on the Ca IR behavior. From the chromosphere kinematics using Van Hoof effect in Sect.~\ref{sect6:vanhoof}, we have shown that H$\alpha$ lines are de-synchronized with an important velocity gradient compared to Ca IR for Cepheids with pulsation period larger than about 10d. We therefore suggest that the same tendency as seen in Fig.~\ref{fig:p_vs_amp} could appear between metallic lines (which is formed at a great optical depth) and Ca IR when observing with the RVS instrument. This de-synchronization between metallic and Ca IR lines was already observed by \cite{wallerstein2015,wallerstein2019} for Cepheids $\delta$ Cep and X~Cyg. Further investigations are necessary to quantify these differences and calibrate, if necessary, the projection factor to be used together with the RVS measurements of Cepheids radial velocities, in the Baade-Wesselink method of distance determination \citep{nardetto17}.

\section{Conclusion }\label{sect:conclusion}
We analyzed H$\alpha$ and Ca IR profile variations to qualitatively describe the chromosphere dynamics over the pulsation period based on a sample of 24 stars with a good period coverage from short- up to long-period Cepheids. We demonstrate that Ca IR are interesting lines for probing dynamics in the lower chromosphere of Cepheids, complementary to H$\alpha$ lines, because their core probe different heights in the atmospheric layers, and it is an unambiguous indicator of chromospheric activity. Our analysis leads to the following conclusions:
\begin{enumerate}

\item We confirm the H$\alpha$ profile variation model made by \cite{gillet14} in the particular case of X~Cygni for a wider sample of long-period Cepheids.
%, although the poor phase coverage of the star sample limits to enter in a more quantitative study.

\item As found by \citet{nardetto08b}, in the case of $\ell$ Car, we identified a H$\alpha$ absorption features centered on 0 km/s in at least 8 long-period Cepheids (TT Aql, RZ Vel, WZ Car, VZ Pup, T Mon, $\ell$ Car, U Car, RS Pup) that we interpret as a static CSEs which obscure the chromosphere and the star. Since the calcium abundance is lower than hydrogen, we expect that the CSE absorption is weaker in Ca IR compared to H$\alpha$. However we do not exclude a possible blend of the calcium emissions if the CSE is denser in some cases. 

\item We demonstrated that at least two mechanisms are pulsationally activating the chromosphere owing to Ca IR diagnostic: (1) An inverse P Cygni profile emission during supersonic in-falling motion and (2) a P Cygni profile emission due to the radiative wake behind the main shock when it is sufficiently detached from the photosphere. In addition, a blended emission during Schwarzschild mechanism is also not excluded due to large amount of kinetic energy to be released when the main shock and the in-falling atmosphere are colliding.

\item The difference in height formation of Ca IR and H$\alpha$ creates a de-synchronization of their associated chromospheric layers due to an outward propagation wave. We observe a weak phase delay and velocity gradient ($\Delta v$ < 5 km/s) in small period Cepheids ($P < 10$ days) while H$\alpha$ lines are delayed  with an important velocity gradient (up to $\Delta v \approx$30 km/s) compared to Ca IR in mid- and long-period Cepheids.

\item This de-synchronization in long-period Cepheids is likely responsible for the Schwarzschild mechanism of Ca IR lines during the beginning of photosphere ascending phase from $\phi$=0.9 to 0.3. This phenomenon is possibly a common feature since we observed this feature among 8 out of 10 long-periods. The collision velocity between in-falling layers of the lower part of the atmosphere and the photospheric region is about 50 up to 80km/s. Since the in-falling motion lasts for at least a quarter cycle we derived the height of Ca IR core formation to be $\approx$50\% of the star radius in long-period Cepheids, which is consistent at the first order with the size of the shell of ionized gas (15\% of the stellar radius) found by \cite{hocde2020}.

%\item Following the {\it Gaia} RVS pipeline procedure from \citet{Sartoretti2018}, we find that the radial velocity curves that will be derived from the analysis of Calcium triplet with the RVS should be different (in amplitude, shape and cycle-averaged velocity) than the cross-correlated radial velocity curves based on the metallic lines that are currently used for the distance determination of Cepheids. If such preliminary results are confirmed with actual RVS {\it Gaia} data it means that dedicated projection factors, used to convert the radial velocity into pulsation velocity \citep{nardetto04, nardetto09, breitfelder16,nardetto17}, will be necessary. Besided, the RVS velocities will be an excellent indicator of the dynamical structure of the dynamical structure and the chromosphere of Cepheids, bringing exquisite constraints on the projection factor and the environment of Cepheids, for a large amount of stars.

%We have shown various and complicated dynamics in Ca~IR profiles over a wide sample of stars. We have demonstrated the impact of such profiles on the radial velocity measurement. Calcium lines lag behind metallic lines around $\phi=0.8$. This effect seems to be stronger for stronger period. This overview shows the importance of understanding Ca~IR profiles for analyzing \textit{Gaia} RVs of Cepheids. 
%\item This study was limited to a qualitative approach and observations with a better phase coverage is required to study more quantitatively the upper chromosphere dynamics of Cepheids.
\end{enumerate}

\bibliographystyle{aa}  % A&A bibliography style file (aa.bst)
%\bibliography{main_draft} % your references in file: Yourfile.bib

\begin{acknowledgements}
%The research leading to these results has received funding from the European Research Council (ERC) under the European Union’s Horizon 2020 research and innovation programme under grant agreement No 695099 (project CepBin). 
The authors acknowledge the support of the French Agence Nationale de la Recherche (ANR), under grant ANR-15-CE31-0012- 01 (project UnlockCepheids). %W.G. and G.P. gratefully acknowledge financial support for this work from the BASAL Centro de Astrofisica y Tecnologias Afines (CATA) AFB-170002.
We acknowledge financial support from ``Programme National de Physique Stellaire'' (PNPS) of CNRS/INSU, France. This project was partially supported by the Polish Ministry of Science grant Ideas Plus. This research made use of the SIMBAD and VIZIER\footnote{Available at \url{http://cdsweb.u- strasbg.fr/}} databases at CDS, Strasbourg (France) and the electronic bibliography maintained by the NASA/ADS system. This research also made use of Astropy, a  community-developed core Python package for Astronomy \citep{astropy2018}.
This research has made use of the SIMBAD database, operated at CDS, Strasbourg, France.  Based on observations made with ESO telescopes at Paranal La  Silla  observatories  under  program  IDs: 098.D-0379(A), 0100.D-0397(A) and 0101.D-0551(A).
\end{acknowledgements}
\begin{appendix}

\section{Results of bi-Gaussian fitting for Schwarzschild mechanism}
\begin{table*}[h!]
\caption{\label{Tab.shock} \small RVs fields in Schwarzschild mechanism in long-period Cepheids. Velocities were calculated only when the characteristic W-shape was clearly identified by-eye. In the thid column of the table, lines 1,2 and 3 refer to Ca $\lambda$8498, 8542 and 8662, respectively. $V_\mathrm{blue}$, $V_\mathrm{red}$ are the blue and the red component velocities respectively of the W shape. $A_\mathrm{shock}$ is the shock amplitude derived by $V_\mathrm{blue}$ -   $V_\mathrm{red}$. The FWHM of blue ($\sigma_{\mathrm{blue}}$) and red ($\sigma_{\mathrm{red}}$) components are also listed. All values are given in (km/s). Stars are ranged by increasing periods from top to bottom.}
\begin{center}
\begin{tabular}{c|c|c|c|c|c|c|c}
\hline
\hline
Star	&	Phase &line	&	$V_\mathrm{blue}$	&	$V_\mathrm{red}$	&	$A_\mathrm{shock}$&$\sigma_{\mathrm{blue}}$	&$\sigma_{\mathrm{red}}$\\%[0.15cm]
\hline	
TT Aql	&	0.947&2	&	-36.7$\pm1.0$	&	+46.2$\pm0.5$	&	82.9$\pm1.1$&42.8$\pm0.9$&31.6$\pm0.7$	\\%[0.15cm]
	    &	&3	&	-29.7$\pm0.7$	&	+47.4$\pm0.6$	&	77.1$\pm0.9$&42.5$\pm0.7$&29.2$\pm0.6$\\%[0.15cm]
\cline{2-8}
&	0.116&2	&	-35.5$\pm0.1$	&	+41.7$\pm0.7$	&77.2$\pm0.7$&41.4$\pm0.7$&15.1$\pm0.8$	\\%[0.15cm]
	    &	&3	&	-31.6$\pm0.1$	&	+44.7$\pm1.2$	&76.3$\pm1.2$&36.1$\pm0.9$&14.2$\pm1.4$	\\%[0.15cm]
\hline	
\hline
RZ Vel	&	0.955 &2	&-43.3$\pm1.2$		&	+36.8$\pm0.6$	&80.1$\pm1.3$&35.0$\pm0.8$&18.5$\pm0.6$		\\%[0.15cm]
&		&3	&	-40.0$\pm0.1$	&	+39.1$\pm0.4$	&	79.1$\pm0.4$&33.6$\pm0.2$&20.4$\pm0.4$	\\%[0.15cm]
\hline	
\hline
	&	&1		&-43.0$\pm0.1$	&	+39.3$\pm0.7$	&	82.3$\pm0.7$&36.3$\pm0.1$&21.7$\pm0.6$	\\%[0.15cm]
WZ Car	&	0.080 &2&	-48.2$\pm0.2$	&	+35.0$\pm0.4$	&	83.2$\pm0.4$&34.3$\pm0.8$&31.6$\pm0.4$	\\%[0.15cm]
&		&3	&	-46.2$\pm0.1$	&	+34.5$\pm0.2$	&	80.7$\pm0.2$&29.6$\pm0.4$&19.8$\pm0.3$	\\%[0.15cm]
\hline
\hline
VZ Pup	&	0.080&2	&	-37.3$\pm0.1$	&	+32.0$\pm0.6$	&	69.3$\pm0.6$&29.8$\pm0.7$&14.0$\pm0.8$	\\%[0.15cm]
	&		&3&	-34.2$\pm0.1$	&	+33.6$\pm1.1$	&	67.8$\pm1.1$&28.5$\pm0.3$&14.2$\pm0.5$	\\%[0.15cm]
\cline{2-8}
	&	0.179&2	&	-28.9$\pm0.1$	&	+32.3$\pm0.4$	&	61.2$\pm0.4$&29.4$\pm0.3$&10.5$\pm0.5$	\\%[0.15cm]
	&		&3&	-26.6$\pm0.1$	&	+37.2$\pm0.4$	&	63.8$\pm0.4$&28.0$\pm0.2$&11.0$\pm0.5$	\\%[0.15cm]
	\hline
	\hline
	&		&1&	-41.6$\pm0.1$	&	+28.2$\pm0.7$	&	69.8$\pm0.8$&35.7$\pm1.1$&16.0$\pm0.5$	\\%[0.15cm]
T Mon	&	0.123&2	&	-49.4$\pm0.2$	&	+33.0$\pm0.8$	&	82.4$\pm0.8$&47.5$\pm2.0$&22.5$\pm0.6$	\\%[0.15cm]%std 47.07 et 22.14[0.15cm]
	&		&3&	-45.8$\pm0.1$ & +28.6$\pm0.1$	&	74.4$\pm0.1$&36.0$\pm0.3$&17.4$\pm0.1$	\\%[0.15cm]
\hline	
\hline
U Car	&	0.197	&2&	-30.9$\pm0.1$	&	+28.2$\pm0.1$	&	59.1$\pm0.4$&30.1$\pm0.5$&17.3$\pm0.5$	\\%[0.15cm]
	&		&3&	-26.8$\pm0.1$	&	+41.5$\pm2.9$	&	68.3$\pm2.9$&20.3$\pm0.3$&6.5$\pm1.6$\\%[0.15cm]
\cline{2-8}
&	0.225	&2&	-27.1$\pm0.2$&	+33.6$\pm1.6$	&	60.7$\pm1.6$&37.0$\pm1.2$&16.5$\pm0.5$	\\%[0.15cm]
	&		&3&	-24.5$\pm0.2$	&+40.6$\pm0.1$	&	65.1$\pm0.2$&37.8$\pm2.3$&18.0$\pm0.9$	\\%[0.15cm]
\hline
\hline
	&	&1	&	-33.6$\pm0.1$	&	+30.4$\pm0.7$	&	64.0$\pm0.7$&33.1$\pm1.3$&14.8$\pm0.5$		\\%[0.15cm]
RS Pup	&	0.280&2	& -41.1$\pm1.0$	&	+40.2$\pm1.4$	&	81.3$\pm1.7$&54.1$\pm3.6$&26.4$\pm1.8$\\%[0.15cm]
	&		&3&	-36.1$\pm0.9$	&	+35.9$\pm0.7$	&	72.0$\pm1.1$&44.0$\pm1.5$&20.2$\pm0.4$\\%[0.15cm]
\hline	
\hline

	&	&1	&	-20.7$\pm0.3$	&	+28.5$\pm0.3$	&	49.2$\pm0.4$&17.2$\pm0.4$&21.6$\pm0.3$	\\%[0.15cm]
V1496 Aql	&	0.625&2	&	-38.5$\pm1.2$	&	+32.8$\pm0.4$	&	71.3$\pm1.3$&28.1$\pm0.6$&32.5$\pm1.2$	\\%[0.15cm]
	&		&3&	-27.2$\pm0.7$	&	+33.4$\pm0.4$	&	60.6$\pm0.8$&26.2$\pm0.8$&24.9$\pm0.6$	\\%[0.15cm]
\hline
\hline
\end{tabular}
\normalsize
\end{center}
\end{table*}

\section{Calcium triplet of Cepheids with UVES \label{Fig.profiles}}
\begin{figure*}
     \centering
          \begin{subfigure}[b]{0.24\textwidth}
         \centering
         \includegraphics[width=\textwidth]{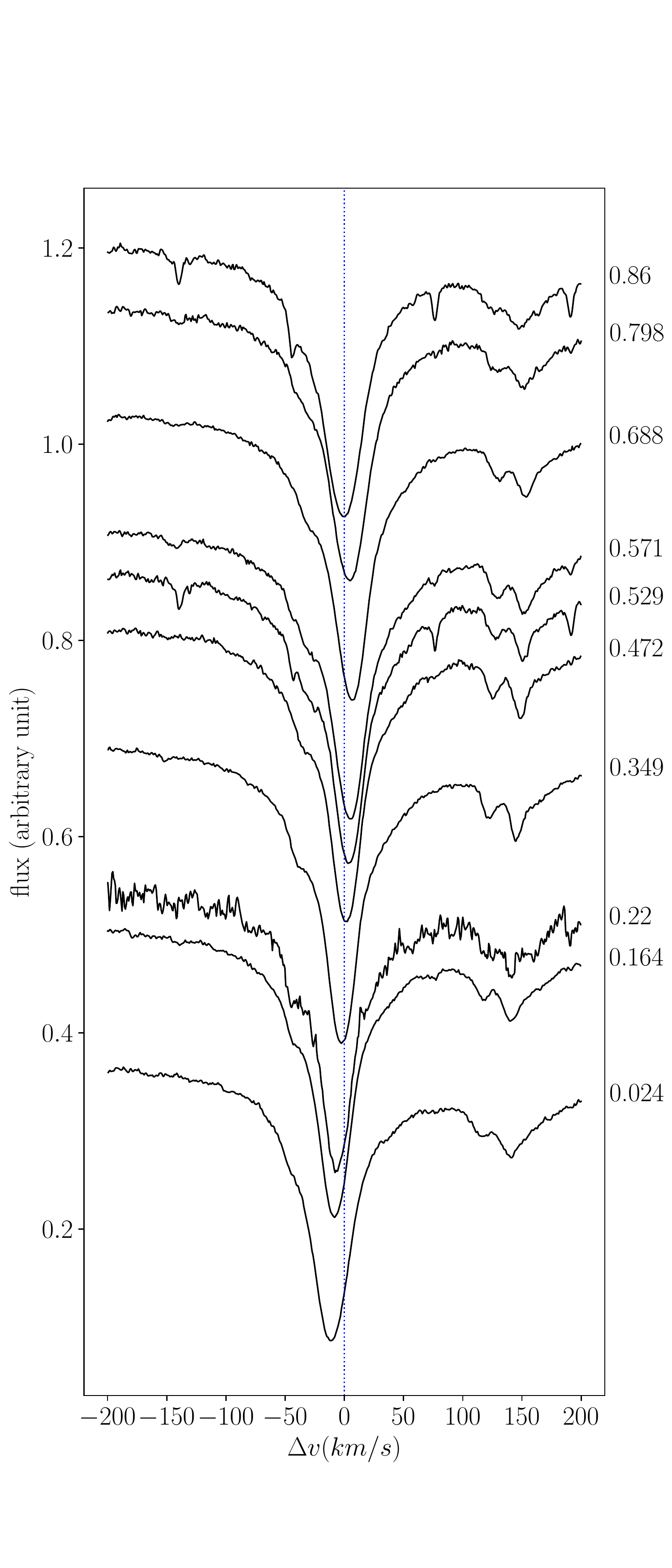}
         \caption{$\lambda$8498}

     \end{subfigure}
     \hfill
     \begin{subfigure}[b]{0.24\textwidth}
         \centering
         \includegraphics[width=\textwidth]{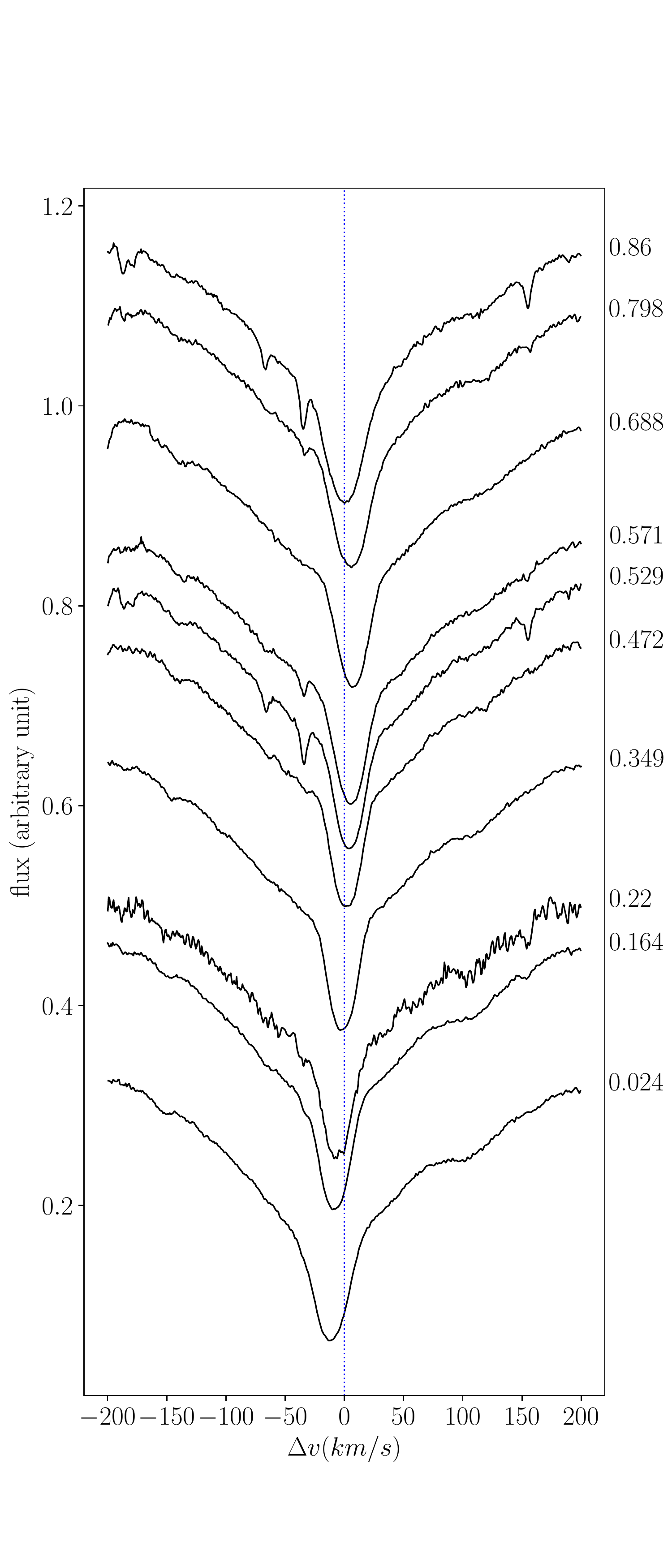}
         \caption{$\lambda$8542}
     \end{subfigure}
     \hfill
     \begin{subfigure}[b]{0.24\textwidth}
         \centering
         \includegraphics[width=\textwidth]{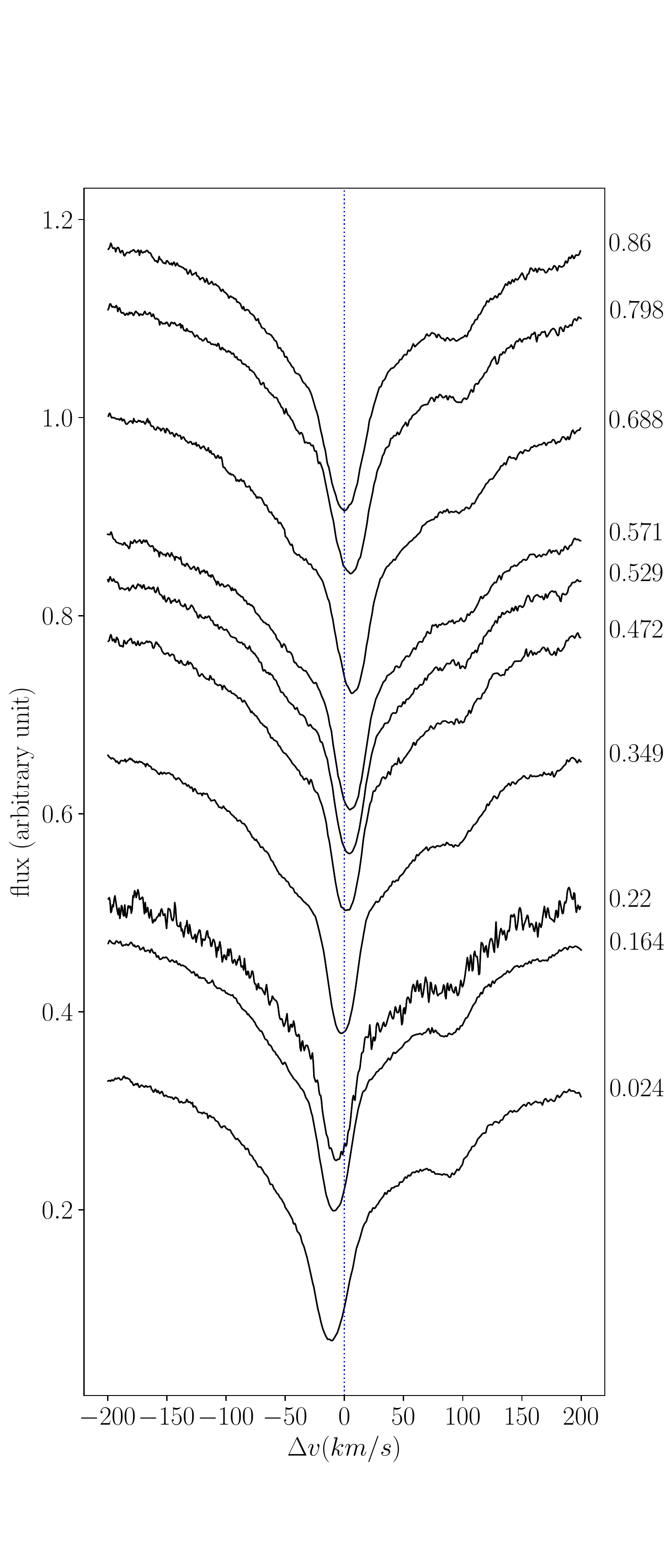}
         \caption{$\lambda$8662}
     \end{subfigure}
     \hfill
     \begin{subfigure}[b]{0.24\textwidth}
         \centering
         \includegraphics[width=\textwidth]{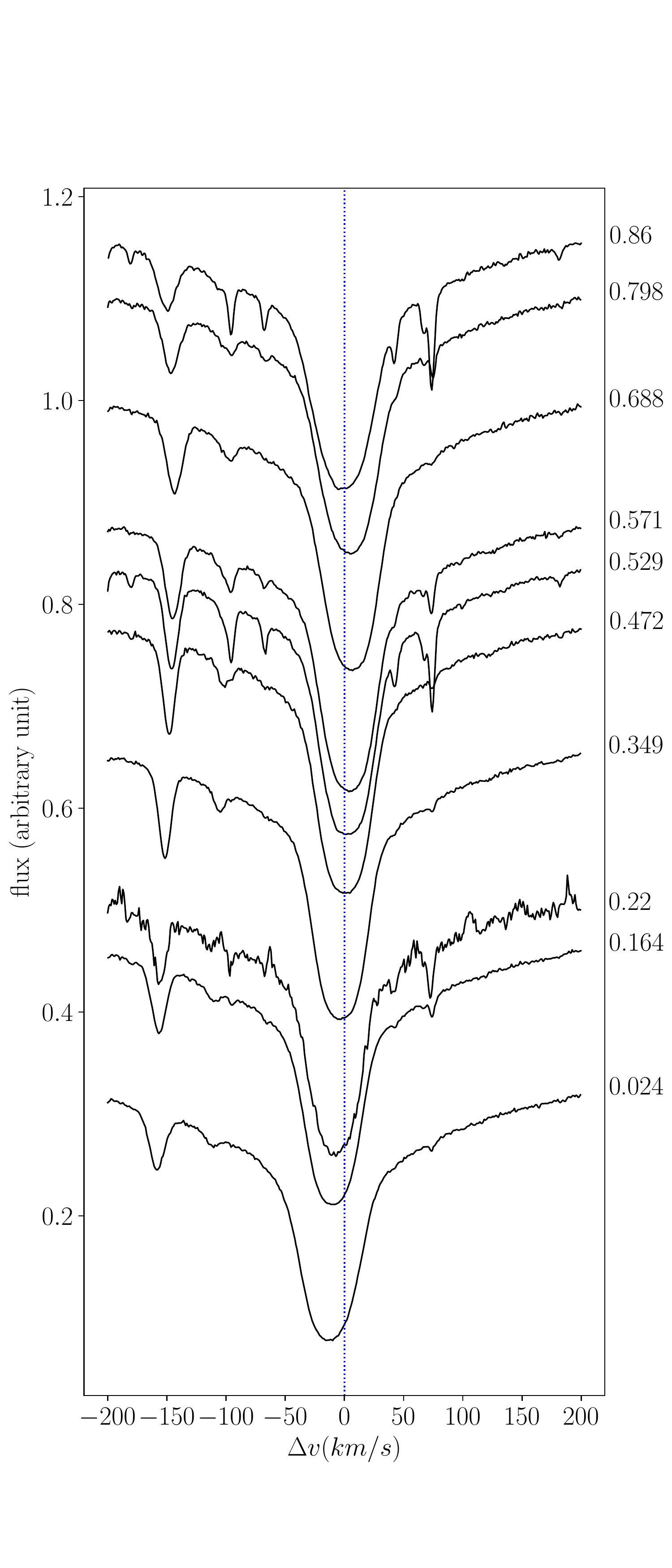}
         \caption{H$\alpha$}
     \end{subfigure}
        \caption{AV Cir, 3.06d}

\end{figure*}

\begin{figure*}
     \centering
     \begin{subfigure}[b]{0.24\textwidth}
         \centering
         \includegraphics[width=\textwidth]{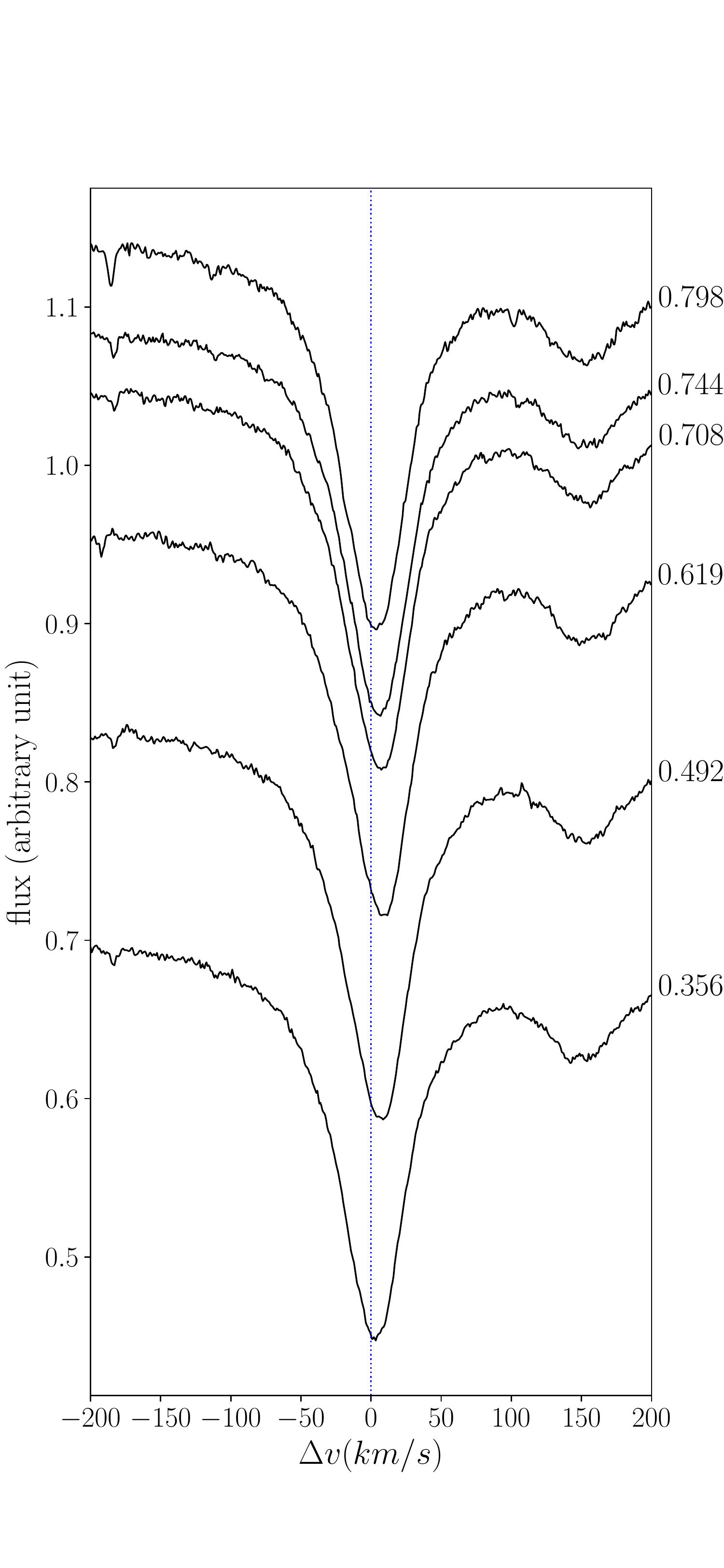}
         \caption{$\lambda$8498}
     \end{subfigure}
     \begin{subfigure}[b]{0.24\textwidth}
         \centering
         \includegraphics[width=\textwidth]{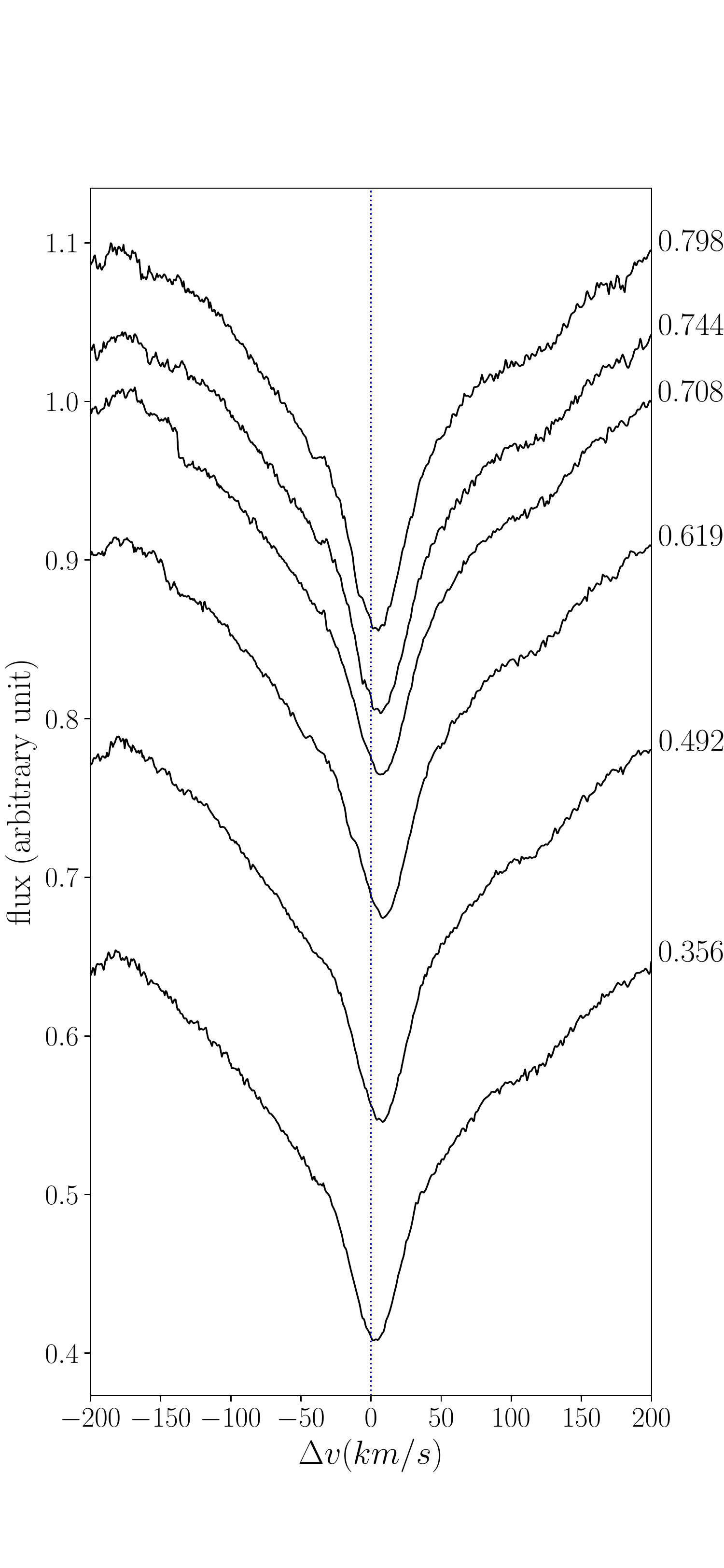}
         \caption{$\lambda$8542}
     \end{subfigure}
     \hfill
     \begin{subfigure}[b]{0.24\textwidth}
         \centering
         \includegraphics[width=\textwidth]{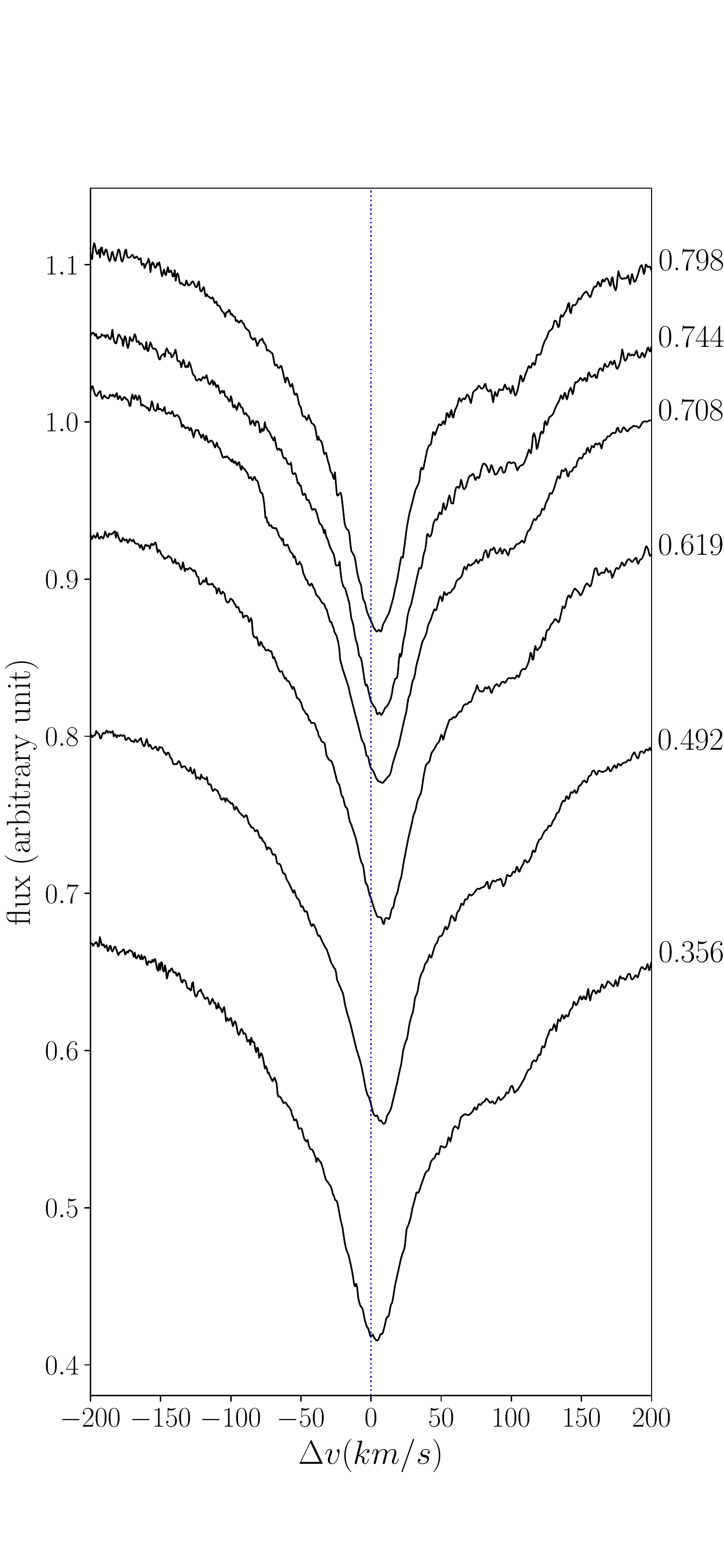}
         \caption{$\lambda$8662}

     \end{subfigure}
     \hfill
     \begin{subfigure}[b]{0.24\textwidth}
         \centering
         \includegraphics[width=\textwidth]{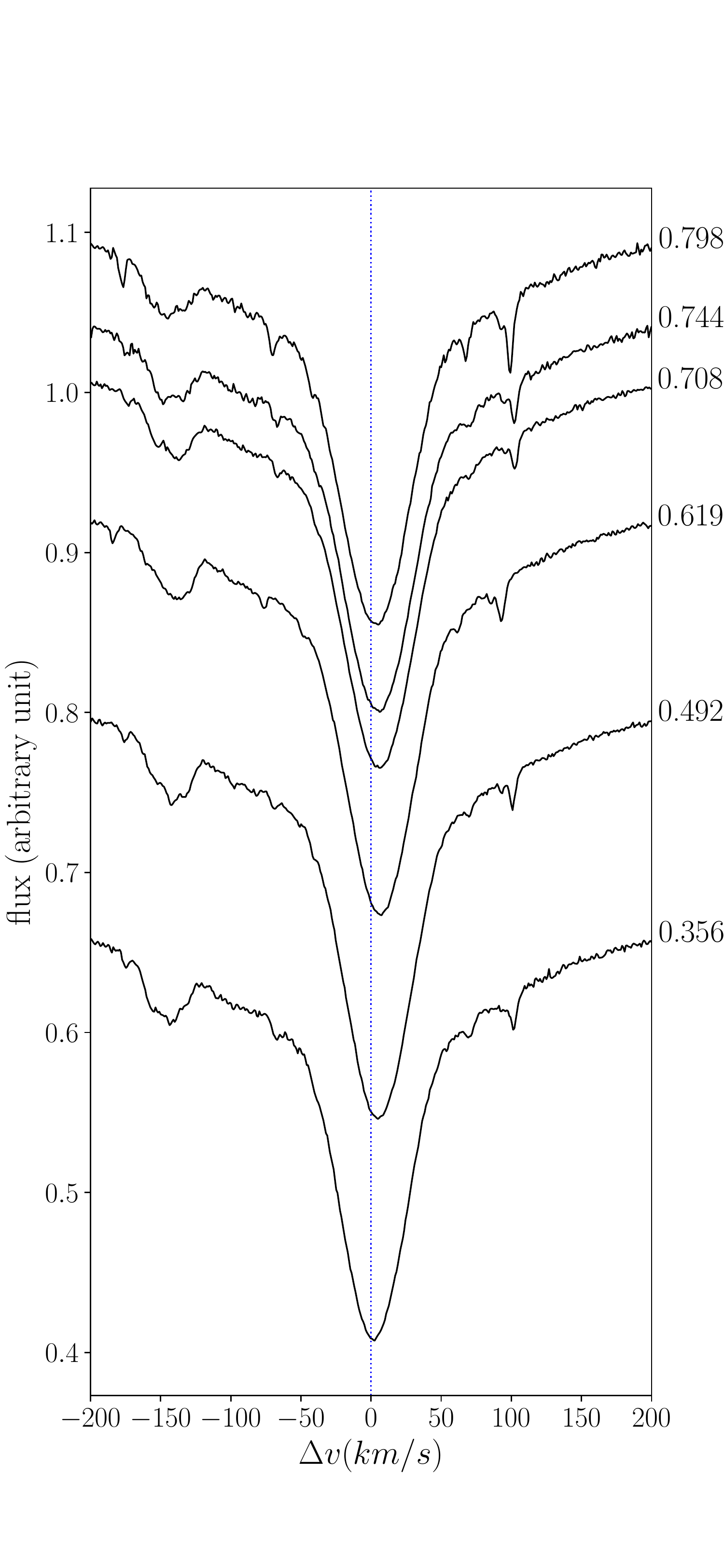}
         \caption{H$\alpha$}
     \end{subfigure}
        \caption{BG Cru, 3.34d}
\end{figure*}

\begin{figure*}
     \centering
         \begin{subfigure}[b]{0.24\textwidth}
         \centering
         \includegraphics[width=\textwidth]{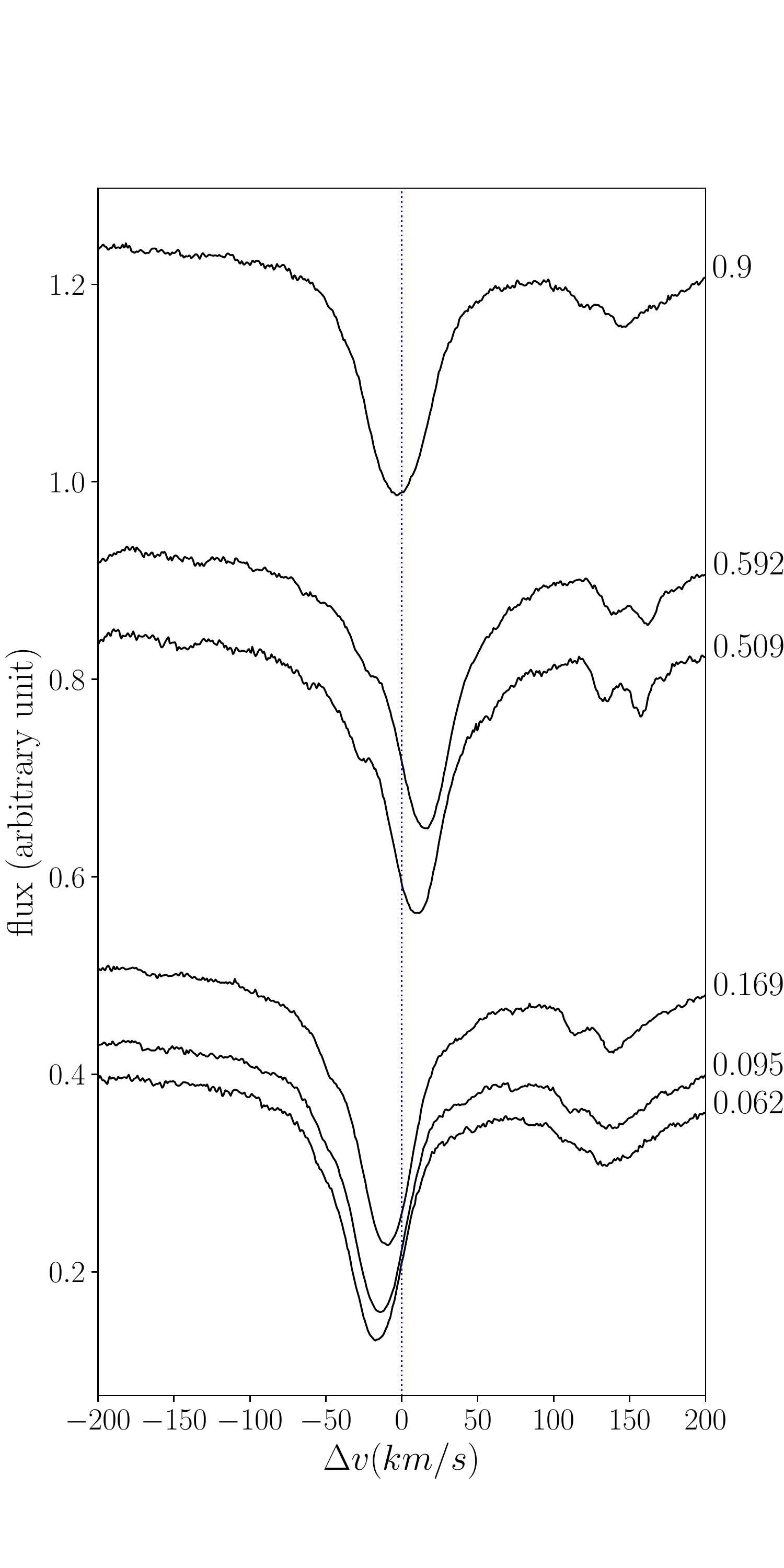}
         \caption{$\lambda$8498}
     \end{subfigure}
     \hfill
     \begin{subfigure}[b]{0.24\textwidth}
         \centering
         \includegraphics[width=\textwidth]{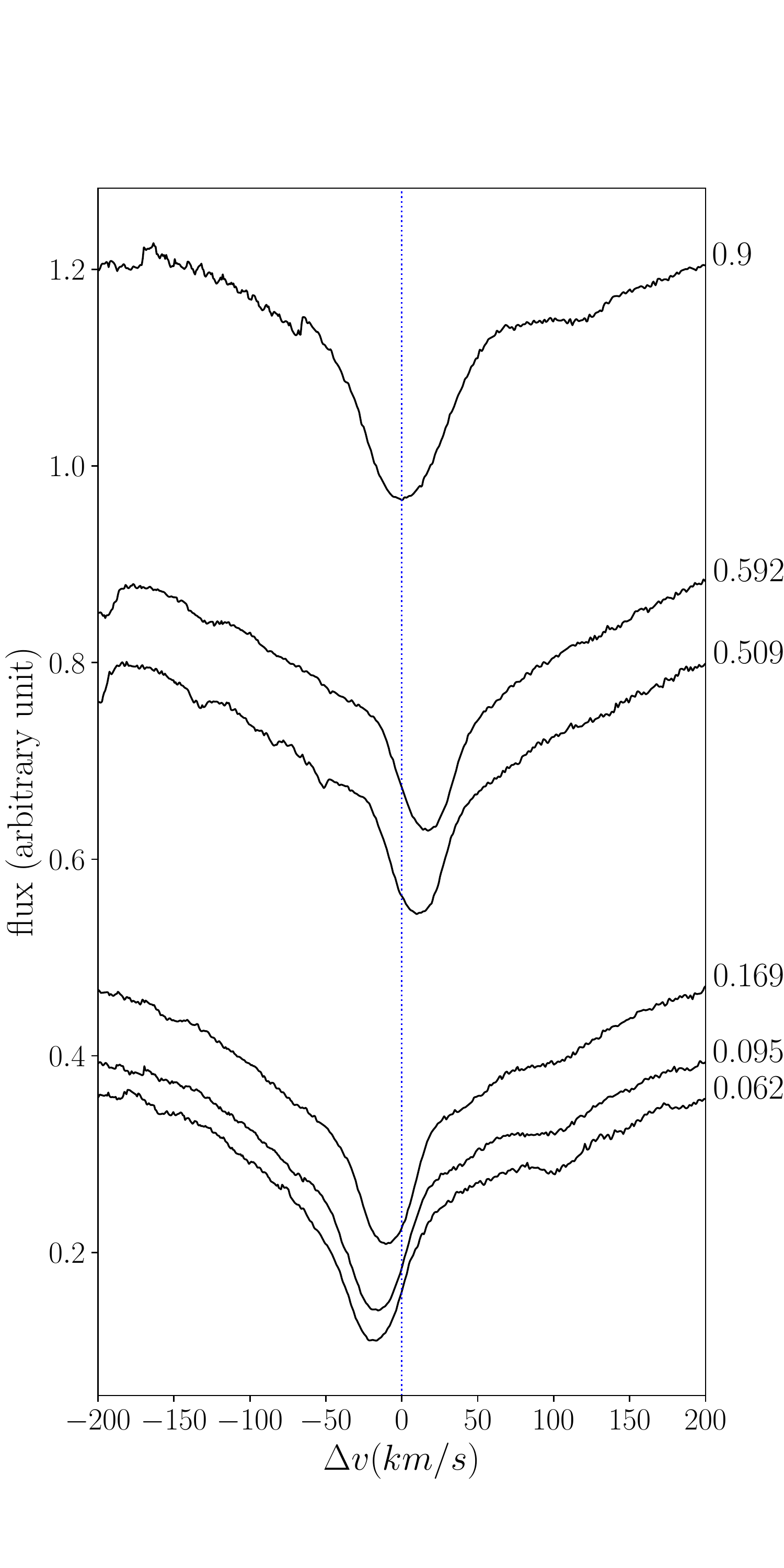}
         \caption{$\lambda$8542}
     \end{subfigure}
     \hfill
     \begin{subfigure}[b]{0.24\textwidth}
         \centering
         \includegraphics[width=\textwidth]{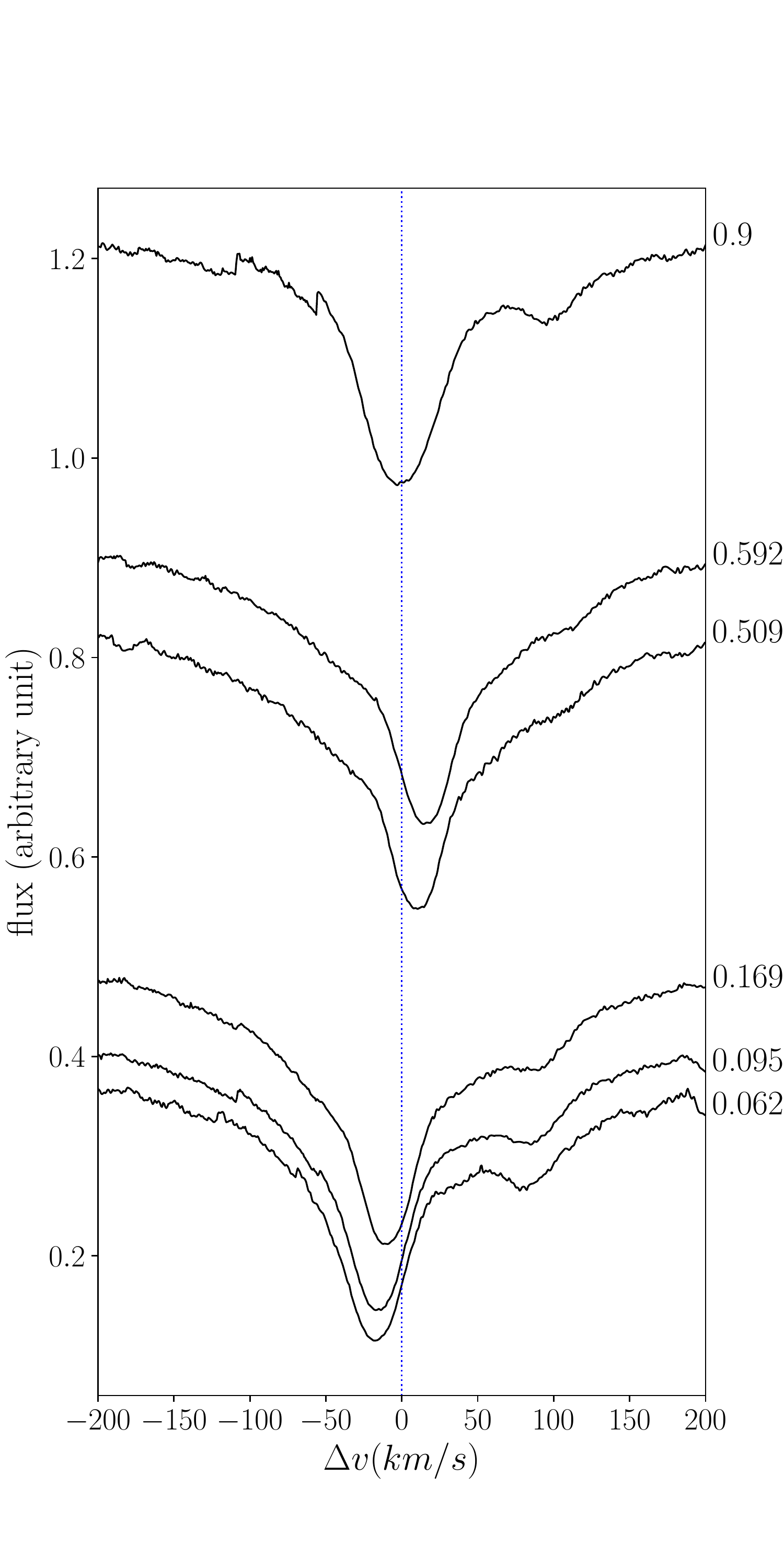}
         \caption{$\lambda$8662}
     \end{subfigure}
     \hfill
     \begin{subfigure}[b]{0.24\textwidth}
         \centering
         \includegraphics[width=\textwidth]{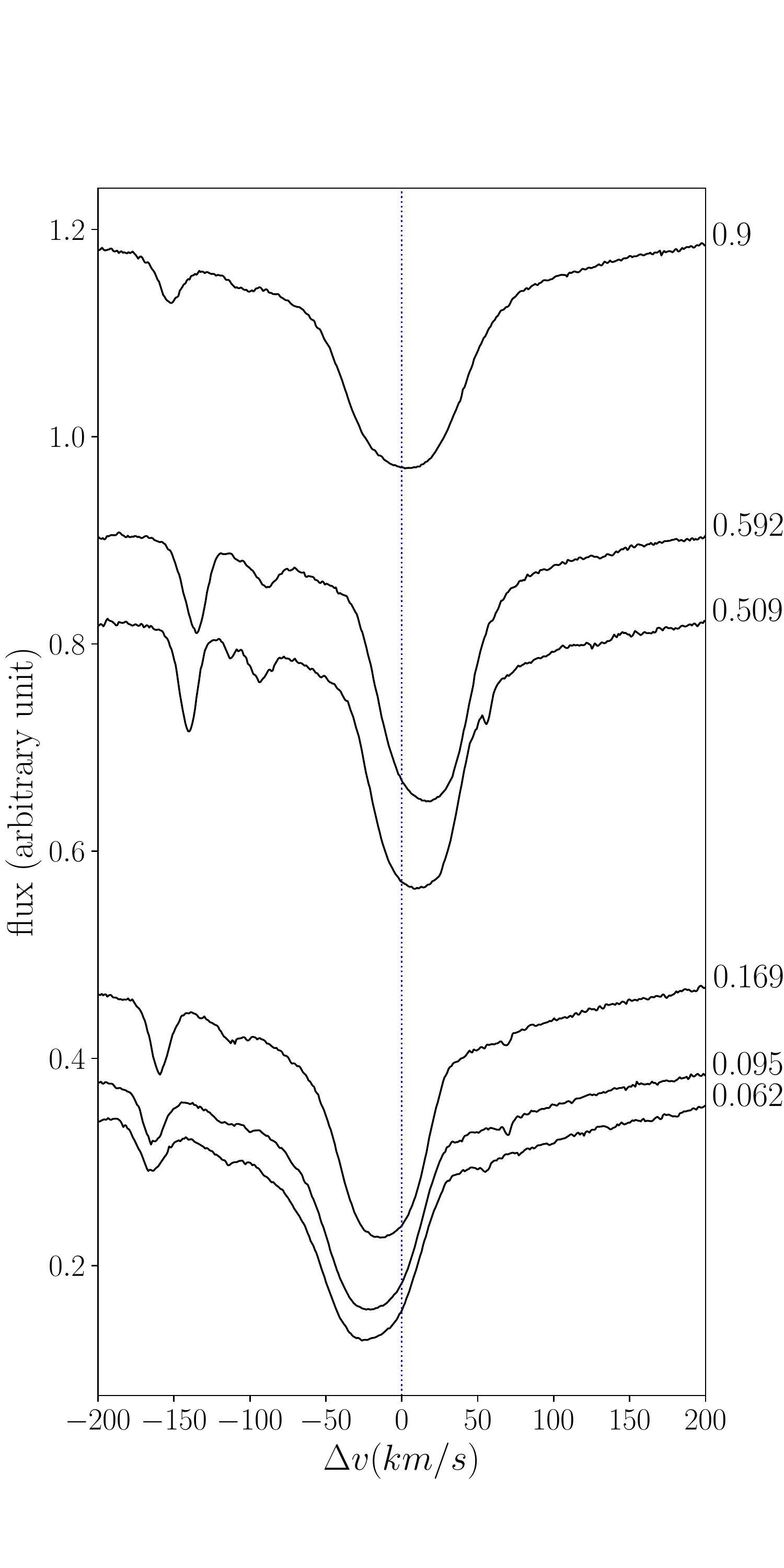}
         \caption{H$\alpha$}
     \end{subfigure}
        \caption{RT Aur, 3.74d}
\end{figure*}

\begin{figure*}
     \centering
     \begin{subfigure}[b]{0.24\textwidth}
         \centering
         \includegraphics[width=\textwidth]{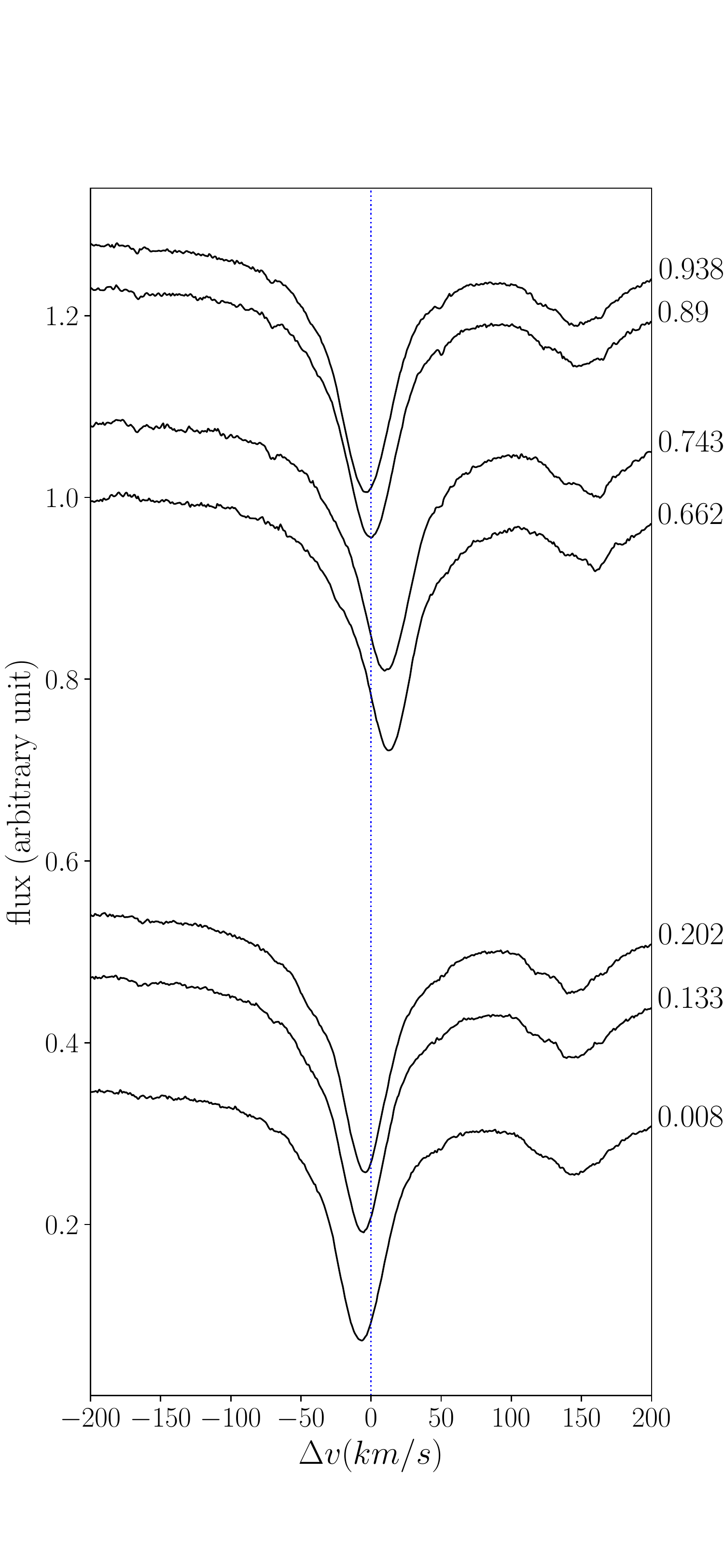}
         \caption{$\lambda$8498}
     \end{subfigure}
     \begin{subfigure}[b]{0.24\textwidth}
         \centering
         \includegraphics[width=\textwidth]{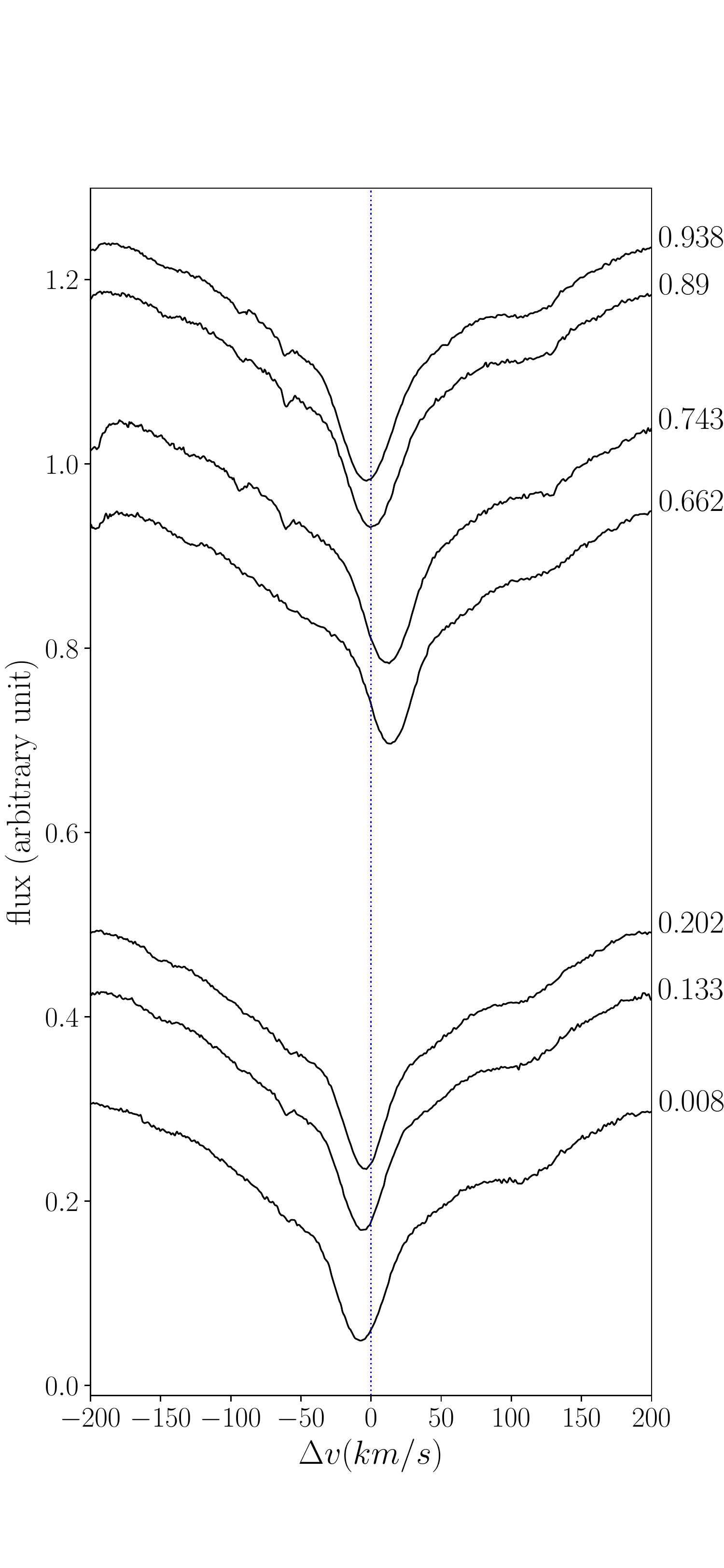}
         \caption{$\lambda$8542}
     \end{subfigure}
     \hfill
     \begin{subfigure}[b]{0.24\textwidth}
         \centering
         \includegraphics[width=\textwidth]{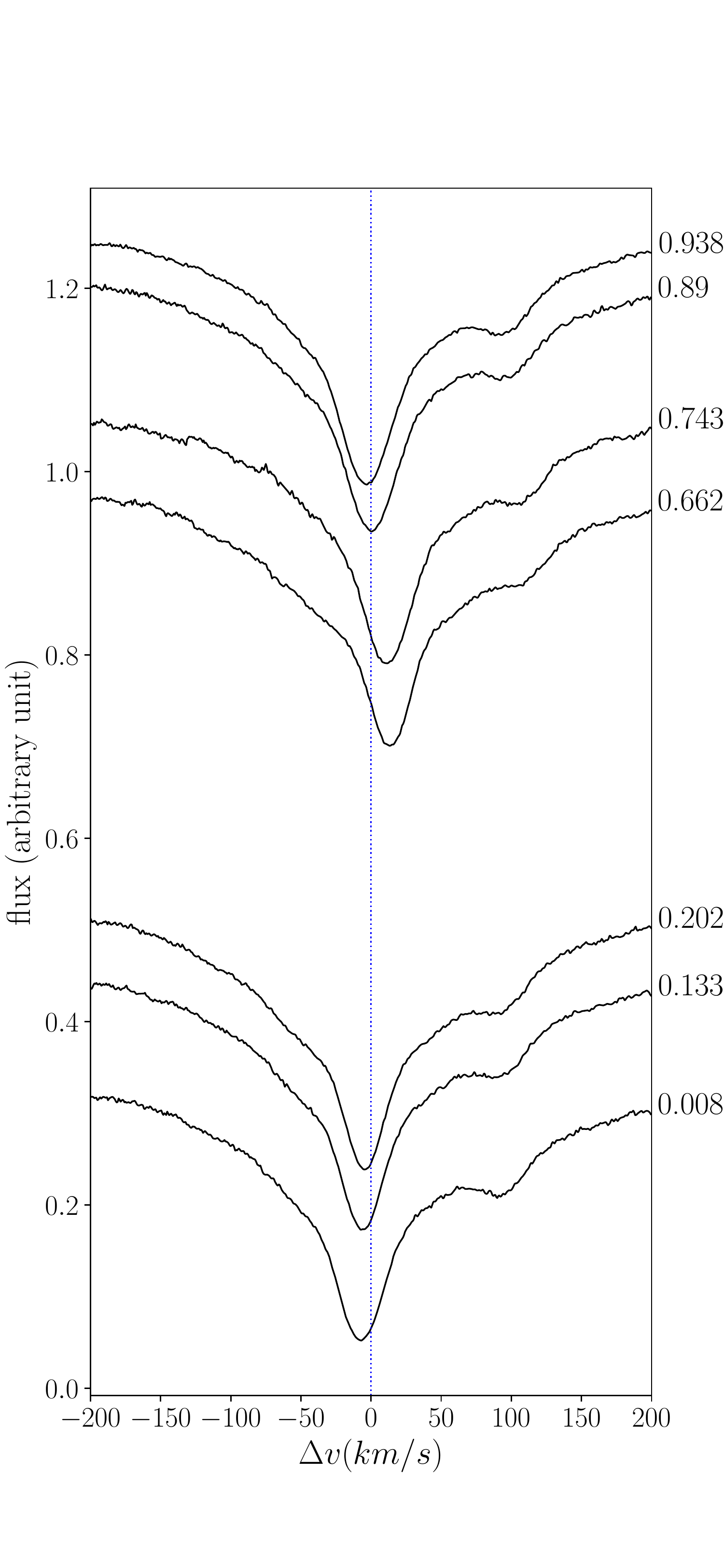}
         \caption{$\lambda$8662}
     \end{subfigure}
     \hfill
     \begin{subfigure}[b]{0.24\textwidth}
         \centering
         \includegraphics[width=\textwidth]{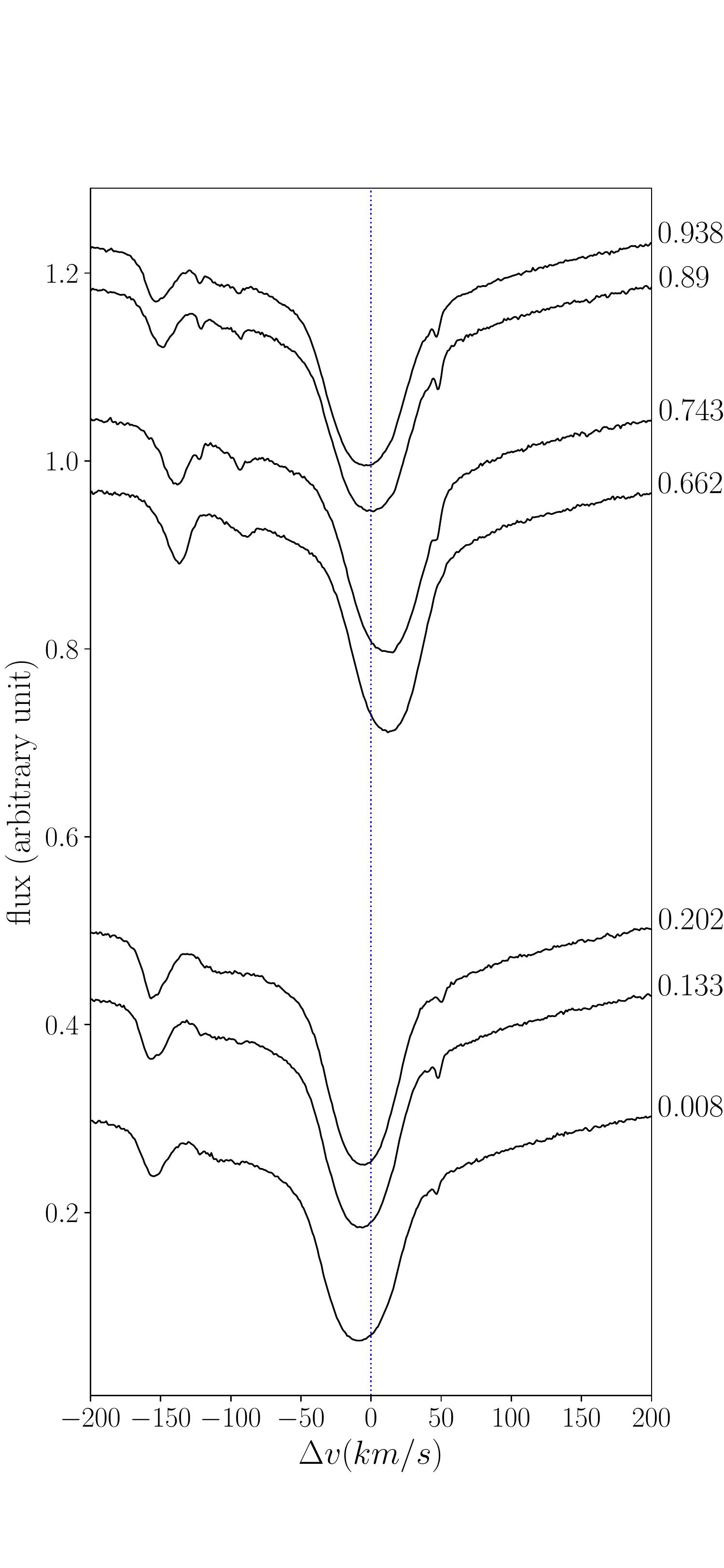}
         \caption{H$\alpha$}
     \end{subfigure}
        \caption{AH Vel, 4.22d}
\end{figure*}

\begin{figure*}
     \centering
     \begin{subfigure}[b]{0.24\textwidth}
         \centering
         \includegraphics[width=\textwidth]{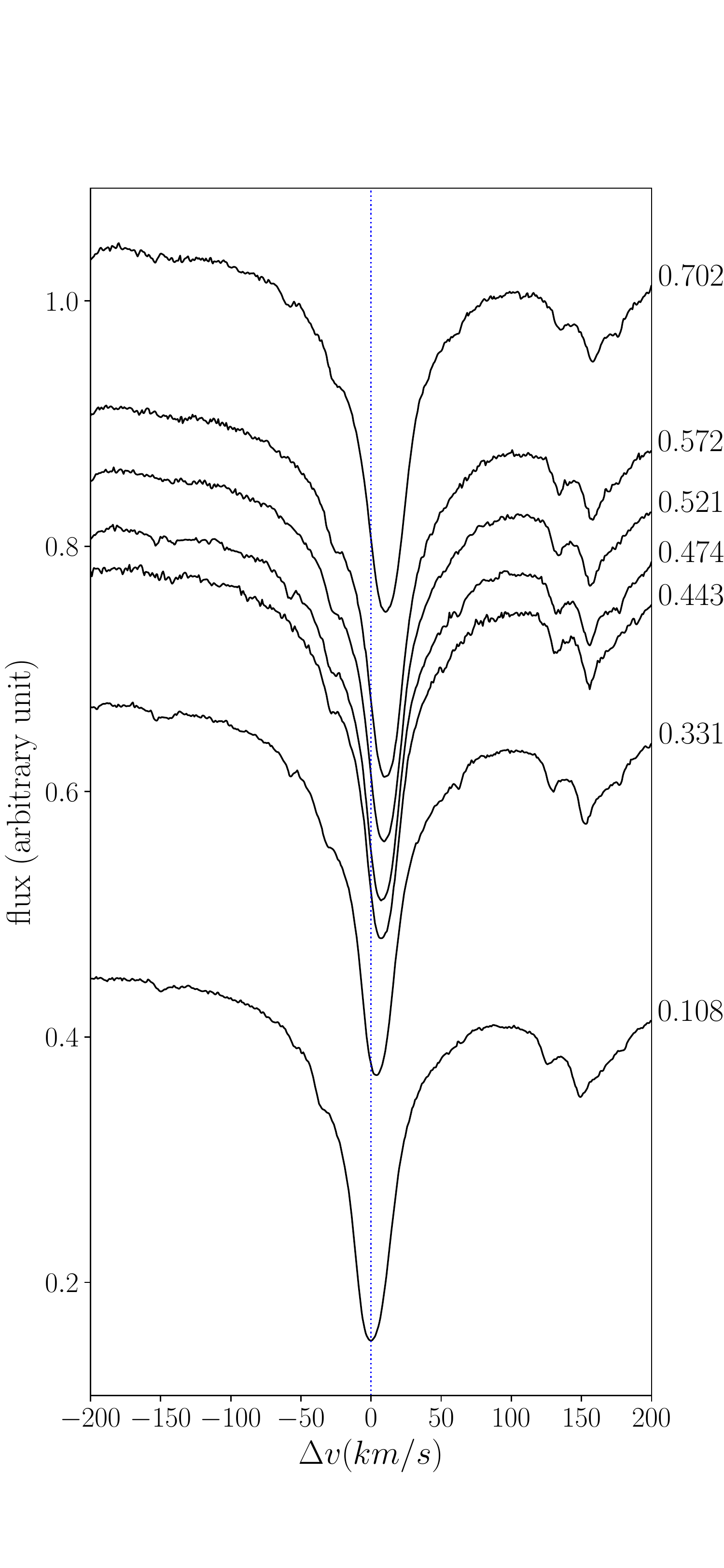}
         \caption{$\lambda$8498}
     \end{subfigure}
     \hfill
     \begin{subfigure}[b]{0.24\textwidth}
         \centering
         \includegraphics[width=\textwidth]{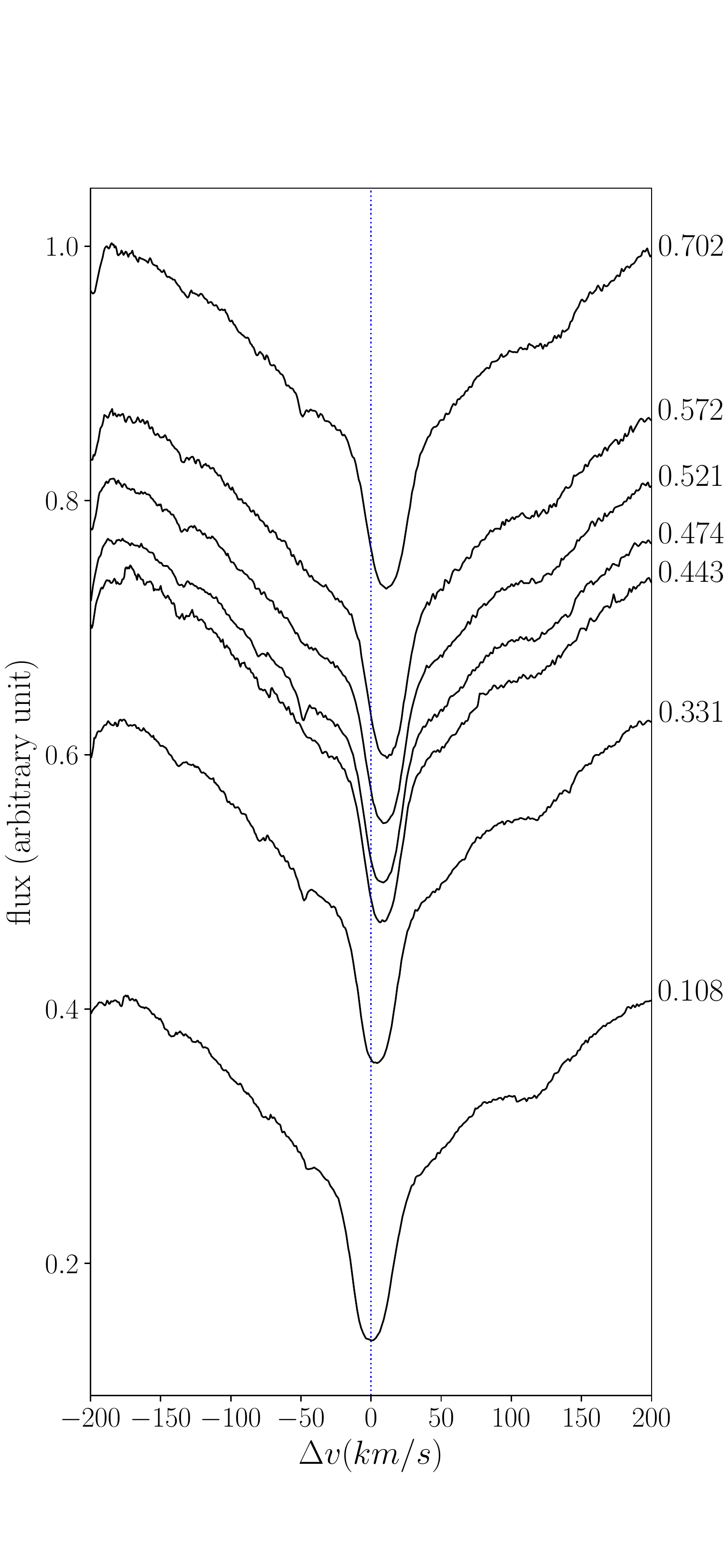}
         \caption{$\lambda$8542}
     \end{subfigure}
     \hfill
     \begin{subfigure}[b]{0.24\textwidth}
         \centering
         \includegraphics[width=\textwidth]{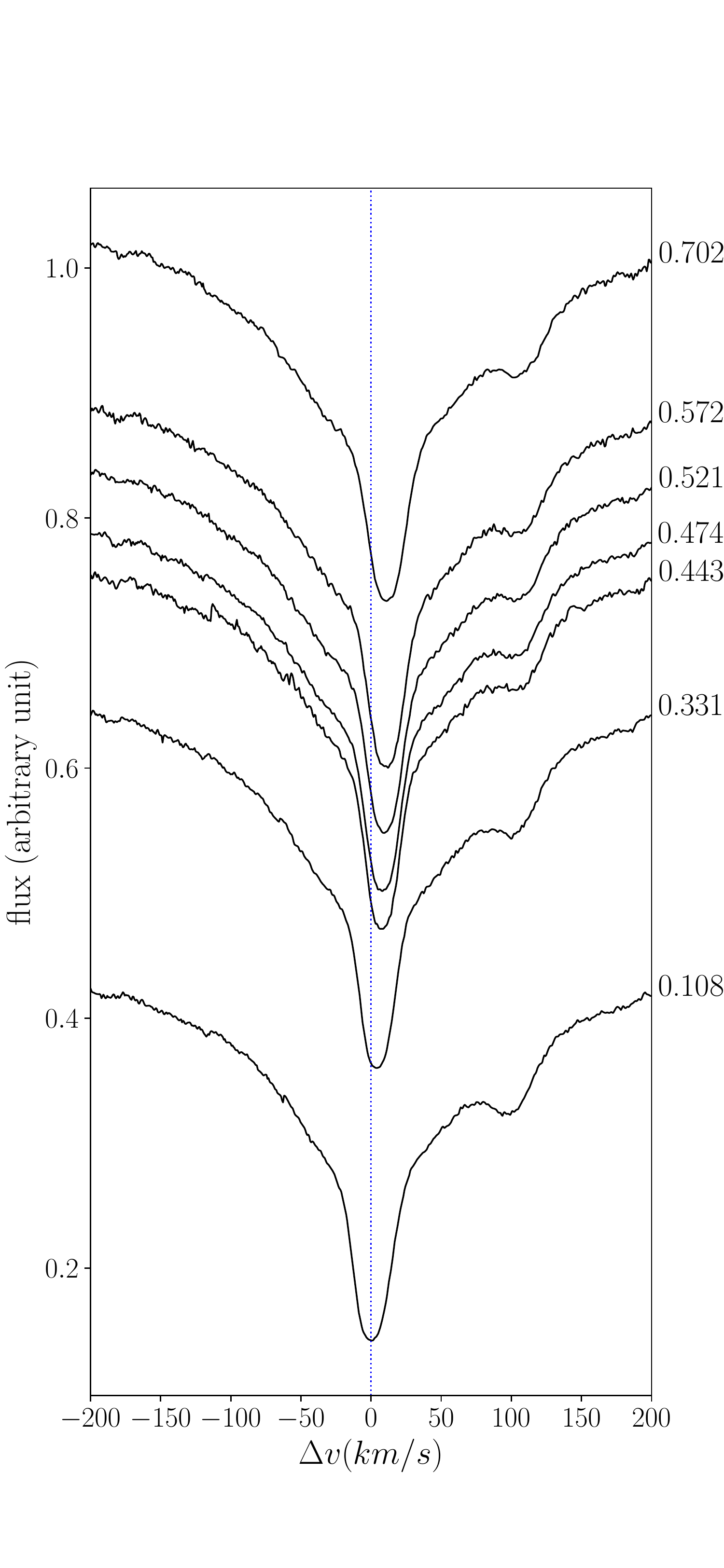}
         \caption{$\lambda$8662}
     \end{subfigure}
     \hfill
     \begin{subfigure}[b]{0.24\textwidth}
         \centering
         \includegraphics[width=\textwidth]{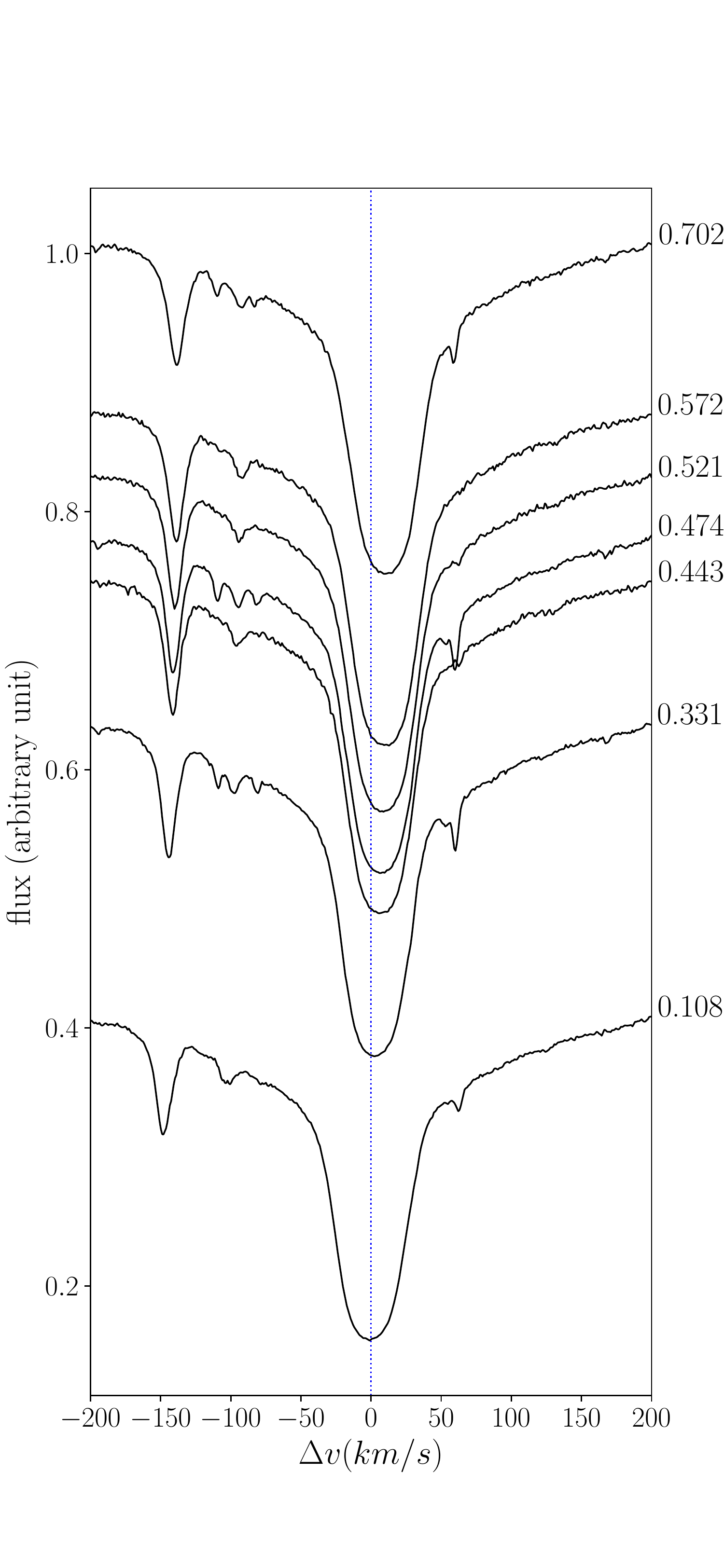}
         \caption{H$\alpha$}
     \end{subfigure}
        \caption{MY Pup, 5.69d}
\end{figure*}

\begin{figure*}[h!]
     \centering
     \begin{subfigure}[b]{0.24\textwidth}
         \centering
         \includegraphics[width=\textwidth]{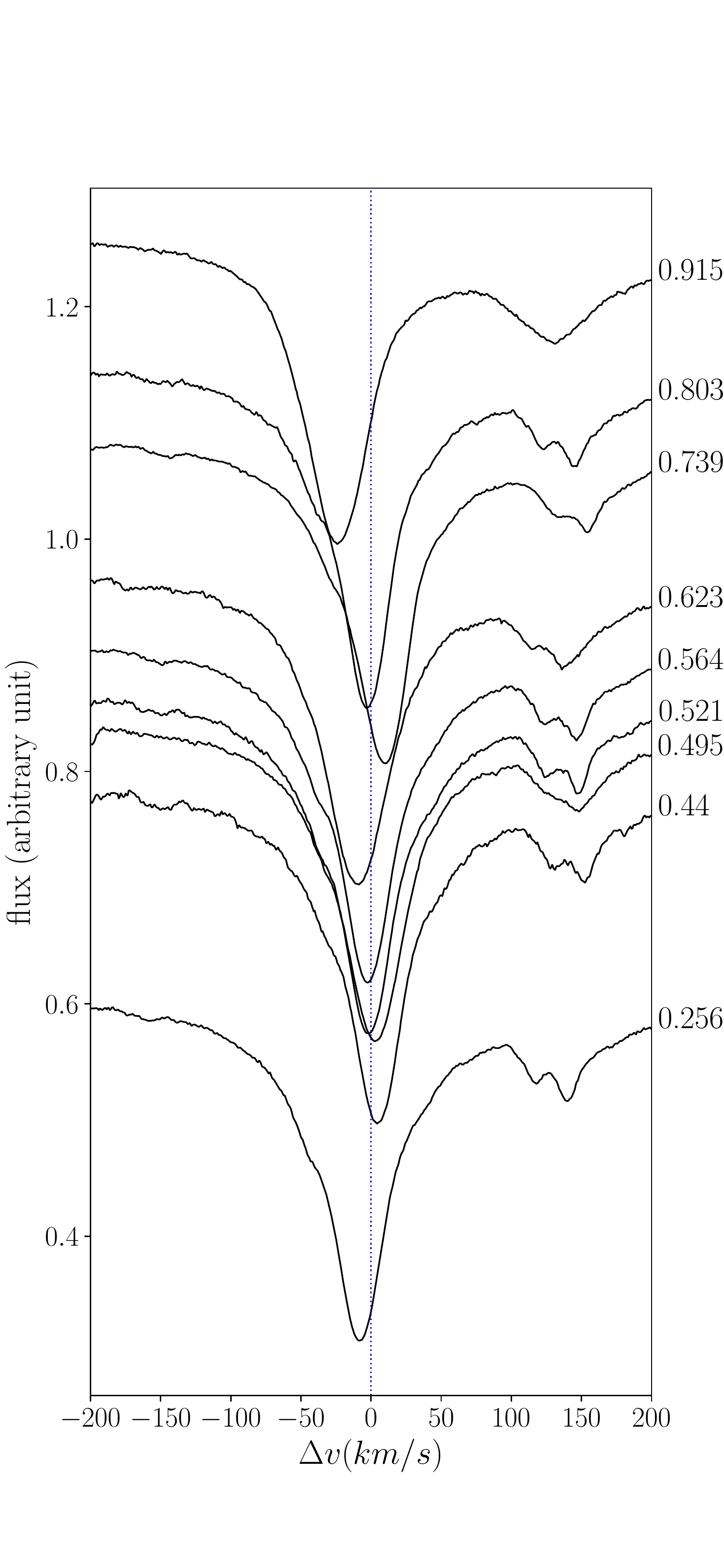}
         \caption{$\lambda$8498}
  
     \end{subfigure}
     \hfill%\vskip\baselineskip
          \begin{subfigure}[b]{0.24\textwidth}
         \centering
         \includegraphics[width=\textwidth]{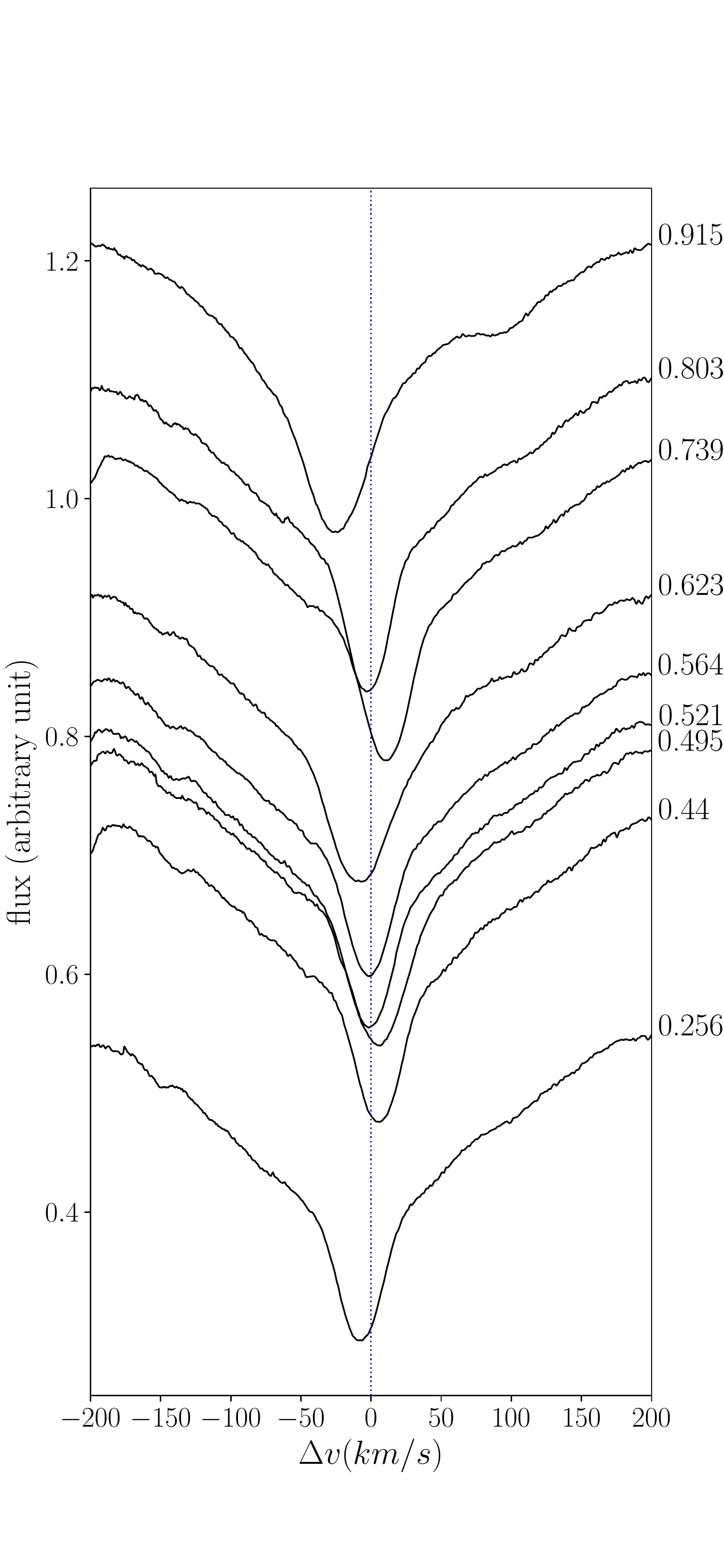}
         \caption{$\lambda$8542 \label{fig:l2_ew_sct}}
     \end{subfigure}
     \hfill
     \begin{subfigure}[b]{0.24\textwidth}
         \centering
         \includegraphics[width=\textwidth]{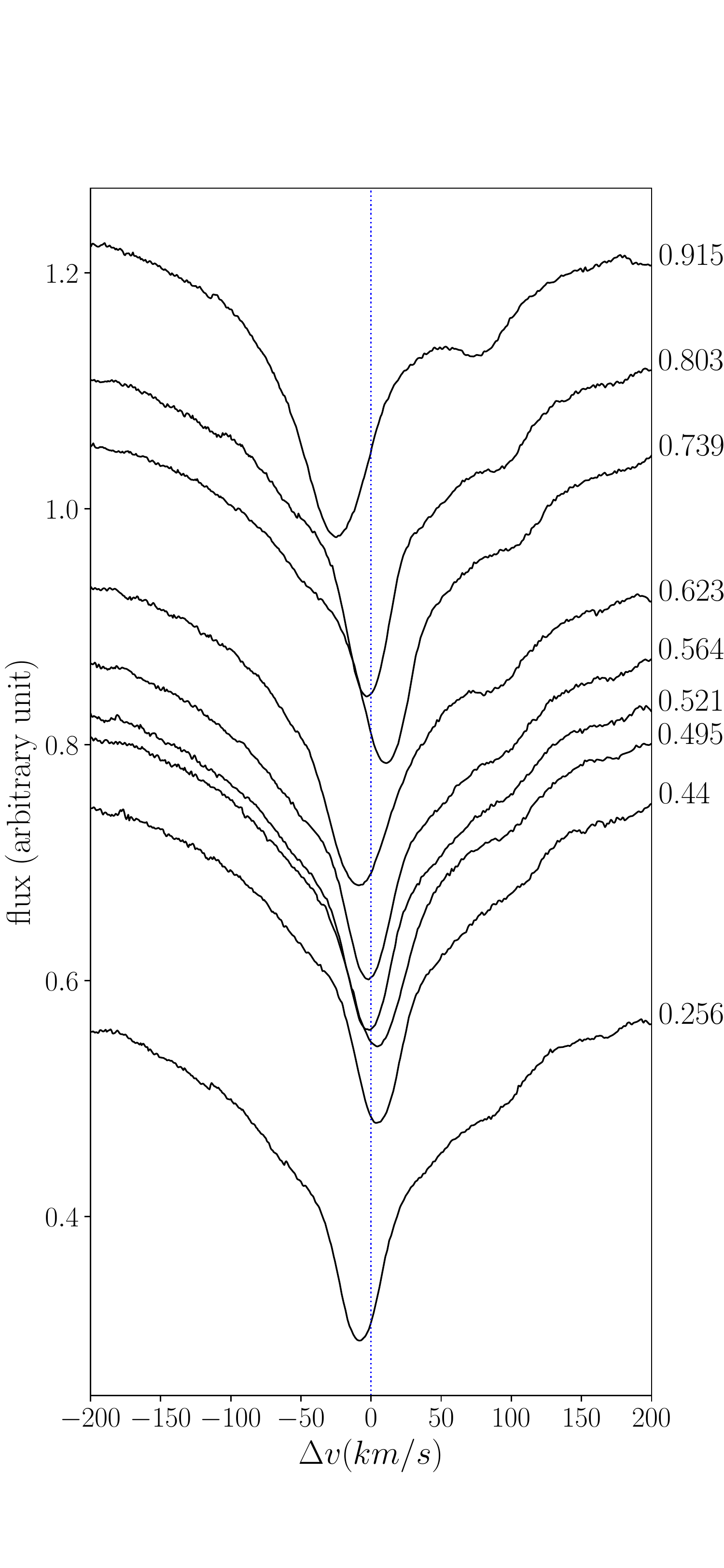}
        \caption{$\lambda$8662  \label{fig:l3_ew_sct}}
     \end{subfigure}
     \hfill
     \begin{subfigure}[b]{0.24\textwidth}
         \centering
         \includegraphics[width=\textwidth]{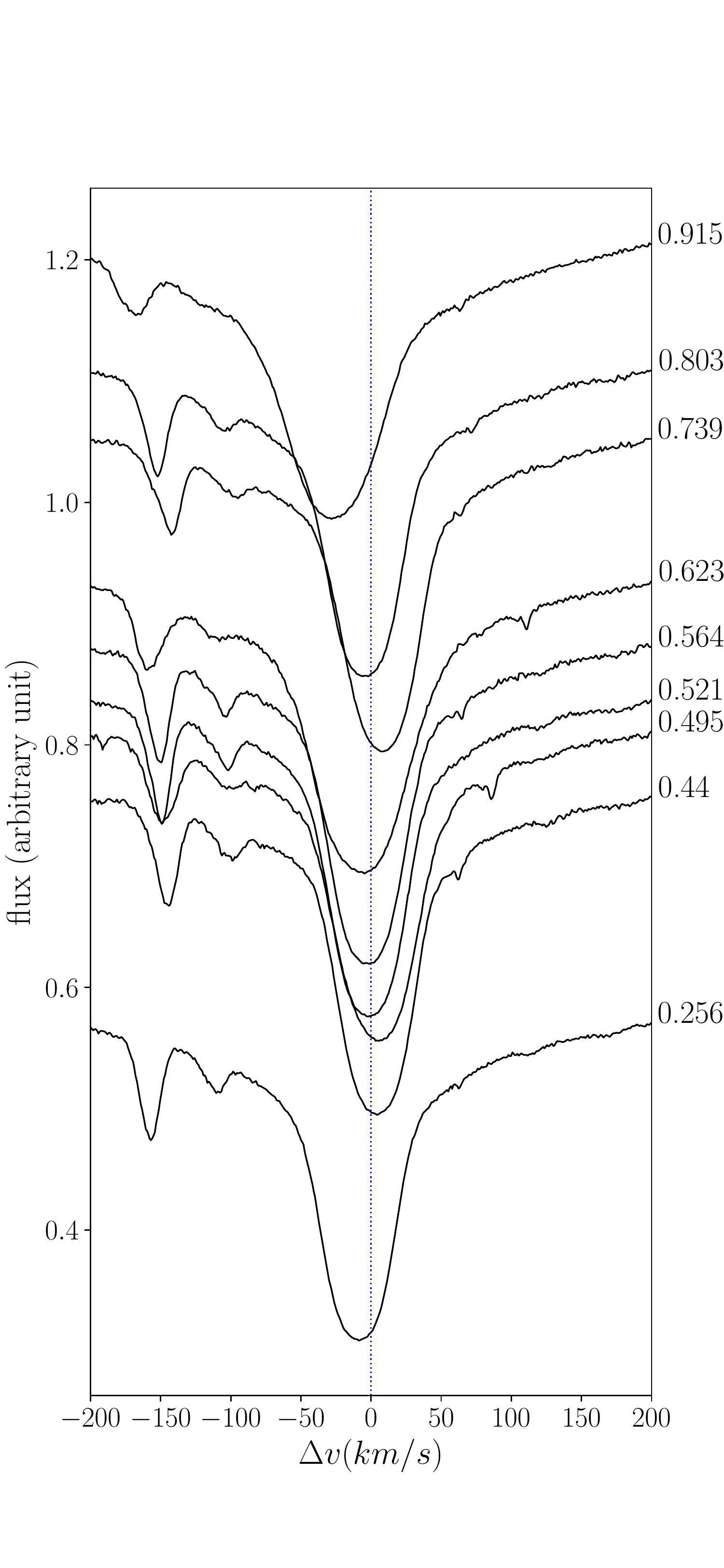}
         \caption{H$\alpha$ \label{fig:Ha_ew_sct}}
     \end{subfigure}
     \caption{EW Sct, 5.82d}
     \end{figure*}

\begin{figure*}
     \centering
         \begin{subfigure}[b]{0.24\textwidth}
         \centering
         \includegraphics[width=\textwidth]{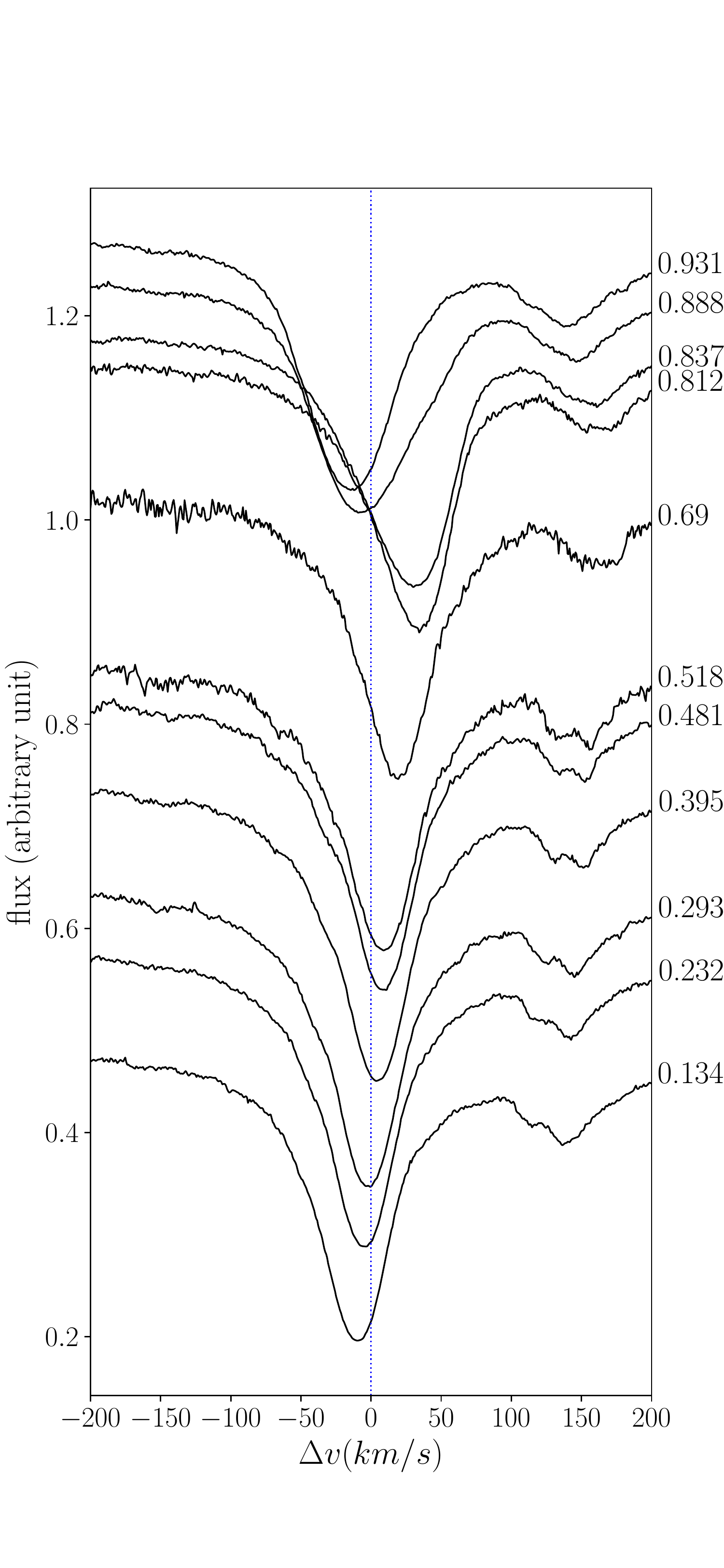}
         \caption{$\lambda$8498}
     \end{subfigure}
     \hfill
     \begin{subfigure}[b]{0.24\textwidth}
         \centering
         \includegraphics[width=\textwidth]{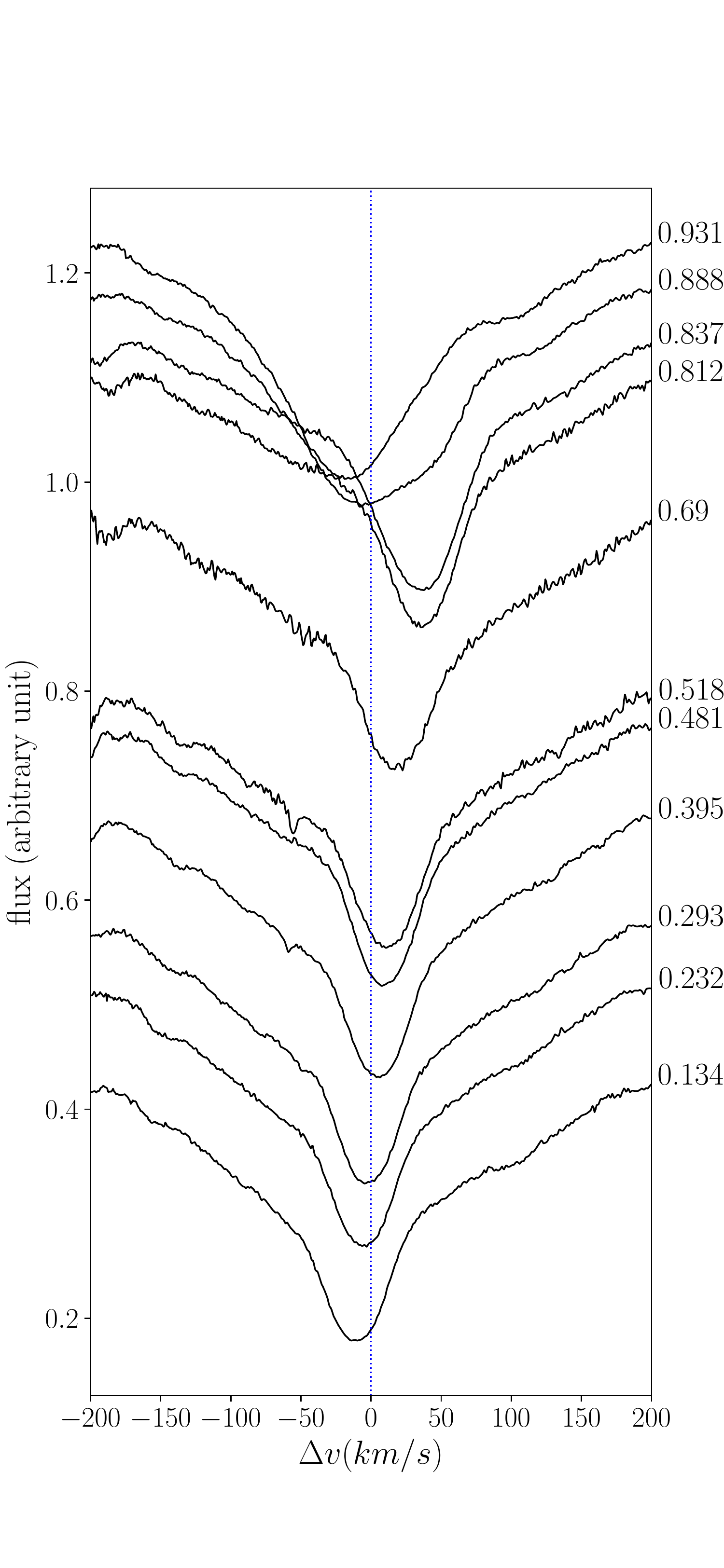}
         \caption{$\lambda$8542}
     \end{subfigure}
     \hfill
     \begin{subfigure}[b]{0.24\textwidth}
         \centering
         \includegraphics[width=\textwidth]{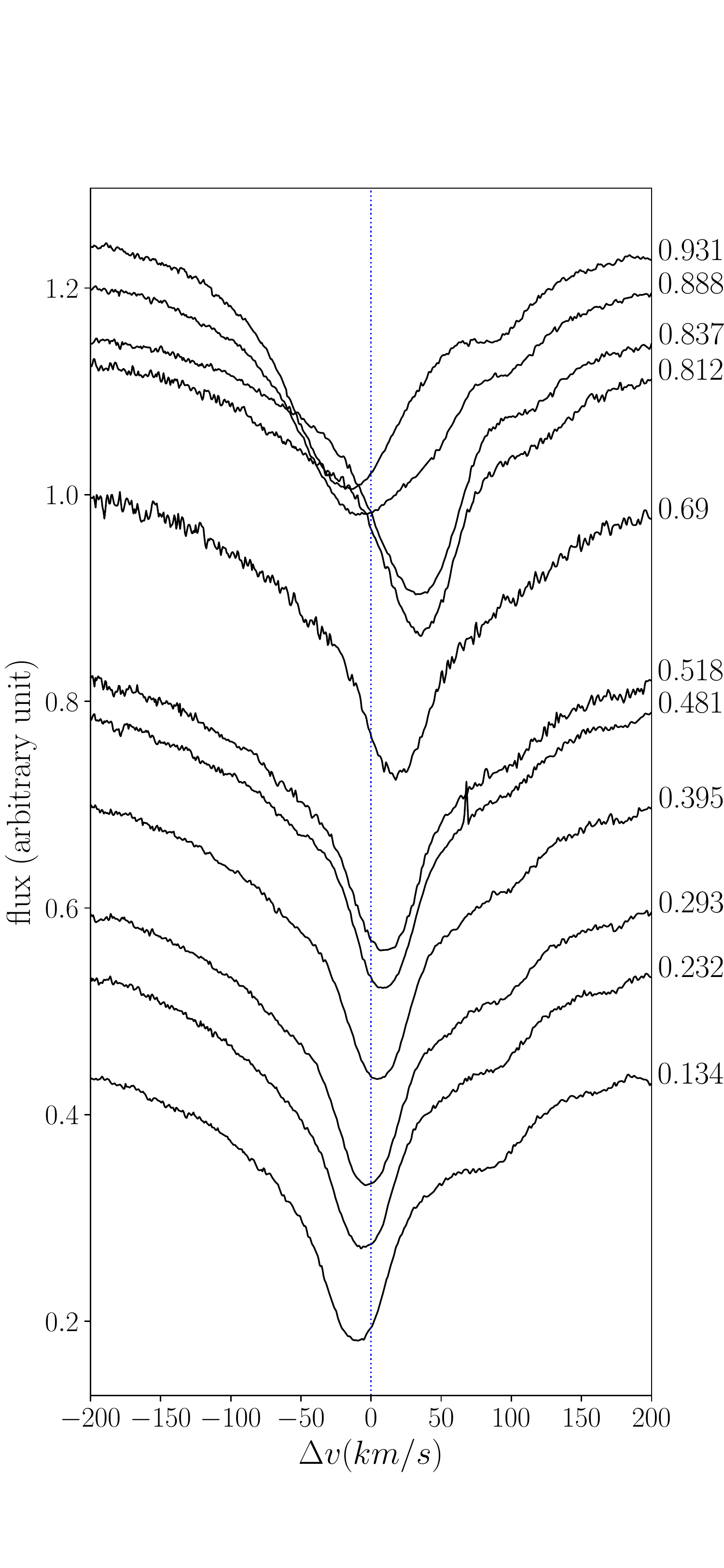}
         \caption{$\lambda$8662}
     \end{subfigure}
     \hfill
     \begin{subfigure}[b]{0.24\textwidth}
         \centering
         \includegraphics[width=\textwidth]{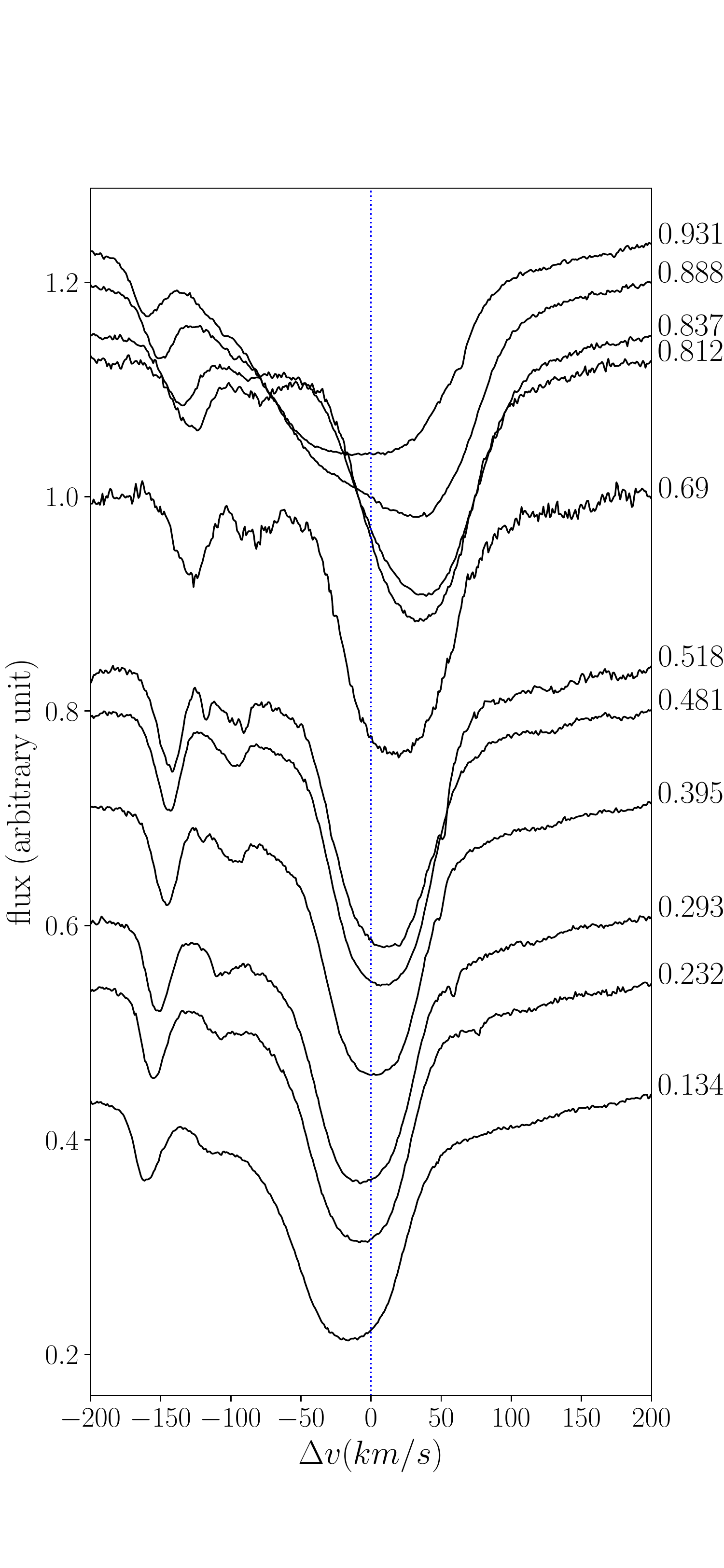}
         \caption{H$\alpha$}
     \end{subfigure}
        \caption{U Sgr, 6.75d}
        \label{fig:u_sgr}
\end{figure*}

\begin{figure*}
     \centering
          \begin{subfigure}[b]{0.24\textwidth}
         \centering
         \includegraphics[width=\textwidth]{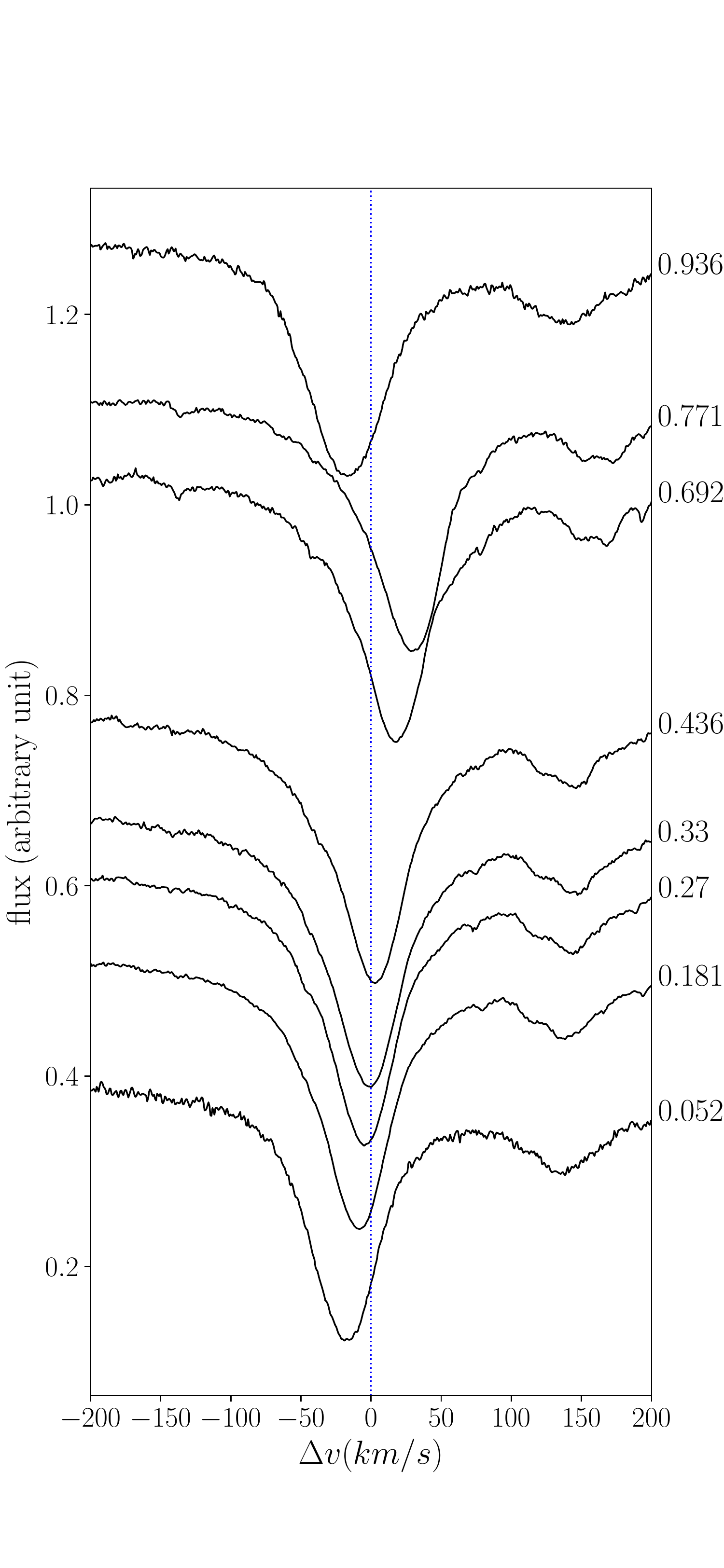}
         \caption{$\lambda$8498}
     \end{subfigure}
     \hfill
     \begin{subfigure}[b]{0.24\textwidth}
         \centering
         \includegraphics[width=\textwidth]{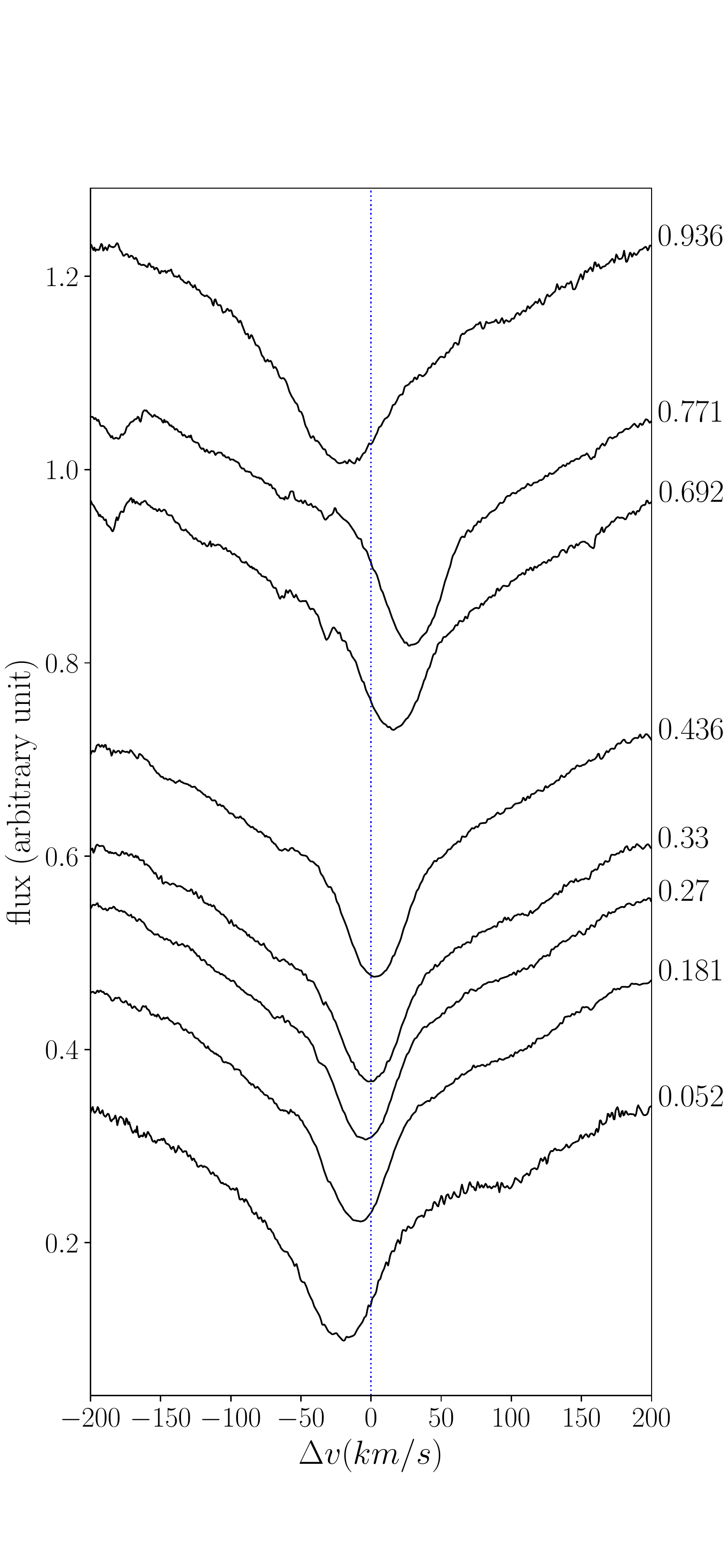}
         \caption{$\lambda$8542}
     \end{subfigure}
     \hfill
     \begin{subfigure}[b]{0.24\textwidth}
         \centering
         \includegraphics[width=\textwidth]{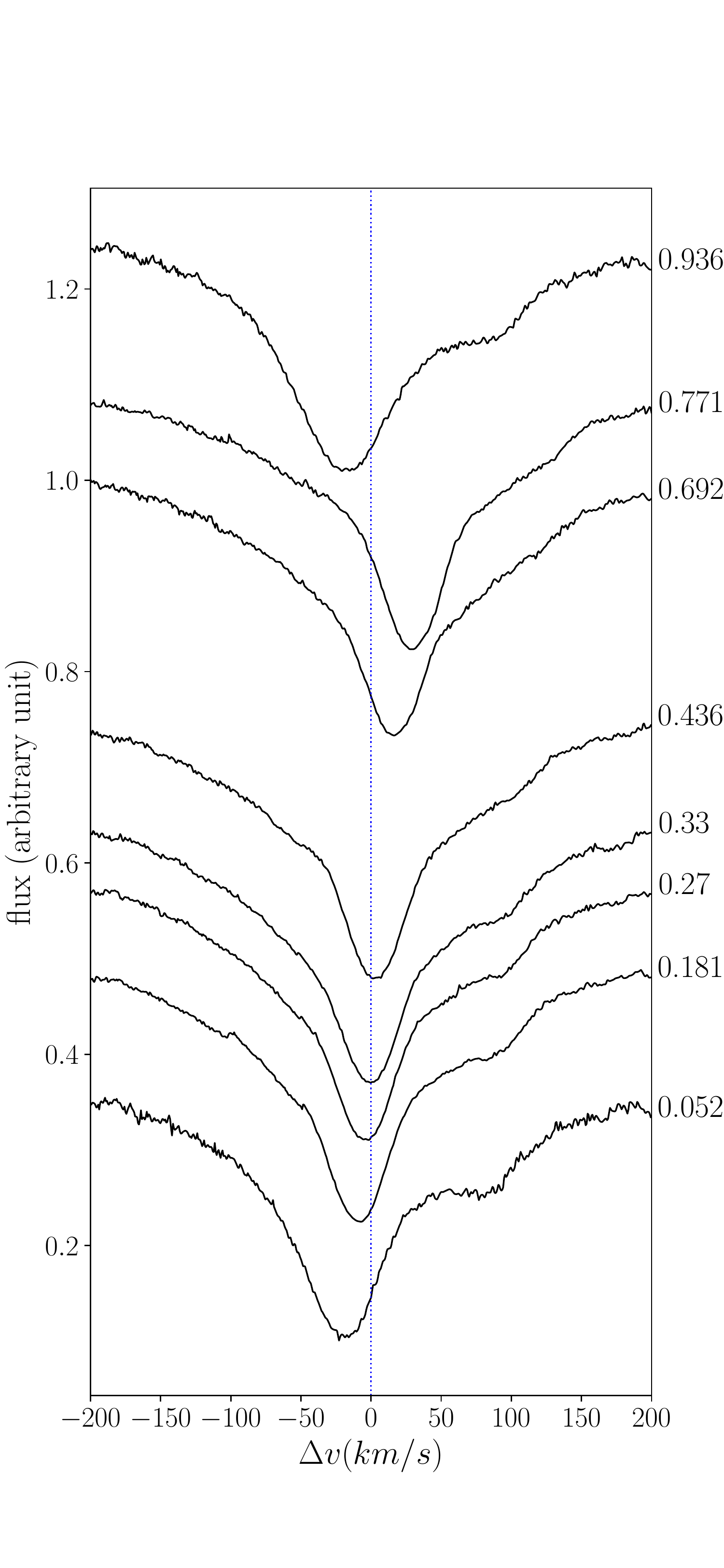}
         \caption{$\lambda$8662}
     \end{subfigure}
     \hfill
          \begin{subfigure}[b]{0.24\textwidth}
         \centering
         \includegraphics[width=\textwidth]{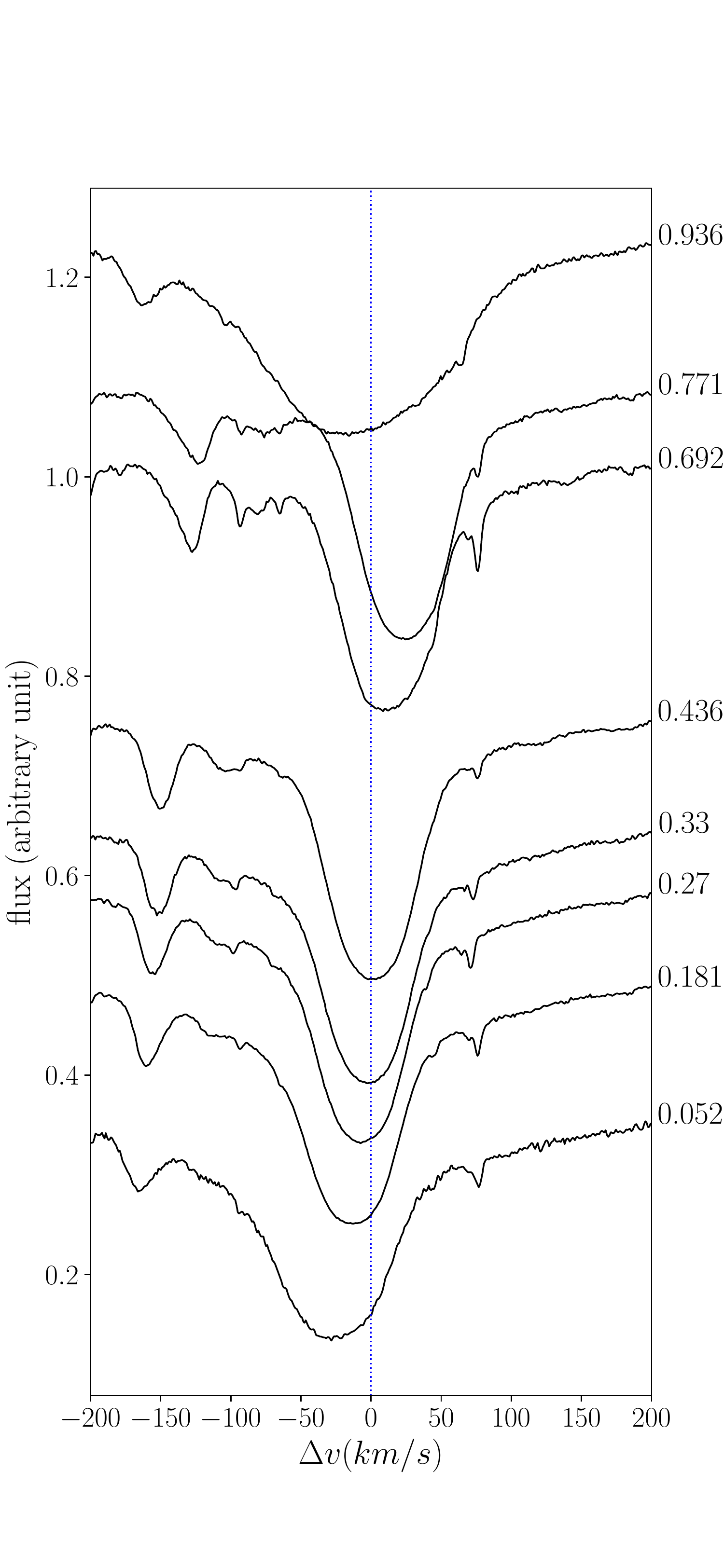}
         \caption{H$\alpha$}
     \end{subfigure}
        \caption{R Mus, 7.51d}
        \label{fig:three graphs}
\end{figure*}

\begin{figure*}
     \centering
     \begin{subfigure}[b]{0.24\textwidth}
         \centering
         \includegraphics[width=\textwidth]{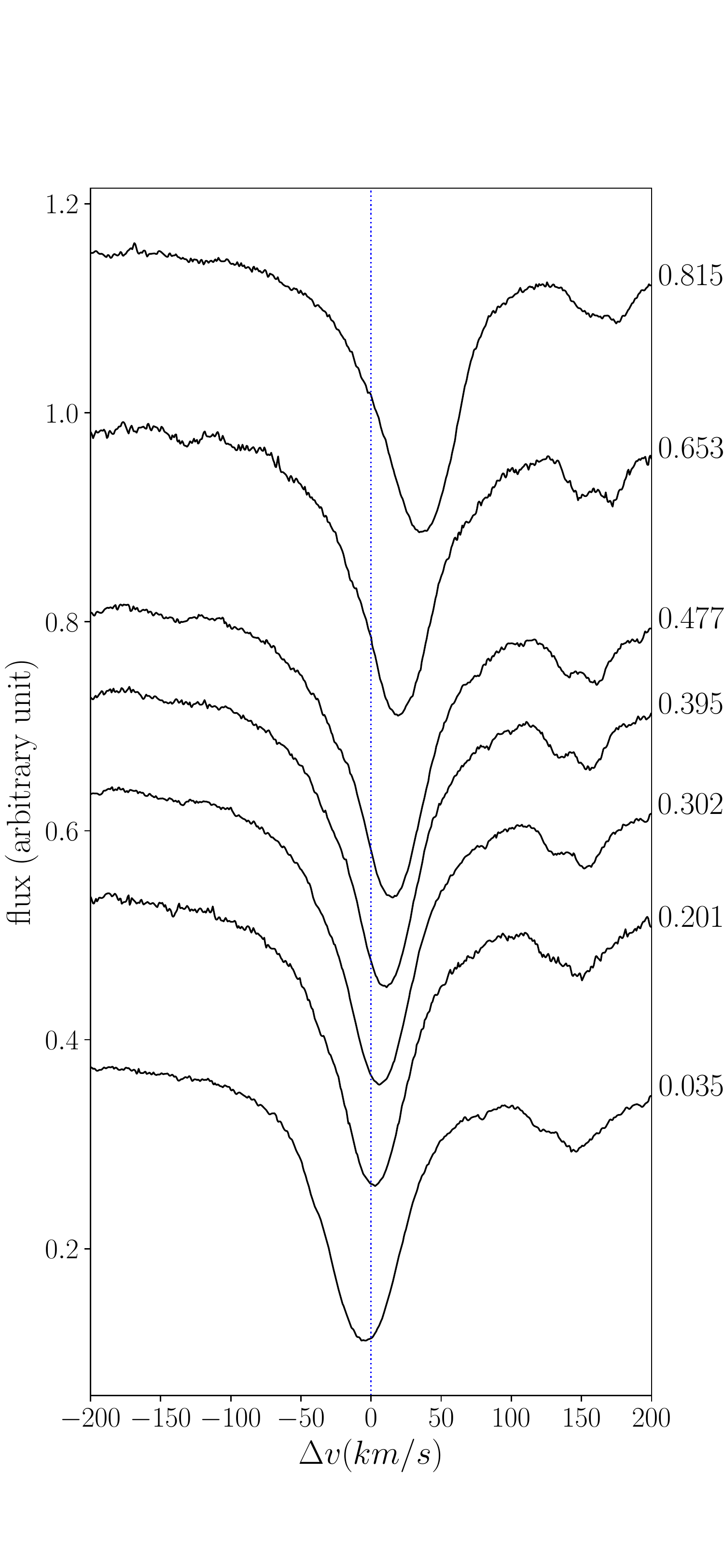}
         \caption{$\lambda$8498}
     \end{subfigure}
     \hfill
     \begin{subfigure}[b]{0.24\textwidth}
         \centering
         \includegraphics[width=\textwidth]{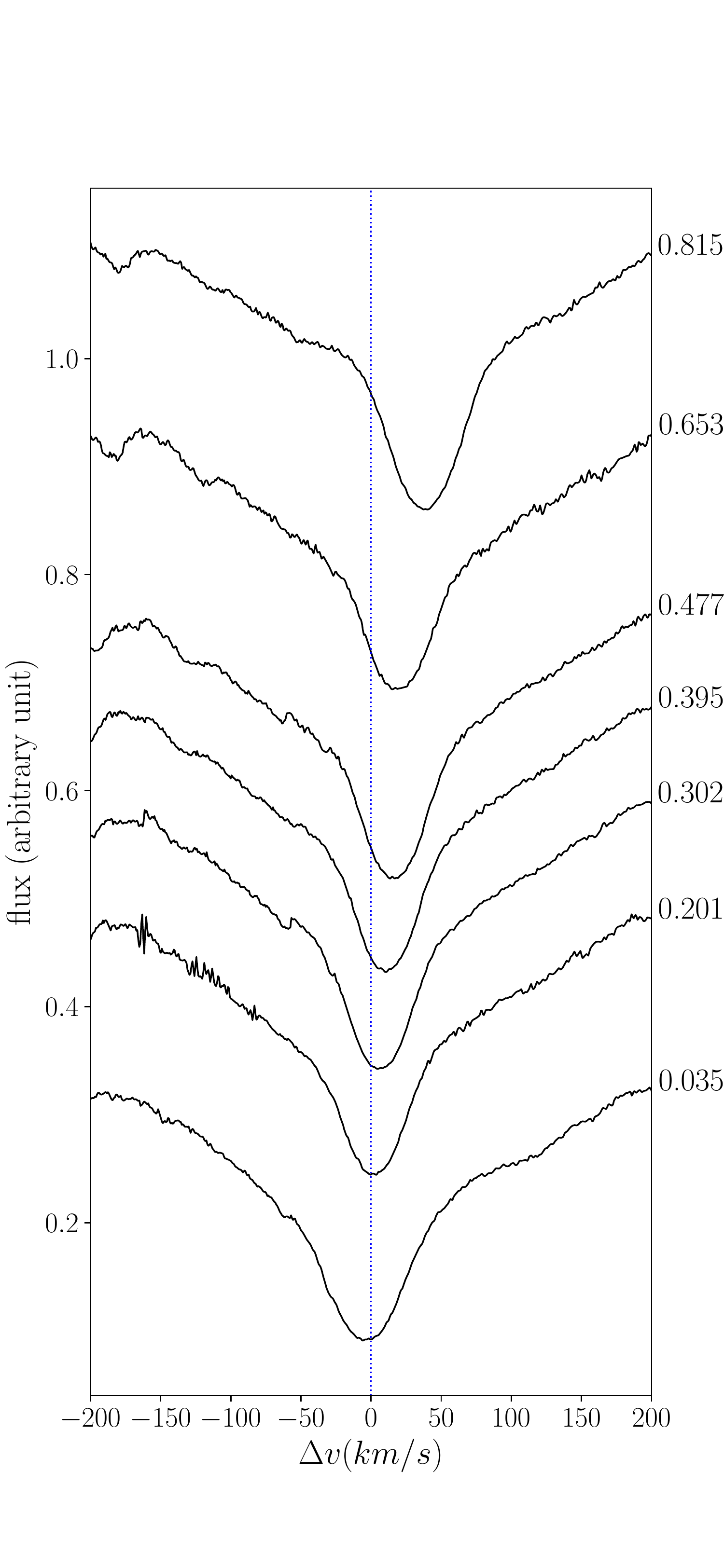}
         \caption{$\lambda$8542}
     \end{subfigure}
     \hfill
     \begin{subfigure}[b]{0.24\textwidth}
         \centering
         \includegraphics[width=\textwidth]{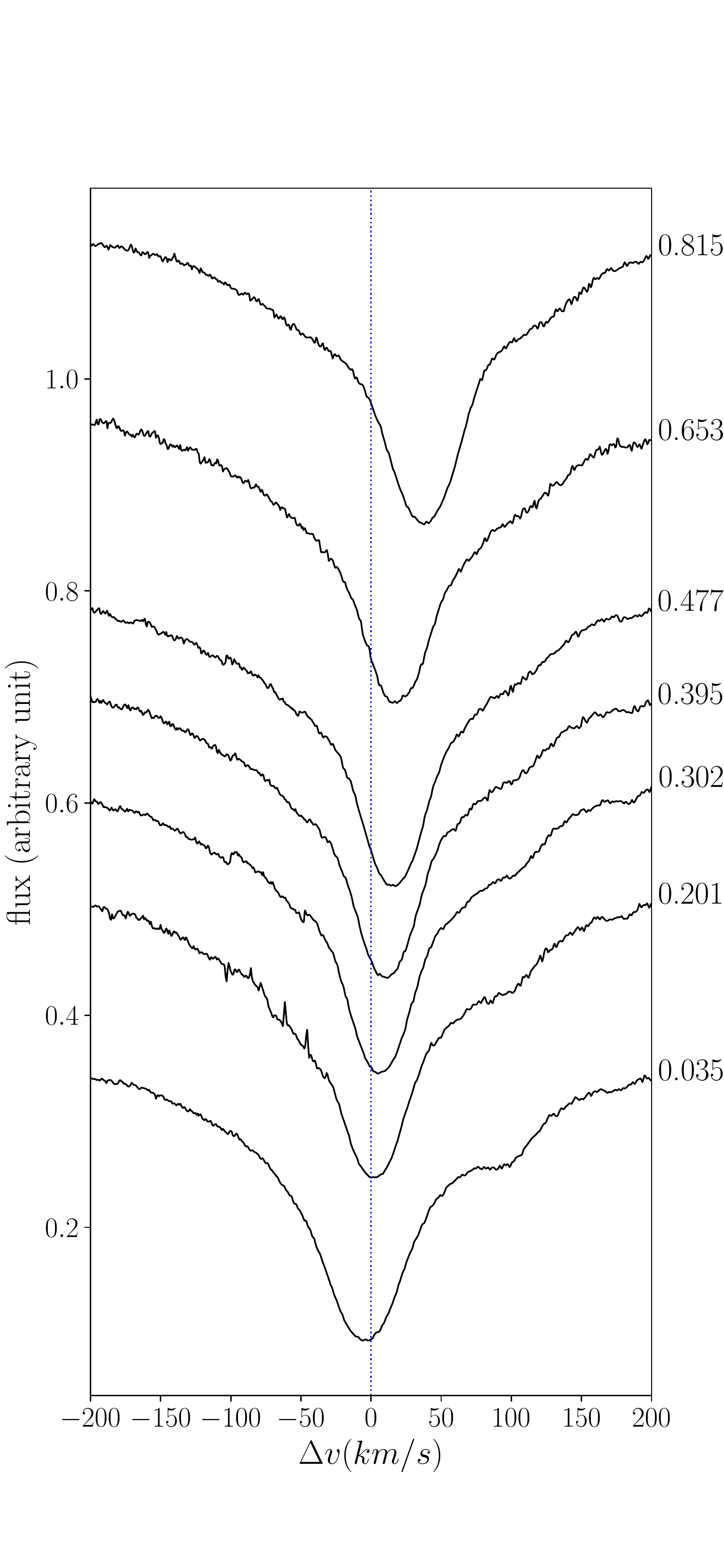}
         \caption{$\lambda$8662}
     \end{subfigure}
     \hfill
     \begin{subfigure}[b]{0.24\textwidth}
         \centering
         \includegraphics[width=\textwidth]{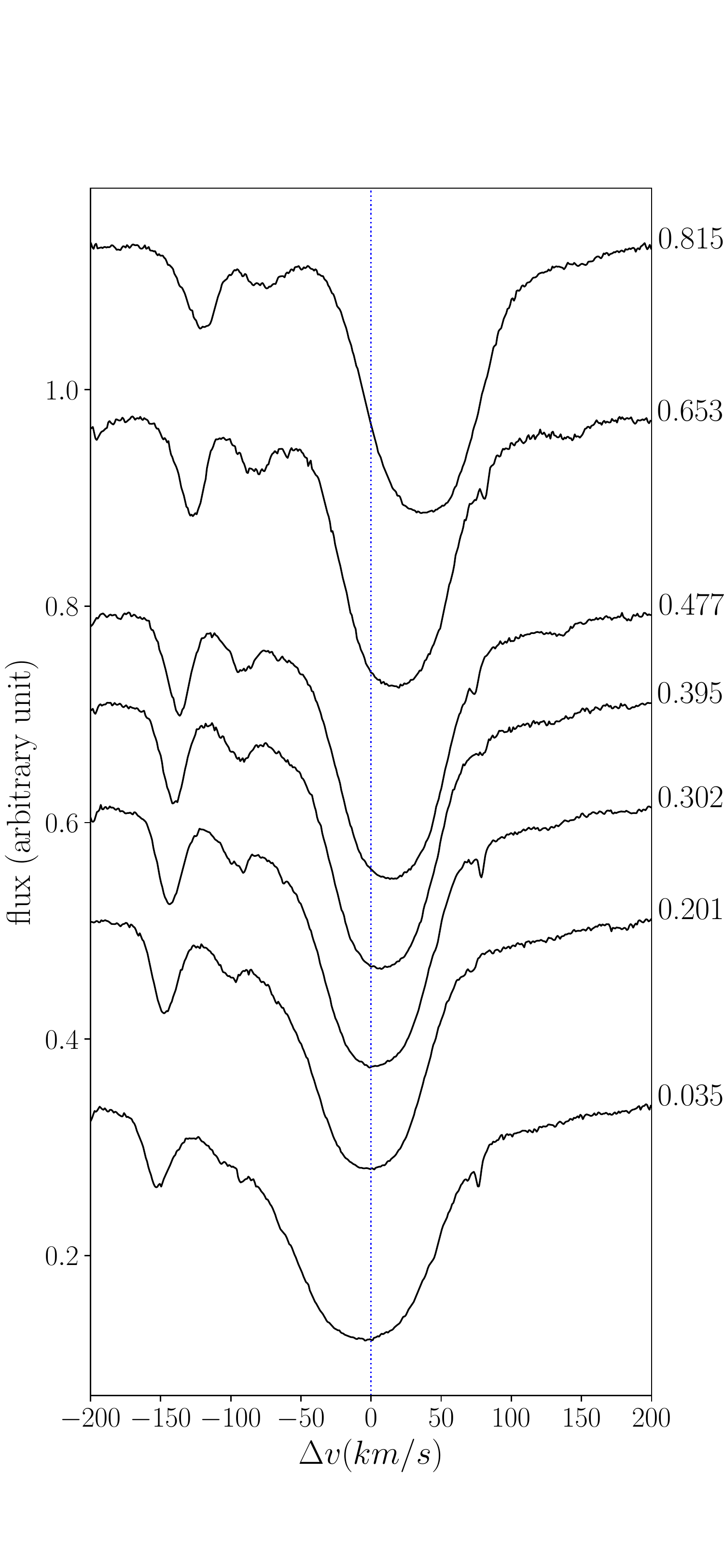}
         \caption{H$\alpha$}
     \end{subfigure}
        \caption{V636 Sco, 7.8d}
\end{figure*}

\begin{figure*}
     \centering
         \begin{subfigure}[b]{0.24\textwidth}
         \centering
         \includegraphics[width=\textwidth]{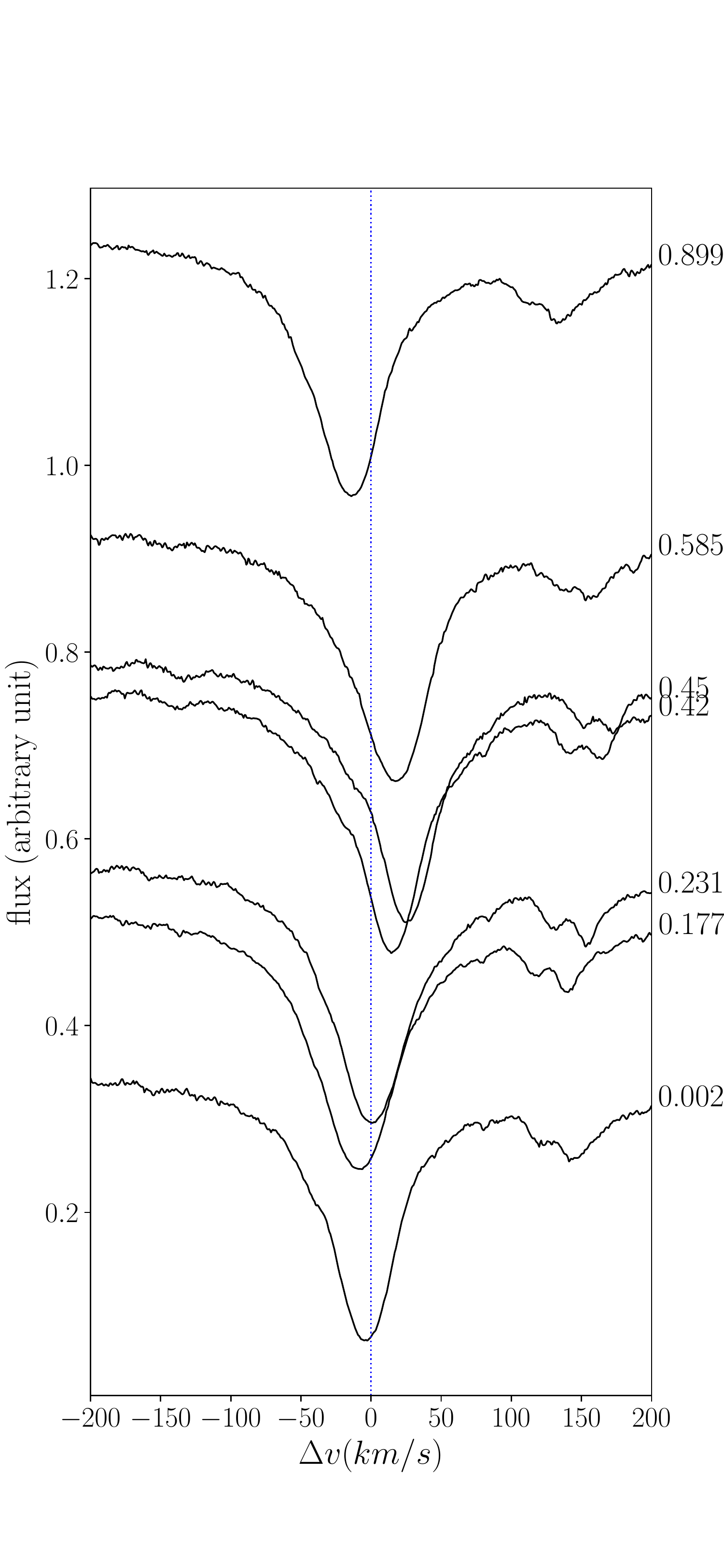}
         \caption{$\lambda$8498}
     \end{subfigure}
     \hfill
     \begin{subfigure}[b]{0.24\textwidth}
         \centering
         \includegraphics[width=\textwidth]{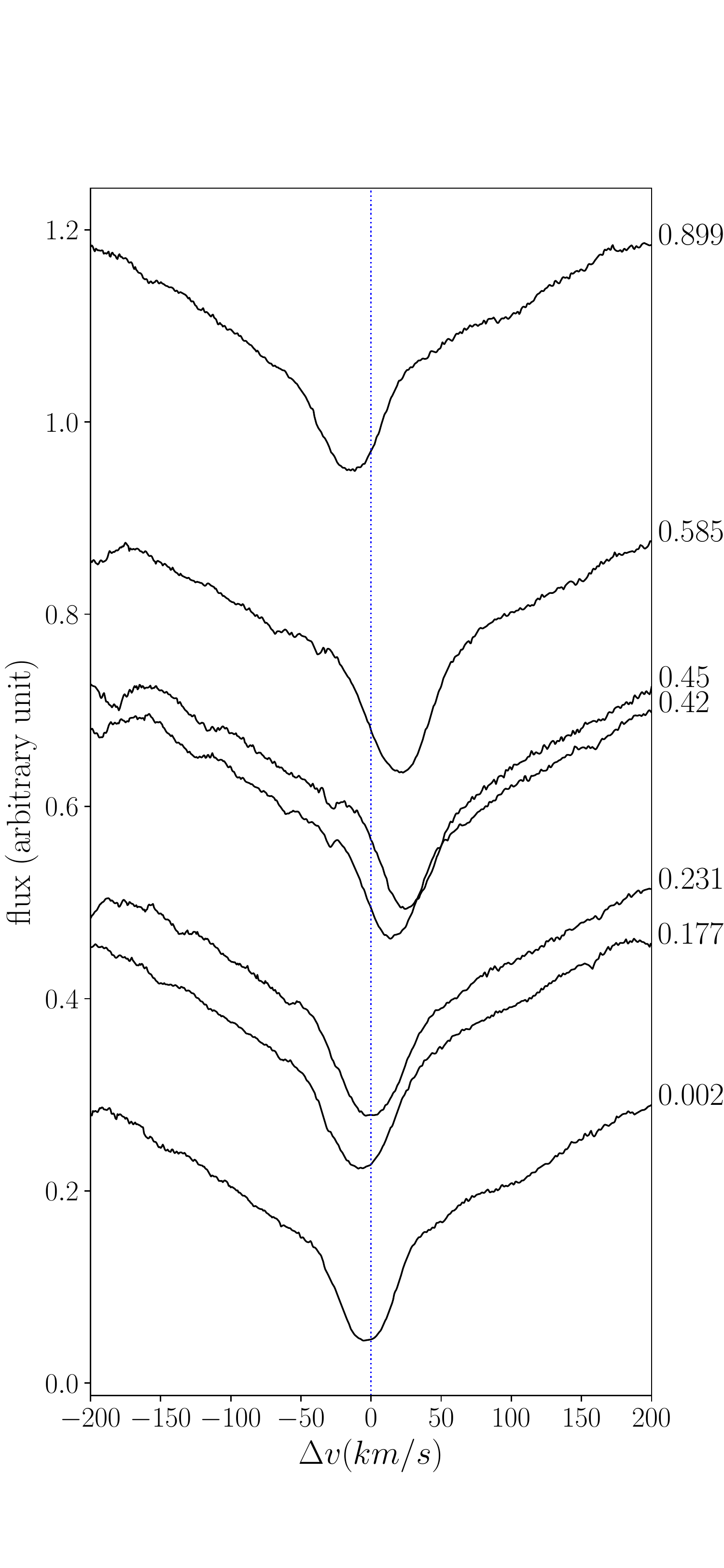}
         \caption{$\lambda$8542}
     \end{subfigure}
     \hfill
     \begin{subfigure}[b]{0.24\textwidth}
         \centering
         \includegraphics[width=\textwidth]{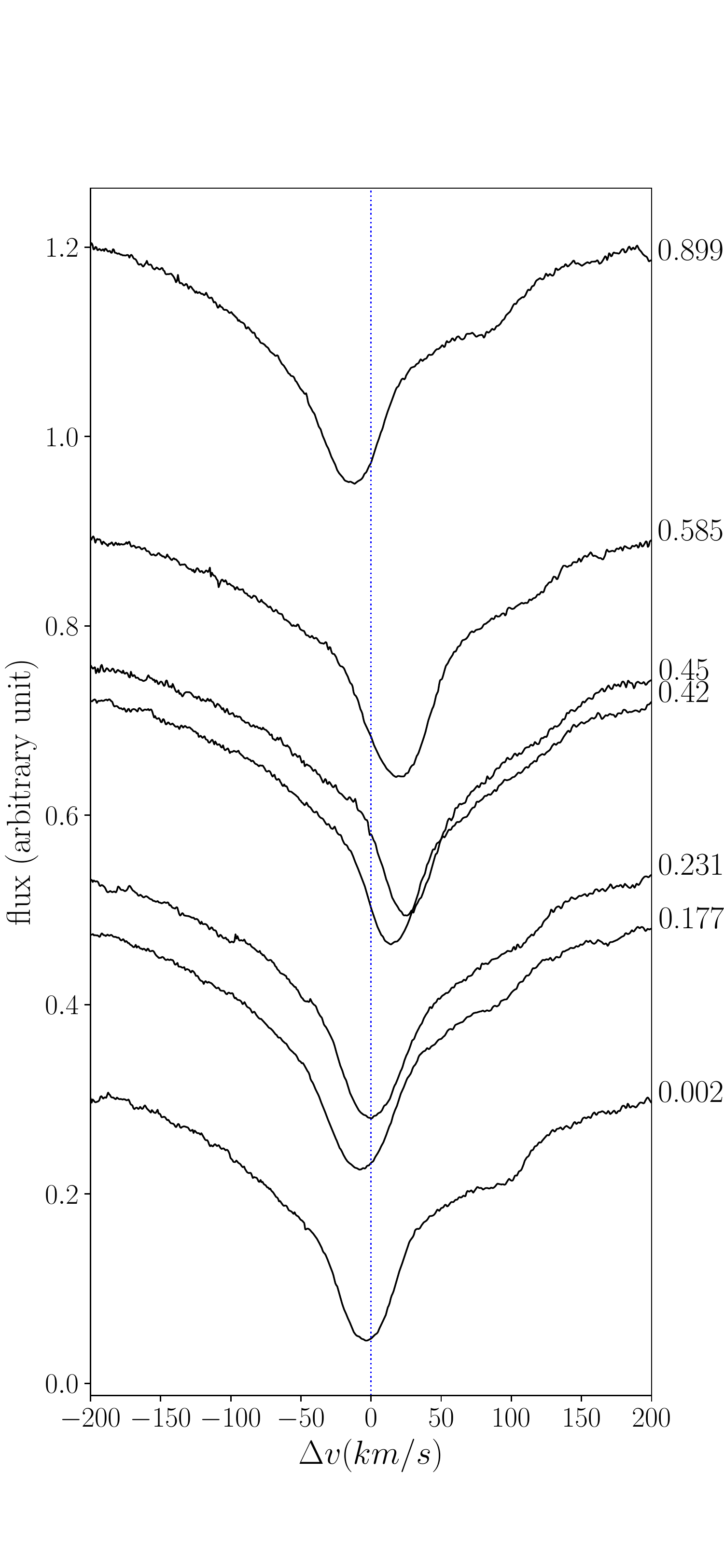}
         \caption{$\lambda$8662}
     \end{subfigure}
     \hfill
          \begin{subfigure}[b]{0.24\textwidth}
         \centering
         \includegraphics[width=\textwidth]{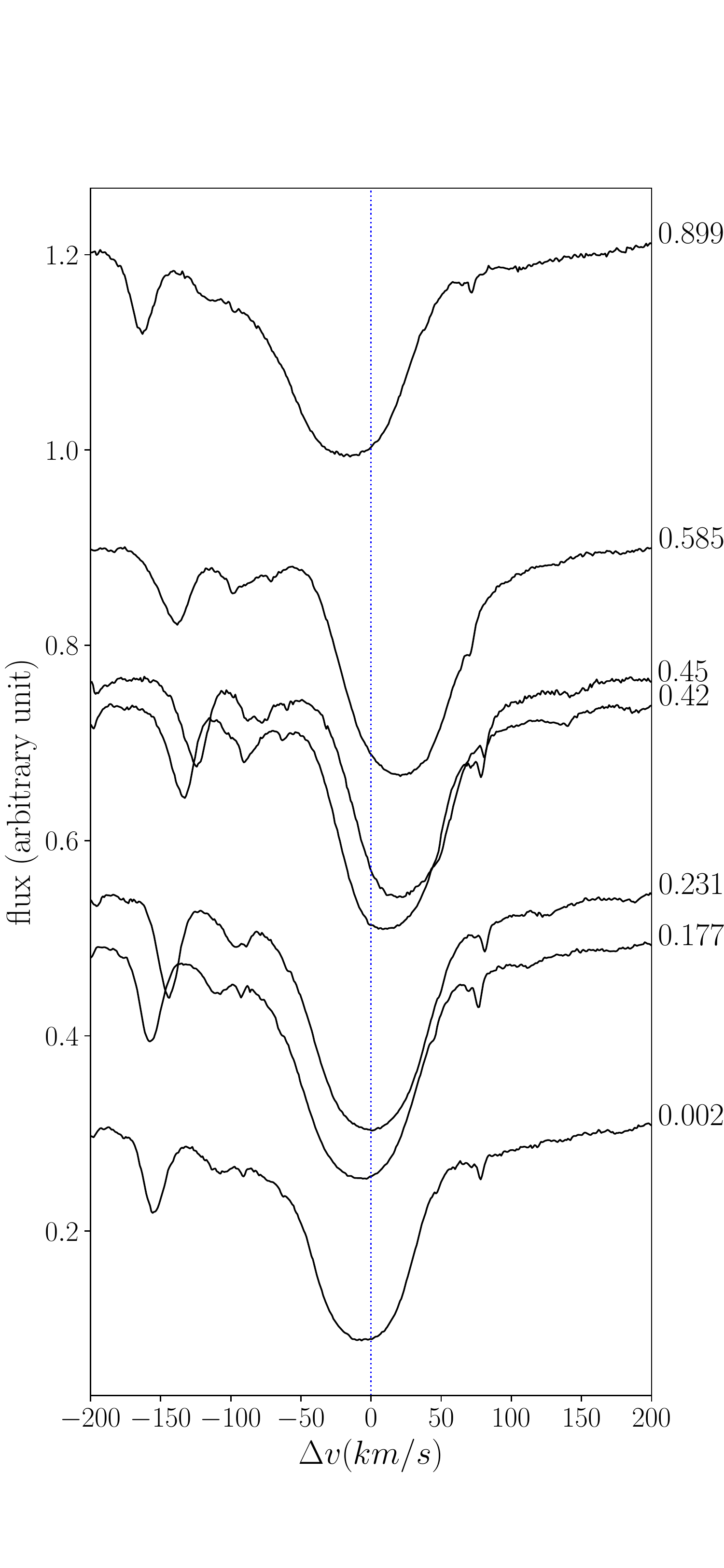}
         \caption{H$\alpha$}
     \end{subfigure}
        \caption{S Mus, 9.66d}
\end{figure*}

\begin{figure*}
     \centering
     \begin{subfigure}[b]{0.24\textwidth}
         \centering
         \includegraphics[width=\textwidth]{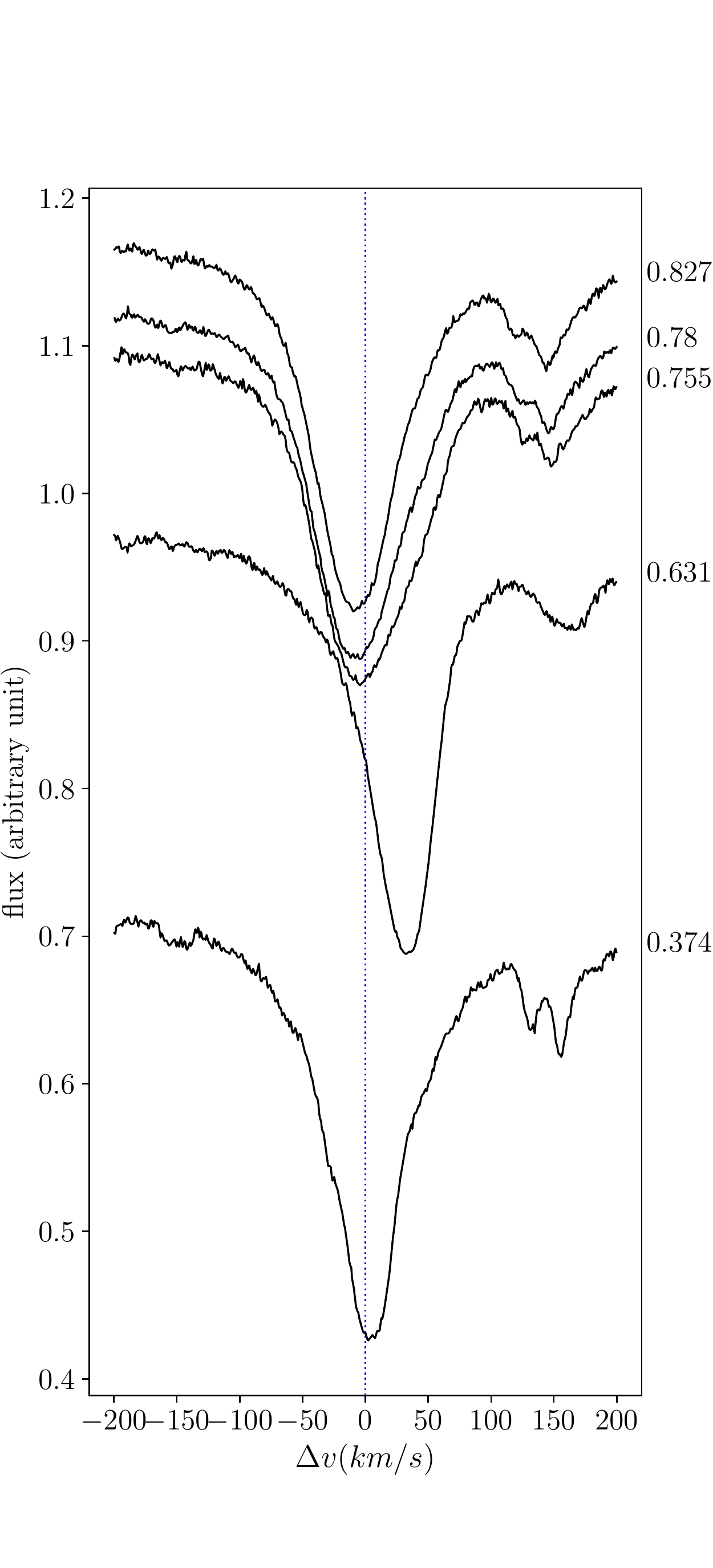}
         \caption{$\lambda$8498}
     \end{subfigure}
     \begin{subfigure}[b]{0.24\textwidth}
         \centering
         \includegraphics[width=\textwidth]{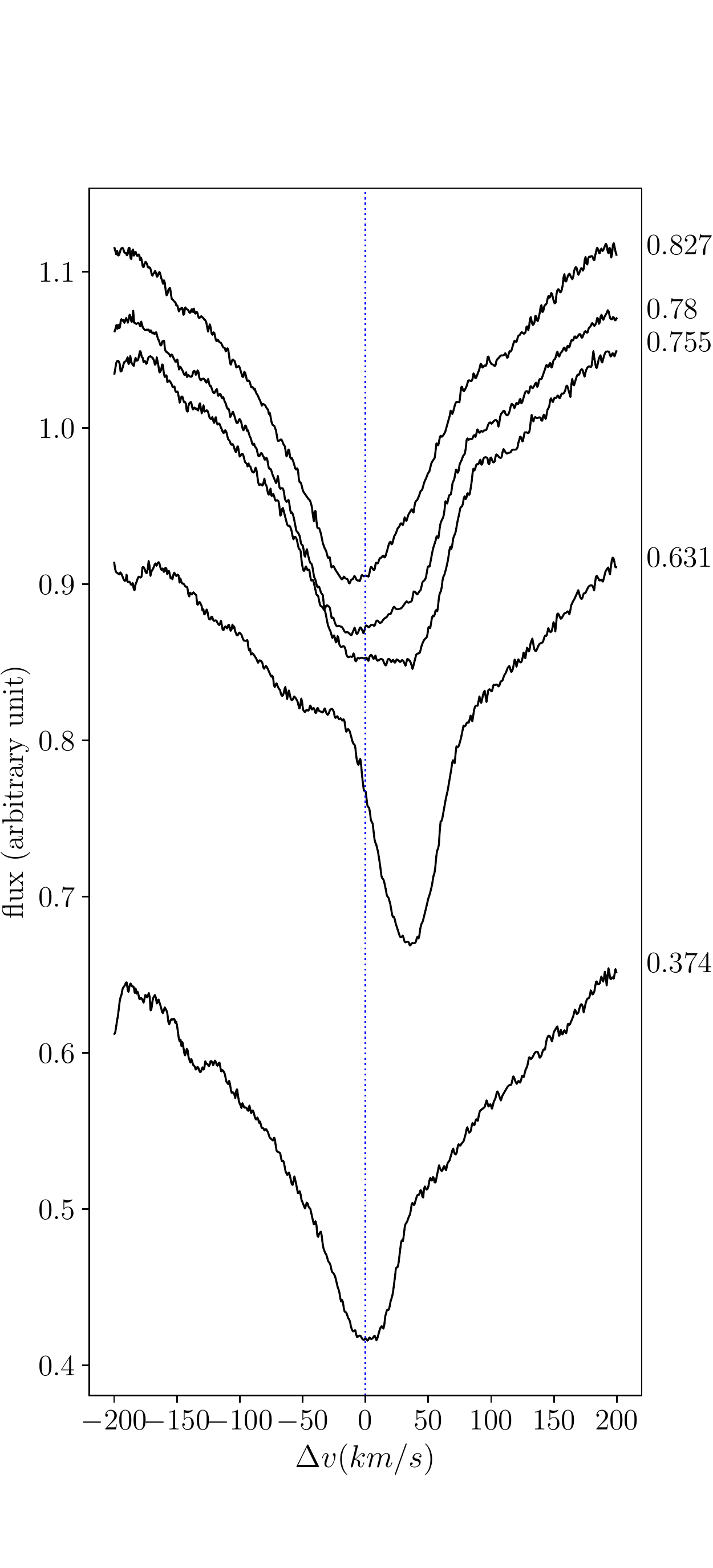}
         \caption{$\lambda$8542}
     \end{subfigure}
     \hfill
     \begin{subfigure}[b]{0.24\textwidth}
         \centering
         \includegraphics[width=\textwidth]{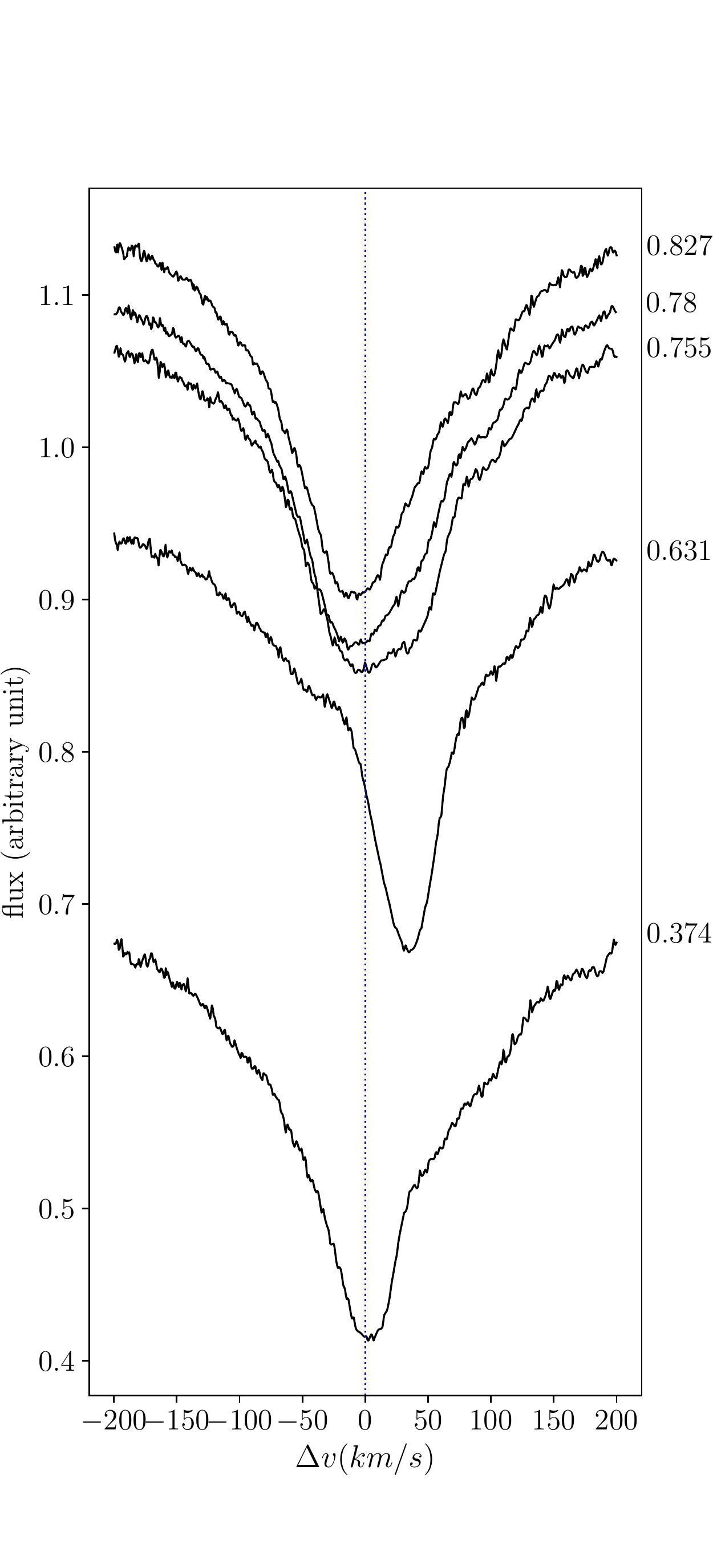}
         \caption{$\lambda$8662}
     \end{subfigure}
     \hfill
     \begin{subfigure}[b]{0.24\textwidth}
         \centering
         \includegraphics[width=\textwidth]{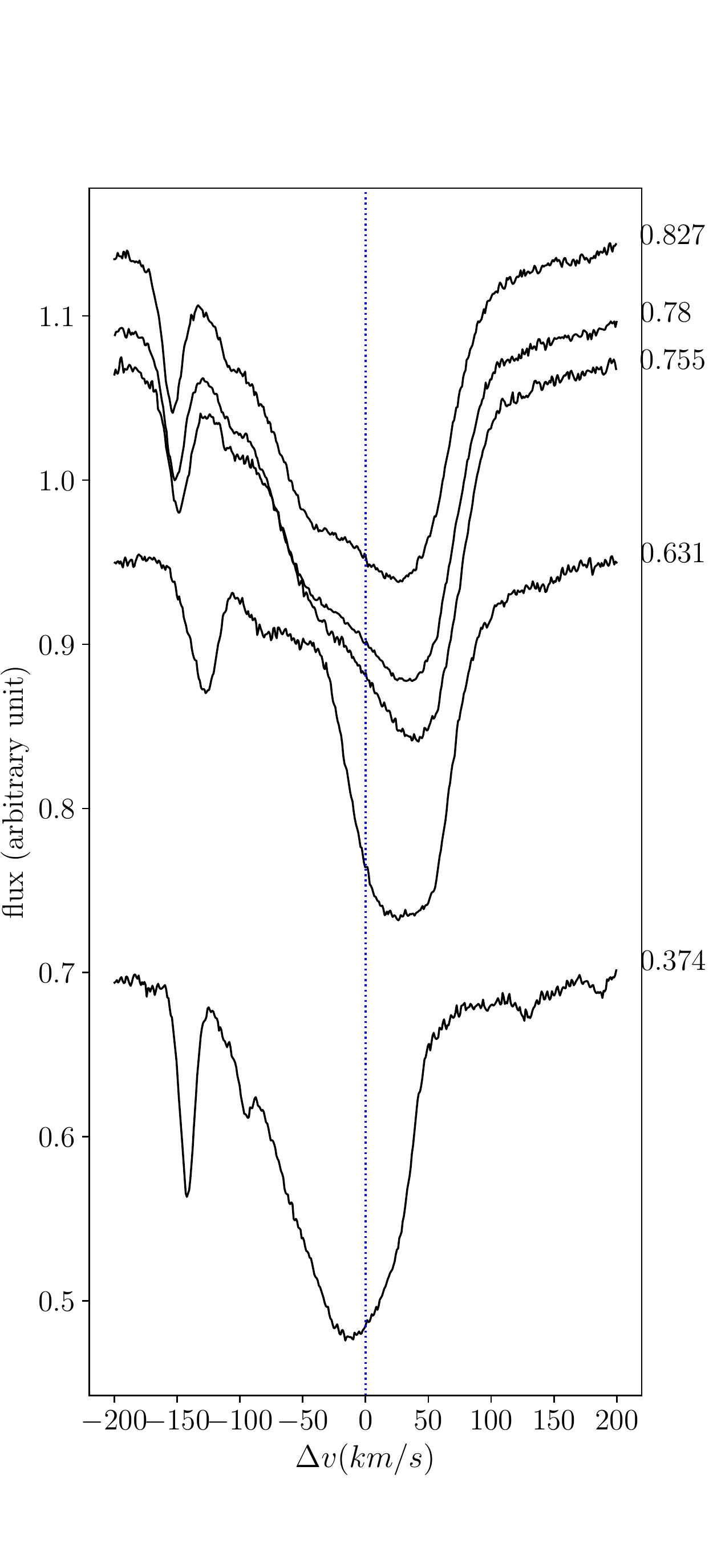}
         \caption{H$\alpha$}
     \end{subfigure}
        \caption{$\beta$ Dor, 9.84d          \label{fig:beta_dor}}
\end{figure*}

\begin{figure*}
     \centering
     \begin{subfigure}[b]{0.24\textwidth}
         \centering
         \includegraphics[width=\textwidth]{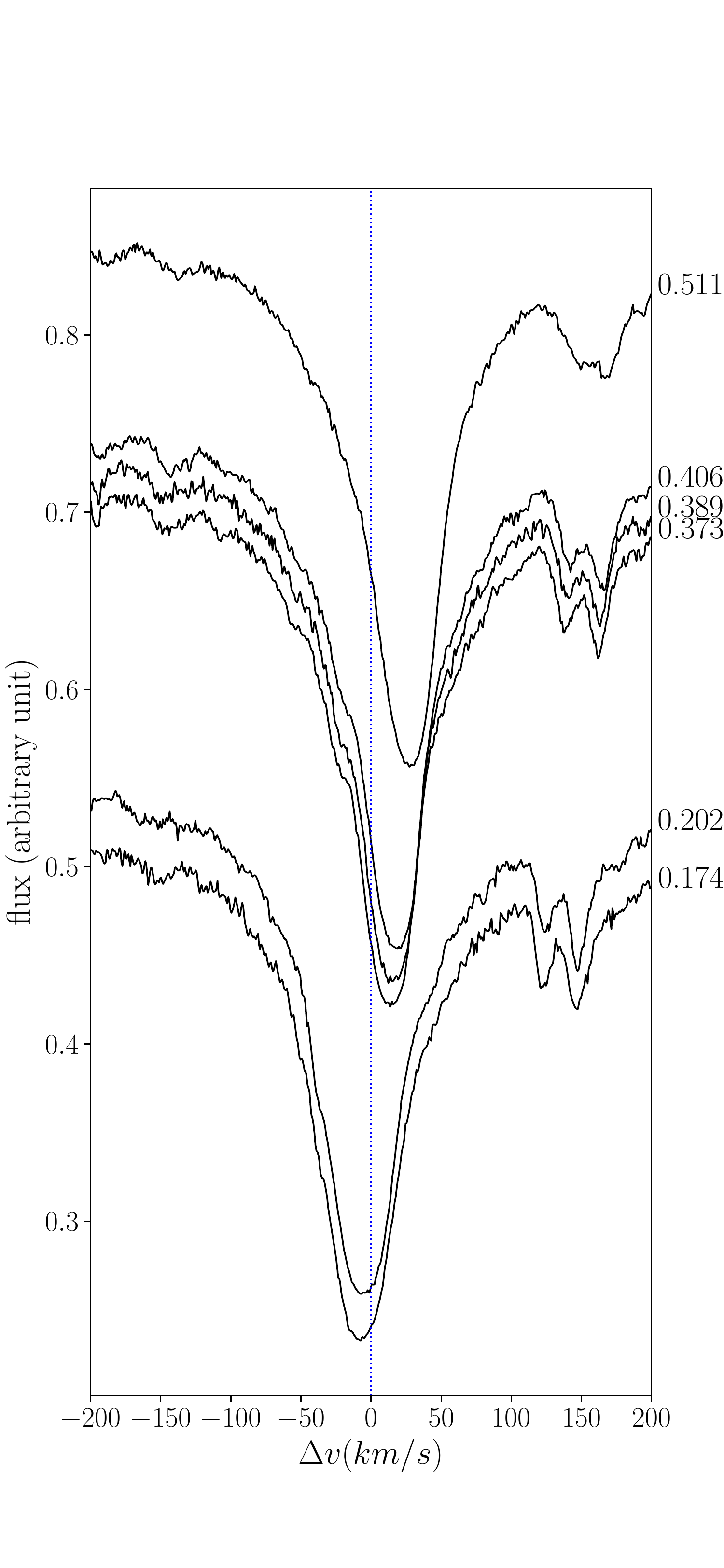}
         \caption{$\lambda$8498}

     \end{subfigure}
     \begin{subfigure}[b]{0.24\textwidth}
         \centering
         \includegraphics[width=\textwidth]{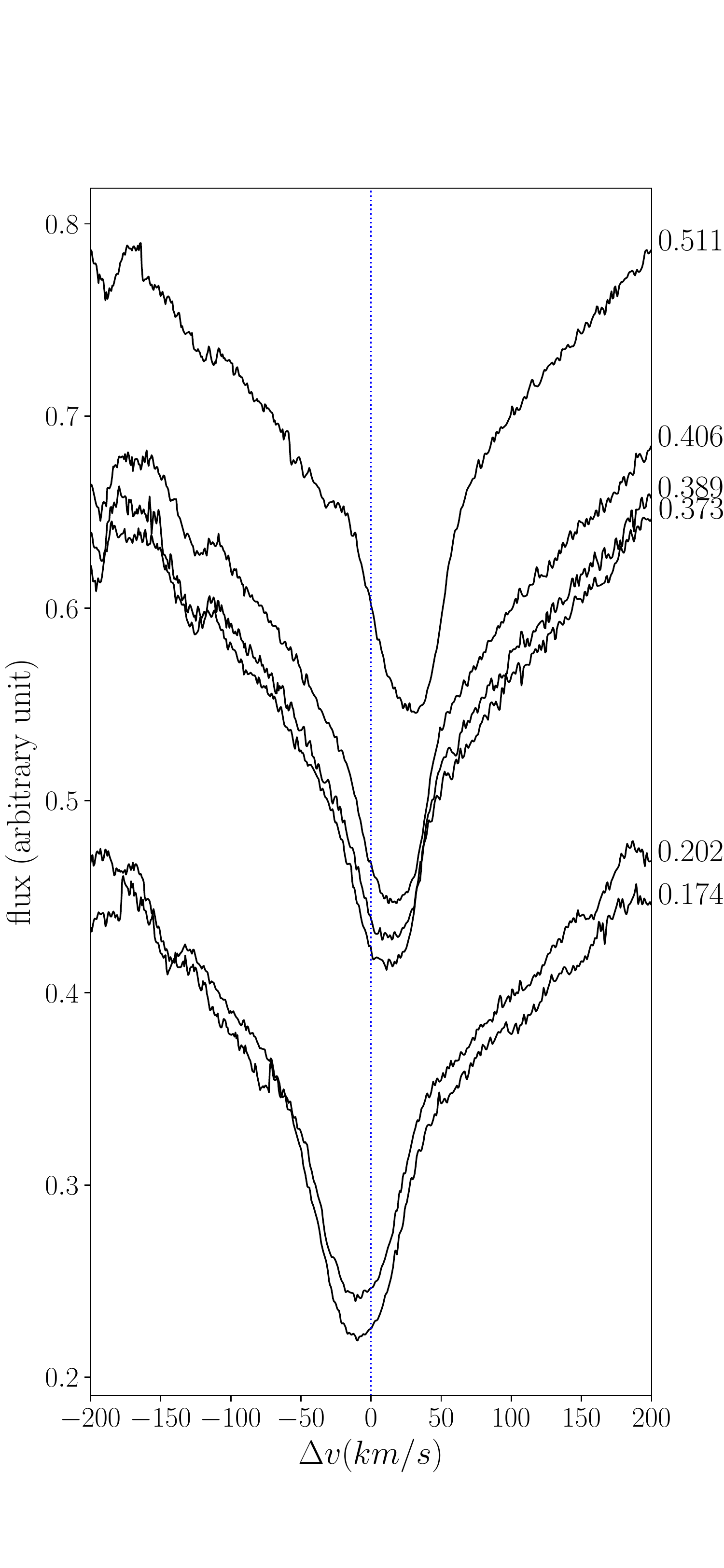}
         \caption{$\lambda$8542}

     \end{subfigure}
     \hfill
     \begin{subfigure}[b]{0.24\textwidth}
         \centering
         \includegraphics[width=\textwidth]{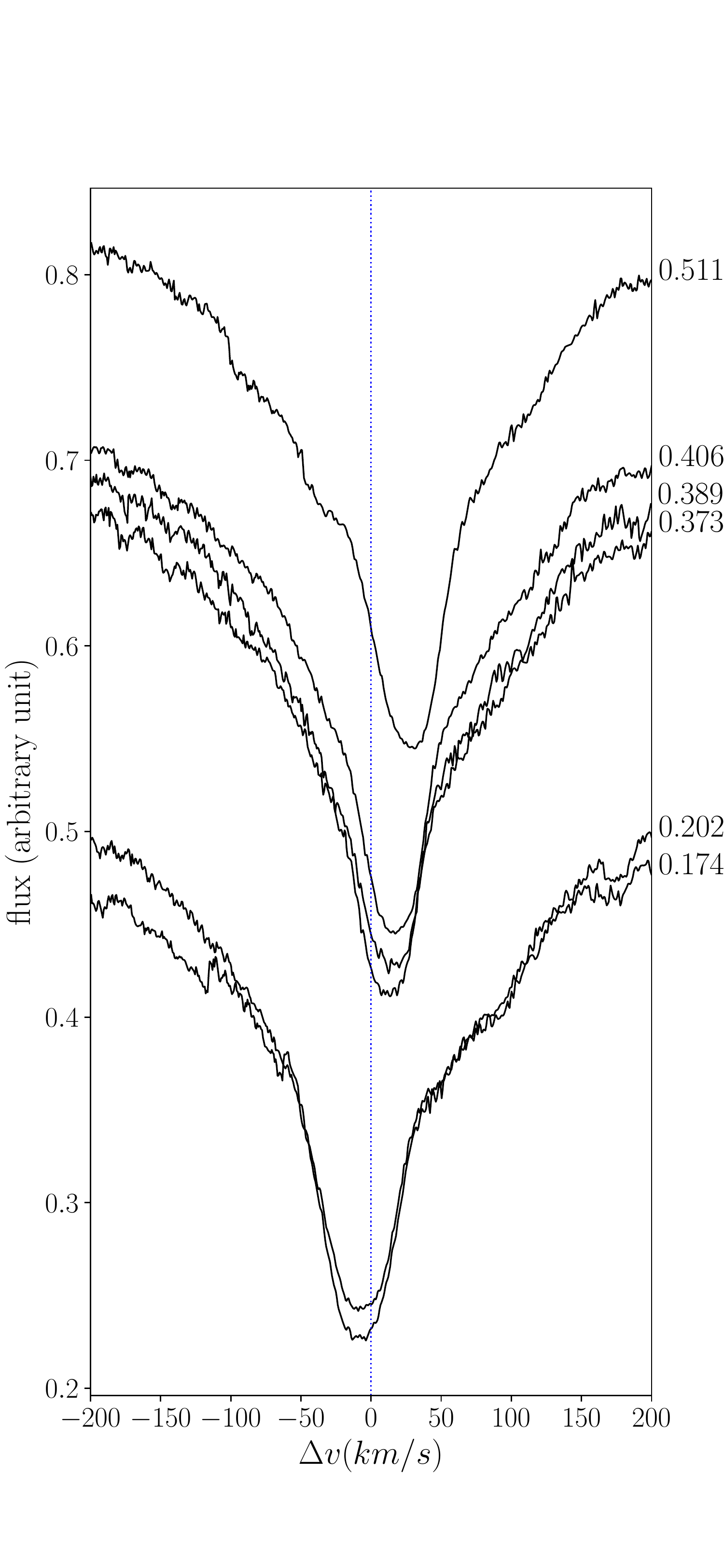}
         \caption{$\lambda$8662}
     \end{subfigure}
     \hfill
     \begin{subfigure}[b]{0.24\textwidth}
         \centering
         \includegraphics[width=\textwidth]{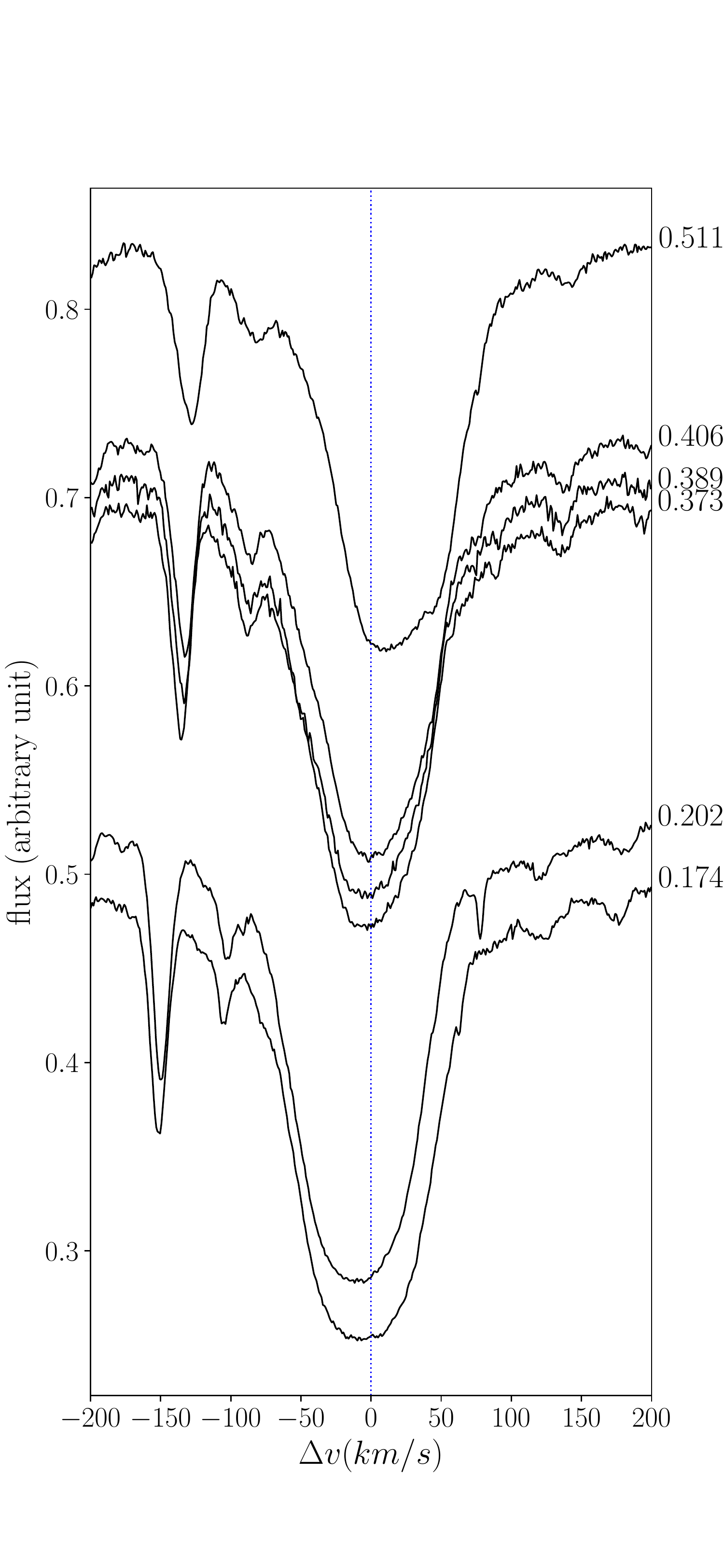}
         \caption{H$\alpha$}
     \end{subfigure}
        \caption{$\zeta$ Gem, 10.15d \label{fig:zeta_gem}}
\end{figure*}

\begin{figure*}
     \centering
         \begin{subfigure}[b]{0.24\textwidth}
         \centering
         \includegraphics[width=\textwidth]{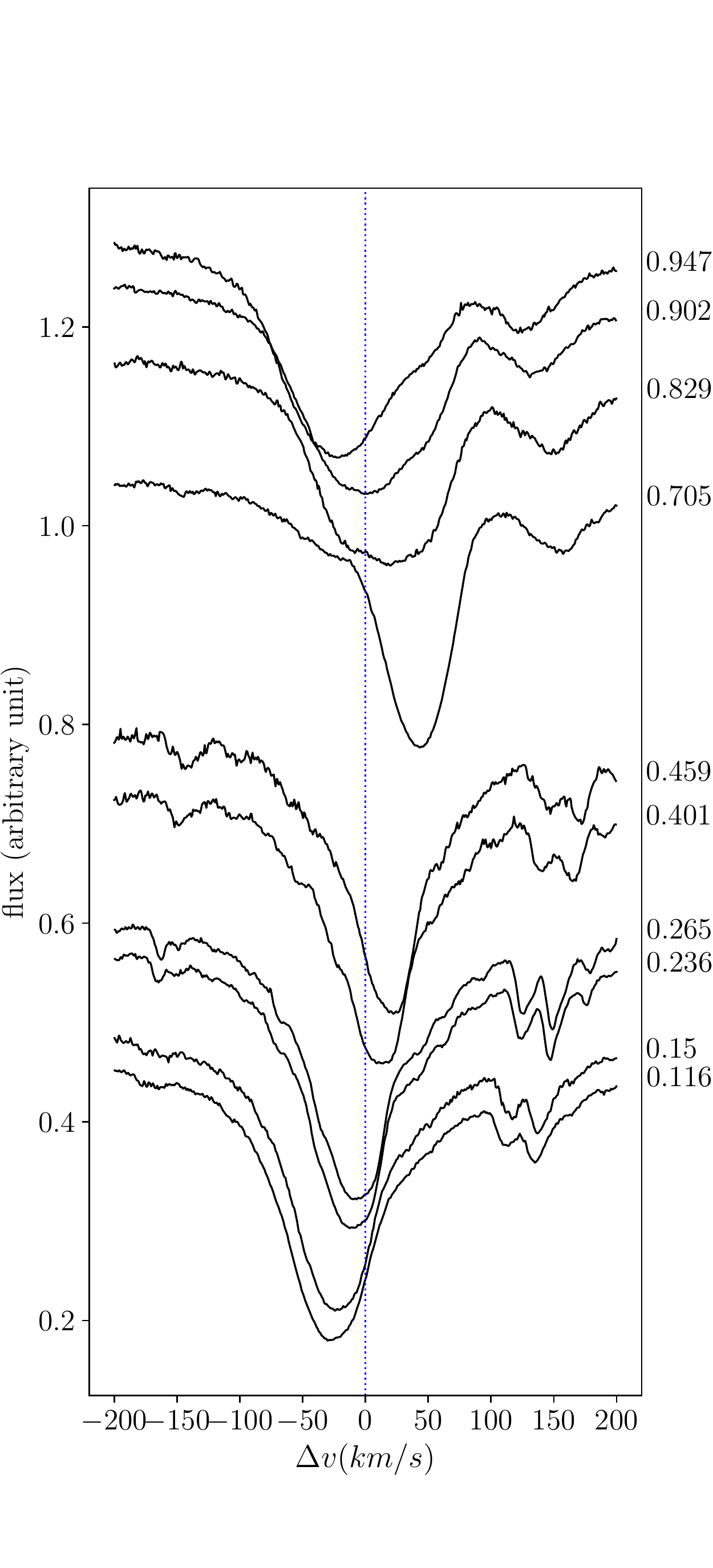}
         \caption{$\lambda$8498}
     \end{subfigure}
     \hfill
     \begin{subfigure}[b]{0.24\textwidth}
         \centering
         \includegraphics[width=\textwidth]{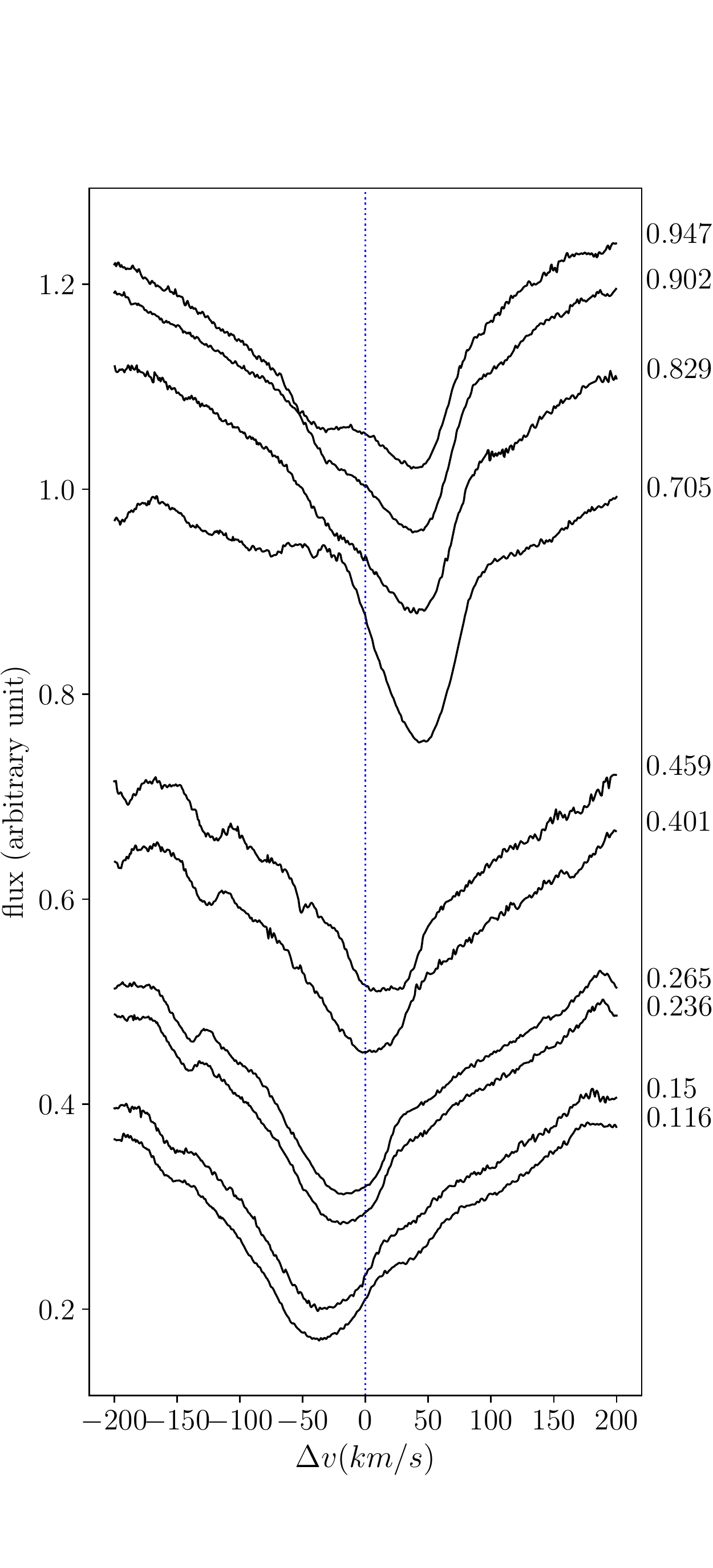}
         \caption{$\lambda$8542}
     \end{subfigure}
     \hfill
     \begin{subfigure}[b]{0.24\textwidth}
         \centering
         \includegraphics[width=\textwidth]{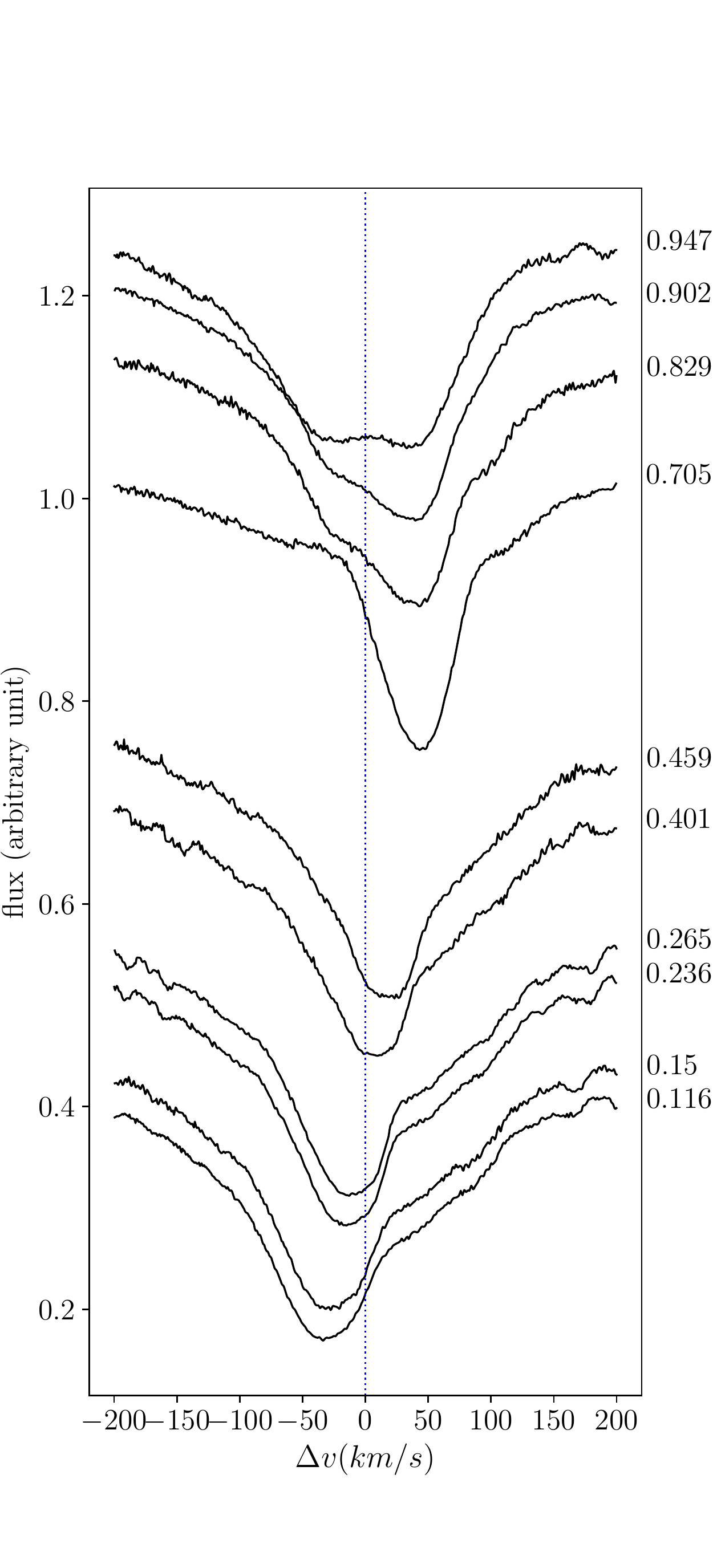}
         \caption{$\lambda$8662}
     \end{subfigure}
     \hfill
     \begin{subfigure}[b]{0.24\textwidth}
         \centering
         \includegraphics[width=\textwidth]{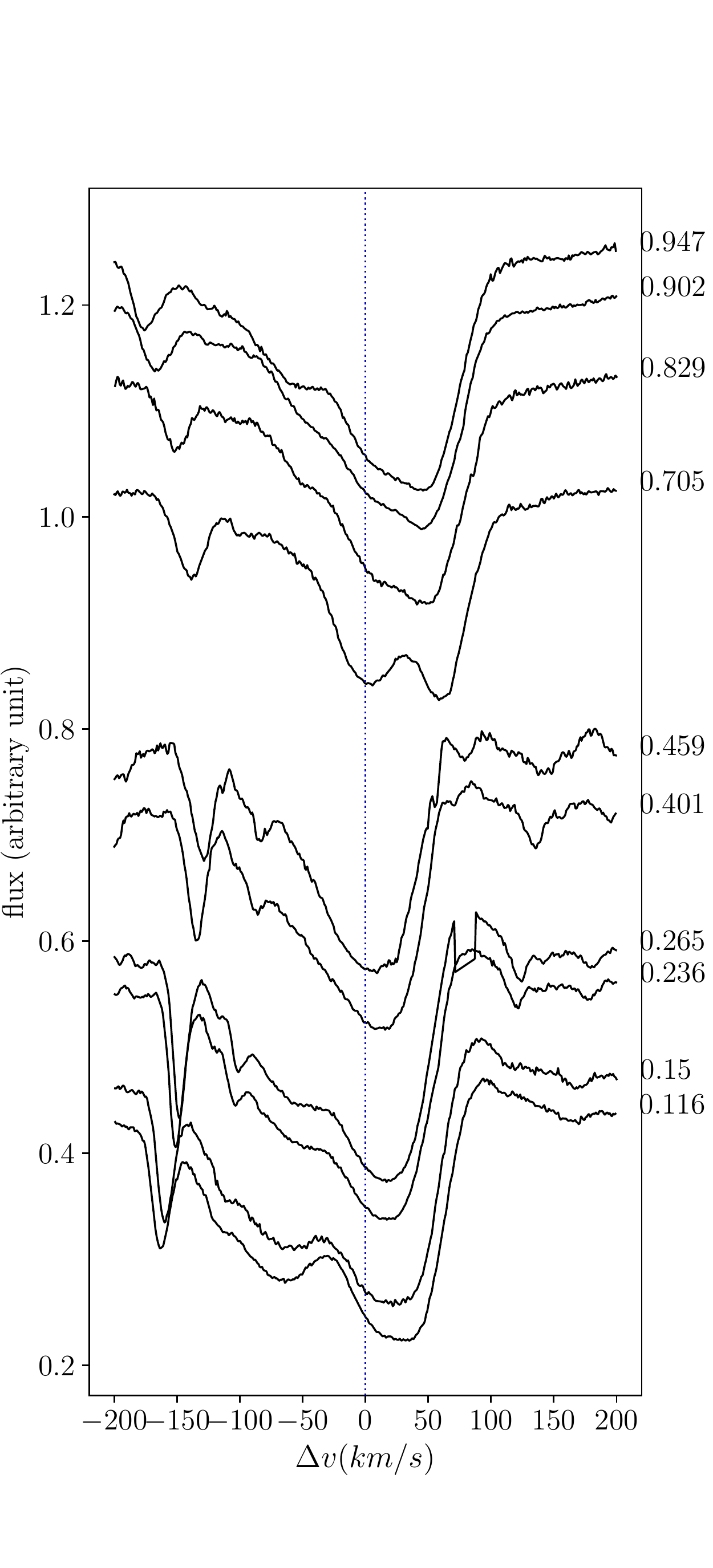}
         \caption{H$\alpha$}
     \end{subfigure}
        \caption{TT Aql, 13.75d}
        \label{fig:tt_aql}
\end{figure*}

\begin{figure*}
     \centering
          \begin{subfigure}[b]{0.24\textwidth}
         \centering
         \includegraphics[width=\textwidth]{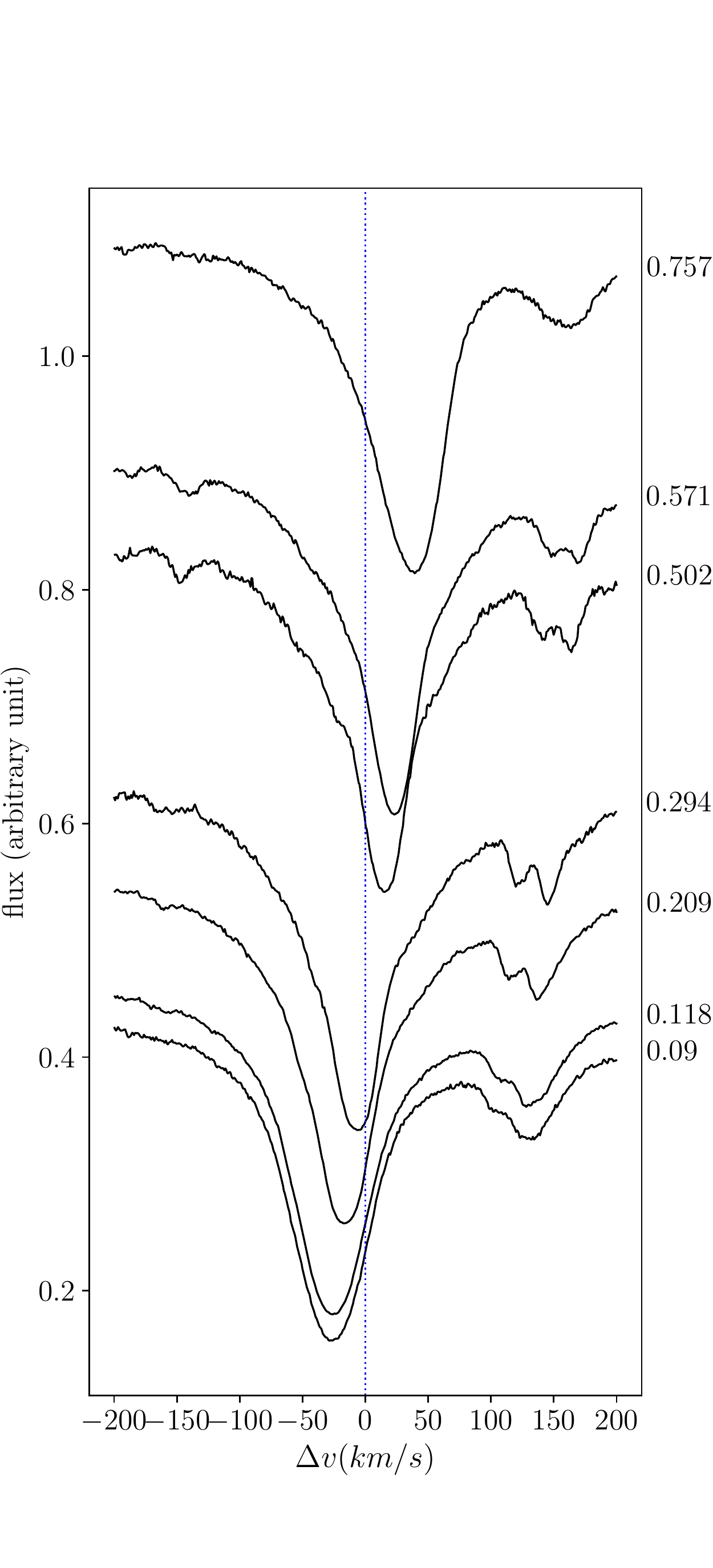}
         \caption{$\lambda$8498}
     \end{subfigure}
     \hfill
     \begin{subfigure}[b]{0.24\textwidth}
         \centering
         \includegraphics[width=\textwidth]{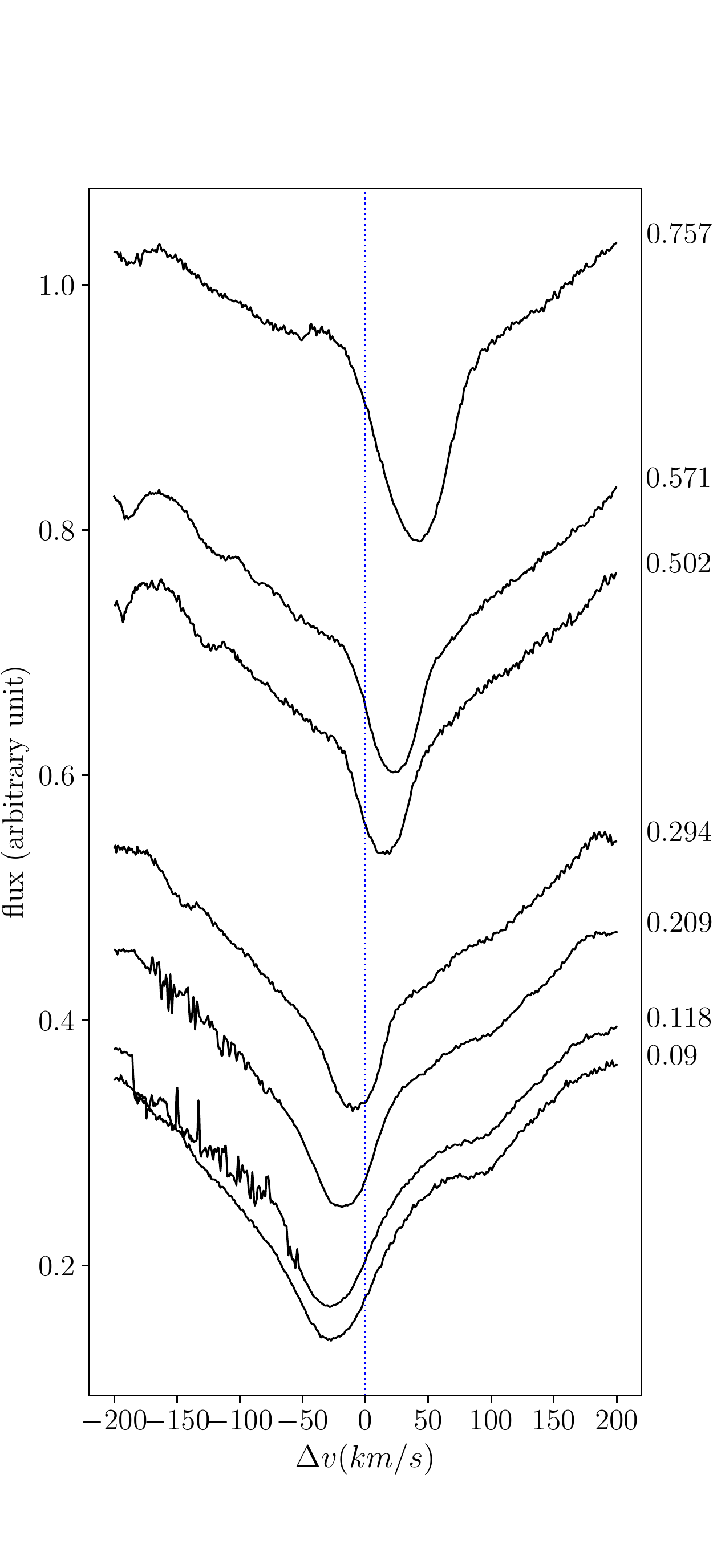}
         \caption{$\lambda$8542}
     \end{subfigure}
     \hfill
     \begin{subfigure}[b]{0.24\textwidth}
         \centering
         \includegraphics[width=\textwidth]{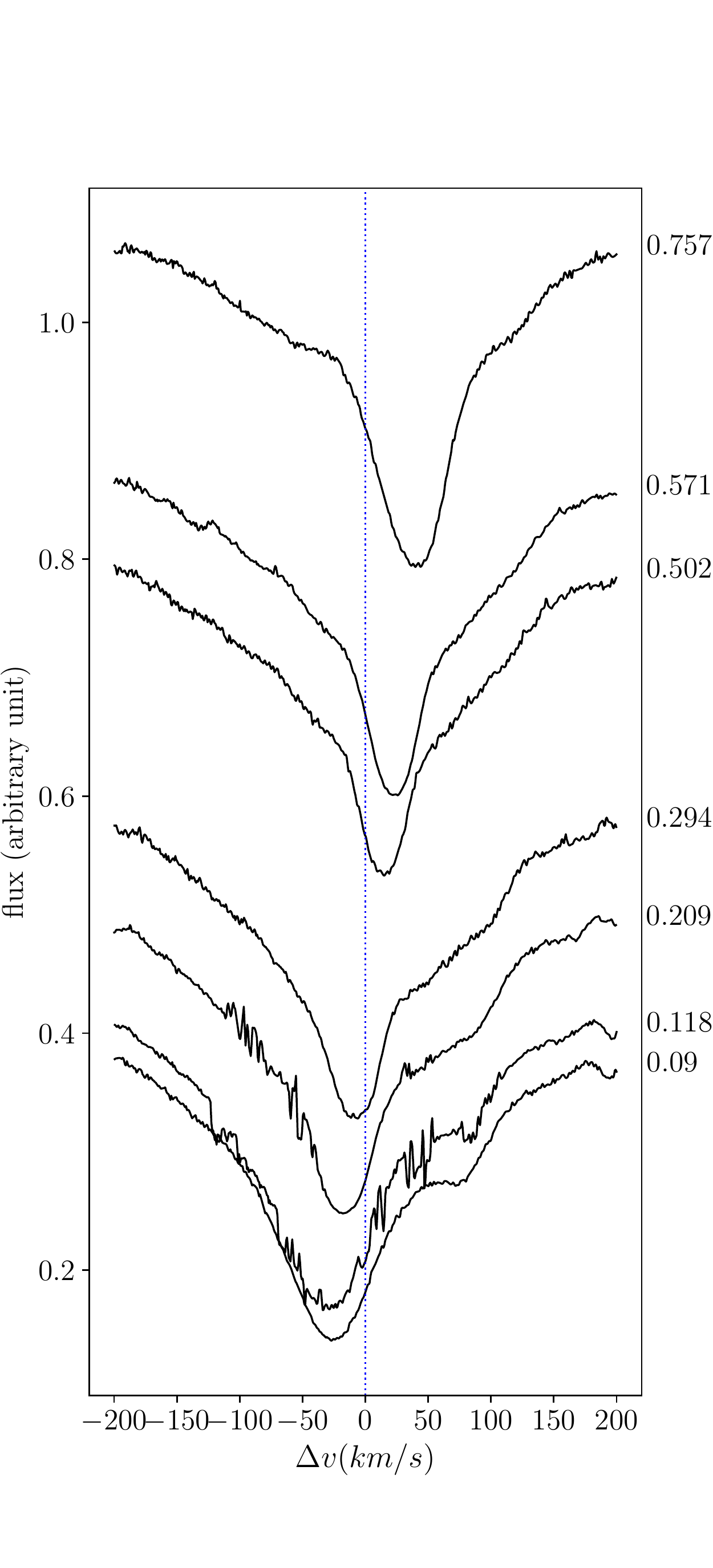}
         \caption{$\lambda$8662}
     \end{subfigure}
     \hfill
         \begin{subfigure}[b]{0.24\textwidth}
         \centering
         \includegraphics[width=\textwidth]{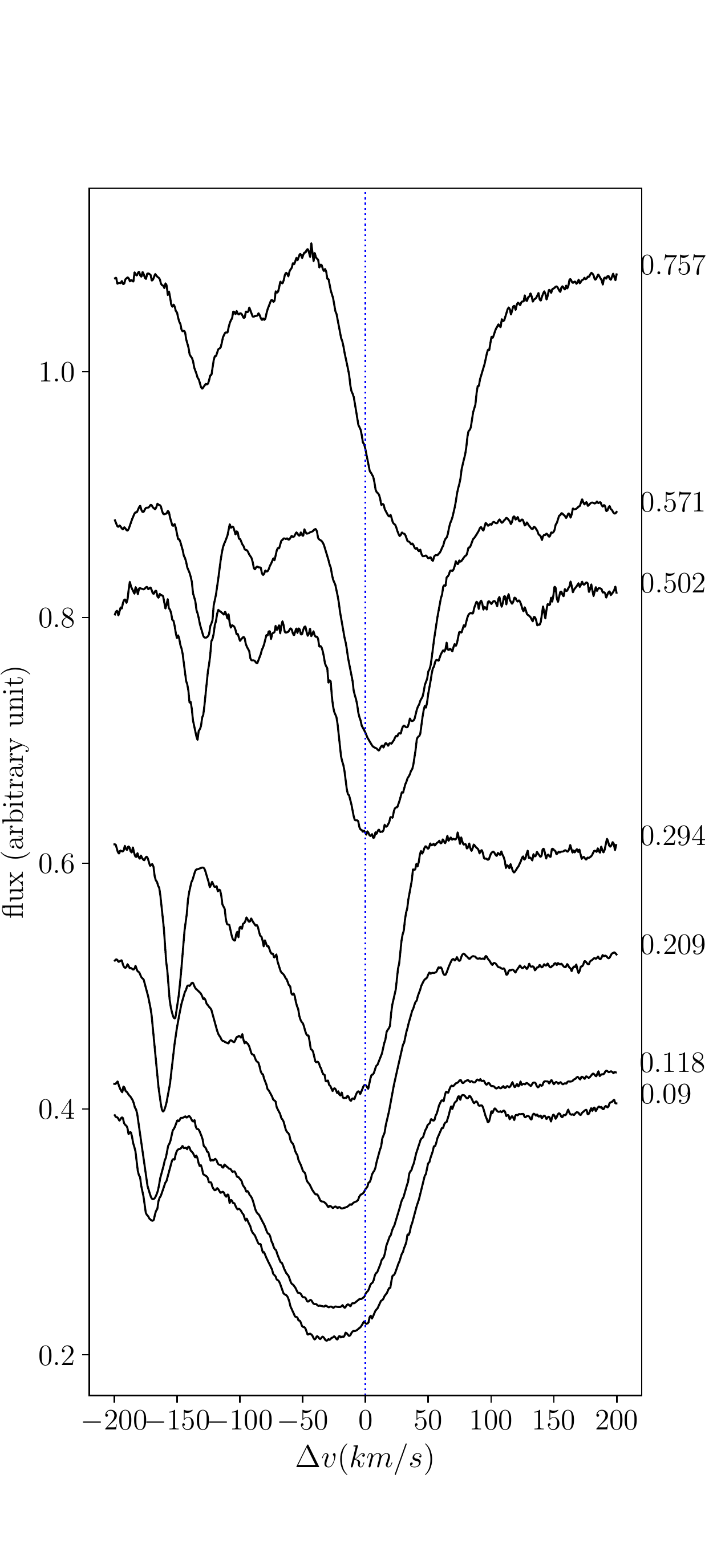}
         \caption{H$\alpha$}
         \label{fig:Ha_ru_sct}
     \end{subfigure}
        \caption{RU Sct, 19.70d}\label{fig:ru_sct}
\end{figure*}

\begin{figure*}
     \centering

          \begin{subfigure}[b]{0.24\textwidth}
         \centering
         \includegraphics[width=\textwidth]{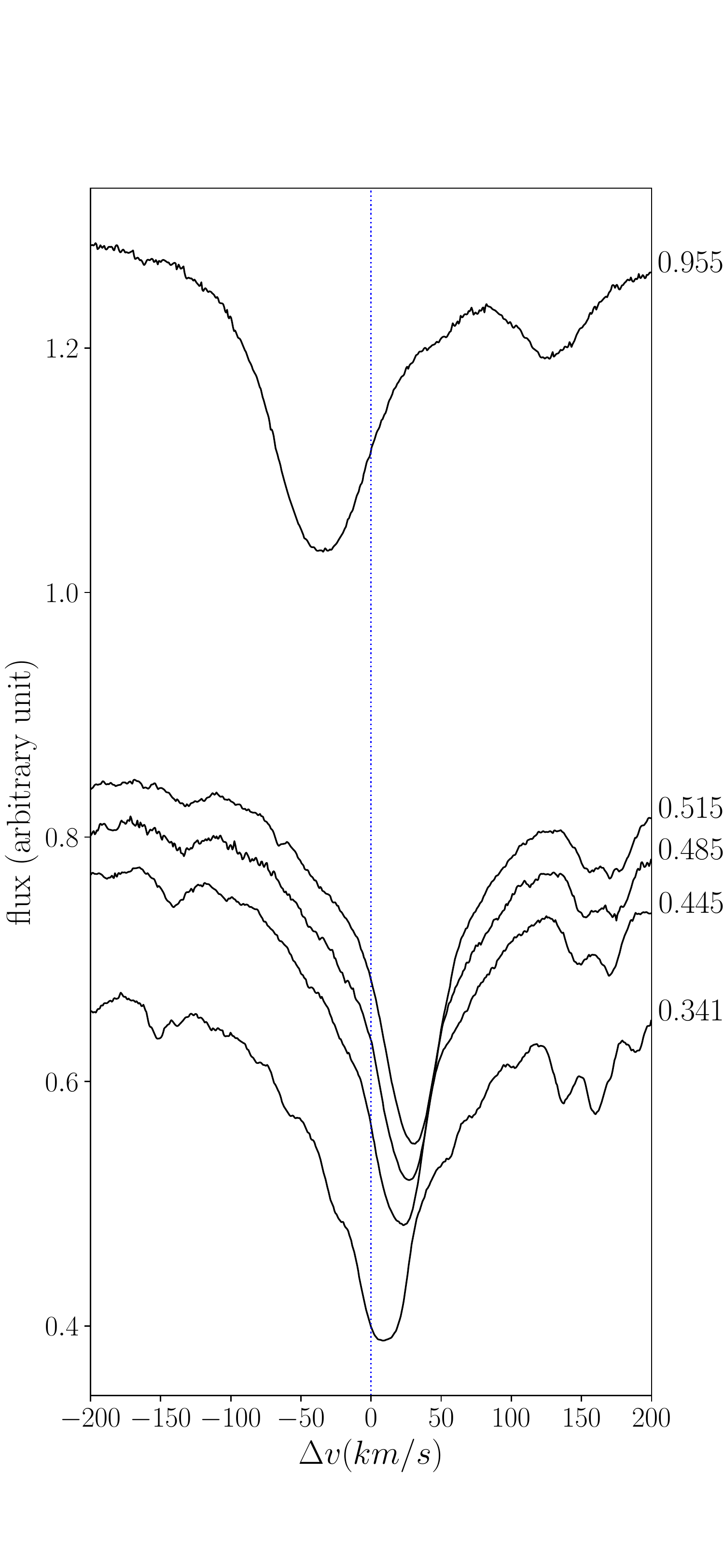}
         \caption{$\lambda$8498}
     \end{subfigure}
     \hfill
     \begin{subfigure}[b]{0.24\textwidth}
         \centering
         \includegraphics[width=\textwidth]{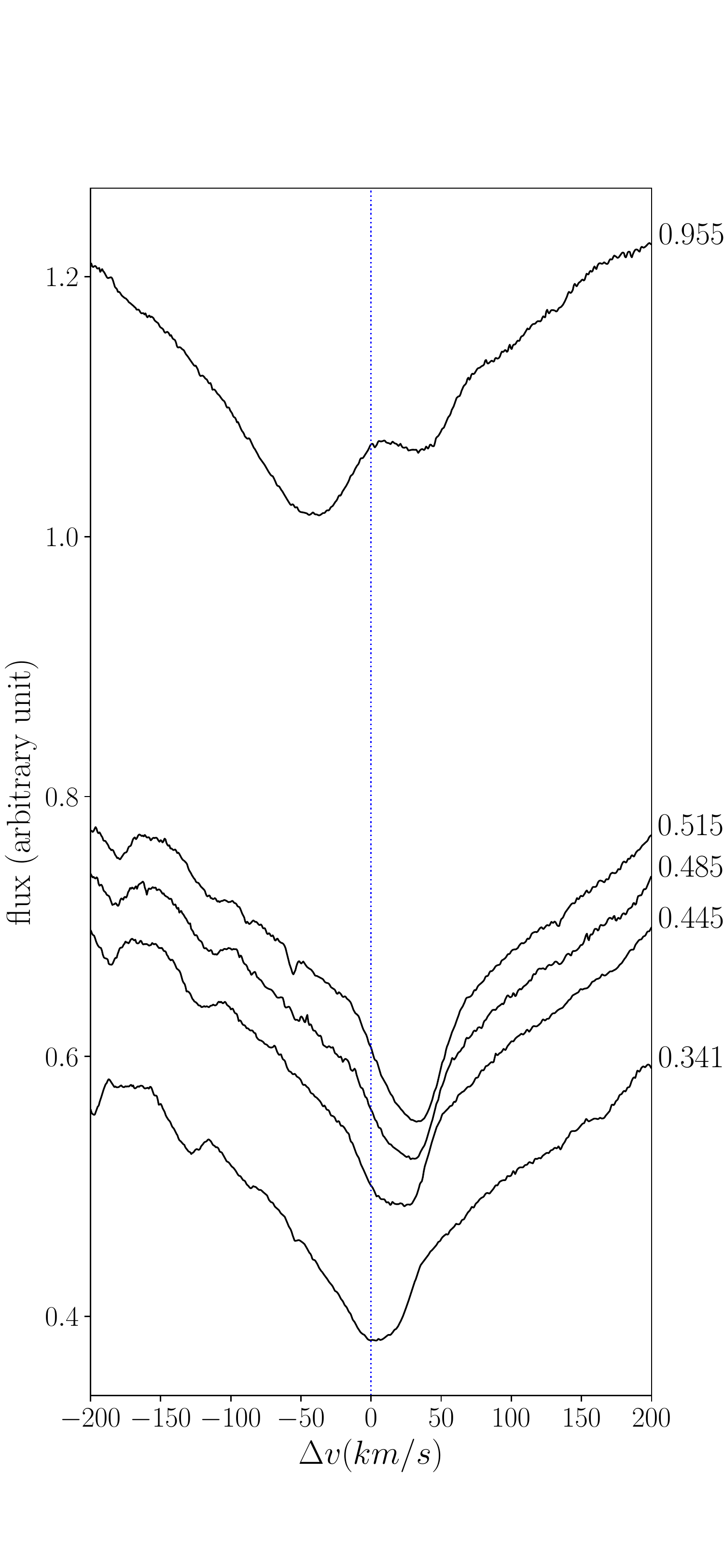}
         \caption{$\lambda$8542}
     \end{subfigure}
     \hfill
     \begin{subfigure}[b]{0.24\textwidth}
         \centering
         \includegraphics[width=\textwidth]{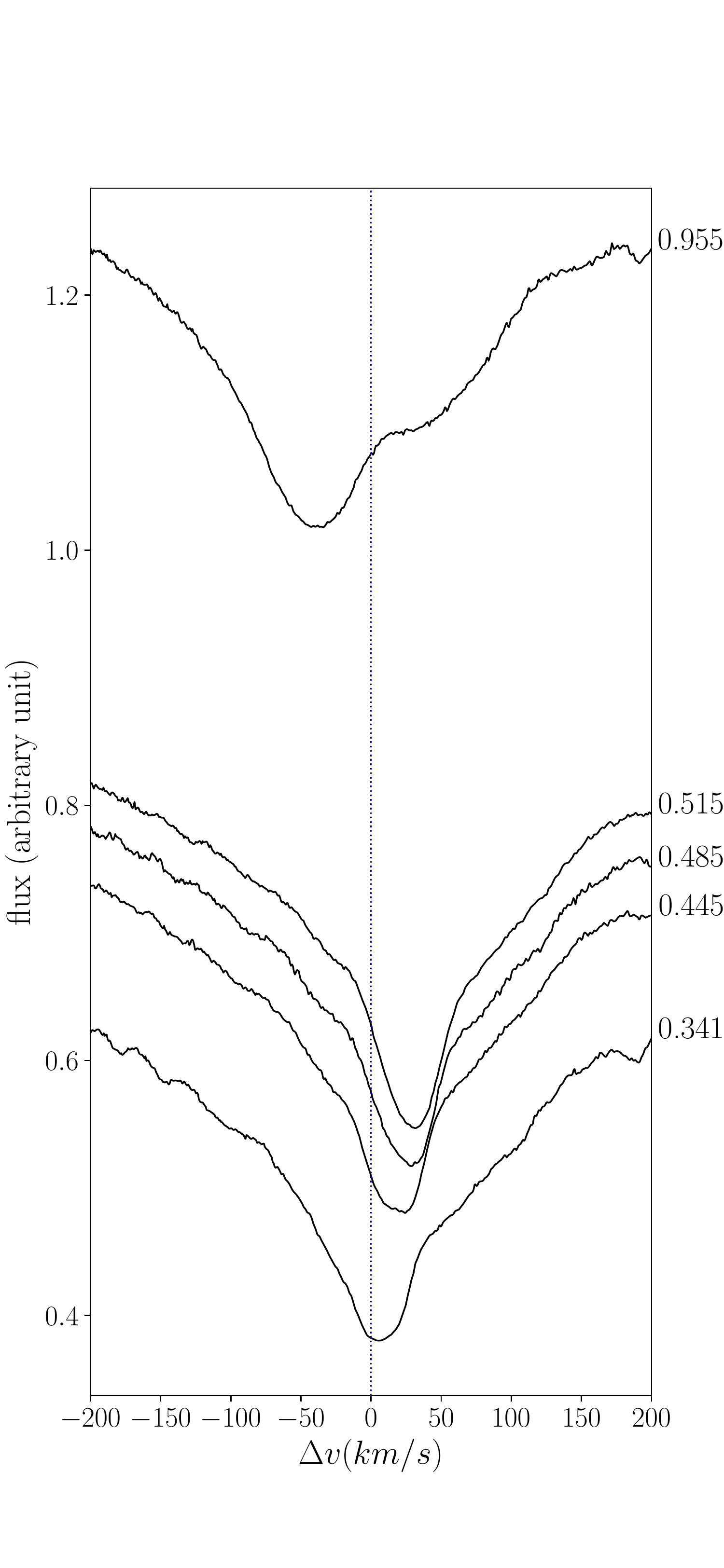}
         \caption{$\lambda$8662}
     \end{subfigure}
     \begin{subfigure}[b]{0.24\textwidth}
         \centering
         \includegraphics[width=\textwidth]{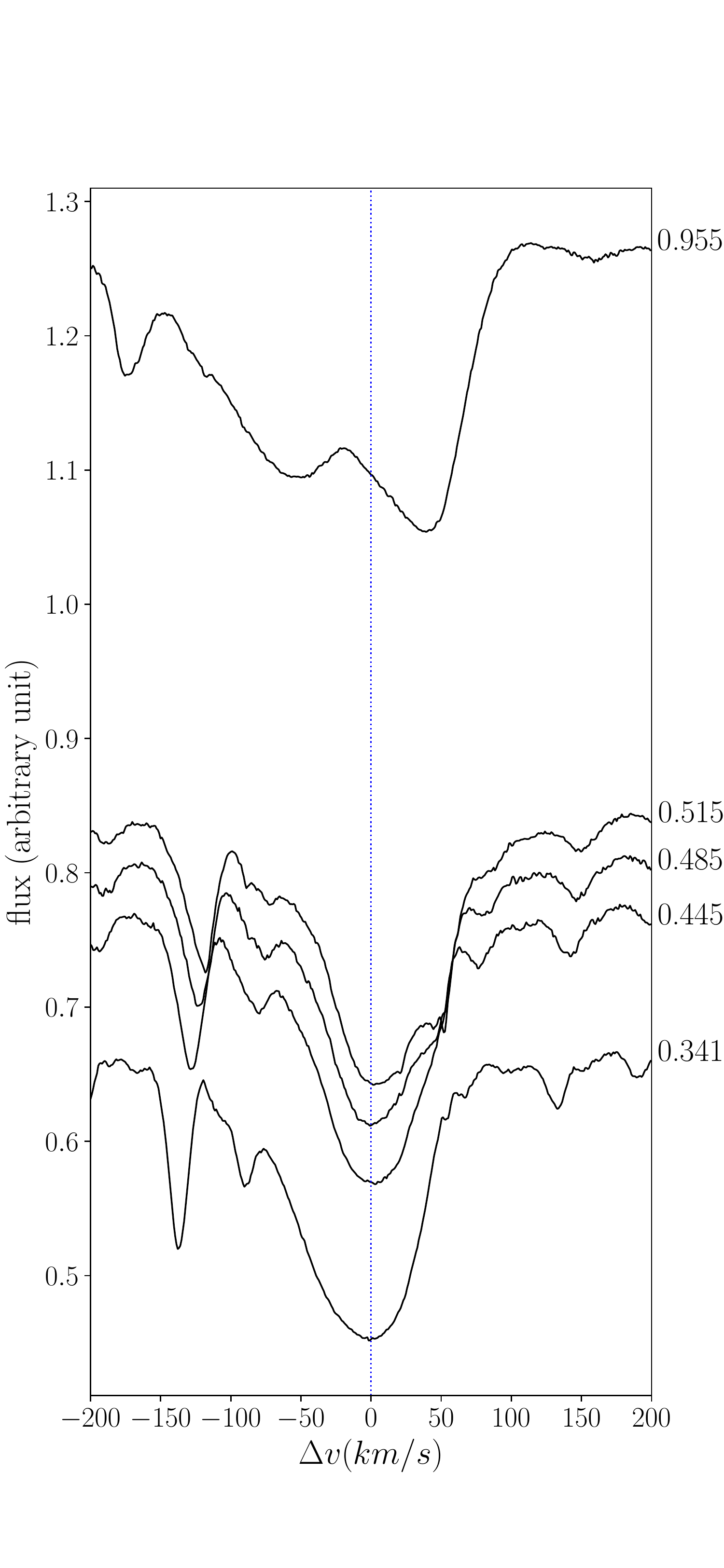}
         \caption{H$\alpha$}
     \end{subfigure}
        \caption{RZ Vel, 20.39d}
\end{figure*}

\begin{figure*}
     \centering
          \begin{subfigure}[b]{0.24\textwidth}
         \centering
         \includegraphics[width=\textwidth]{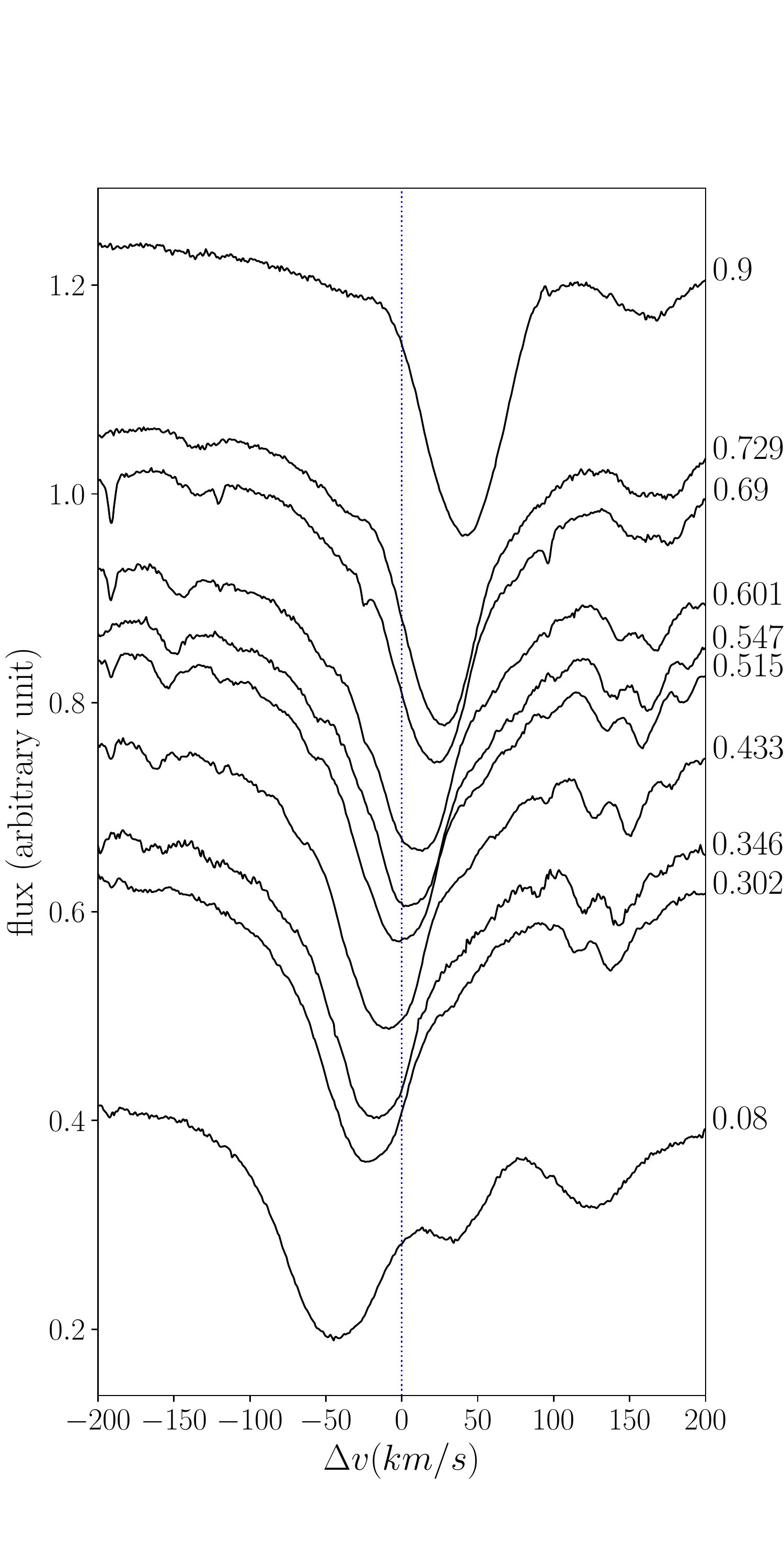}
         \caption{$\lambda$8498}
     \end{subfigure}
     \hfill
     \begin{subfigure}[b]{0.24\textwidth}
         \centering
         \includegraphics[width=\textwidth]{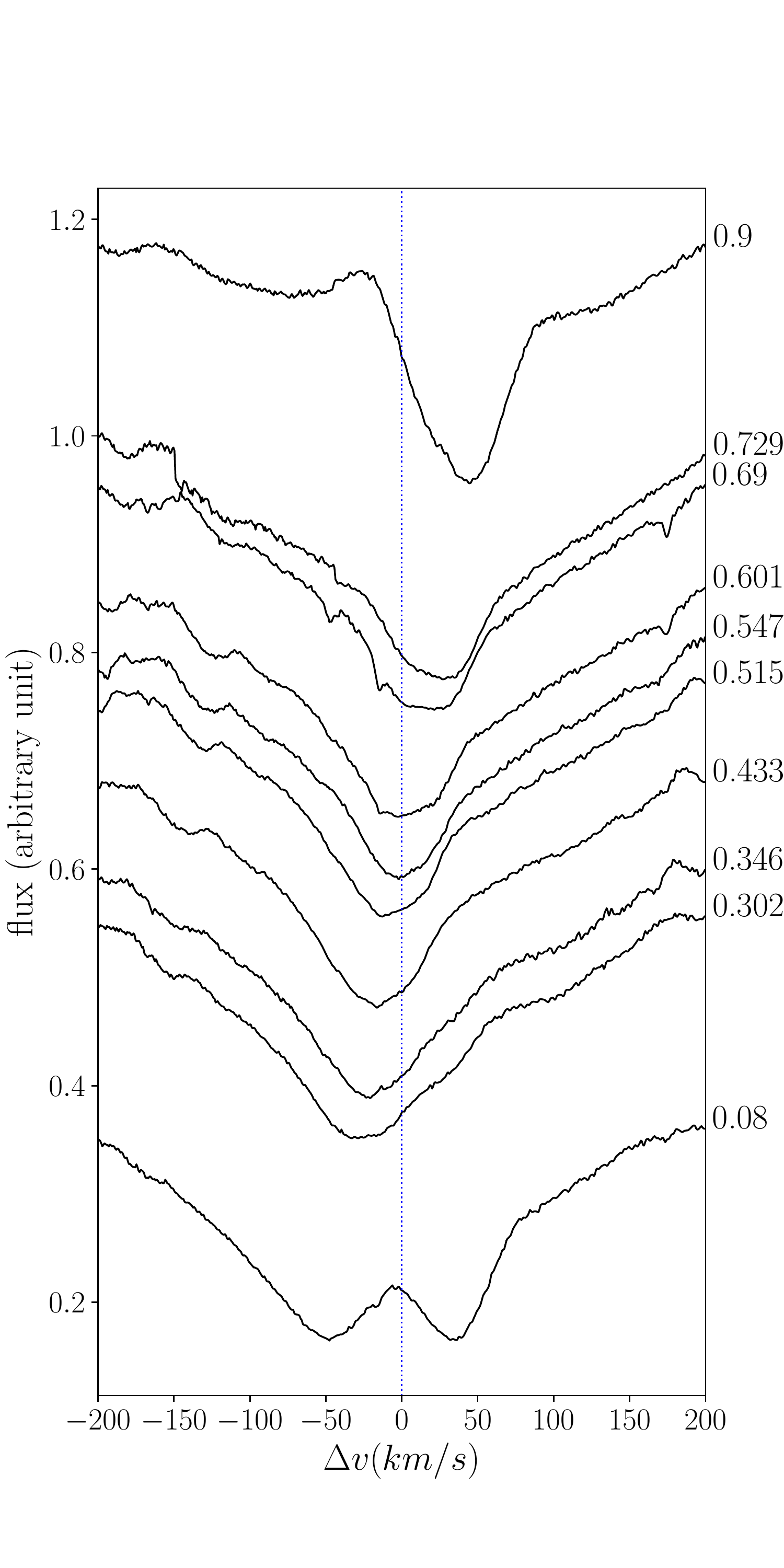}
         \caption{$\lambda$8542}
     \end{subfigure}
     \hfill
     \begin{subfigure}[b]{0.24\textwidth}
         \centering
         \includegraphics[width=\textwidth]{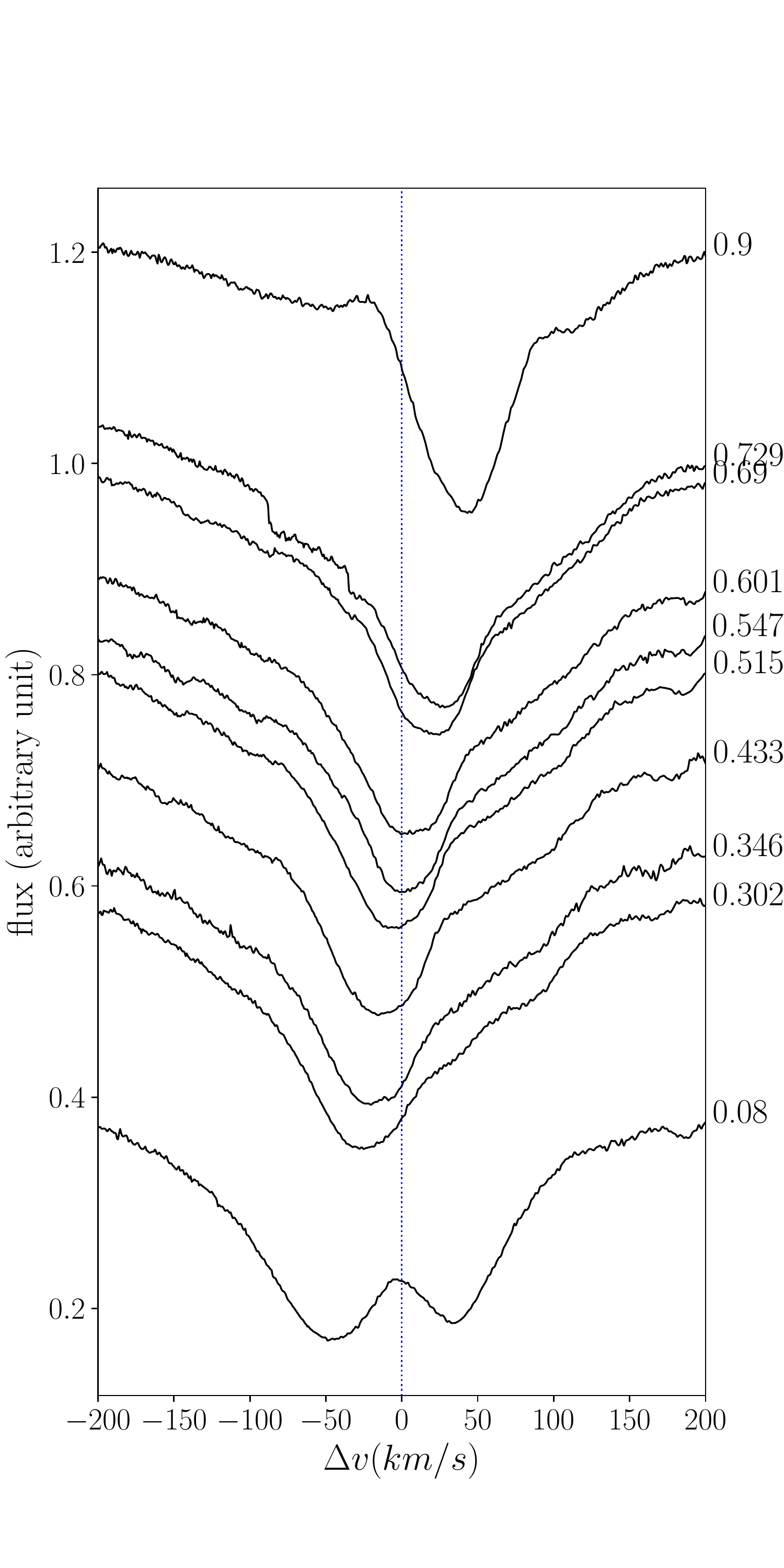}
         \caption{$\lambda$8662}
     \end{subfigure}
     \hfill
         \begin{subfigure}[b]{0.24\textwidth}
         \centering
         \includegraphics[width=\textwidth]{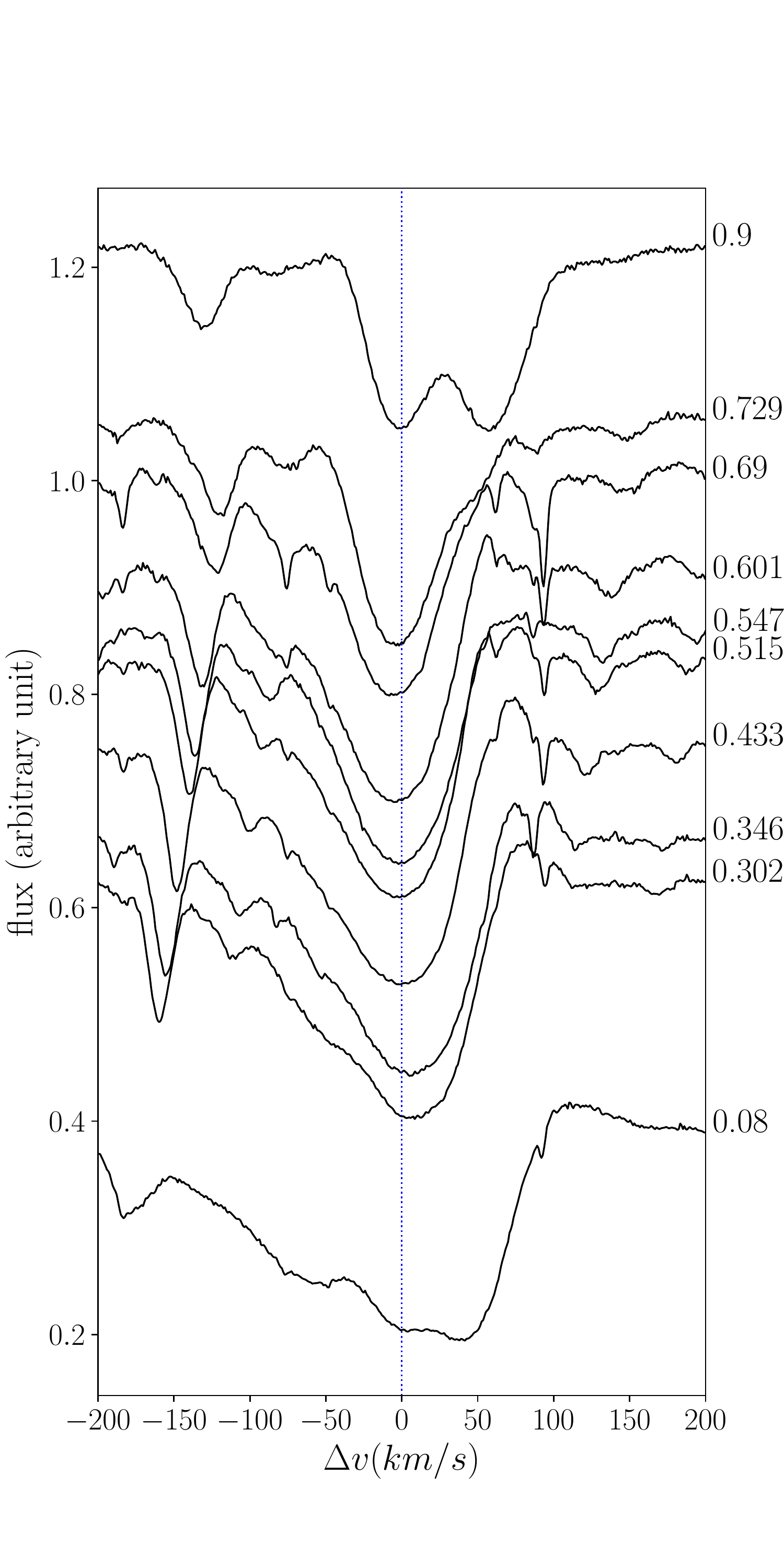}
         \caption{H$\alpha$}
     \end{subfigure}
        \caption{WZ Car, 23.01d}\label{fig:wz_car}
\end{figure*}

\begin{figure*}
     \centering
         \begin{subfigure}[b]{0.24\textwidth}
         \centering
         \includegraphics[width=\textwidth]{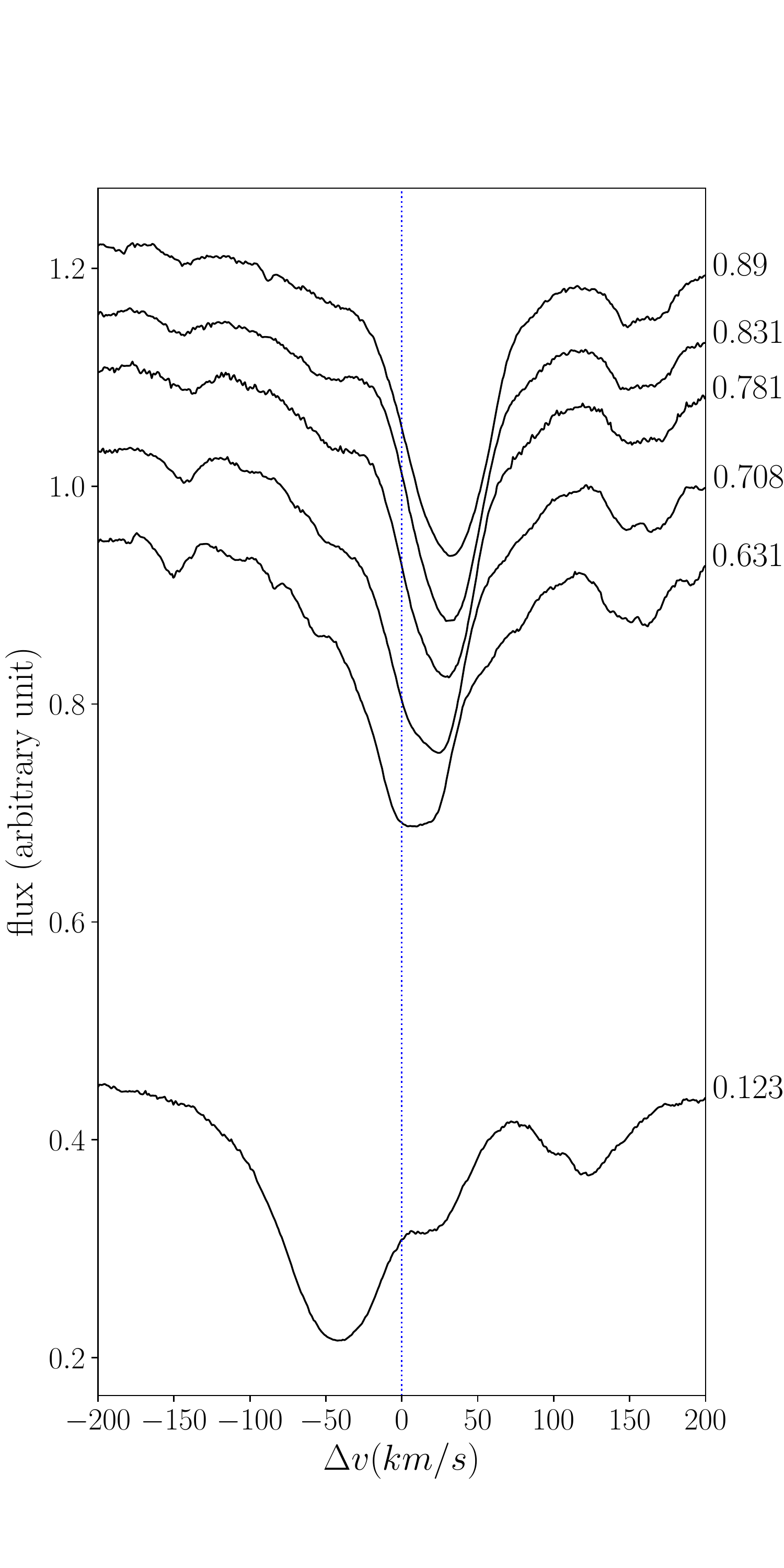}
         \caption{$\lambda$8498}
     \end{subfigure}
     \hfill
     \begin{subfigure}[b]{0.24\textwidth}
         \centering
         \includegraphics[width=\textwidth]{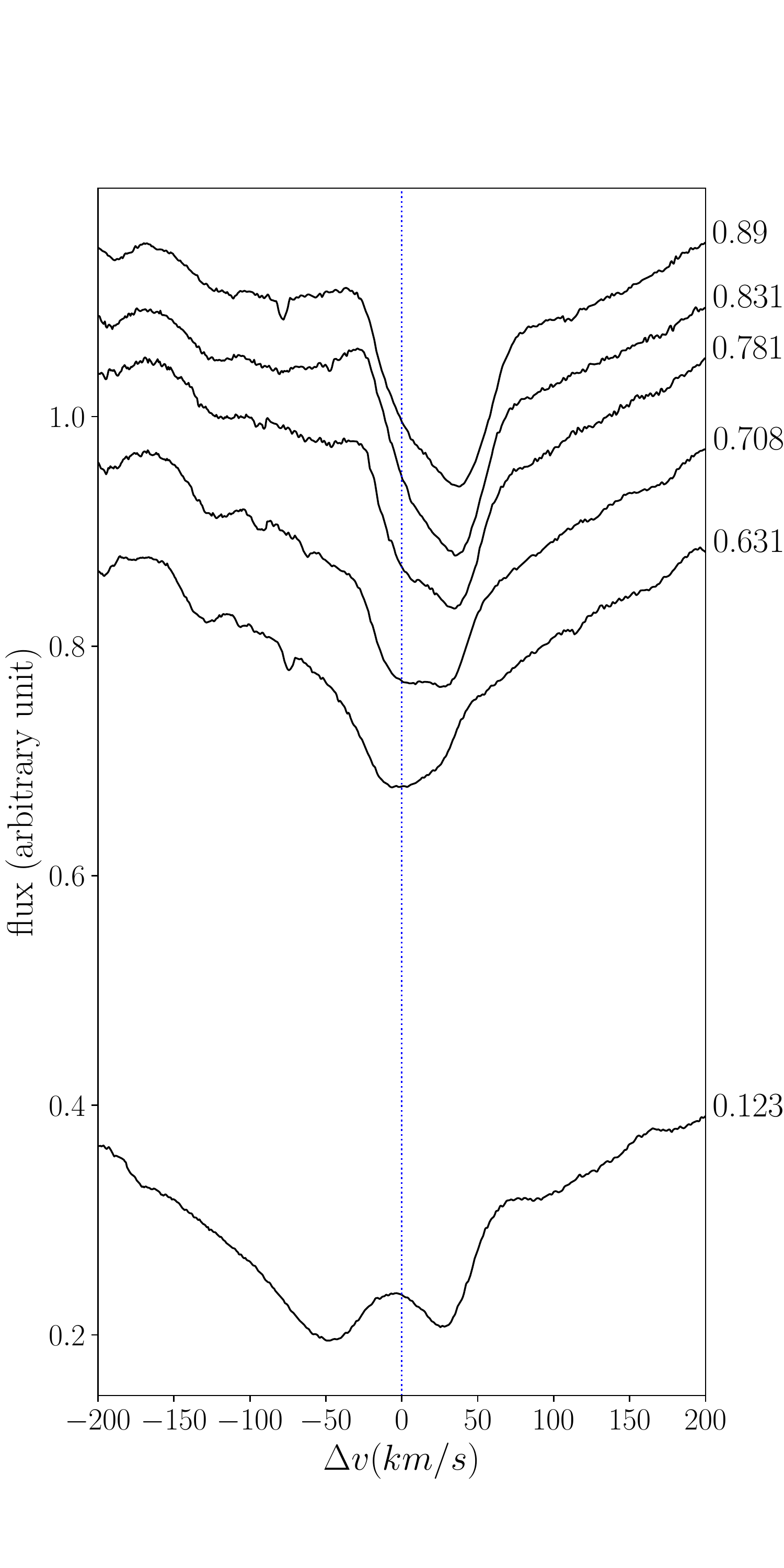}
         \caption{$\lambda$8542}
     \end{subfigure}
     \hfill
     \begin{subfigure}[b]{0.24\textwidth}
         \centering
         \includegraphics[width=\textwidth]{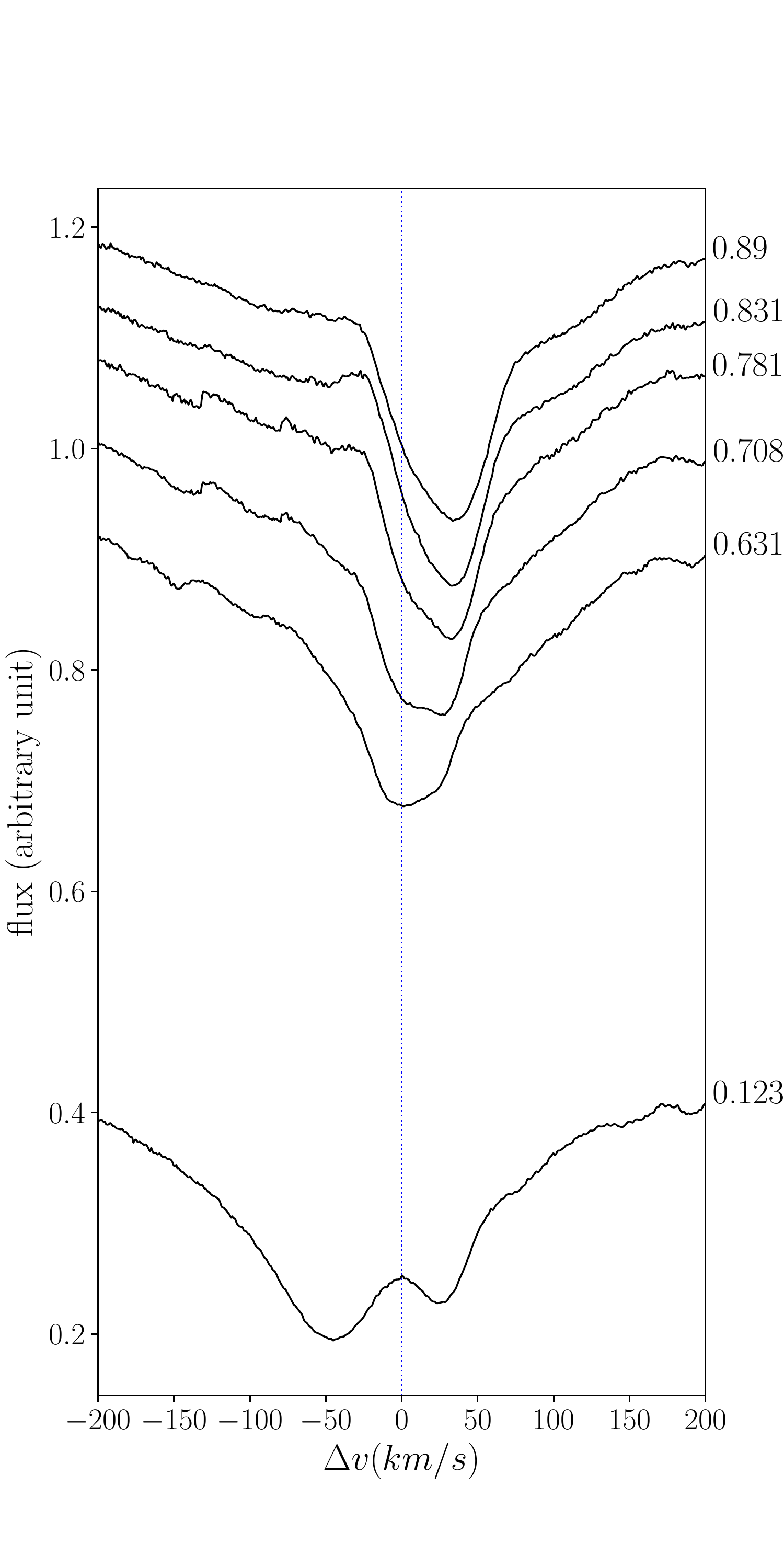}
         \caption{$\lambda$8662}
     \end{subfigure}
     \hfill
     \begin{subfigure}[b]{0.24\textwidth}
         \centering
         \includegraphics[width=\textwidth]{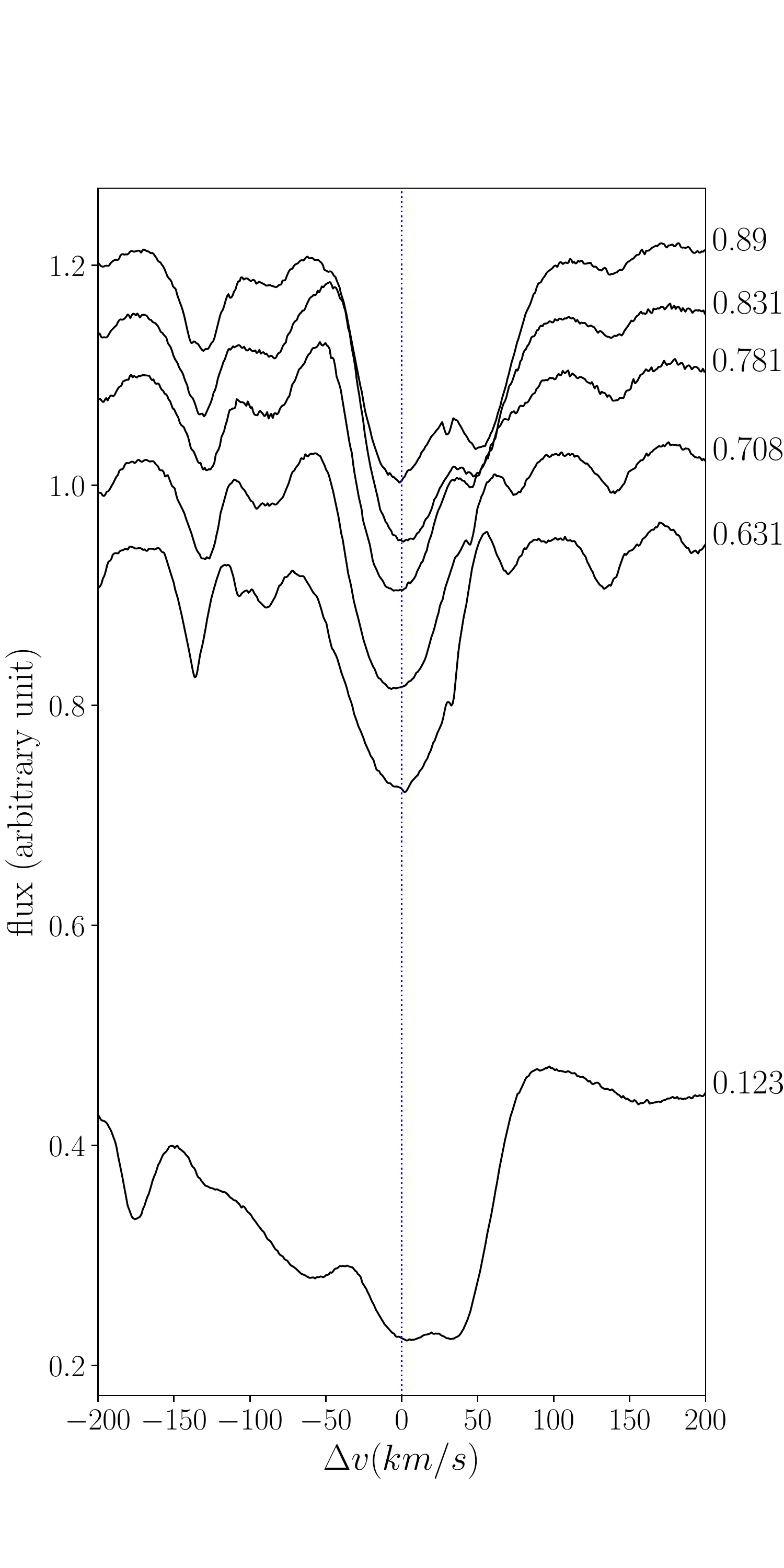}
         \caption{H$\alpha$}
     \end{subfigure}
        \caption{T Mon, 27.02d}
\end{figure*}

\begin{figure*}
     \centering

     \begin{subfigure}[b]{0.24\textwidth}
         \centering
         \includegraphics[width=\textwidth]{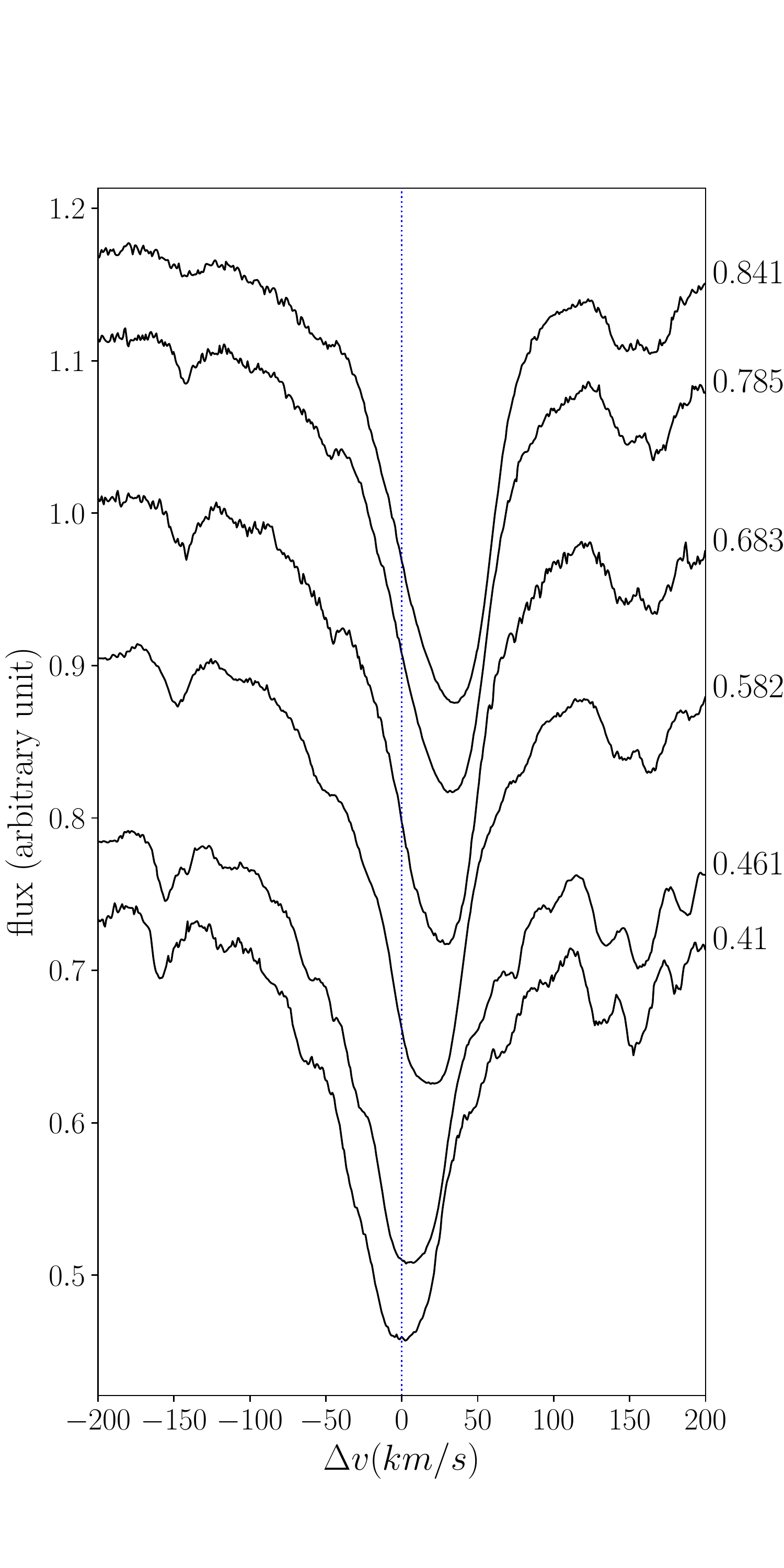}
         \caption{$\lambda$8498}
     \end{subfigure}
          \hfill
     \begin{subfigure}[b]{0.24\textwidth}
         \centering
         \includegraphics[width=\textwidth]{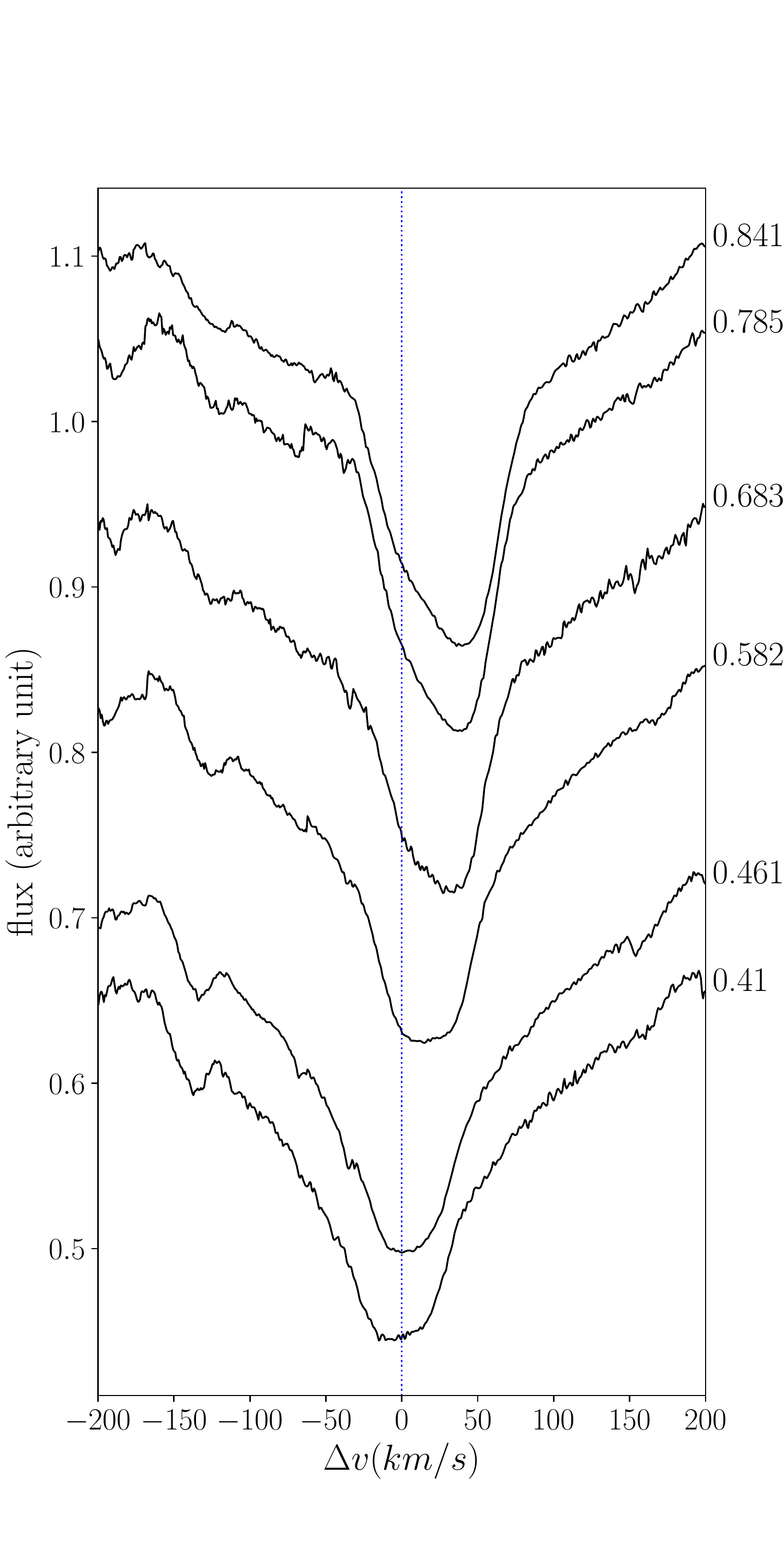}
         \caption{$\lambda$8542}
     \end{subfigure}
     \hfill
     \begin{subfigure}[b]{0.24\textwidth}
         \centering
         \includegraphics[width=\textwidth]{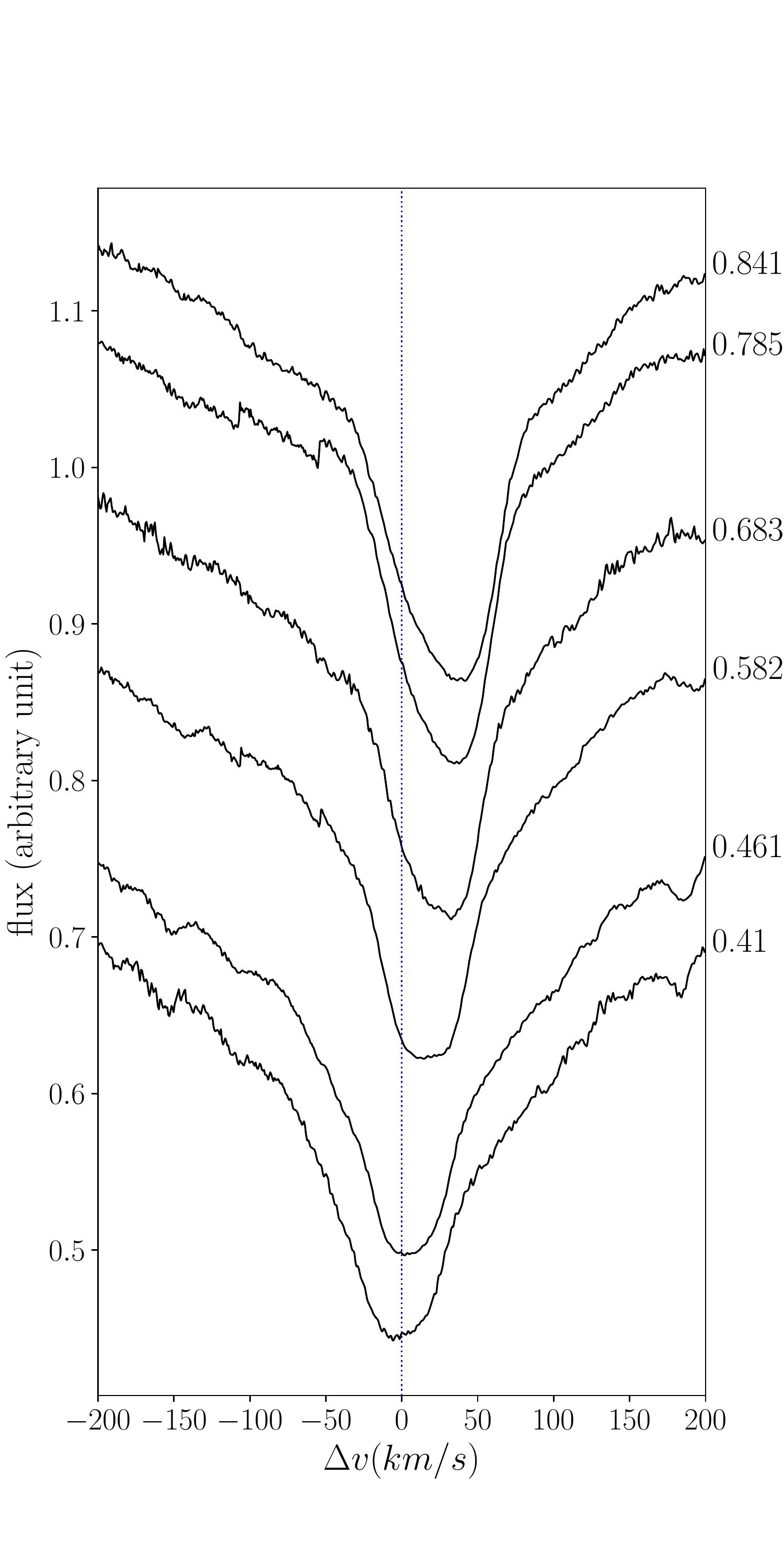}
         \caption{$\lambda$8662}
     \end{subfigure}
     \hfill
     \begin{subfigure}[b]{0.24\textwidth}
         \centering
         \includegraphics[width=\textwidth]{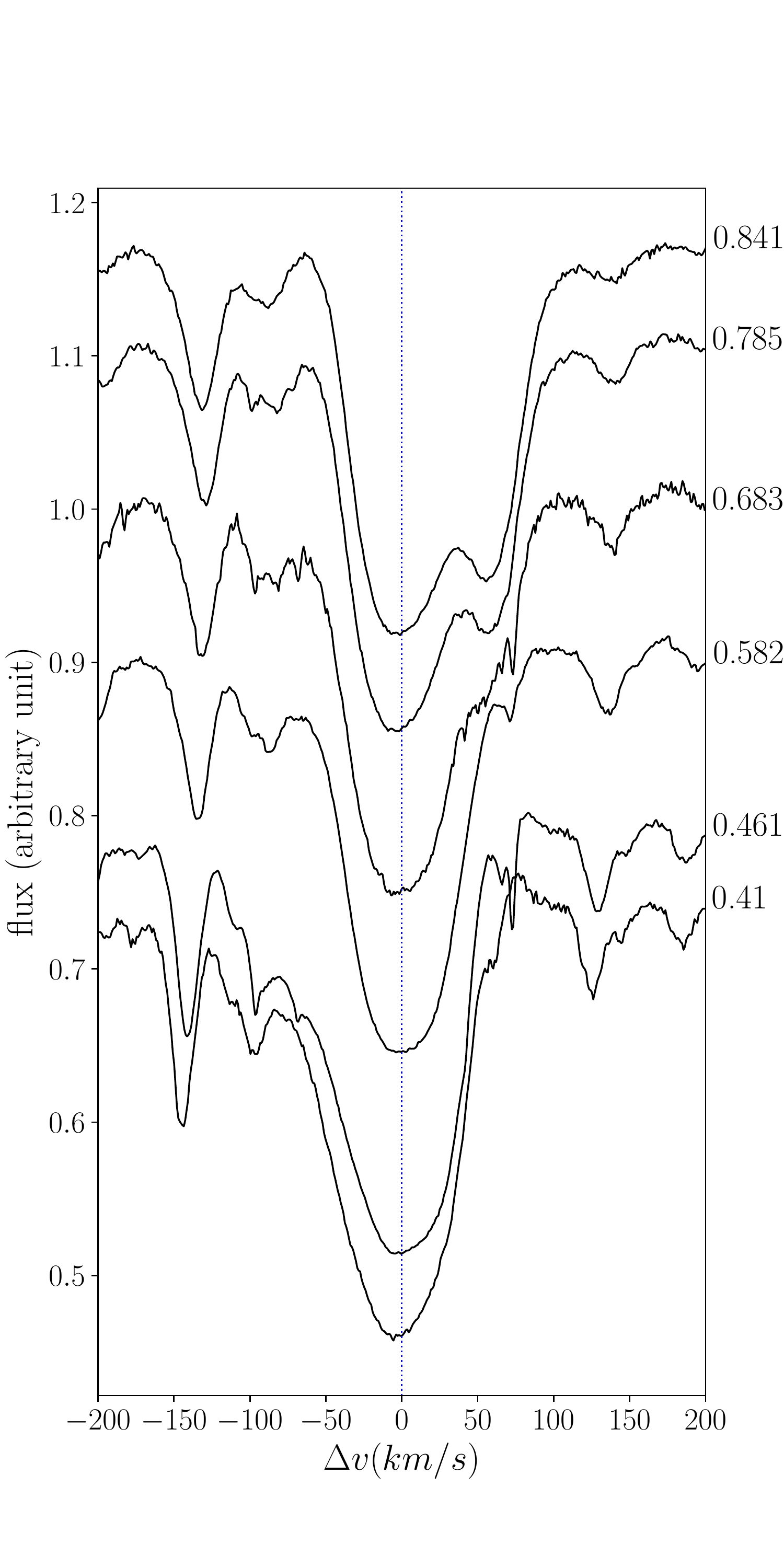}
         \caption{H$\alpha$}
     \end{subfigure}

        \caption{$\ell$ Car, 35.55d}\label{fig:l_car}
\end{figure*}

\begin{figure*}
     \centering
     \begin{subfigure}[b]{0.24\textwidth}
         \centering
         \includegraphics[width=\textwidth]{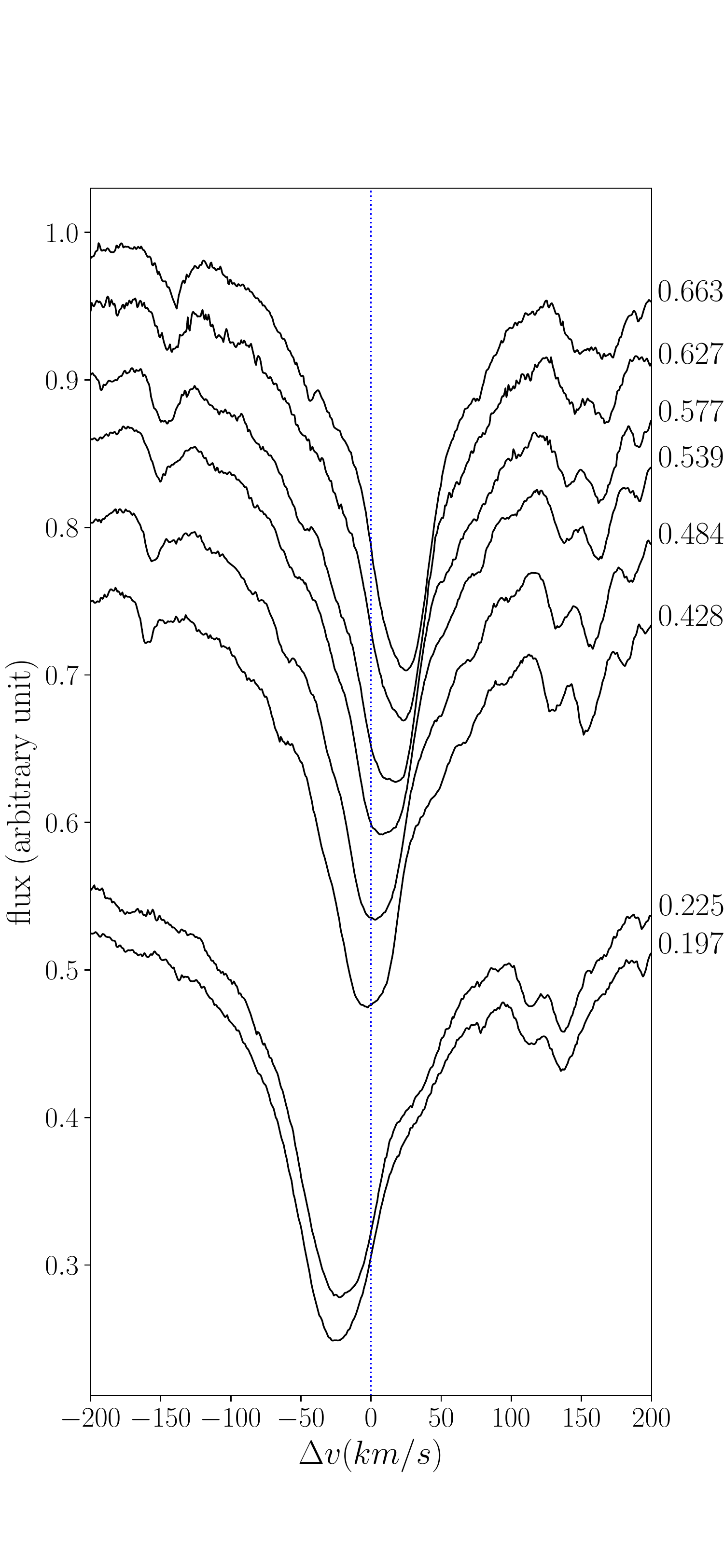}
         \caption{$\lambda$8498}
     \end{subfigure}
     \hfill
     \begin{subfigure}[b]{0.24\textwidth}
         \centering
         \includegraphics[width=\textwidth]{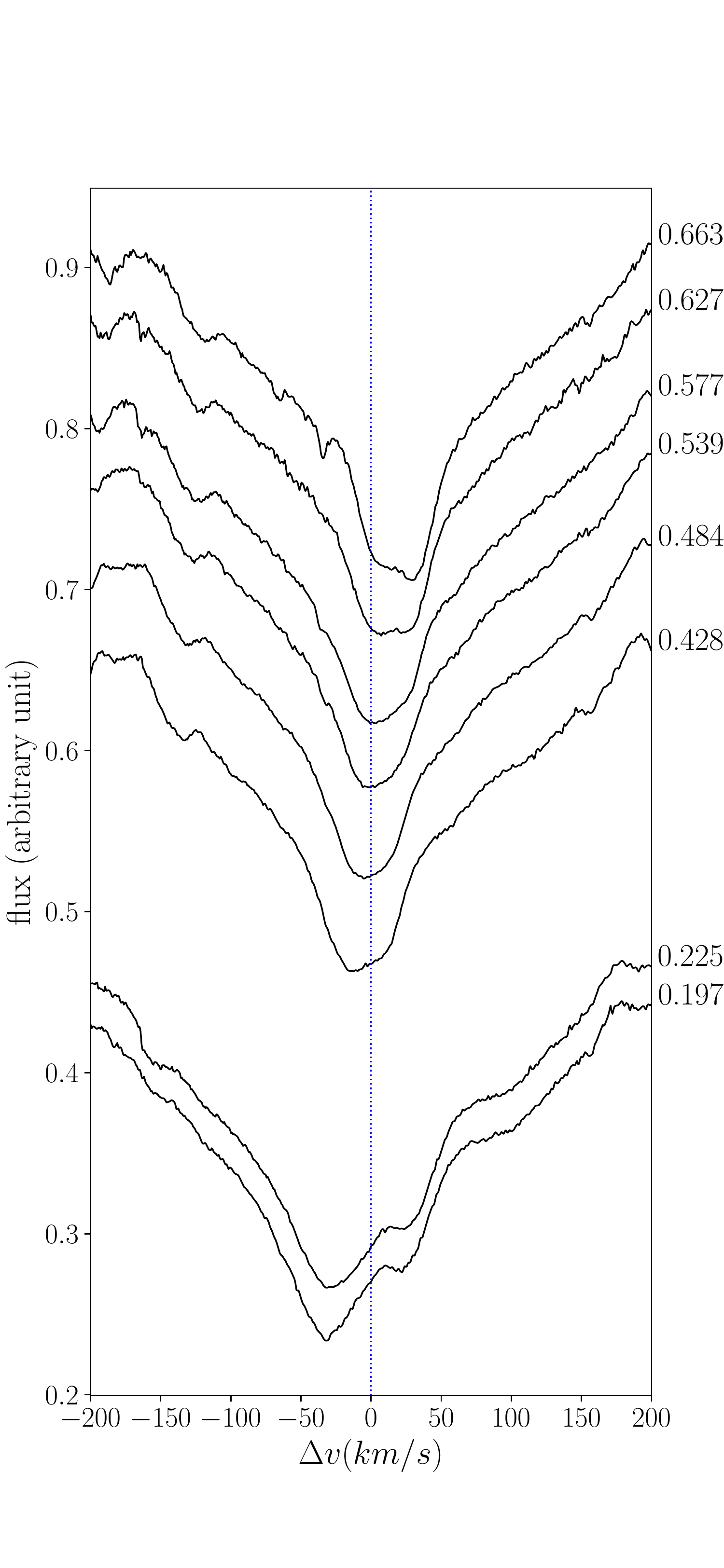}
         \caption{$\lambda$8542}
     \end{subfigure}
     \hfill
     \begin{subfigure}[b]{0.24\textwidth}
         \centering
         \includegraphics[width=\textwidth]{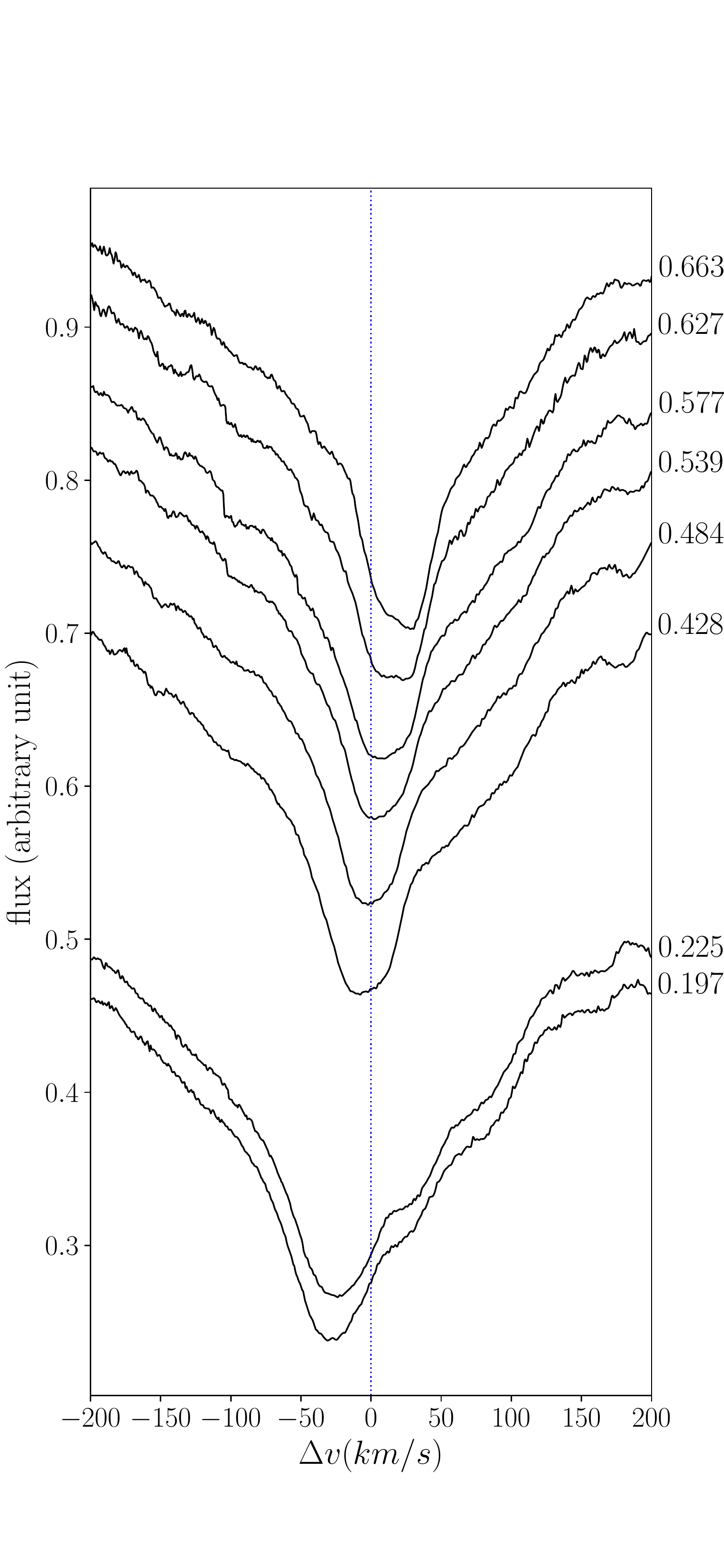}
         \caption{$\lambda$8662}
     \end{subfigure}
     \hfill
          \begin{subfigure}[b]{0.24\textwidth}
         \centering
         \includegraphics[width=\textwidth]{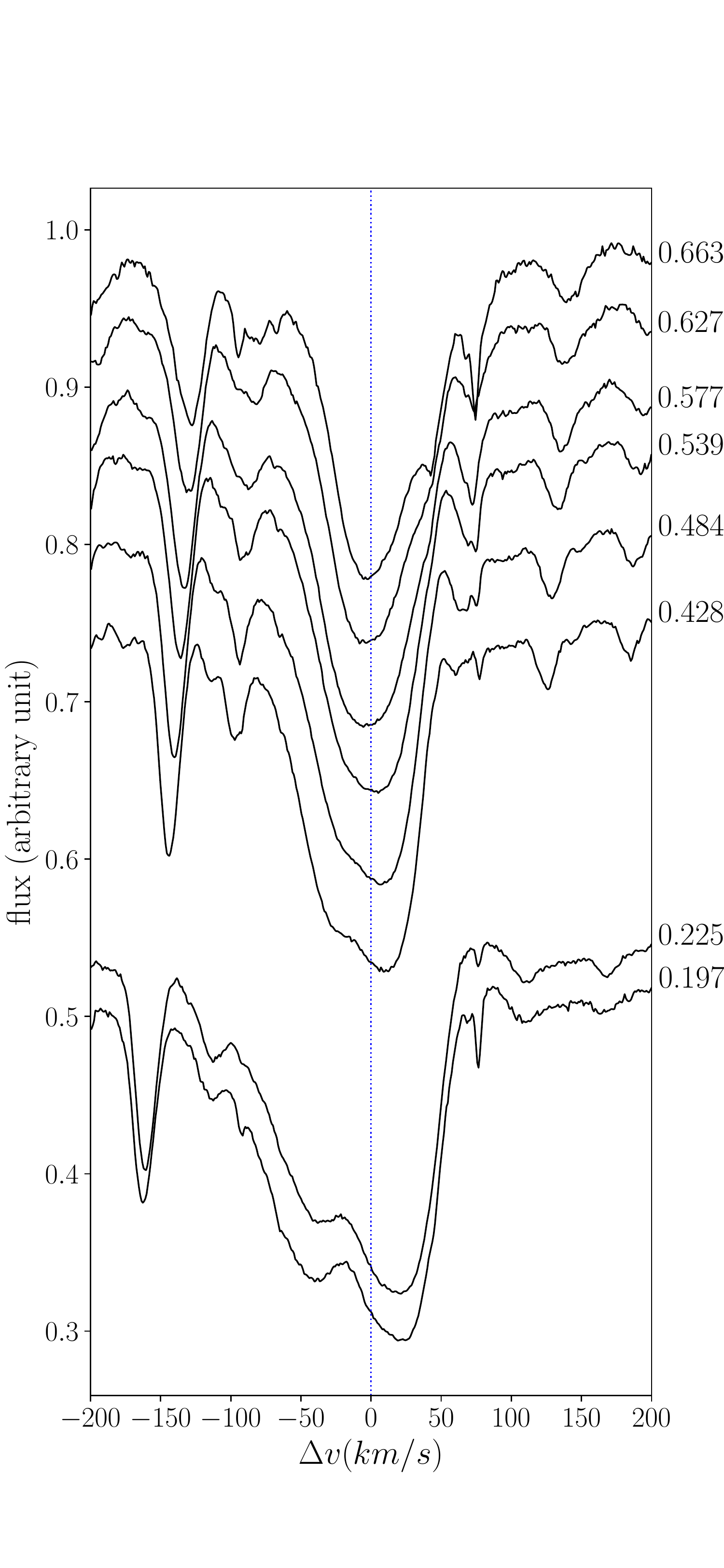}
         \caption{H$\alpha$}
     \end{subfigure}
        \caption{U Car, 38.80d}\label{fig:u_car}
\end{figure*}

\begin{figure*}
     \centering

          \begin{subfigure}[b]{0.24\textwidth}
         \centering
         \includegraphics[width=\textwidth]{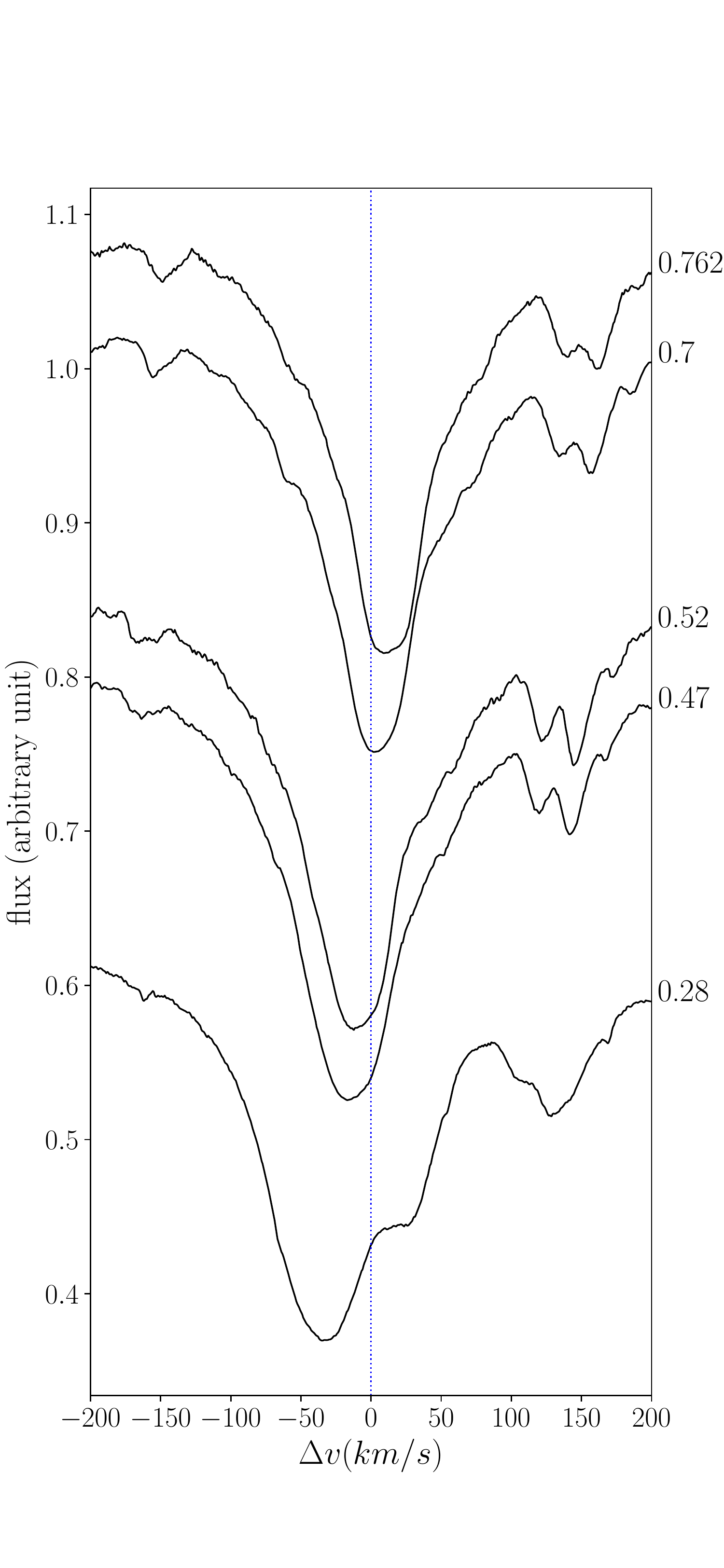}
         \caption{$\lambda$8498}
     \end{subfigure}
     \hfill
     \begin{subfigure}[b]{0.24\textwidth}
         \centering
         \includegraphics[width=\textwidth]{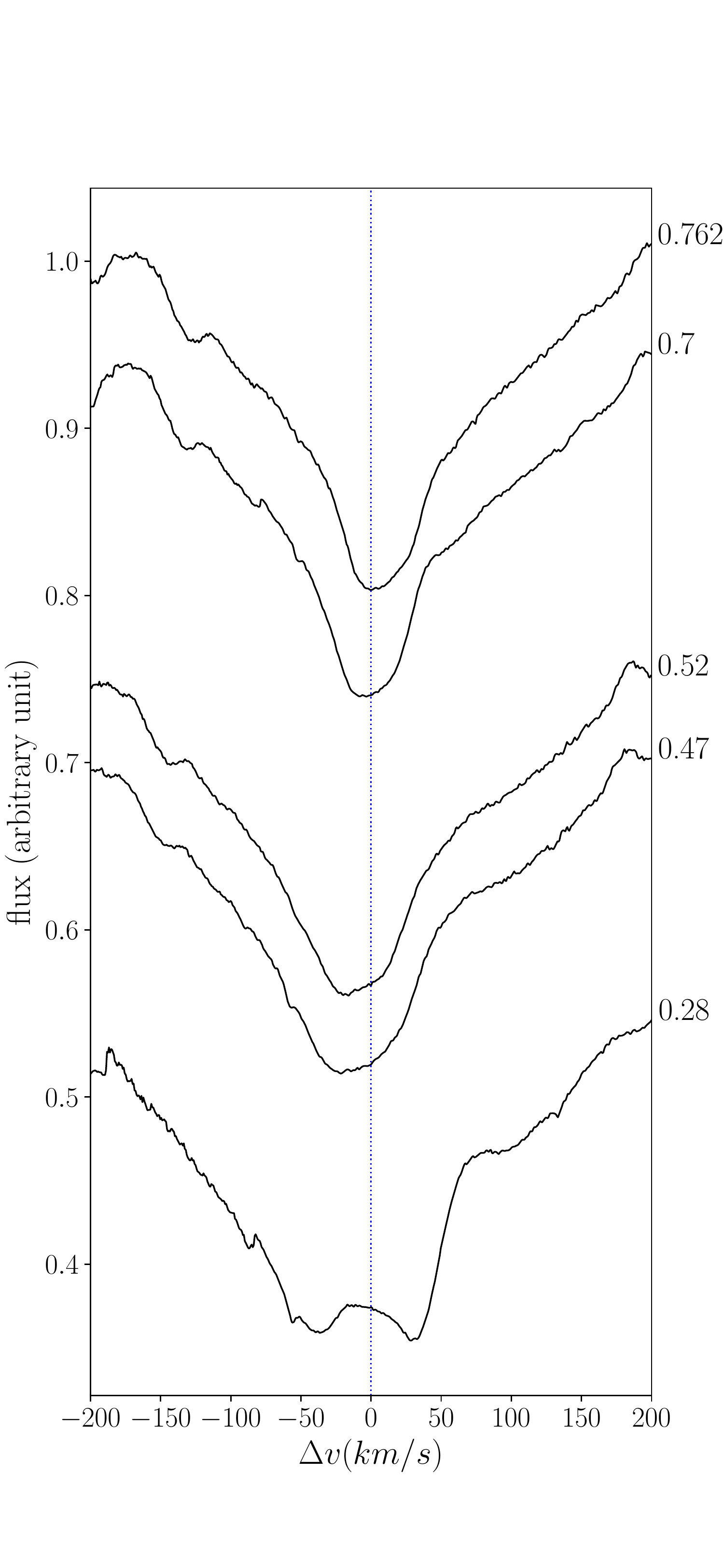}
         \caption{$\lambda$8542}
     \end{subfigure}
     \hfill
     \begin{subfigure}[b]{0.24\textwidth}
         \centering
         \includegraphics[width=\textwidth]{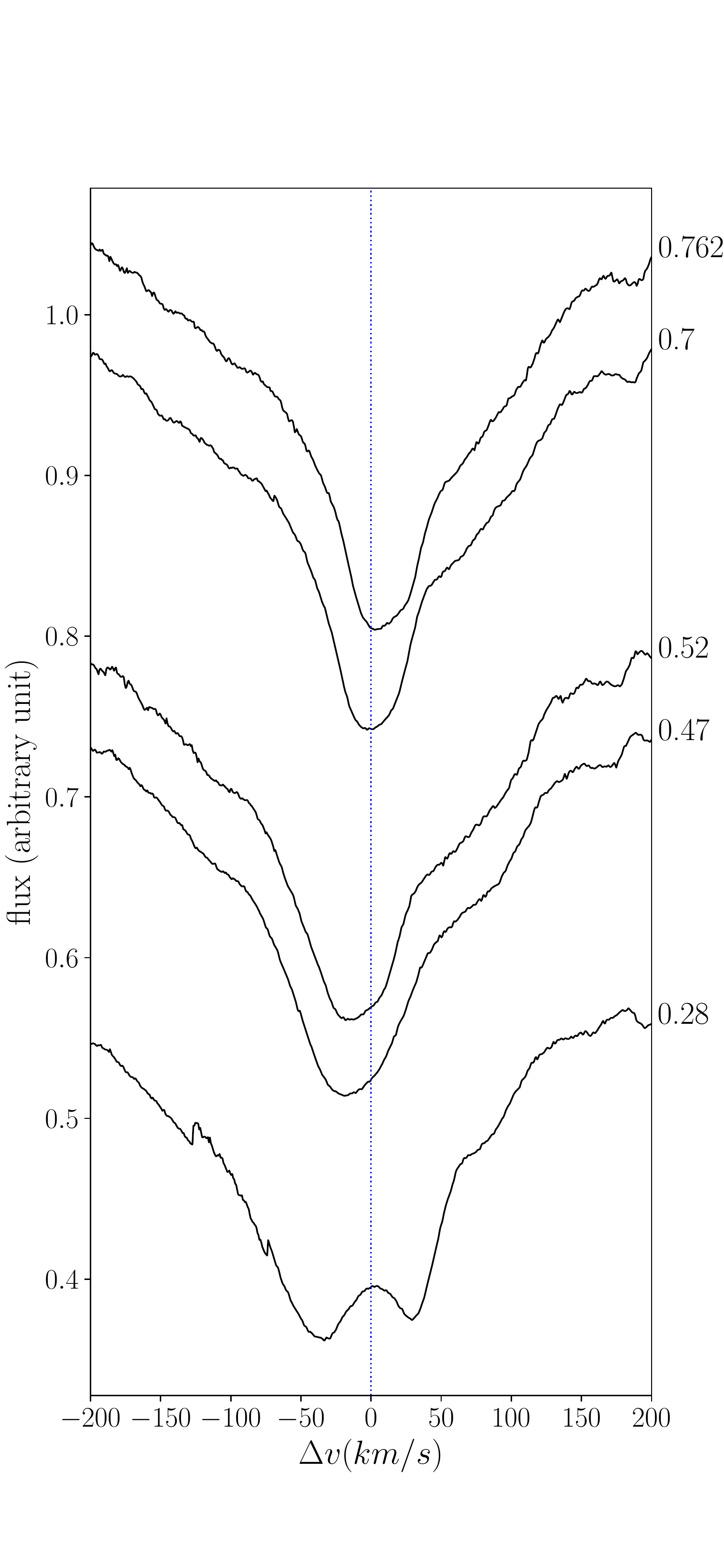}
         \caption{$\lambda$8662}
     \end{subfigure}
     \hfill
          \begin{subfigure}[b]{0.24\textwidth}
         \centering
         \includegraphics[width=\textwidth]{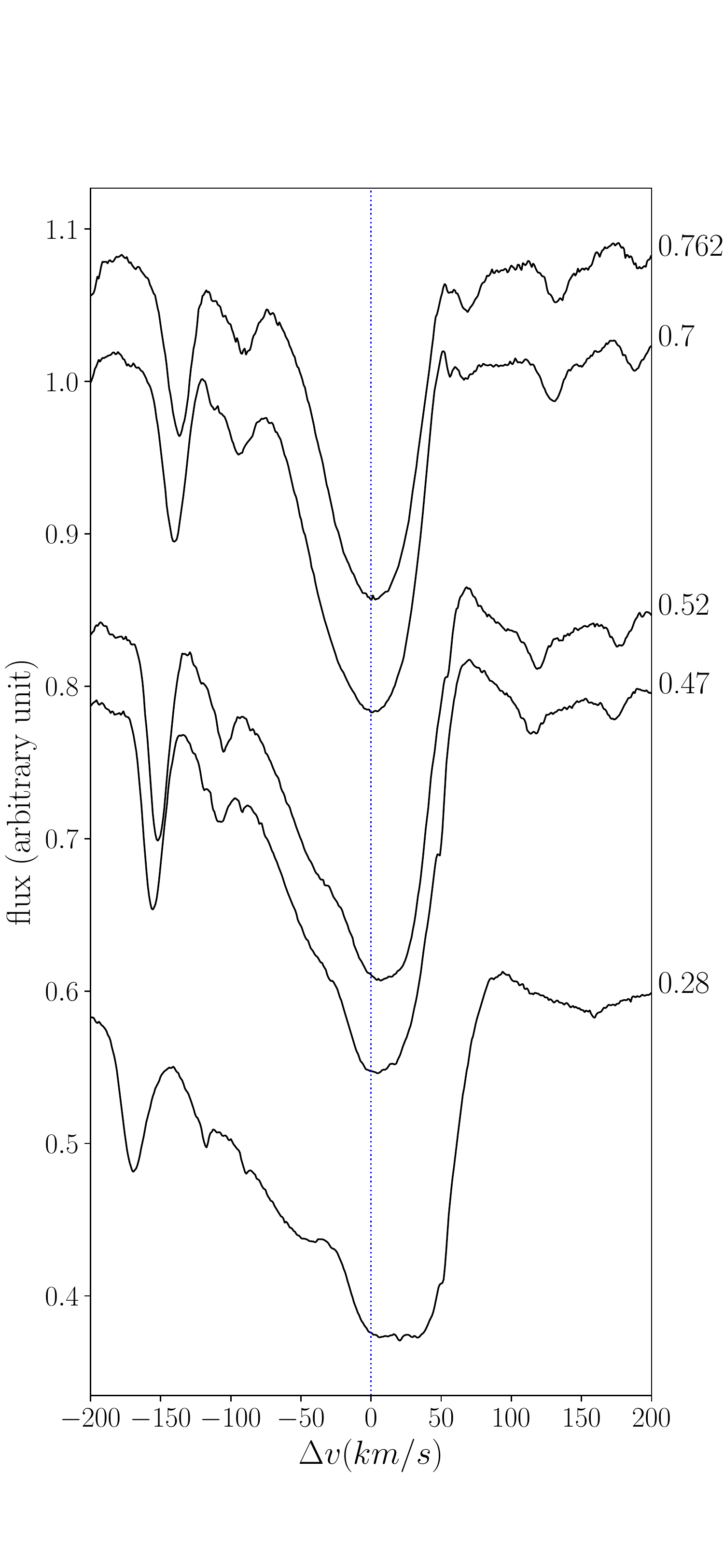}
         \caption{H$\alpha$}
     \end{subfigure}
        \caption{RS Pup, 41.46d}
\end{figure*}

\begin{figure*}
     \centering

          \begin{subfigure}[b]{0.24\textwidth}
         \centering
         \includegraphics[width=\textwidth]{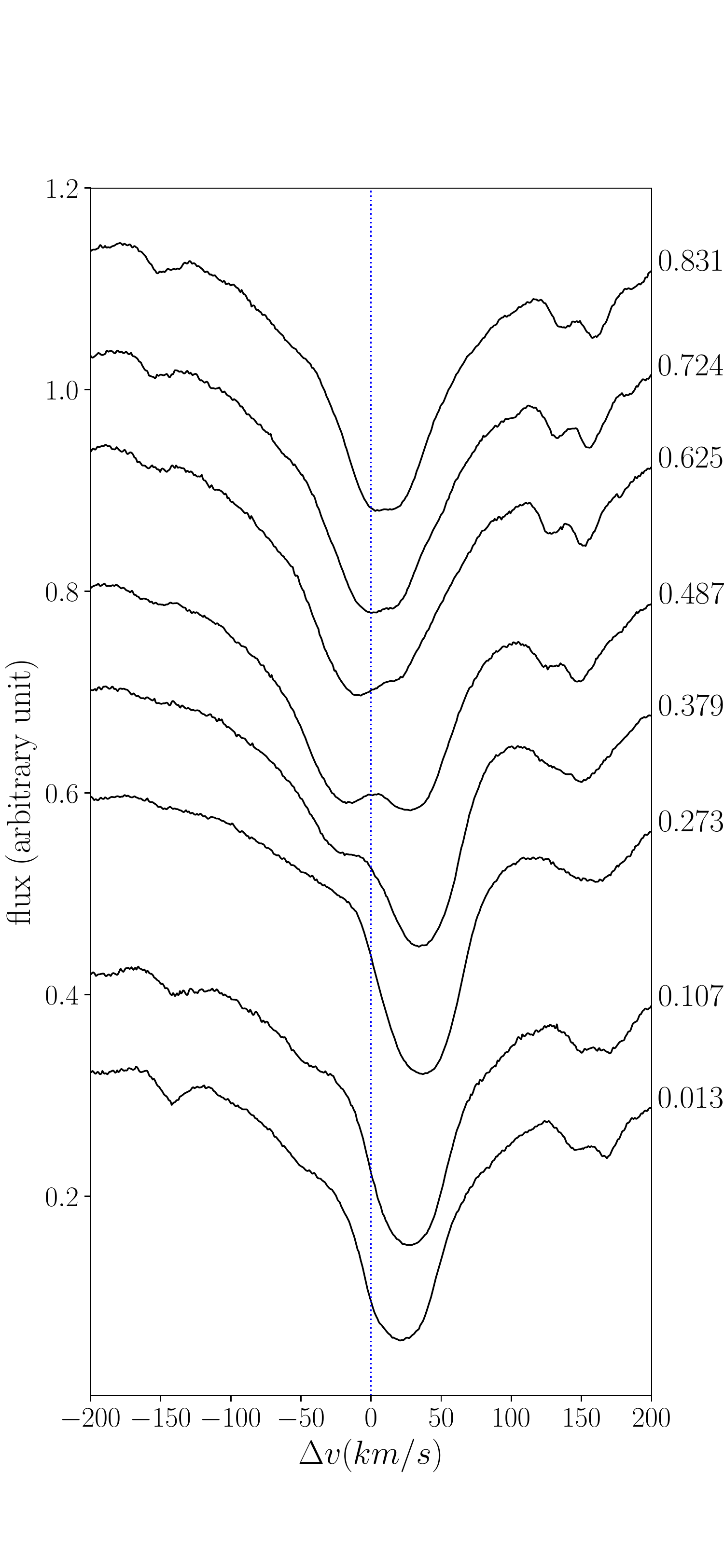}
         \caption{$\lambda$8498}
     \end{subfigure}
     \hfill
     \begin{subfigure}[b]{0.24\textwidth}
         \centering
         \includegraphics[width=\textwidth]{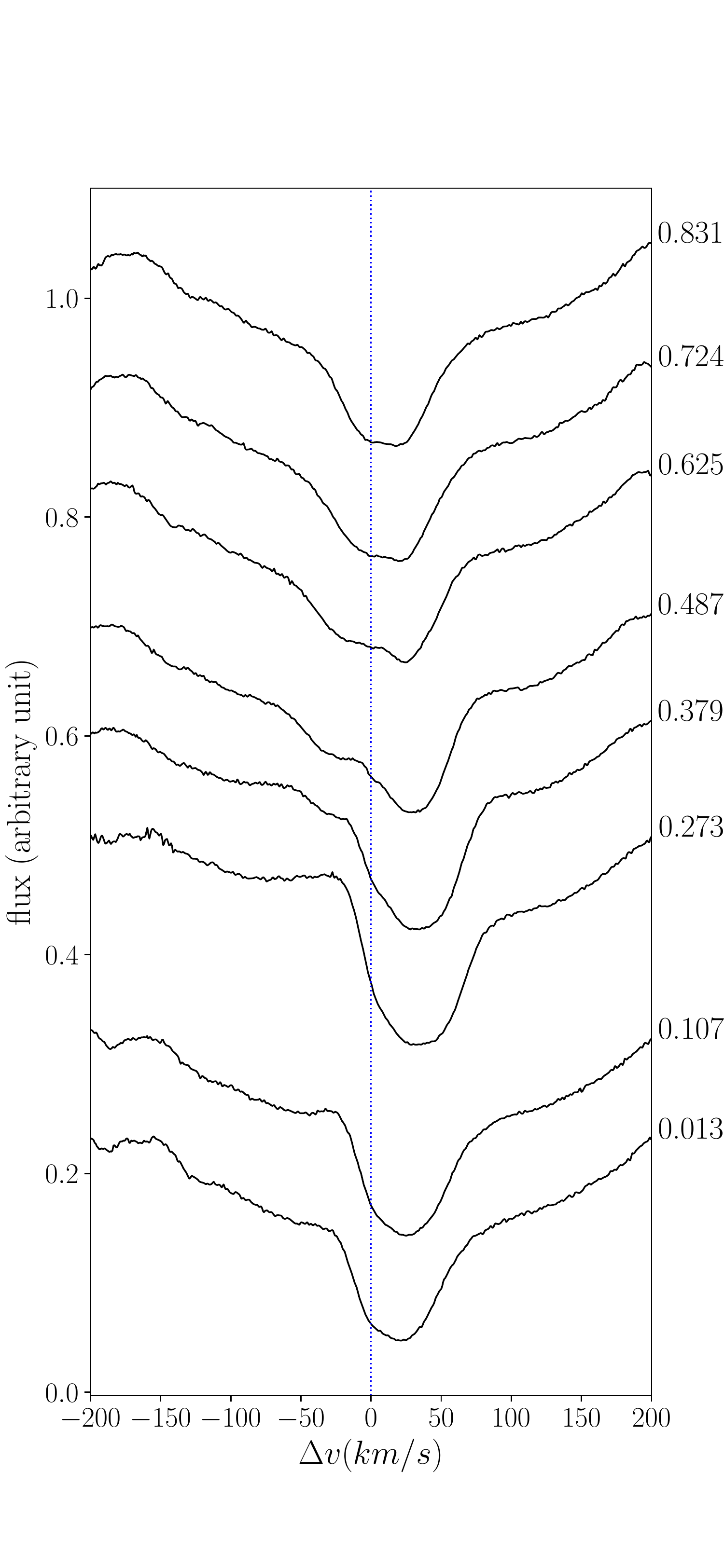}
         \caption{$\lambda$8542}
     \end{subfigure}
     \hfill
     \begin{subfigure}[b]{0.24\textwidth}
         \centering
         \includegraphics[width=\textwidth]{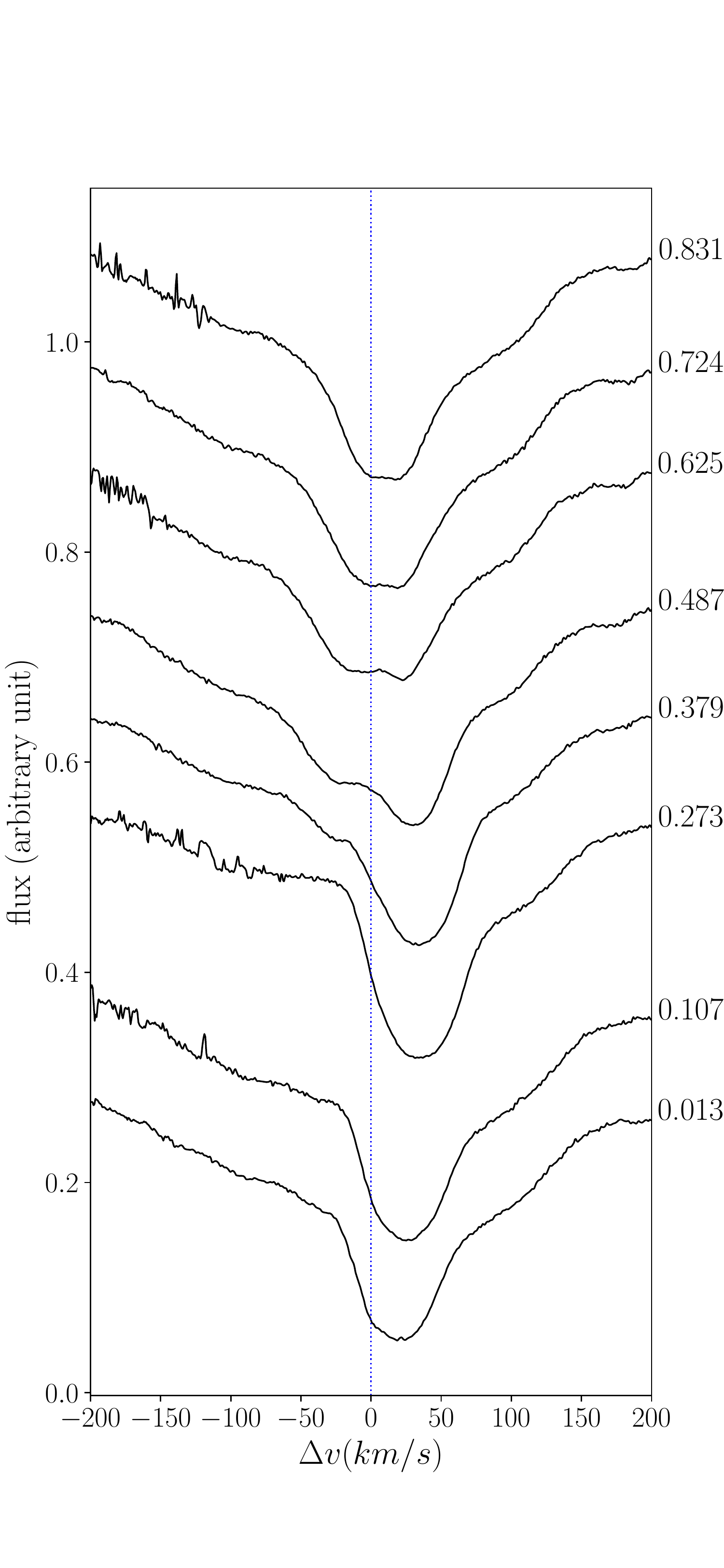}
         \caption{$\lambda$8662}
     \end{subfigure}
     \hfill
          \begin{subfigure}[b]{0.24\textwidth}
         \centering
         \includegraphics[width=\textwidth]{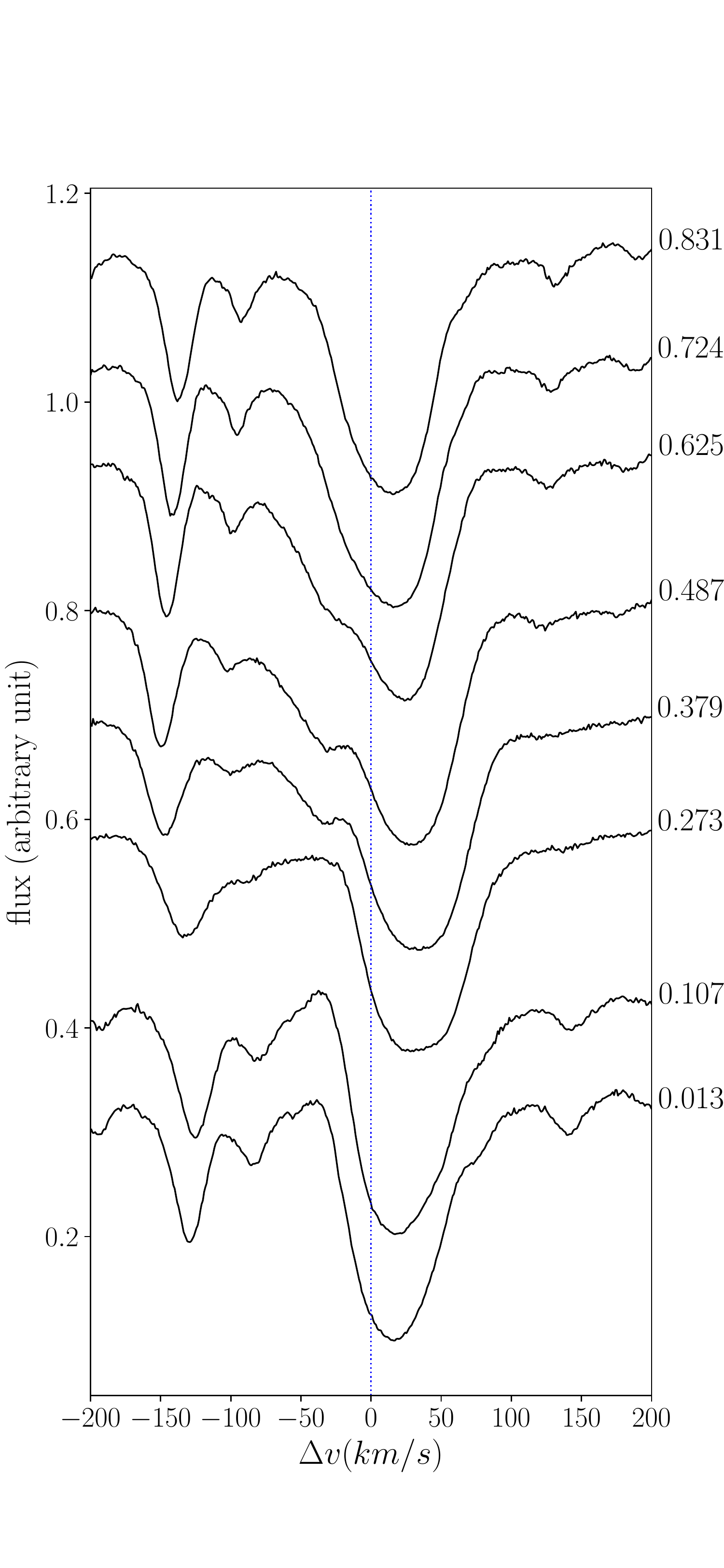}
         \caption{H$\alpha$}
         \label{fig:Ha_v1496_aql}
     \end{subfigure}
        \caption{V1496 Aql, 65.37d}
\end{figure*}

\end{appendix}

\end{document}